\documentclass[12pt,oneside,openany,letterpaper,final]{report}

\usepackage{setspace}
\usepackage{fancyhdr}
\usepackage{amssymb}
\usepackage{latexsym}
\usepackage{amsmath}
\usepackage{graphics}
\usepackage{graphicx}
\usepackage{subfigure}
\usepackage{rotating}
\usepackage{cite}
\usepackage{epsfig}
\usepackage{mathrsfs}
\usepackage{array}
\usepackage{layout}
\usepackage{calc}
\usepackage{bm}
\usepackage[usenames,dvipsnames]{color}
\definecolor{forest-green}{RGB}{0,150,0}
\definecolor{dark-red}{RGB}{195,0,0}
\definecolor{navy-blue}{RGB}{0,0,195}
\usepackage{multirow}
\usepackage{booktabs}
\usepackage{url}
\usepackage{hyperref}
\usepackage{booktabs}

\hypersetup
{
    pdfauthor={James D. Maxwell},%
    pdfsubject={Proton Spin Structure},%
    pdftitle={ Measurement of the Proton Spin Structure Function $g_2$ \,\,at Four-momentum Transfer of 2 to 6 GeV$^2$},%
    pdfkeywords={Nuclear Spin Structure, Jefferson Lab, Polarized Targets},%
    colorlinks=true,
    citecolor=forest-green, %
    linkcolor=navy-blue, 
    pdfborder={0 0 0}
}

\usepackage[vcentering]{geometry}
\newcommand{\degrees}{\ensuremath{^\circ}}

\begin{document} 
%
\pagestyle{fancy}
\fancyhf{} 
\fancyhead[LO]{}
\fancyhead[RO]{\thepage}
\fancyhead[LE]{\thepage} 
\fancyhead[RE]{}
\fancypagestyle{plain}{ \fancyhead[LO]{} \fancyhead[RO]{\thepage} \fancyhead[LE]{\thepage} \fancyhead[RE]{}}

\newpage
\thispagestyle{empty}
\singlespacing
\noindent
\begin{minipage}[c][1in-\topmargin-\headheight-\headsep][t]{\textwidth}
\ \ \ \ \ 
\end{minipage}
\noindent
\begin{minipage}[c][0.4in][t]{\textwidth}
\begin{center}
   \begin{large}
\textsc{Probing Proton Spin Structure: a Measurement of $g_2$\\ at Four-momentum Transfer of 2 to 6 GeV$^2$}\end{large}
\end{center}
\end{minipage}

\vspace{0.65in}

\noindent
\begin{minipage}[c][0.2in][c]{\textwidth}
\begin{center}
James Davis Maxwell
\end{center}
\end{minipage}

\noindent
\begin{minipage}[c][0.2in][c]{\textwidth}
\begin{center}
Poquoson, Virginia
\end{center}
\end{minipage}

\vspace{0.55in}

\noindent
\begin{minipage}[c][0.2in][c]{\textwidth}
\begin{center}
B.S. Physics, Mathematics, University of Virginia, 2004\\
M.A. Physics, University of Virginia, 2010
\end{center}
\end{minipage}



\vspace{0.95in}

\noindent
\begin{minipage}[c][0.2in][c]{\textwidth}
\begin{center}
     A Dissertation presented to the Graduate Faculty
\end{center}
\end{minipage}

\noindent
\begin{minipage}[c][0.2in][c]{\textwidth}
\begin{center}
    of the University of Virginia in Candidacy for the Degree of
\end{center}
\end{minipage}

\noindent
\begin{minipage}[c][0.2in][c]{\textwidth}
\begin{center}
    Doctor of Philosophy
\end{center}
\end{minipage}

\vspace{0.4in}

\noindent
\begin{minipage}[c][0.2in][c]{\textwidth}
\begin{center}
    Department of Physics
\end{center}
\end{minipage}

\vspace{0.35in}

\noindent
\begin{minipage}[c][0.2in][c]{\textwidth}
\begin{center}
     University of Virginia
\end{center}
\end{minipage}

\noindent
\begin{minipage}[c][0.2in][c]{\textwidth}
\begin{center}
December, 2011
\end{center}
\end{minipage}

\vspace{1.7in}
\noindent
\begin{minipage}[c][0.4in][c]{\textwidth}
\begin{flushright}
  \includegraphics{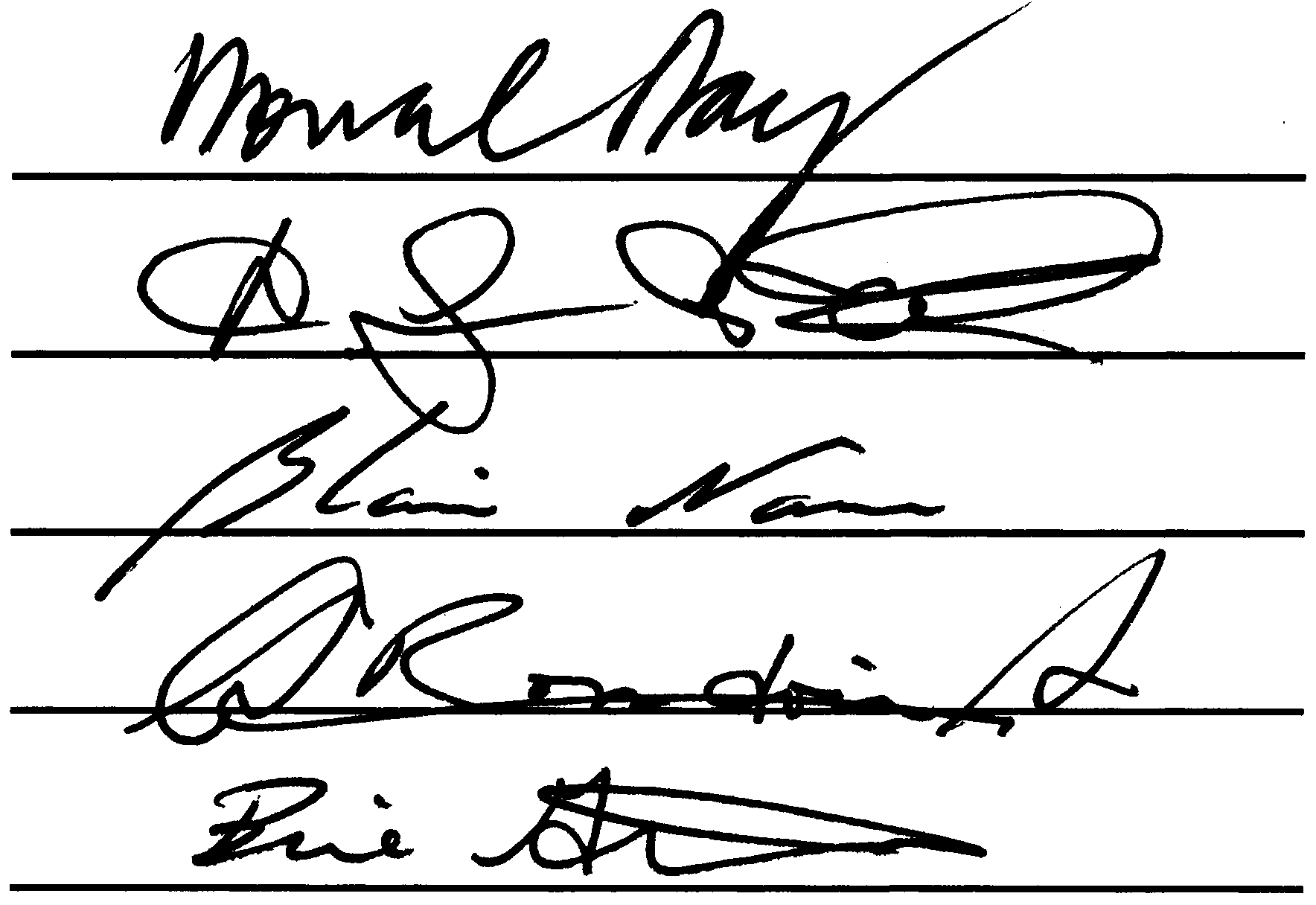}

\end{flushright}
\end{minipage}

\clearpage
\newpage
\thispagestyle{empty}
\singlespacing
\noindent
\begin{minipage}[c][8in-\topmargin-\headheight-\headsep][t]{\textwidth}
\ \ \ \ \ 
\end{minipage}
\noindent
\begin{minipage}[c][0.8in][t]{\textwidth}
\begin{center}
    \copyright\  Copyright by \\
    James Davis Maxwell \\
    All Rights Reserved \\
    December 2011
\end{center}
\end{minipage}

\clearpage
\pagenumbering{roman}
\doublespacing
\begin{abstract}

The Spin Asymmetries of the Nucleon Experiment investigated the spin structure of the proton via inclusive electron scattering at the Continuous Electron Beam Accelerator Facility at Jefferson Laboratory in Newport News, VA.  A double--polarization measurement of polarized asymmetries was performed using the University of Virginia solid polarized ammonia target with target polarization aligned longitudinal and near transverse to the electron beam, allowing the extraction of the spin asymmetries $A_1$ and $A_2$, and spin structure functions $g_1$ and $g_2$.  Polarized electrons of energies of 4.7 and 5.9 GeV were scattered to be viewed by a novel, non-magnetic array of detectors observing a four-momentum transfer range of 2 to 6 GeV$^2$.  This document addresses the extraction of the spin asymmetries and spin structure functions, with a focus on spin structure function $g_2$, which we have measured as a function of $x$ and $W$ in four $Q^2$ bins.
\end{abstract}
\newpage

\clearpage
%
%
\singlespacing
\tableofcontents
\listoffigures
\listoftables
\newpage
\doublespacing
\section*{Acknowledgments}

It feels facile and inadequate to distill my gratitude into a few words on a page quickly skirted in a document so long.  Nevertheless, I hope that the many people who helped make this work possible, a list too large to recount entirely here, know the depth of my appreciation for their support and contribution.

This experiment was a trying one, and all those involved deserve many thanks for persevering in the face of such challenges.   All the SANE collaborators and JLab staff who contributed their expertise and equipment to the planning and execution of the experiment, as well as those who gave their time during shifts, deserve great credit for this work.  In particular, I thank the experiment's spokespersons, O. Rondon, S. Choi, Z. Meziani and M. Jones for tireless work to make SANE possible. My sincere thanks also go out to the JLab target group, lead by C. Keith, who were indefatigable in the installation and continued repair of the target during the experiment.

After 10 years, the Polarized Target Group at UVa seems almost a second home to me. I thank my graduate student colleagues, especially J. Mulholland who labored beside me in SANE, and J. Pierce and N. Fomin who I rightly consider mentors.  I cannot thank the professors of the Target Group enough; I have benefited immeasurably from the guidance of D. Crabb and O. Rondon.  Most of all, I thank my advisor, D. Day, my longtime mentor and steadfast advocate.

Finally, I'd like to thank all those who have shaped me as a scientist and a person; my friends, teachers, and family.  It should go without saying that I owe all the success I meet to my parents, grandparents and the rest of my family; how can a grateful child ever repay his family?  Lastly, I thank my wonderful wife, Ginny, who has been my best friend and vital support all through my graduate career.

 Again, thank you all.

\newpage
\clearpage

\pagenumbering{arabic}

\pagestyle{fancy}
\renewcommand{\chaptermark}[1]{\markboth{#1}{}}
\renewcommand{\sectionmark}[1]{\markright{\thesection.\ #1}}
\fancyhf{} 
\fancyhead[LO]{\sc \rightmark}
\fancyhead[RO]{\thepage}
\fancyhead[RE]{\sc \leftmark}
\fancyhead[LE]{\thepage}
\fancypagestyle{plain}{ \fancyhead[LO]{} \fancyhead[RO]{\thepage} }
%
\chapter{Introduction}
\label{sec:intro}
The investigation of our world naturally leads us to seek the most basic building blocks of creation and to uncover how they interact with one another.  While the early flights of fancy of Democritus and his school struck eerily close to home, it would be another 2,300 years before J.J. Thompson's discovery of the electron\cite{jjthompson} made the first entry into today's roll of elementary particles.  Cataloging these particles warrants the compilation of their intrinsic qualities, so we have endeavored to measure their mass and charges---the magnitudes of their interaction via the known forces.  The measurements of Stern and Gerlach\cite{sterngerlach} in the 1920s, lead to the addition of \textit{spin} to this list of fundamental properties.

The concept of spin is aptly, if perhaps misleadingly, named.  In the electron, we observe a magnetic moment equivalent to that of a rotating charged particle, but how can a particle of no spatial extent rotate?  Spin looks identical to angular momentum, but with the startling caveat that it is unrelated to any motion of the particle in space.  We must abandon our intuition and accept spin as an fundamental quality; the electron is a spin-$\frac{1}{2}$ particle.  

In 1927, Dennison established that the proton was also a spin-$\frac{1}{2}$ entity.  When Stern and Estermann approached the measurement of the proton's magnetic moment in 1933\cite{stern}, the study of spin offered a seminal insight.  The proton was observed to have an \textit{anomalous} magnetic moment which was far larger than could be expected for a point particle of spin-$\frac{1}{2}$. This was the first clue to the internal structure of nucleons---protons and neutrons---and began the inquiry into the nature and behavior of their constituents that continues today.

\section{Leptons, Quarks and Bosons}
\label{sec:standardmodel}

The Standard Model provides only three types of elementary particles, two of which have corresponding antiparticles.  There are six known leptons: the electron, muon and tauon, and their corresponding neutrinos; six known quarks: the up, down, charm, strange, top and bottom; and five known bosons: the gluon of the strong force, and the photon, Z and W$^\pm$ of the electroweak force.  We model the interactions of the spin-$\frac{1}{2}$ quarks and leptons which form matter via dynamical rules involving the exchange of the spin-1 mediating bosons.

Quantum Electrodynamics  (QED) describes the interaction of all electromagnetically charged particles via the photon.  Codified by Feynman, Schwinger and Tomanaga, QED has produced startlingly accurate predictions and represents the crowning achievement of modern Physics.  Measurements of the electron's anomalous magnetic moment agree with QED beyond 10 significant digits\cite{qed}.

Quantum Chromodynamics (QCD) is the attempt to extend the rules and success of QED towards the description of the interaction of gluons and quarks.  Quarks and gluons carry ``color'' charge; the $\pm e$ electromagnetic charges of QED become six charges under QCD: red, anti-red, blue, anti-blue, green and anti-green.  QCD is based upon an SU(3) symmetry group of the three colors, which form a ``color octet'' of gluons and a ``color singlet'' gluon which is not observed in our world\cite{griffiths}.  These 8 gluons are superpositions of color and anti-color charges; for example, a red quark could exchange a red--anti-blue gluon to become blue. 

QCD exhibits two related properties which make it quite different from QED: \textit{confinement} and \textit{asymptotic freedom}.  Confinement requires that naturally occurring particles be colorless. This explains why we don't observe free quarks, only combinations of two (mesons) or three (baryons) in which the colors of the quarks add up to white---as in red--anti-red or red--green--blue, for example.  

Asymptotic freedom arises from the fact that gluons carry color charge and can thus couple to themselves.  In QED, we observe ``charge-screening'' in which particle--antiparticle pair loops produced in the vacuum around an electron, for instance, serve to lessen the apparent charge of the electron as the distance from the electron increases. But in QCD, we have not only particle--antiparticle loops, but also gluon loops.  

Since the gluon itself carries color charge, a red charge will beget more red charge in the vacuum around it, creating an \textit{anti}-screening.  As the distance from a color charge increases, the charge appears \textit{larger}.  Thus color charges in close proximity have a low coupling constant and are essentially free, but as they move away the coupling strength becomes greater and greater.  As we will see later, this vanishing coupling strength at short distances enables a perturbative description of quark--gluon interactions at high energies.

\section{Scattering Experiments}

Scattering experiments have been the mainstay of elementary particle studies beginning with Rutherford's seminal experiments in 1911.  Rutherford, Geiger and Marsden\cite{rutherford,geiger} scattered alpha particles through thin gold foil, and were able to discern the nucleus of the atom as a compact entity with a charge a multiple of the electron charge.  The advance of experimental technology continues to expand the reach of scattering probes of nuclear structure.

\begin{figure}[htb]
  \begin{center}
   \includegraphics[width=2.7in]{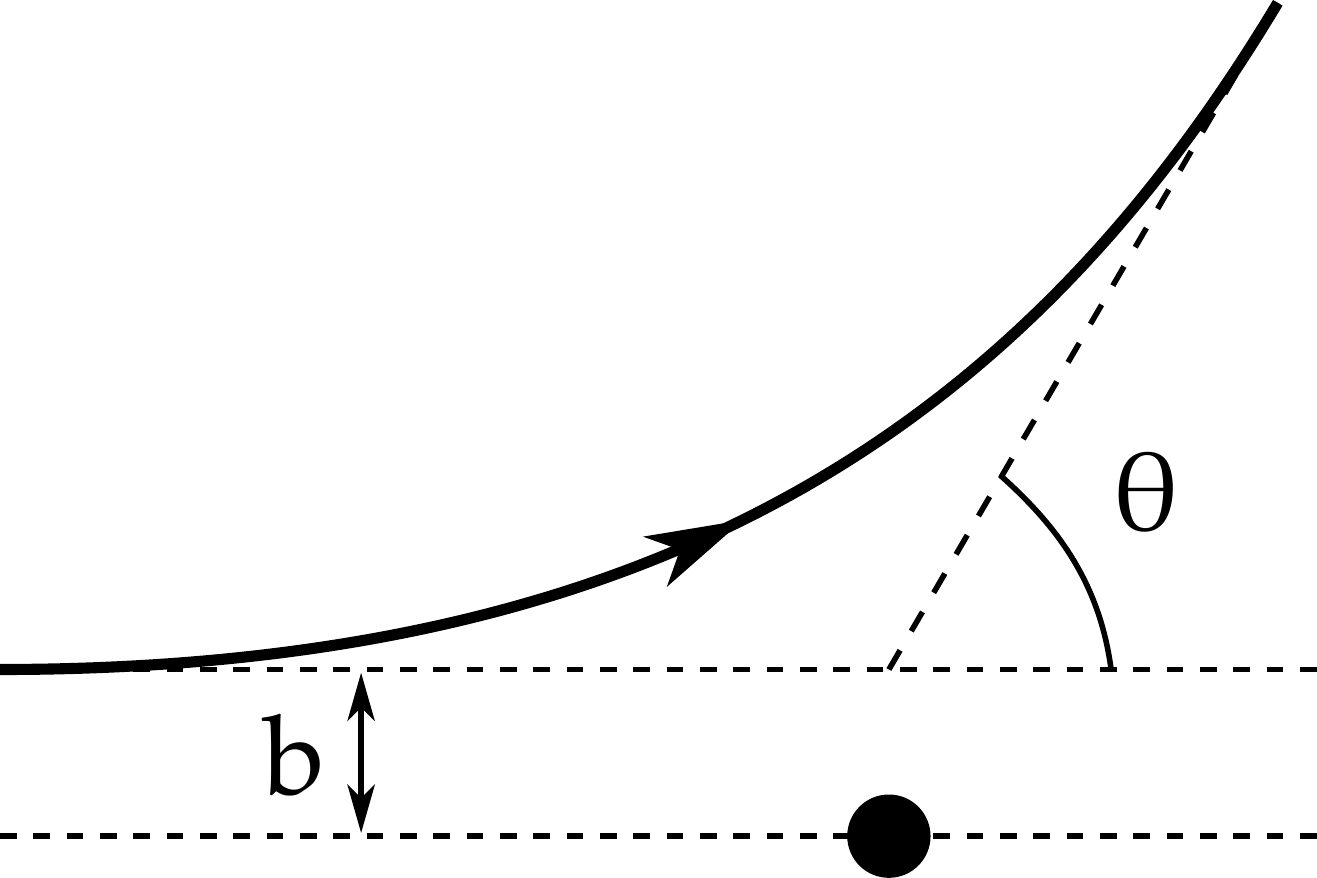}
  \end{center}
  \caption{Scattering from a fixed potential in which the scattering particle is repulsed from the target particle. }
  \label{fig:scatdef}
\end{figure}

The fundamental measured quantity in scattering experiments is the cross section.  We first define two quantities, seen in figure \ref{fig:scatdef}\footnote{A note on the diagrams in this document.  Unless otherwise noted, they are my own, most produced as vector graphics in Inkscape.  They are available for free use with attribution.}:  for an incoming particle approaching a target particle, the distance by which it would have missed the target had it continued on its original path is called the \textit{impact parameter} $b$, and the angle of the final trajectory from the initial is the \textit{scattering angle} $\theta$.  More generally, for a infinitesimal area around $b$, $d\sigma$, the particle will scatter into a solid angle around $\theta$, $d\Omega$.  We will see that we can use the ratio $d\sigma/d\Omega$ to connect experimental observation of scattering processes to theoretical prediction.

\subsection{Variables}
\label{sec:variables}

Before embarking on a discussion of the formalism of scattering processes, we will quickly establish a lexicon of commonly used variables.  For an electron of four momentum $k^{\mu} = (E,\vec{k})$ interacting with a target particle of four momentum $p^{\mu} = (\epsilon,\vec{p})$, as in figure \ref{fig:em}, a single virtual photon is exchanged at leading order, scattering the electron at angle $\theta$ and resulting in final state four momenta of the particles $k'^{\mu} = (E',\vec{k}')$ and $p'^{\mu}$.  The virtual photon four momentum is $q^{\mu} = (\nu,\vec{q})$, which for a space-like virtual photon has $q^2<0$, and includes an energy component $\nu$ , the energy loss of the electron.  We thus define $-Q^2 \equiv q^2 = (k-k')^2 = (p-p')^2$, the four-momentum transfer squared of the process. 

It is useful in inclusive experiments, where only the final electron state is observed, to define the invariant mass of the final state $W = \sqrt{(p+q)^2}$, as well as the invariant scalar $x = Q^2/(2 p \cdot q)$, whose significance will be explained later. 
  In the laboratory frame, where $p^{\mu} = (M,0)$, we have the following kinematic relations\footnote{We will be using natural units, in which $\hbar = c = 1$, unless otherwise noted.}:
\begin{equation}
\begin{split}
\nu &= E-E' \\
Q^2 &= 4EE'\sin^2 \frac{\theta}{2}\\
W^2 &= M^2 + 2M\nu - Q^2\\
x &= \frac{Q^2}{2M\nu}.
\end{split}
\end{equation}

\section{Inclusive Electron Scattering}

	We can construct the transition probability of a particular process using the invariant  amplitude, or so-called ``matrix element,'' $\mathfrak{M}$ for the process, and the differential phase space available:
\begin{equation}
\textrm{transition rate} = \frac{2\pi}{\hbar}\vert\mathfrak{M}\vert^2 \times \textrm{(phase space)}
\end{equation}	
This is known as Fermi's ``Golden Rule.''  The amplitude contains the dynamical information on the process, which we build using the Feynman calculus, while the phase space is simply the kinematical ``room to maneuver'' from the initial to final states.

In the context of scattering, we want to develop an expression for the differential cross section $d\sigma$ to relate to measured scattering angles and energies:
\begin{equation}
d\sigma = \frac{\vert\mathfrak{M}\vert^2}{F} \times dQ
\end{equation}	
for Lorentz invariant phase space $dQ$ and a flux factor $F$ \cite{Halzen}.

\begin{figure}[htb]
  \begin{center}
   \includegraphics[width=2in]{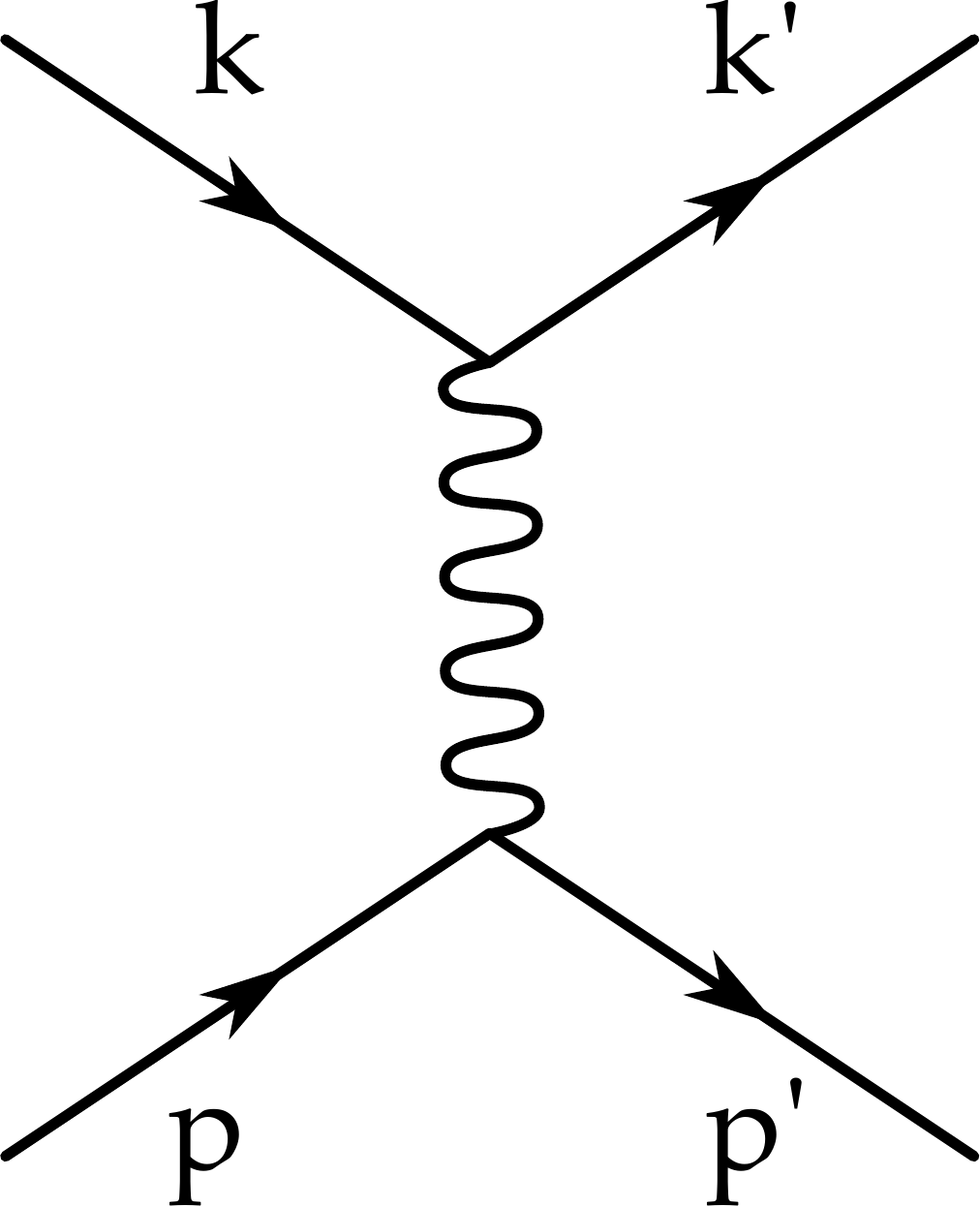}
  \end{center}
  \caption{Leading order Feynman diagram for lepton--lepton scattering. }
  \label{fig:em}
\end{figure}
To build an invariant amplitude for a scattering process such as the one shown in figure \ref{fig:em} for lepton--lepton scattering, the Feynman calculus\footnote{See \cite{Halzen} table 6.2 or \cite{griffiths} section 7.5.} prescribes the factors to collect based on features of our diagram\footnote{Feynman diagrams in this document will generally show space-time proceeding from left to right.}.  For each line leaving the diagram, we include an external line factor such as $u(k)$ or $\bar{u}(k')$ for an incoming or outgoing electron.  This $u$, and its adjoint $\bar{u}$, represent solutions to the momentum space Dirac equation $(\gamma^{\mu}p_{\mu} -mc)u = 0$.  Each vertex adds a $ig_e \gamma^{\mu}$, with $g_e$ representing the coupling strength of the vertex, here the charge of the electron $e$. We then need factors for internal line propagation, which in this case is a photon: $-ig_{\mu\nu}/q^2$.  

After including delta function factors to ensure conservation of momentum, we have an integral over internal momenta
\begin{equation}
(2\pi)^4 \int{[\bar{u}(k')\gamma^{\mu}u(k)]\frac{ig_{\mu\nu}}{q^2}[\bar{u}(p')\gamma^{\mu}u(p)]\delta^4(k-k'-q)\delta^4(p-p'+q)d^4q},
\end{equation}
which we integrate and cancel the delta functions to reach the matrix element
\begin{equation}
\label{eq:mlep}
\mathfrak{M} = -\frac{g^2_e}{(k-k')^2}[\bar{u}(k')\gamma^{\mu}u(k)][\bar{u}(p')\gamma^{\mu}u(p)].
\end{equation}

\subsection{Electron--Muon Scattering}

The matrix element we have achieved in equation \ref{eq:mlep} applies directly to $e^-\mu^+ \rightarrow e^-\mu^+$ scattering.  By proceeding with this example, we illustrate a procedure which will carry over naturally to the case of elastic electron--proton scattering.  

For the time being, we will assume no knowledge of the spin degrees of freedom; to find such a scattering amplitude we need to average over all spin states of $\vert \mathfrak{M}\vert^2$ to get $\overline{\vert \mathfrak{M}\vert^2}$, which we can compare with measurement.

Squaring our matrix element we have:
\begin{equation}
\label{eq:m1}
\vert{\mathfrak{M}\vert^2} = \frac{e^4}{(k-k')^4}[\bar{u}(k')\gamma^{\mu}u(k)][\bar{u}(p')\gamma_{\mu}u(p)][\bar{u}(k')\gamma^{\nu}u(k)]^*[\bar{u}(p')\gamma_{\nu}u(p)]^*.
\end{equation}
As we produce the spin average, it is convenient to separate the sums over the electron and muon spins such that 
\begin{equation}
\label{eq:L}
\overline{\vert \mathfrak{M}\vert^2} = \frac{e^4}{q^4}L_e^{\mu\nu}L_{\mu\nu}^{\mathrm{muon}},
\end{equation}
with the electron tensor 
\begin{equation}
L_e^{\mu\nu} = \frac{1}{2} \sum_{\mathrm{spins}}[\bar{u}(k')\gamma^{\mu}u(k)][\bar{u}(k')\gamma_{\nu}u(k)]^*,
\end{equation}
and a similar muon tensor.  Using ``Casimir's trick'' we can turn these sums over spins into traces of $4\times4$ matrices, which we then apply trace theorems\footnote{See \cite{Halzen} sections 6.3 and 6.4 or \cite{griffiths} section 7.7.} to simplify and remove the bilinear covariants of the Dirac equation:
\begin{equation}
\begin{split}
L_e^{\mu\nu} &= \frac{1}{2}\textrm{Tr}((\not{k}' + m)\gamma^{\mu}(\not{k} + m)\gamma^{\nu})\\
 &= 2\left(k'^{\mu}k^{\nu} + k'^{\nu}k^{\mu} - (k'\cdot k - m^2)g^{\mu\nu}\right).
\end{split}
\end{equation}
Now plugging these electron and muon tensor expressions back into \ref{eq:L}, we have the following expression, with $m$ the mass of the electron, and $M$ of the muon:
\begin{equation}
\overline{\vert \mathfrak{M}\vert^2} = \frac{8e^4}{q^4} \left[ (k' \cdot p')(k\cdot p) + (k' \cdot p)(k\cdot p') -m^2p' \cdot p - M^2 k' \cdot k + 2m^2M^2   \right].
\end{equation}

\begin{figure}[htb]
  \begin{center}
   \includegraphics[width=3in]{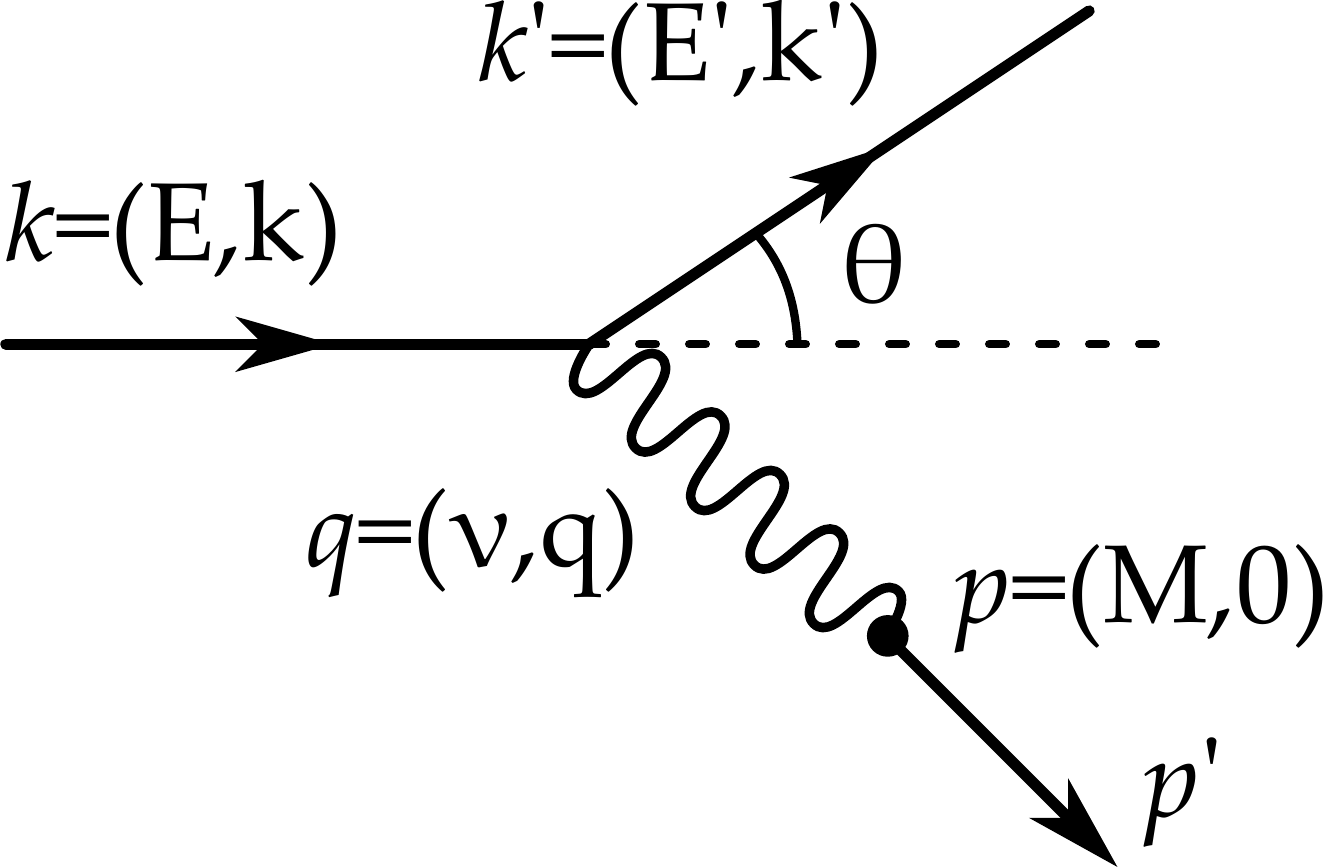}
  \end{center}
  \caption{Lab frame electron scattering from a stationary muon target. }
  \label{fig:mulab}
\end{figure}

Armed with this expression, we can construct a differential cross section for scattering in the laboratory frame.  For a stationary muon as shown in figure \ref{fig:mulab}, and neglecting the electron mass, we recall the relations of section \ref{sec:variables} to get
\begin{equation}
\label{eq:m2}
\begin{split}
\overline{\vert \mathfrak{M}\vert^2} &= \frac{8e^4}{q^4} \left[-\frac{1}{2}q^2M(E-E') + 2EE'M^2 + \frac{1}{2}M^2q^2   \right]\\
&= \frac{8e^4}{q^4}2M^2EE'\left\{ \cos^2\frac{\theta}{2} - \frac{q^2}{2M^2}\sin^2\frac{\theta}{2} \right\}.
\end{split}
\end{equation}

Now we apply the golden rule to build a differential cross section, still neglecting the electron mass:
\begin{equation}
\label{eq:sig1}
d\sigma = \frac{1}{4ME}\frac{\overline{\vert \mathfrak{M}\vert^2}}{4\pi^2}\frac{1}{2}E'dE'd\Omega\frac{d^3p'}{2p'_0}\delta^4(p+q-p')
\end{equation}

Finally, we arrive at a result, combining equations \ref{eq:m2} and \ref{eq:sig1}:
\begin{equation}
\label{eq:emscat}
\left(\frac{d\sigma}{d\Omega}\right)_{lab} = \left( \frac{\alpha^2}{4E^2\sin^4\frac{\theta}{2}} \right) \frac{E'}{E} \left\{  \cos^2\frac{\theta}{2} - \frac{q^2}{2M^2}\sin^2\frac{\theta}{2}\right\},
\end{equation}
with the factor $E'/E =  1/(1+2E/M\sin^2(\theta/2))$ arising from the target's recoil, and the fine structure constant $\alpha \approx 1/137$. 

If we have a condition where the mass of the target particle is much larger than the scattering energy $(M\gg q)$ in equation \ref{eq:emscat}, we recognize a familiar result from experiment---the Mott cross section of spin coupled Coulomb scattering:
\begin{equation}
\left(\frac{d\sigma}{d\Omega}\right)_{Mott} = \frac{\alpha^2}{4E^2} \left(\frac{\cos^2\frac{\theta}{2}}{\sin^4\frac{\theta}{2}} \right) \frac{E'}{E}.
\end{equation}

\subsection{Elastic Electron--Proton Scattering}
	
\begin{figure}[tb]
  \begin{center}
   \includegraphics[width=2in]{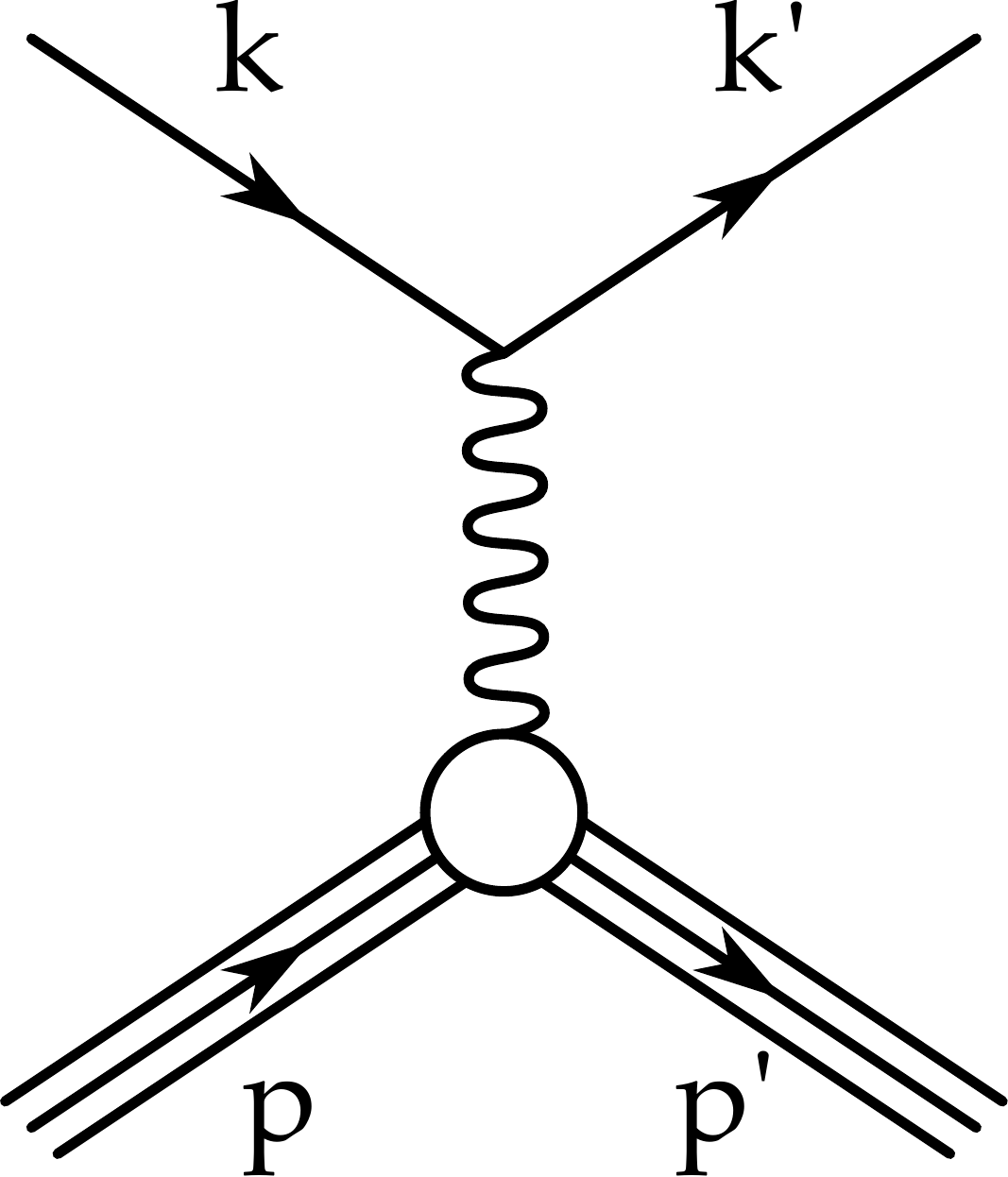}
  \end{center}
  \caption{Leading order diagram for elastic electron--proton scattering.}
  \label{fig:epelastic}
\end{figure}	
	
	Were the proton a point charge $e$ with Dirac magnetic moment $e/2M$, we would have reached our goal at equation \ref{eq:emscat}.  For a proton with internal structure, we need to adjust our matrix element accordingly.  The key is that we can keep our electron tensor as is, carrying over what we know well from quantum electrodynamics and addressing the proton tensor separately:
\begin{equation}
\overline{\vert \mathfrak{M}\vert^2} = \frac{e^4}{q^4}L^{\mu\nu}_{electron}W_{\mu\nu}^{proton}.
\end{equation}	
Taking a step back, we change the matrix element from equation \ref{eq:mlep} accordingly; the $\gamma^{\mu}$ of a spin-$\frac{1}{2}$ point particle doesn't apply to the proton:
\begin{equation}
\mathfrak{M} = -\frac{g^2_e}{(k-k')^2}[\bar{u}(k')\gamma^{\mu}u(k)][\bar{u}(p') \Bigl[\ ? \ \Bigr] u(p)].
\end{equation}	

To fill those square brackets which have taken the place of a $\gamma^{\mu}$, we look for a four-vector to fit between our Dirac spinors.  We naively build a four-vector out of $p$, $p'$, $q$ and bilinear covariants, except $\gamma^5$ which is ruled out by parity conservation.  Following section 8.2 of \cite{Halzen}, without loss of generality, we can insert
\begin{equation}
\left [ f_1(q^2)\gamma^{\mu} + \frac{\kappa}{2M}f_2(q^2)i\sigma^{\mu\nu}q_{\nu} \right],
\end{equation}	
where we have introduced two independent form factors, $f_1(q^2)$ and $f_2(q^2)$, and the anomalous magnet moment $\kappa$.  These two form factors parametrize the unknown behavior shown by the open circle in figure \ref{fig:epelastic}.  In practice, these form factors are written so that no interference terms appear in the cross section:
\begin{equation}
\begin{split}
G_E &\equiv f_1 + \frac{\kappa q^2}{4M^2}f_2\\
G_M &\equiv f_1 + \kappa f_2
\end{split}
\end{equation}
Now, for elastic $e$--$p$ scattering, equation \ref{eq:emscat} becomes
\begin{equation}
\left(\frac{d\sigma}{d\Omega}\right)_{lab} = \left( \frac{\alpha^2}{4E^2\sin^4\frac{\theta}{2}} \right) \frac{E'}{E} \left\{  \frac{G_E^2+\tau G_M^2}{1+\tau}\cos^2\frac{\theta}{2} + 2\tau G_M^2\sin^2\frac{\theta}{2}\right\}
\end{equation}
with $\tau \equiv -q^2/4M^2$.  This is the Rosenbluth cross section, with the Sachs form factors $G_E(q^2)$ and $G_M(q^2)$.  We can think of the form factors as the \textit{extent} of the electric and magnetic charge, and are rightly the Fourier transforms of the charge distributions.  Differences between the ratios of these form factors from measurements using polarization transfer and Rosenbluth separation techniques continue to prompt inquiry
\cite{afan2phot,arr2phot}.  An overview of these electromagnetic form factors can be found in reference \cite{dejager}. 

\subsection{Deep Inelastic Electron--Proton Scattering}
\label{sec:dis}
\begin{figure}[tb]
  \begin{center}
   \includegraphics[width=2.2in]{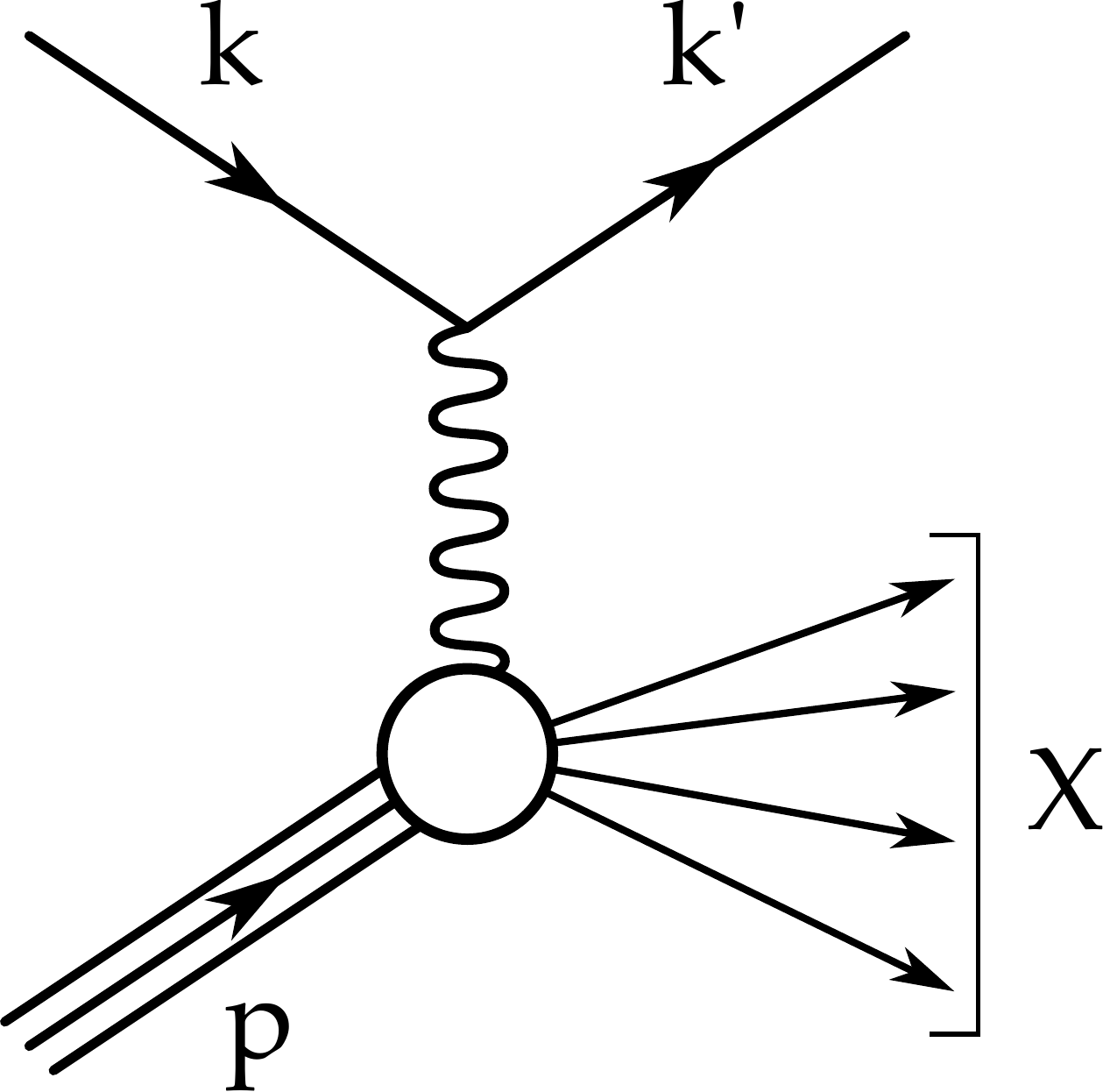}
  \end{center}
  \caption{Leading order diagram for inelastic $e p \rightarrow e' X$ scattering.}
  \label{fig:epinel}
\end{figure}

	As we peer deeper into the proton using a virtual photon of smaller wavelength, the increased energy of the scattering interaction will tear apart the proton.  In elastic scattering $e p \rightarrow e' p$, the final state of the proton could be represented by the Dirac $\bar{u}$ entry into the matrix element.  As we break up the proton $e p \rightarrow e' X$, shown in figure \ref{fig:epinel}, we need a new formalism for the final state.  
	
\begin{figure}[tb]
  \begin{center}
   \includegraphics[width=4.0in]{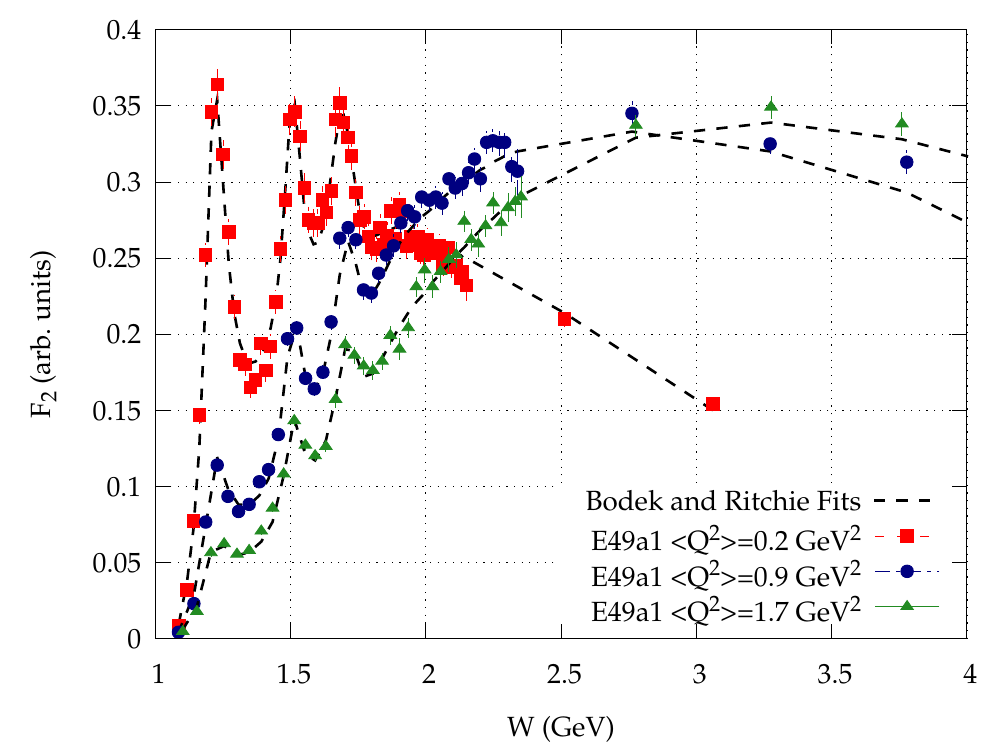}
  \end{center}
  \caption[Resonance peaks in the cross section vs. W]{Resonance peaks in the e--p cross section vs. W, reproduced from SLAC E49a1 data \cite{e49a1}.  Plotted is $F_2$, which is directly proportional to the cross section for these fixed angle, low $\theta$ examples.  The fits shown are from Bodek and Ritchie\cite{Bodek}.}
  \label{fig:reson}
\end{figure}

	In inelastic scattering, the invariant mass of the final state $W$, or the ``missing'' mass in inclusive scattering, becomes a quantity of interest.  With increasing $q^2$, peaks emerge in the spectrum of $d^2\sigma/(d\Omega dE')$ versus the missing mass $W = M^2 + 2M\nu + q^2$.  The first, at $W$ equal to the proton mass, is the \textit{elastic peak} in which the proton does not break up.  At higher $W$ are resonance peaks in which the target is excited into resonant baryon states, such as the $\Delta$ at mass 1232 MeV (see figure \ref{fig:reson}).  Beyond the resonances is the smooth curve made up of the many complicated multi-particle states of deep inelastic scattering.

As in the case of elastic $e$--$p$ scattering, to proceed to form an expression for this scattering we separate the matrix element into an electron tensor and a proton tensor:
\begin{equation}
\frac{d^2\sigma}{d\Omega dE'} = \frac{\alpha^2}{2Mq^4}\frac{E'}{E} L^{\mu\nu}_{e}W_{\mu\nu}.
\end{equation}
We recognize the electron tensor, now dealing with the spins explicitly:
\begin{equation}
\begin{split}
L^{\mu\nu}_{e} &= \frac{1}{2} \sum_{\mathrm{spins}}\bar{u}(k,s)\gamma^{\mu}u(k',s')\bar{u}(k',s')\gamma_{\nu}u(k,s)\\
&= k'^{\mu}k^{\nu} + k'^{\nu}k^{\mu} - g^{\mu\nu}k\cdot k' + [i\epsilon^{\mu\nu\lambda\sigma}q_{\lambda}s_{\sigma}],
\end{split}
\end{equation}
after summing over spins, where here we have enclosed the part which is antisymmetric under $\mu \nu$ interchange in brackets, which includes the spin vector for the electron $s$.

As we look to the proton tensor $W_{\mu\nu}$, we must be even more general in our formulation than in the elastic case as we can't even rely on Dirac $u$.  Taking into account parity conservation, Lorentz invariance, gauge invariance, and standard discrete symmetries of the strong force, we can maintain generality while parameterizing $W_{\mu\nu}$ in four dimensionless structure functions\cite{Drell:1963ej}, two symmetric in $\mu$, $\nu$ interchange (superscript $(S)$) and two antisymmetric (superscript $(A)$):
\begin{equation}
W_{\mu\nu}(q;p,S) = W_{\mu\nu}^{(S)}(q;p) + iW_{\mu\nu}^{(A)}(q;p,S)
\end{equation}
with
\begin{equation}
\begin{split}
\label{eq:hadronic}
\frac{1}{2M}W_{\mu\nu}^{(S)}(q;p) &= \left( -g_{\mu\nu} + \frac{q_{\mu}q_{\nu}}{q^2}\right) W_1(p\cdot q,q^2) \\ &+ \left(	p_{\mu} - \frac{p\cdot q}{q^2}q_{\mu}\right)\left( p_{\nu} - \frac{p\cdot q}{q^2}q_{\nu} \right)\frac{W_2(p\cdot q,q^2)}{M^2}\\
\frac{1}{2M}W_{\mu\nu}^{(A)}(q;p,S) &= \epsilon_{\mu\nu\alpha\beta}q^{\alpha}MS^{\beta}G_1(p\cdot q,q^2) \\ &+\epsilon_{\mu\nu\alpha\beta}q^{\alpha}[(p\cdot q)S^{\beta}+(S\cdot q)p^{\beta}]\frac{G_2(p\cdot q,q^2)}{M}  .
\end{split}
\end{equation}
Here we have used the proton spin vector $S$.  We notice the symmetric portion of the hadronic tensor $W_{\mu\nu}$ consists of two spin-independent structure functions, $W_1$ and $W_2$, while the spin-dependent, antisymmetric portion gives us structure functions, $G_1$ and $G_2$.  

As we measure experimental cross sections, we access different structure functions depending on our control of the spin degrees of freedom\cite{Anselmino19951}.  For instance, unpolarized electron--proton scattering results in a cross section which is proportional to the symmetric terms:
\begin{equation}
\frac{d^2\sigma^{unpol}}{d\Omega dE'}(k,p;k') = \frac{\alpha^2}{Mq^4}\frac{E'}{E}L_{\mu\nu}^{(S)}W^{\mu\nu(S)}.
\end{equation} 
Or, if we take a difference of cross sections of opposite target spin polarizations, still summing over electron spins, we can measure the antisymmetric terms:
\begin{equation}
\sum_{s'}\left[\frac{d^2\sigma}{d\Omega dE'}(k,s,p,-S;k',s')-\frac{d^2\sigma}{d\Omega dE'}(k,s,p,S;k',s')\right] = \frac{2\alpha^2}{MQ^4}\frac{E'}{E}L_{\mu\nu}^{(A)}W^{\mu\nu(A)}.
\end{equation}
We will present explicit expressions for the structure functions in terms of cross sections of different spin orientations in section \ref{sec:measure}.  We can now focus our interest in these structure functions to continue our investigation of the structure of the nucleon.

\section{Bjorken Scaling}

\begin{figure}[htb]
  \begin{center}
   \includegraphics[width=4in]{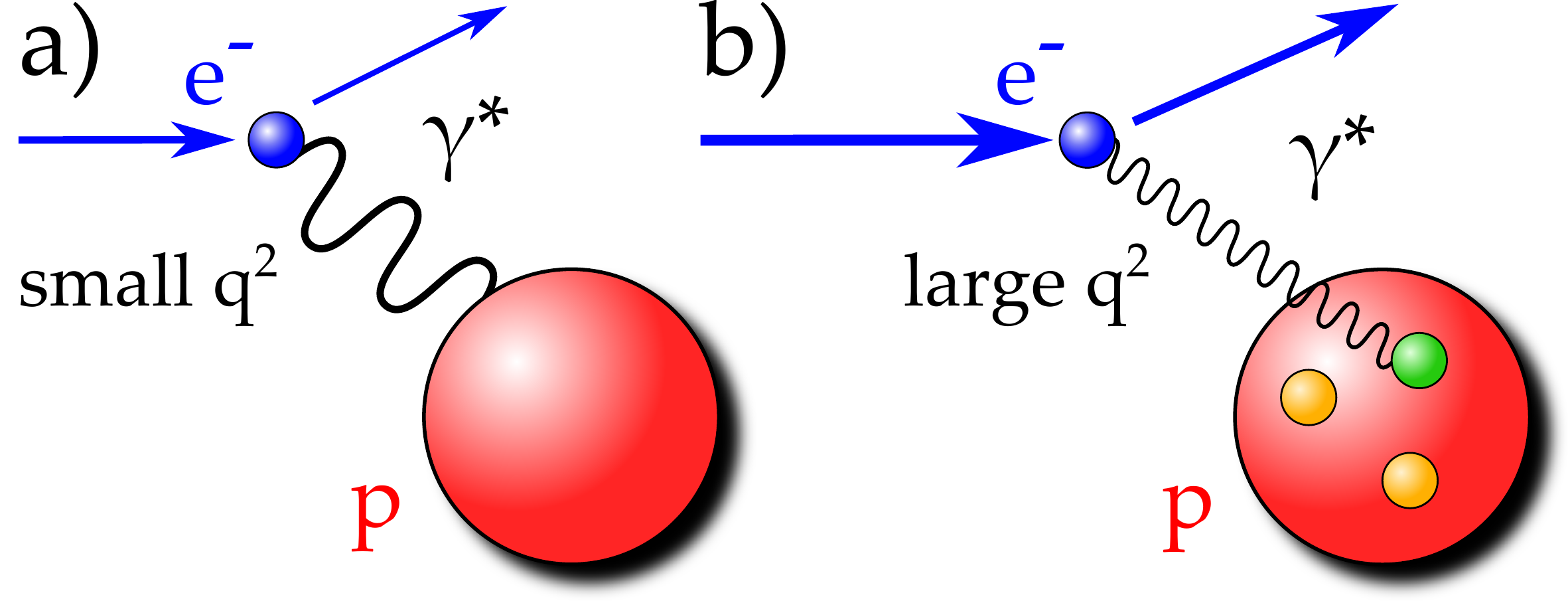}
  \end{center}
  \caption[Diagram of the resolving power of the virtual photon.]{Diagram of the resolving power of the virtual photon in elastic (a) and deep inelastic (b) electron--proton scattering.}
  \label{fig:scaling}
\end{figure}

We have seen that as we increase the momentum transfer of our scattering interaction, the proton ceases to behave like a point particle, revealing internal structure. At yet higher $-q^2$, we begin to suspect the presence of point particles, or \textit{partons}, inside the proton (figure \ref{fig:scaling}) as the first two proton structure functions simplify to
\begin{equation}
\begin{split}
\label{bjorkenfunc}
2mW_1^{point}(\nu ,Q^2) &= \frac{Q^2}{2m\nu}\delta\left( 1 - \frac{Q^2}{2m\nu}\right)\\
\nu W_2^{point}(\nu ,Q^2) &= \delta\left( 1 - \frac{Q^2}{2m\nu} \right).
\end{split}
\end{equation}
Here we notice these functions depend only on the dimensionless ratio $Q^2/2m\nu$, where mass $m$ is of that of the constituent particle inside the proton \cite{Halzen}.

\begin{figure}[htbp]
  \begin{center}
   \includegraphics[width=4in]{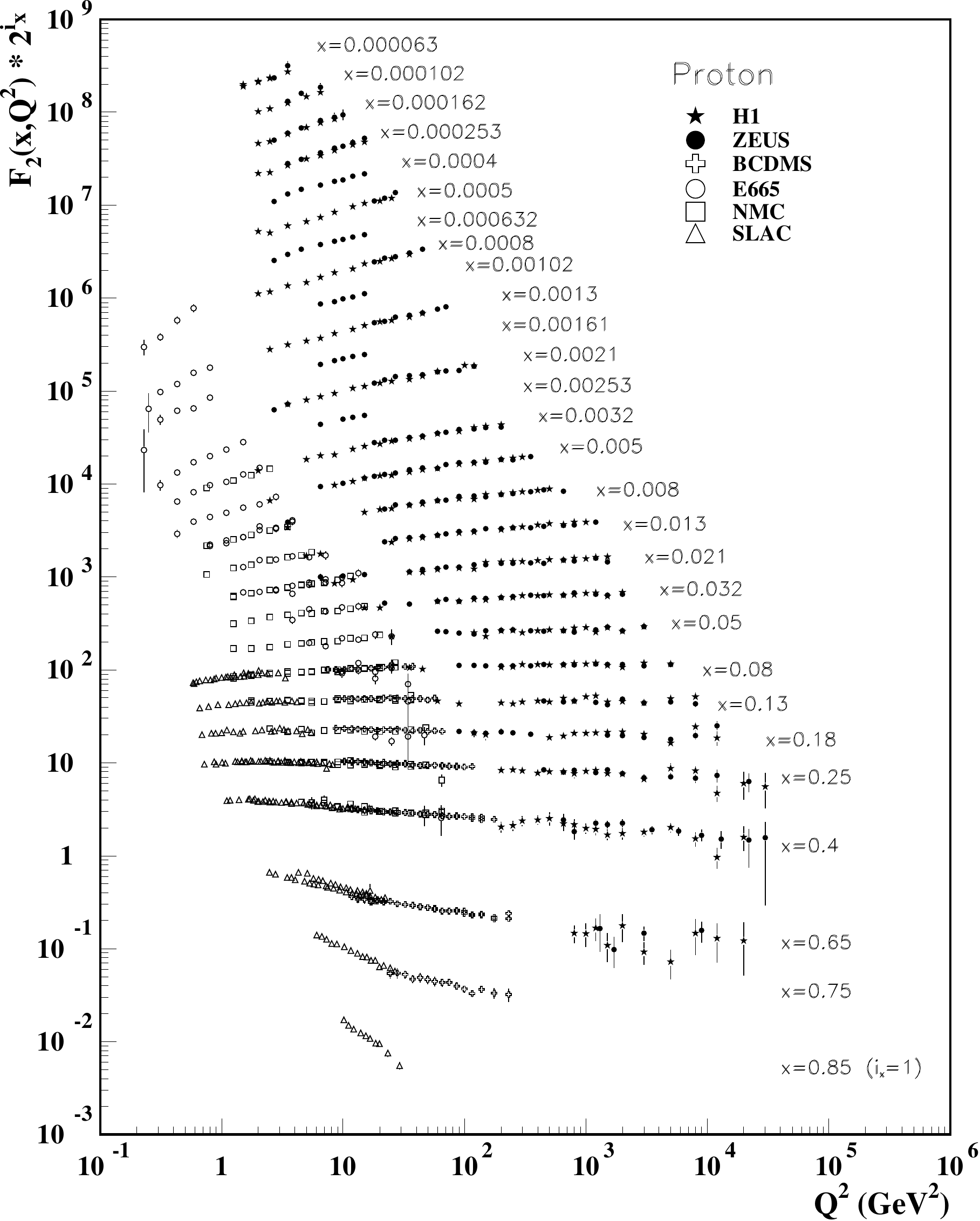}
  \end{center}
  \caption[Scaling in structure function $F_2$.]{Scaling in the structure function $F_2$, where the structure function is roughly constant in $Q^2$ for most values of $x$. Figure from reference \cite{pdb}.}
  \label{fig:f2scale}
\end{figure}

With this in mind we define the deep inelastic regime in the Bjorken limit:
\begin{equation}
\begin{split}
-q^2 &\equiv Q^2 \rightarrow \mathrm{large}, \\
\nu &= E-E' \rightarrow \infty, \\
x &= \frac{Q^2}{2p \cdot q} = \frac{Q^2}{2M\nu} \quad \textrm{constant}.
\end{split}
\end{equation}
In the Bjorken limit, the proton structure functions, which depend on $\nu$ and $Q^2$, become dependent only upon the dimensionless Bjorken $x$, a sign that the \textrm{partons} themselves have no internal structure.  Figure \ref{fig:f2scale} shows an example of scaling behavior for $F_2$.

Thus, in the Bjorken limit we can give the structure functions as
\begin{equation}
\begin{split}
\label{scalingfunc}
MW_1(\nu,Q^2) &\equiv F_1(x,Q^2) \xrightarrow[\mathrm{large}\ Q^2]{} F_1(x),\\
\nu W_2(\nu,Q^2) &\equiv F_2(x,Q^2) \xrightarrow[\mathrm{large}\ Q^2]{} F_2(x),\\
\frac{(p\cdot q)^2}{\nu}G_1(\nu,Q^2) &\equiv g_1(x,Q^2) \xrightarrow[\mathrm{large}\ Q^2]{} g_1(x),\\
\nu(p\cdot q)G_2(\nu,Q^2) &\equiv g_2(x,Q^2) \xrightarrow[\mathrm{large}\ Q^2]{} g_2(x).
\end{split}
\end{equation}

\begin{figure}[htbp]
  \begin{center}
   \includegraphics[width=4in]{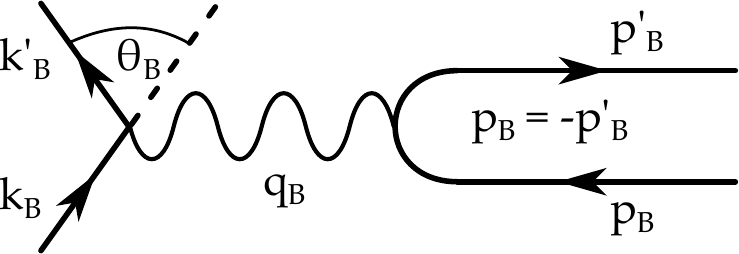}
  \end{center}
  \caption{Diagram of scattering in the Breit frame.}
  \label{fig:breit}
\end{figure}

We have now bundled up all the inner workings of the proton into these four \textit{scaling} structure functions which are functions only of $x$ in the Bjorken limit.  Bjorken $x$ can be thought of as the fraction of the proton's momentum which was carried by the struck constituent particle.  Obviously, in the lab frame the proton is stationary; this definition applies in the Breit frame of reference, where the outgoing momentum of the proton is equal but opposite the incoming momentum, shown in figure \ref{fig:breit}.

From equations \ref{bjorkenfunc} and \ref{scalingfunc}, we also see a useful relation between the unpolarized structure functions:
\begin{equation}
\label{eq:callan}
F_2(x) = 2xF_1(x),
\end{equation}
known as the Callan-Gross relation.  Looking at figure \ref{fig:f2scale}, the scaling behavior falls off at high and low $x$, hinting at the effects of the constituents' interactions.  The change, or so-called ``evolution'', of the structure functions in $Q^2$ is described by the Dokshitzer--Gribov-–Lipatov-–Altarelli-–Parisi (DGLAP) equations\cite{Altarelli, Gribov}.

\section{Compton Scattering \& Inclusive $ep \rightarrow eX$}
\label{sec:compton}
Before moving on to a deeper discussion of the spin structure functions, it is worthwhile to take a brief aside to show another way to look at the hadronic tensor $W_{\mu\nu}$ and thus $F_1$, $F_2$, $g_1$, and $g_2$.  As the hadronic tensor deals with the virtual photon's intersection with the proton, the connection with virtual Compton scattering $\gamma p \rightarrow \gamma p$ is not entirely unintuitive.

\begin{figure}[htb]
  \begin{center}
   \includegraphics[width=2.3in]{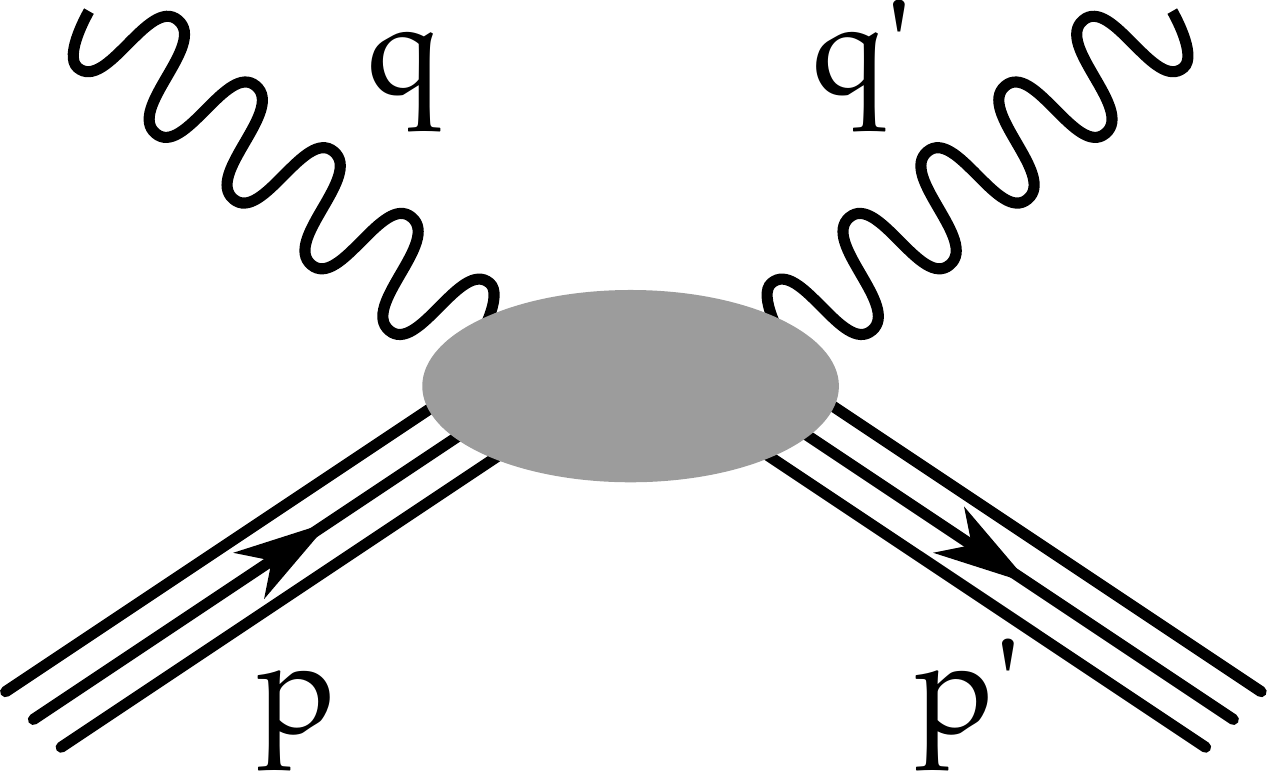}
  \end{center}
  \caption{Compton scattering diagram.}
  \label{fig:compton}
\end{figure}

If we consider virtual ($Q^2<0$) forward ($q=q'$) Compton scattering seen in figure \ref{fig:compton}, we can express the scattering amplitude in terms of the electromagnetic current $J_{\mu}$ as 
\begin{equation}
\label{eq:compton}
T_{\mu\nu}(q;p,s)= i \int d^4 z e^{iq\cdot z} \langle p,s \vert \mathcal{T}(J_{\mu}(z)J_{\nu}(0)) \vert p,s\rangle
\end{equation}
with the time ordering operator $\mathcal{T}$\cite{bass-book,Anselmino19951}.

The hadronic tensor can be similarly expressed as the Fourier transform of the matrix elements of the commutator of electromagnetic currents in inclusive $e$--$p$ scattering:
\begin{equation}
\label{eq:hadcur}
W_{\mu\nu}(q;p,s)=  \frac{1}{2\pi}\int d^4 z e^{iq\cdot z} \langle p,s \vert [(J_{\mu}(z),J_{\nu}(0)] \vert p,s\rangle
\end{equation}  

With equations \ref{eq:compton} and \ref{eq:hadcur}, the relation between the forward virtual Compton tensor and the inclusive hadronic tensor, properly a result of the \textit{optical theorem}, is apparent:
\begin{equation}
W_{\mu\nu}(\nu,Q^2) = \frac{1}{\pi}\mathrm{Im}T_{\mu\nu}(\nu,Q^2).
\end{equation}
The hadronic tensor is proportional to the imaginary (or absorptive) part of the forward virtual Compton tensor\cite{Anselmino19951,bass}.

One of the results of this relation is the connection between virtual photon absorption asymmetries $A_1$ and $A_2$, and the $e$--$p$ structure functions.  Asymmetries $A_1$ and $A_2$ are defined in terms of virtual photon absorption cross sections for polarized photons and nucleons; these 4 cross sections are labeled by the spin sum, anti-parallel $(J_z = 1 + \frac{1}{2} = \mathbf{\frac{3}{2}})$ or parallel $(J_z = 1 - \frac{1}{2} = \mathbf{\frac{1}{2}})$, and L or T for a longitudinal or transverse photon\cite{leaderpredazzi}.
\begin{equation}
\begin{split}
A_1 &= \frac{\sigma_{1/2}^T-\sigma_{3/2}^T}{\sigma_{1/2}^T+\sigma_{3/2}^T}\\
A_2 &= \frac{2\sigma_{1/2}^{TL}}{\sigma_{1/2}^T+\sigma_{3/2}^T}
\end{split}
\end{equation}
The spin structure functions are expressed in terms of these asymmetries and the structure function $F_1$ as
\begin{equation}
\label{eq:virtcomp}
\begin{split}
g_1 &= \frac{F_1}{1+\gamma^2}\left(A_1 + \gamma A_2 \right)\\
g_2 &= \frac{F_1}{1+\gamma^2}\left(\frac{A_2}{\gamma} -  A_1 \right) 
\end{split}
\end{equation}
for $\gamma^2 = 4x^2M^2/Q^2$.


\chapter{Proton Spin Structure}

In the previous chapter we established a framework for studying nucleon structure through lepton scattering experiments, parameterizing the proton's unknown behavior in four structure functions.  In this chapter we will endeavor to interpret physical meaning from these structure functions, detail a methodology to measure them, and review existing measurements.  We take advantage of excellent review papers on the study of nucleon spin structure in this chapter, references \cite{Filippone,leaderpredazzi,Anselmino19951,bass,bass-book,Melnitchouk,Jaffe:1996zw,DonnellyRaskin,Manohar}.

\section{Partons}

Faced with Bjorken scaling, we look for a model of the proton with point particle constituents.  The parton model put forward by Feynman in 1969 \cite{feynmanparton} does just this, describing a nucleon made up of different kinds of point particles, partons, which were later recognized as quarks and gluons.

In this model, we consider the constituent partons to be semi-free and point-like.  We can begin to put together a picture of how the spin of these partons might contribute to the spin of the proton, as in this non-relativistic wave function for a proton made of up ($u$) and down ($d$) quarks
\cite{Filippone}:

\begin{equation}
\vert p^{\uparrow} \rangle = \frac{1}{\sqrt{6}}(2\vert u^{\uparrow}u^{\uparrow}d^{\downarrow} \rangle - \vert u^{\uparrow}u^{\downarrow}d^{\uparrow} \rangle  -\vert u^{\downarrow}u^{\uparrow}d^{\uparrow} \rangle),
\end{equation}
where the superscript arrows represent the spin state of the quarks as aligned or anti-aligned with the proton spin.  Here the quarks carry all of the proton's spin.

	\subsection{Structure Functions in the Parton Model} 
	\label{sec:partonmodel}

Armed with a model of a proton made of semi-free partons, we return to deep inelastic electron--proton scattering to formulate our structure functions, recalling the hadronic tensor $W_{\mu\nu}$.   Following references \cite{leader,Anselmino19951}, if we let $n_q(x',s;S)dx'$ be the number of partons $q$ with charge $e_q$, momentum fraction $x'$, and spin vector $s$, inside a nucleon of momentum $P$ and spin vector $S$, we can express our hadronic tensor as
\begin{equation}
\begin{split}
\label{eq:partonw}
W_{\mu\nu}(q;P,S) &= W_{\mu\nu}^{(S)}(q;P)+ iW_{\mu\nu}^{(A)}(q;P,S)\\
 &= \sum_{q,s} e_q^2 \frac{1}{2P\cdot q}\int_0^1 \frac{dx'}{x'} \delta(x'-x) n_q(x',s;S){w}_{\mu\nu}(x',q,s).
\end{split}
\end{equation}	
The sum $q$ goes over all quarks and anti-quarks. Here the $e_q^2{w}_{\mu\nu}(x',q,s)\delta(x'-x)$ can been seen as the analogue of the hadronic tensor for the case of photon interacting with a ``free'' parton. 

\begin{figure}[htb]
  \begin{center}
   \includegraphics[width=3in]{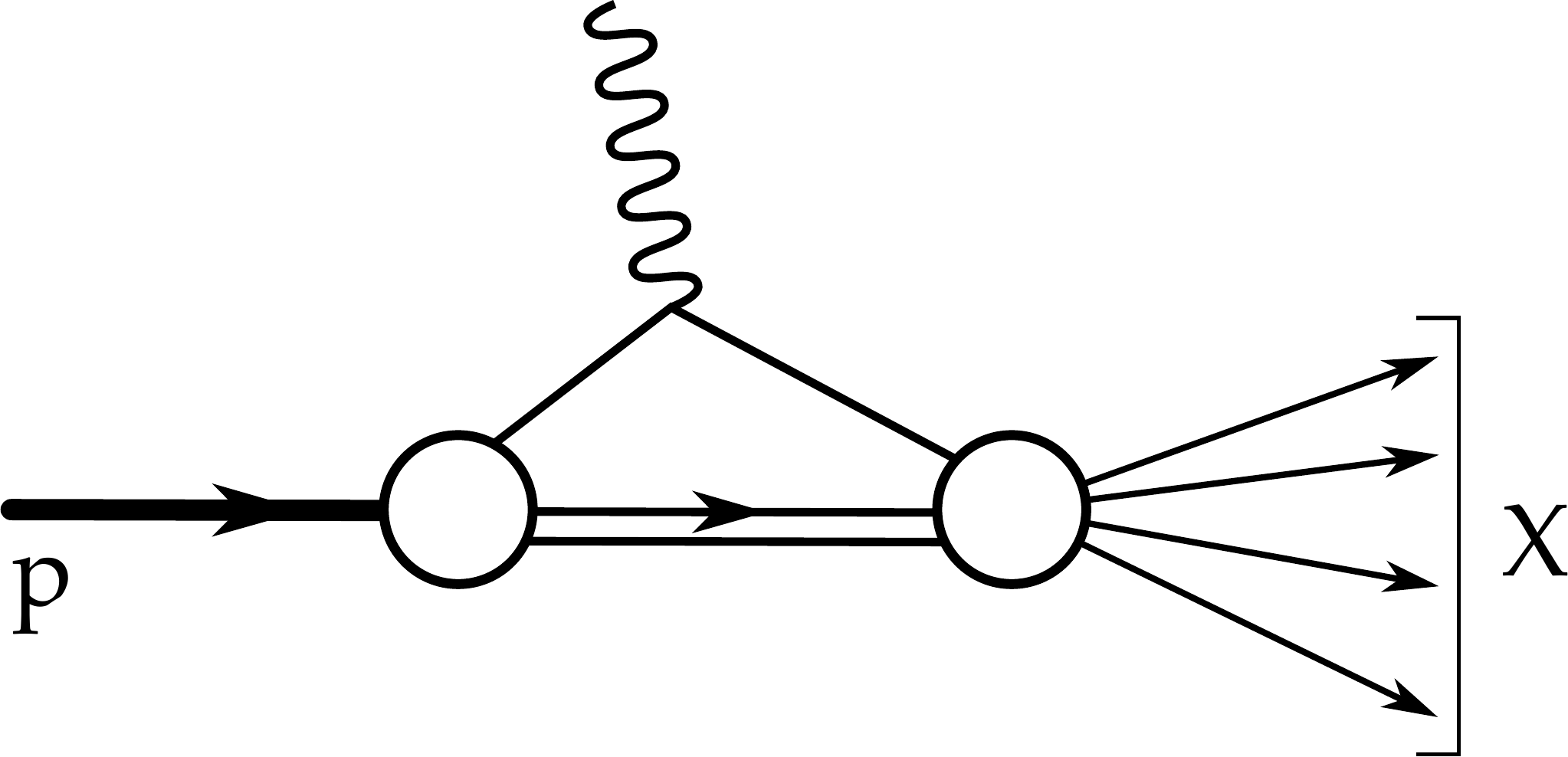}
  \end{center}
  \caption[Parton model interpretation of the $\gamma$--$p$ vertex from $e$--$p$ scattering.]{Parton model interpretation of the $\gamma$--$p$ vertex from $e$--$p$ scattering, based on \cite{leaderpredazzi} 16.1.}
  \label{fig:parton}
\end{figure}
As we see in figure \ref{fig:parton}, we have now simplified the photon--proton interaction to a photon--parton vertex with the parton as a point, charged fermion. We can thus calculate  ${w}_{\mu\nu}$ using QED, leaving the strong interaction dynamics in the number density function.  Treating  ${w}_{\mu\nu}$ as we did $L_{\mu\nu}$, but with replacements $k^{\mu} \rightarrow xP^{\mu}$ and $k'^{\mu} \rightarrow xP^{\mu} + q^{\mu}$, we have:
\begin{equation}
{w}_{\mu\nu} = {w}_{\mu\nu}^{(S)} + i{w}_{\mu\nu}^{(A)}
\end{equation}
with
\begin{equation}
\begin{split}
 {w}_{\mu\nu}^{(S)} &= 2[2x^2P_{\mu} P_{\nu} + xP_{\mu}q_{\nu} + xq_{\mu}P_{\nu} - x(p \cdot q)g^{\mu\nu}] \\
  {w}_{\mu\nu}^{(A)} &= 2 m_q \epsilon_{\mu\nu\alpha\beta}s^{\alpha}q^{\beta}.
\end{split}
\end{equation}
Before we move forward, we condense our notation so that the parton number densities are 
\begin{equation}
q_{\lambda} \equiv P \int d^2p_{\perp}n_q(p,\lambda;\Lambda=1/2),
\end{equation}
so that $q_{\pm}$ represents the number density of quarks with momentum $p \rightarrow xP$, helicity $\lambda = \pm 1/2$ in a proton of momentum $P$ and helicity $\Lambda = 1/2$.  We can now create the unpolarized number density $q(x)$ and difference of spin-dependent quark distribution functions $\Delta q(x)$:
\begin{equation}
\begin{split}
q(x) &= q_+(x) + q_-(x),\\
\Delta q(x) &= q_+(x) - q_-(x).
\end{split}
\end{equation}

Integrating over the assumed small transverse momentum and comparing with equation \ref{eq:hadronic}, we combine ${w}_{\mu\nu}^{(S)}$ with the above equations to arrive at predictions for our structure functions in this quark-parton interpretation:
\begin{equation}
 2xF_1(x,Q^2) = F_2(x,Q^2) = \frac{1}{2} \sum_q e_q^2 q(x),
\end{equation}
where $e_q$ are the charges of these quark flavors and we have used the Callan--Gross relation of equation \ref{eq:callan}.  Likewise, plugging in ${w}_{\mu\nu}^{(A)}$ gives us expressions of the spin structure functions
\begin{equation}
\begin{split}
\label{eq:gparton}
g_1(x,Q^2) &= \frac{1}{2} \sum_q e_q^2\Delta q(x),\\
g_2(x,Q^2) &= 0.
\end{split}
\end{equation}
	
In the zero result for $g_2$, we begin to see cracks in the so-called \textit{naive} quark-parton model. The hard-photon, free-quark interaction is not sensitive to $g_2$ in which transverse spin is important.  Non-zero values of $g_2$ can be obtained by adding transverse momentum to the model, which we have neglected above, but these formulations have an extreme sensitivity to the quark mass.  To access $g_2$ we abandon our simplistic model in favor of the more robust formulation of QCD in DIS.

\section{pQCD and Duality}

Quantum Chromodynamics moves beyond the naive model of semi-free partons to tackle the color charge interactions between the quarks via mediating gluons.  However, the study of semi-free quarks was not entirely wasted.  Due to the property of \textit{asymptotic freedom} discussed in section \ref{sec:standardmodel}, quarks in the nucleon actually do appear to be nearly free at small enough distance scales.  This means at high $Q^2$ we can treat the processes perturabtively, in what is aptly named \textit{perturbative QCD}, or pQCD.

At large $Q^2$, pQCD describes experimental findings quite well.  pQCD correctly predicts the logarithmic $Q^2$ violations of Bjorken scaling in the structure function $F_2$, which comes from gluon production and quark--anti-quark pair creation.  Due to pQCD, we can expect the structure function expression in terms of parton distribution functions from section \ref{sec:partonmodel} to hold at high $Q^2$. However at low $Q^2$, as the interactions between quarks and gluons become important, pQCD predictions should break down.

\subsection{Quark--Hadron Duality}

At lower $Q^2$, approaching the region where resonance production dominates the cross section, a peculiar property was discovered which extends the usefulness of pQCD.  In 1970, Bloom and Gilman \cite{bloom1,bloom2} saw that when the structure function $F_2(\nu,Q^2)$ was measured in the resonance region, it roughly averaged out to the value of $F_2(x)$ expected from the scaling limit. 
 
\begin{figure}[htb]
  \begin{center}
   \includegraphics[width=3.5in]{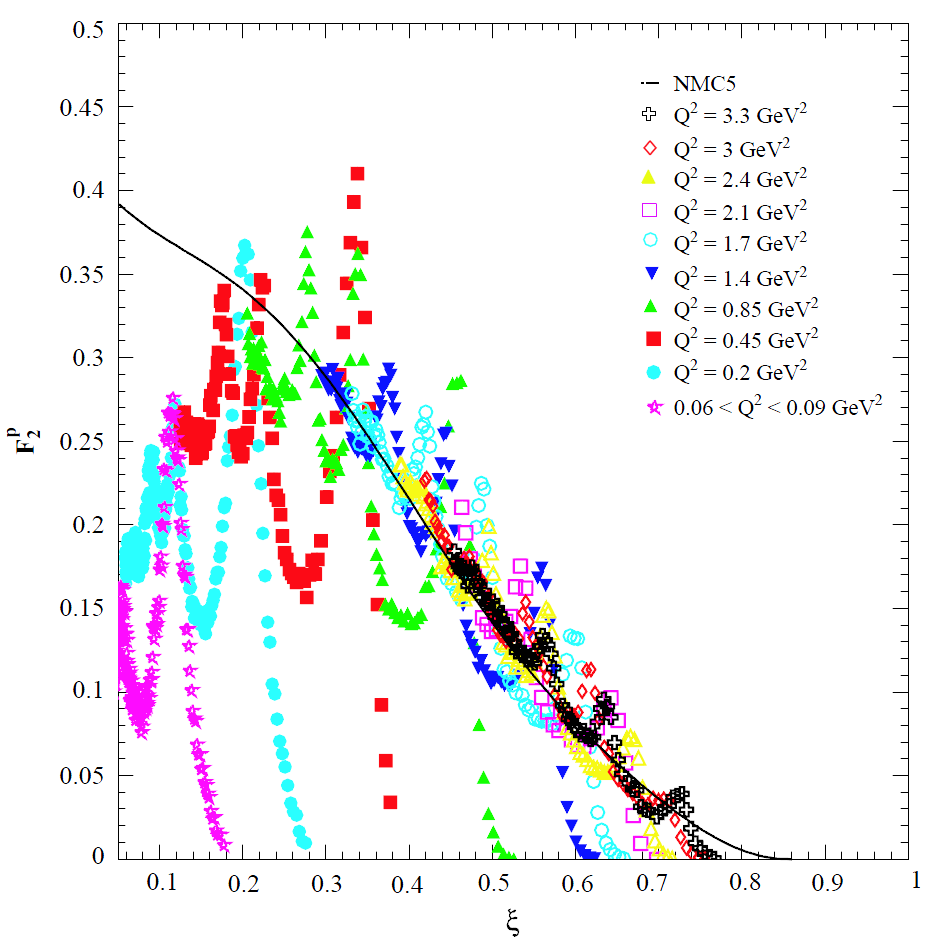}
  \end{center}
  \caption[Duality in Structure Function $F_2^p$.]{Structure function $F_2^p$ from SLAC and JLab data in the resonance region.  Duality can be clearly seen as the resonances sit on the curve to DIS data at the same $\xi$, but higher $Q^2$, from \cite{F2duality}.  As $Q^2$ drops below 1 GeV$^2$, duality ceases to hold.  Plot from reference \cite{Melnitchouk}.}
  \label{fig:duality}
\end{figure}

Defining the Nachtmann scaling variable 
\begin{equation}
\xi = \frac{2x}{1+\sqrt{1+\frac{4M^2x^2}{Q^2}}} \ \xrightarrow[ Q^2 \rightarrow \infty]{} \ x
\end{equation}
attempts to generalize Bjorken $x$ to take into account target mass corrections, counteracting the troublesome sensitivity to the quark mass.  
Plotting $F_2(\xi)$ gives a convincing view of duality.  As $Q^2$ increases, the resonance peaks can be seen sliding along the curve of $F_2(\xi)$ at high $Q^2$, as seen in figure \ref{fig:duality}.  When an individual resonance follows duality in a given $Q^2$ region, we call it ``local'' duality.  In ``global'' duality, this averaging is satisfied over all resonances.  Duality thus extends the results of pQCD into regions of $Q^2$ far lower than might be expected, for certain quantities\cite{Melnitchouk}.  

\section{Moments and Twist}

When evaluating the behavior of structure functions as they evolve in $Q^2$, it is useful to define moments, or $x$-weighted integrals, of the structure functions.  We define the $n$th moment of $F_1$ and $F_2$ as
\begin{equation}
\begin{split}
M_1^{(n)}(Q^2) &= \int_0^1 dx\, x^{n-1} F_1(x,Q^2)\\
M_2^{(n)}(Q^2) &= \int_0^1 dx\, x^{n-2} F_2(x,Q^2).
\end{split}
\end{equation}
These are the Cornwall--Norton moments \cite{CornwallNorton}.  For $n=1$ of $F_1$ we have an effective count of quark charges, while $n=2$ of $F_2$ gives the momentum sum rule.  Likewise, the spin structure function moments are 
\begin{equation}
\label{eq:cnmom}
\begin{split}
\Gamma_1^{(n)}(Q^2) &= \int_0^1 dx\, x^{n-1} g_1(x,Q^2)\\
\Gamma_2^{(n)}(Q^2) &= \int_0^1 dx\, x^{n-2} g_2(x,Q^2).
\end{split}
\end{equation}

\subsection{Operator Product Expansion}
\label{sec:ope}
To describe quark--hadron duality, as well as the spin structure function $g_2$ in QCD, we turn to the \textit{operator product expansion}.  The ``OPE'' was introduced in 1968 by K. Wilson \cite{wilson} as a way to understand the $Q^2$ behavior of moments in DIS, and remains useful after the formulation of QCD to evaluate calculations outside the perturbative region.  In the case of inclusive DIS, the OPE lets us express the products of operators in the asymptotic limit.  The operators we are interested in are the electromagnetic currents as discussed in section \ref{sec:compton}.

In the OPE, as the spatial four-vector $z$ goes to zero, the product of operators $\mathcal{O}_a(z)	$ and $\mathcal{O}_b(0)$ can be expressed as the series
\begin{equation}
\label{eq:ope}
\lim_{z\rightarrow0}\mathcal{O}_a(z)\mathcal{O}_b(0) = \sum_k C_{abk}(z)\mathcal{O}_k(0)
\end{equation}
The key here is that the so-called Wilson coefficients $C_{abk}(z)$ contain all the spatial dependence in the sum.  The equivalence holds as long as the external states of the process have momenta which are small compared to the separation $z$.  Since our coupling constant in QCD is small at short distances due to asymptotic freedom, we can calculate the coefficients in the perturbative range\cite{Manohar}. Thus pQCD calculations can be used to understand our operators in other regimes.

To apply the OPE for the spin structure functions, we start with the expression for the hadronic tensor in terms of the commutator of electromagnetic currents (equation \ref{eq:hadcur}):
\begin{equation}
\label{eq:hadcur2}
W_{\mu\nu}(q;p,s)=  \frac{1}{2\pi}\int d^4 z e^{iq\cdot z} \langle p,s \vert [(J_{\mu}(z),J_{\nu}(0)] \vert p,s\rangle.
\end{equation}
Taking the Fourier transform of \ref{eq:ope} gives us the momentum space version of the OPE, which we can apply to \ref{eq:hadcur2}:
\begin{equation}
\lim_{z\rightarrow0}\int d^4 z e^{iq\cdot z}\mathcal{O}_a(z)\mathcal{O}_b(0) = \sum_k C_{abk}(q)\mathcal{O}_k(0).
\end{equation}

The product of our electromagnetic currents in equation \ref{eq:hadcur2} can now be expanded as a sum of local operators times coefficients which are functions of $q$.  These expansion operators are quark and gluon operators with arbitrary dimension $d$ and spin $n$.  The contribution of any operators to $W_{\mu\nu}L^{\mu\nu}$ is of order
\begin{equation}
(p\cdot q)^n\left(\frac{Q}{M}\right)^{2+n-d} = (p\cdot q)^n\left(\frac{Q}{M}\right)^{2-\tau},
\end{equation}	
where we now define the \textit{twist} of the operator $\tau$ as $\tau = d - n$.	

The lowest, or leading twist, twist-2, contributes the largest in the Bjorken limit, with higher twist contributions suppressed by powers of $M/Q$.  Using dispersion relations, we can apply the OPE to equation \ref{eq:hadcur2} to arrive at expressions for the odd moments of our structure functions. Ignoring contributions beyond twist-3, we have 
\begin{equation}
\begin{split}
\label{eq:moments}
\int^1_0 x^{n-1} g_1(x,Q^2)dx &= \frac{1}{2}a_{n-1}; \quad \mathrm{for}\ n = 1,3,5...\\
\int^1_0 x^{n-1} g_2(x,Q^2)dx &= \frac{n-1}{2n}(d_{n-1}-a_{n-1}); \quad \mathrm{for}\ n = 3,5...
\end{split} 
\end{equation}
	where $a_{n-1}$ and $d_{n-1}$ are matrix elements of the quark and gluon operators for twist-2 and twist-3, respectively.

	\subsection{Burkhardt--Cottingham Sum Rule}
	\label{sec:bcsum}
The OPE has nothing to say about the $n=1$ term of the $g_2$ expression in equation \ref{eq:moments}, but the Burkhardt--Cottingham sum, which addresses the first moment of $g_2$, is not entirely unexpected \cite{Burkhardt}:
\begin{equation}
\label{eq:bcsum}
\Gamma_2(Q^2) = \int_0^1 dx g_2(x,Q^2) = 0.
\end{equation}
This result was first derived from the asymptotic behavior of the virtual Compton helicity amplitude which is proportional to $g_2$.

If this B.C. sum rule is violated, it is likely due to one of two circumstances, according to reference \cite{jaffe-ji}:
\begin{enumerate}
\item $g_2$ is so singular that $\int_0^1 dx g_2(x,Q^2)$ does not exist.
\item $g_2$ has a delta function singularity at $x=0$.
\end{enumerate}
	
 	\subsection{Wandzura--Wilczek Relation}
 	
By combining the two equations in \ref{eq:moments}, we can cancel the leading twist terms to achieve an expression for $g_1$ and $g_2$:
\begin{equation}
\int_0^1 x^{n-1}dx\left ( g_1(x,Q^2) + \frac{n}{n+1}g_2(x,Q^2)\right) = \frac{d_n}{2}
\end{equation}
for $n$ an integer greater or equal to 3.  After performing Mellin transforms, which relate the product of moments of two functions to the moment of their convolution, we arrive at the following result:
\begin{equation}
\label{eq:ww}
g_2^{WW}(x,Q^2) + g_1(x,Q^2) = \int_0^1 \frac{dy}{y} g_1(y,Q^2)
\end{equation}
where here we have set the twist-3 $d_n$ terms to zero.  We've labeled $g_2$ in this equation as $g_2^{WW}$ to designate that this expression ignores higher twist terms.  As it stands, this expression, known as the Wandzura--Wilczek relation\cite{Wandzura1977195}, allows us to determine the leading twist portion of $g_2$ using knowledge of $g_1$, which in turn allows its expression in terms of the parton model.  It should be noted that the OPE does not cover the $n=1$ term of the $g_2$ expansion, so this definition assumes validity of the Burkhardt--Cottingham sum rule.

With our definition of $g_2^{WW}$, we have relegated the higher twist contribution to $g_2$ into the portion here called $\bar{g}_2$:
\begin{equation}
g_2(x,Q^2) = g_2^{WW}(x,Q^2) + \bar{g}_2(x,Q^2)
\end{equation}
While Wandzura and Wilczek went further to hazard that $\bar{g}_2(x,Q^2)$ is zero, we can think of it as the \textit{interesting} part of $g_2$\cite{Jaffe:1996zw}.  The moments
\begin{equation}
\int _0^1 dx x^n \bar{g}_2(x,Q^2) = \frac{n}{4(n+1)}d_n(Q^2)
\end{equation}
are of twist--3 and thus access quark--gluon correlations\cite{Shuryak}.

This $\bar{g}_2$ can itself be split into multiple terms, following \cite{cortes}:
\begin{equation}
\bar{g}_2(x,Q^2) = -\int_x^1\frac{dy}{y}\frac{\partial }{\partial y}\left(\frac{m_q}{M} h_T(y,Q^2) + \xi(y,Q^2)\right)
\end{equation}
where we introduce $\xi$, the twist--3 contribution, and $h_T$, the ``transversity'' distribution from transverse quark polarization, which is a twist--2 term suppressed by the  ratio of the quark to target nucleon mass\cite{bass}.
	
 \subsection{Twist--Three and $g_2$}
While the operator product expansion has given us a foundation to express $g_2$ in the form of higher-order twist, with twist we are left with a mathematical construct from which it is difficult to draw physical meaning.  To understand higher-twist, we must consider parton correlations initially present in the participating hadrons.

\begin{figure}[htb]
  \begin{center}
   \includegraphics[width=2.5in]{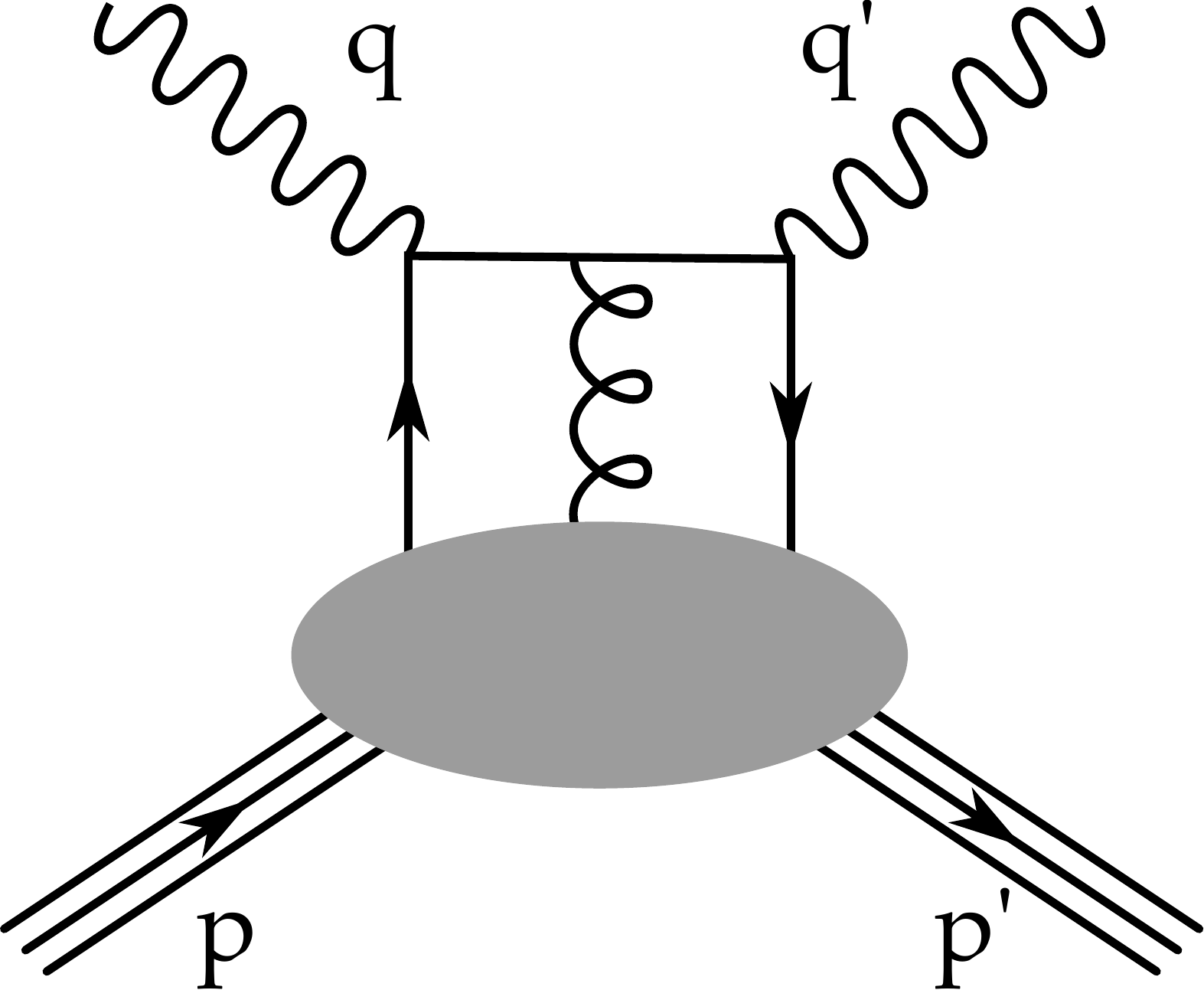}
  \end{center}
  \caption{Twist--3 deep inelastic e-p scattering diagram.}
  \label{fig:twist}
\end{figure}

Higher-twist processes can be thought of as involving \textit{more than one} parton of the hadron in the scattering process, such as in the example in figure \ref{fig:twist}.  We can see the influence of other partons through helicity exchange which is necessary to allow the process.  This exchange can happen in two ways in QCD: through single quark scattering in which the quark carries angular momentum though its transverse axis; or through quark scattering with a transverse-polarized gluon from the hadron \cite{Filippone}.
	
Twist--3 represents the first of the higher-order terms, and therefore gives the greatest contribution to $g_1$ and $g_2$, after leading-order, of course.  In twist--3 we see quark--gluon--quark correlations; instead of viewing only a bare quark we are beginning to probe how the quarks and gluons interact in the context of the nucleon!  With this in mind, $g_2$, which offers the most direct view of these correlations, becomes an attractive quantity to measure.

\section{Measuring Spin Structure Functions}
\label{sec:measure}
As we asserted in section \ref{sec:dis}, we can access the antisymmetric portion of the hadronic tensor via deep inelastic electron--proton scattering by taking a difference of cross sections of opposite polarizations.  In this section, we'll develop expressions to obtain the structure functions $g_1$ and $g_2$ using measurements of asymmetries of cross sections, from a polarized electron beam upon a polarized proton target, anticipating the measurements of SANE.

\begin{figure}[htb]
  \begin{center}
   \includegraphics[width=3in]{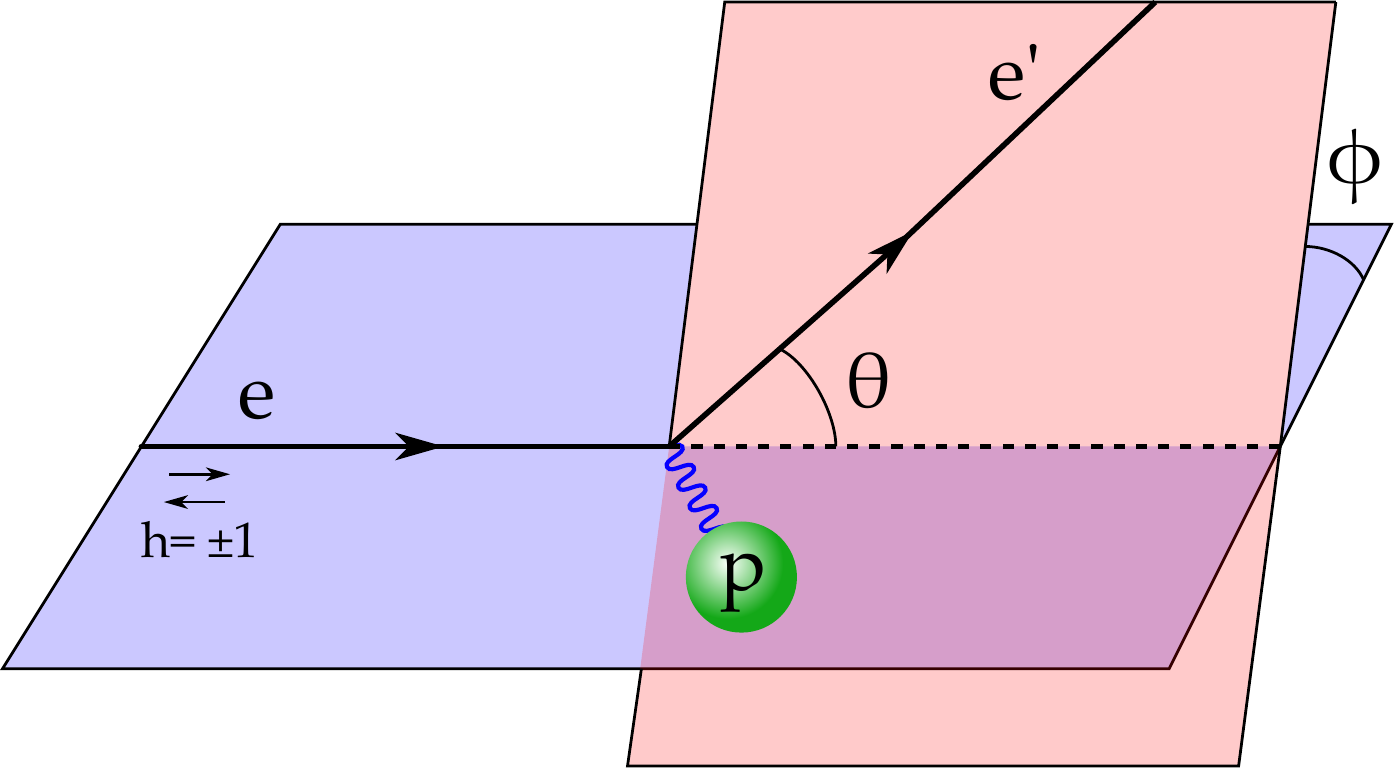}
  \end{center}
  \caption{Electron--proton scattering angle definitions.}
  \label{fig:angle1}
\end{figure}

To save space, we define the difference of cross sections $\Delta\sigma$ and expand it following the steps of section \ref{sec:dis}:
\begin{equation}
\begin{split}
\Delta \sigma &= \sum_{s'}\left[\frac{d^2\sigma}{d\Omega dE'}(k,s,p,-S;k',s')-\frac{d^2\sigma}{d\Omega dE'}(k,s,p,S;k',s')\right]\\
&= \frac{8m\alpha^2E'}{q^4E}\\
&\quad\times\left\{[(q\cdot S)(q\cdot s)+ Q^2(s\cdot S)]MG_1 + Q^2[(s\cdot S)(P\cdot q) - (q\cdot S)(P \cdot s)]\frac{G_2}{M}   \right\}
\end{split}
\end{equation}
where we maintain the notation of previous chapters; namely $k^\mu$ and $k'^\mu$ represent the incoming and outgoing electron momentum, $S^\mu$ represents the target spin vector, and $s^\mu$ and $s'^\mu$ represent the incoming and outgoing electron spin vector.  We define the scattering planes, and angles $\theta$ and $\phi$ as shown in figure \ref{fig:angle1}.

\begin{figure}[htb]
  \begin{center}
   \includegraphics[width=2.5in]{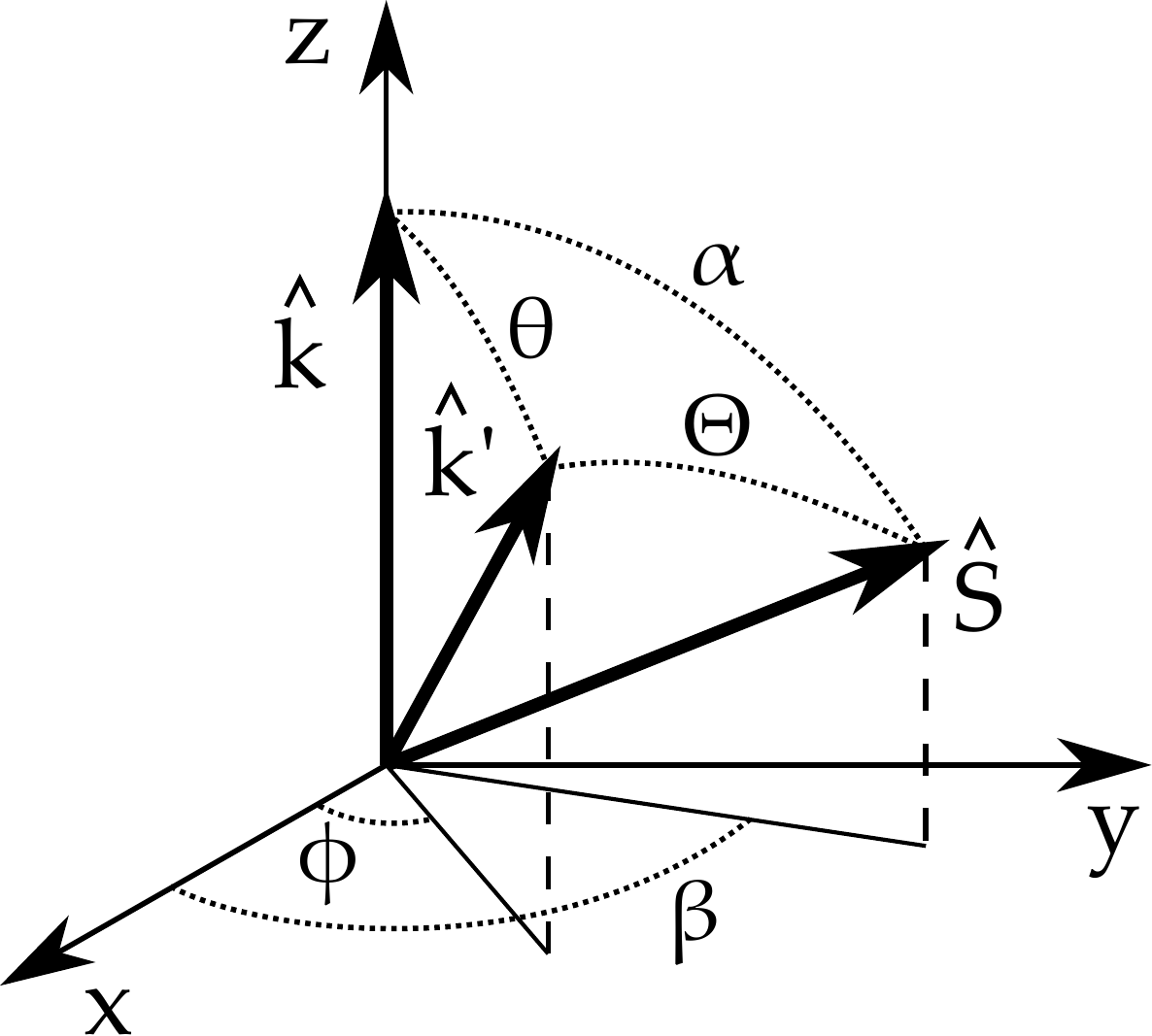}
  \end{center}
  \caption{Scattering coordinate system and angle definitions.}
  \label{fig:angle2}
\end{figure}

The initial electron spin vector is aligned along ($\rightarrow$) or opposite ($\leftarrow$) the momentum $k^\mu$, and for now we let the target spin be polarized along ($\Rightarrow$) or opposite ($\Leftarrow$) an arbitrary direction $\hat{S}$.  If we take the $z$-axis along the incoming electron direction we have
\begin{equation}
\begin{split}
k^{\mu} &= (E,0,0,\vert k \vert) \approx E(1,0,0,1),\\
k'^{\mu} &= (E',k') \approx E'(1,\hat{k'}) = E'(1,\sin\theta\cos\phi,\sin\theta\sin\phi,\cos\theta),\\
S^{\mu} &= (0,\hat{S}) = (0,\sin\alpha\cos\beta,\sin\alpha\sin\beta,\cos\alpha),
\end{split}
\end{equation}
as illustrated in figure \ref{fig:angle2}.
With these definitions we express the difference of cross sections as
\begin{equation}
\frac{d^2\sigma^{\rightarrow\Rightarrow}}{d\Omega dE'}-\frac{d^2\sigma^{\rightarrow\Leftarrow}}{d\Omega dE'} = -\frac{4\alpha^2}{Q^2}\frac{E'}{E}\left([E\cos\alpha + E'\cos\Theta]MG_1 + 2EE'[\cos\Theta - \cos\alpha]G_2\right),
\end{equation}
where we recognize the structure functions $G_1$ and $G_2$\cite{Anselmino19951}.  We will find it useful to have the angle $\Theta$ given in terms of the other angles; after simplification this is
\begin{equation}
\cos\Theta = \sin\theta\cos\phi\sin\alpha + \cos\theta\cos\alpha.
\end{equation}

Looking towards the target polarization orientations used during SANE, 180$\degrees$ and 80$\degrees$ to the incident electron momentum, we can set the angle $\alpha$ accordingly to create differences of cross section for these two cases:
\begin{equation}
\label{eq:diff}
\begin{split}
\Delta\sigma_{180\degrees} &= -\frac{4\alpha^2E'}{Q^2E}[(E+E'\cos\theta)MG_1 - Q^2G_2]\\
\Delta\sigma_{80\degrees} &= -\frac{4\alpha^2E'}{Q^2E}[(E+E'\cos\theta)\cos80\degrees+E'\sin\theta\cos\phi\sin80\degrees)MG_1 \\
&\quad + (2EE'\sin\theta\cos\phi\sin80\degrees - Q^2\cos80\degrees)G_2].
\end{split}
\end{equation}

To create expressions for our measured asymmetries, we'll also need the sum of cross sections, which comes simply from the unpolarized cross section from section \ref{sec:dis}:
\begin{equation}
\label{eq:unpol}
\frac{d^2\sigma^{\rightarrow\Rightarrow}}{d\Omega dE'} + \frac{d^2\sigma^{\rightarrow\Leftarrow}}{d\Omega dE'} = 2\frac{d^2\sigma^{\mathrm{unp}}}{d\Omega dE'} = \frac{8\alpha^2E'^2}{q^4}\left[2W_1\sin^2\frac{\theta}{2}+W_2\cos^2\frac{\theta}{2}\right].
\end{equation}
We'll label $\frac{d^2\sigma^{\mathrm{unp}}}{d\Omega dE'}$ as $\sigma^{\mathrm{unp}}$ for convenience.

Using the expressions for the difference of cross sections and unpolarized cross section, we can now put together a measured spin asymmetry:
\begin{equation}
\label{eq:asym}
A = \frac{\dfrac{d^2\sigma^{\rightarrow\Rightarrow}}{d\Omega dE'} - \dfrac{d^2\sigma^{\rightarrow\Leftarrow}}{d\Omega dE'}}{\dfrac{d^2\sigma^{\rightarrow\Rightarrow}}{d\Omega dE'} + \dfrac{d^2\sigma^{\rightarrow\Leftarrow}}{d\Omega dE'}} = \frac{\Delta\sigma}{2\sigma^{\mathrm{unp}}}.
\end{equation}

Combining equations \ref{eq:diff}, \ref{eq:unpol}, and \ref{eq:asym}, we have our spin structure functions in terms of the measured asymmetries from SANE:
\begin{equation}
\begin{split}
\label{eq:measasyms}
A_{180\degrees} &= -\frac{D'}{W_1}[(E+E'\cos{\theta})MG_1-Q^2G_2],\\
A_{80^\circ} &= \frac{-D'}{W_1}  [(E+E'\cos{\theta})\cos{80^\circ}+
E'\sin{\theta}\cos{\phi}\sin{80^\circ}]MG_1 \\
 &\quad+ (2EE'\sin{\theta}\cos{\phi}\sin{80^\circ} - Q^2\cos{80^\circ})G_2.
\end{split}
\end{equation}
These measured asymmetries can now be used to produce spin structure functions, provided knowledge of the unpolarized structure function $W_1$.  Here we have introduced variable $D'$
\begin{equation}
D' = \frac{1-\epsilon}{1+\epsilon R},
\end{equation}
which contains the virtual photon polarization $\epsilon = 1/(1+2(1+\nu^2/Q^2)\tan^2(\theta/2))$ and $R = \sigma_l/\sigma_T$ the ratio of longitudinal and transverse Compton cross sections\cite{aparaper}.  

We now solve equations \ref{eq:measasyms} for $G_1$ and $G_2$ to get:
\begin{equation}
\label{eq:ga}
\begin{split}
&\quad\quad\frac{MG_1}{W_1} = -\frac{A_{180\degrees}(Q^2\cos{80^\circ}-
2EE'\sin{\theta}\cos{\phi}\sin{80^\circ})+Q^2A_{80^\circ}}{D'E'\sin{\theta}\cos{\phi}
\sin{80^\circ}[2E(E+E'\cos{\theta})+Q^2]},\\
&\frac{G_2}{W_1} = -\frac{[(E+E'\cos{\theta})\cos{80^\circ} + E'\sin{\theta}\cos{\phi}\sin{80^\circ}]A_{180\degrees}
+(E+E'\cos{\theta})A_{80^\circ}} {D'E'\sin{\theta}\cos{\phi}
\sin{80^\circ}[2E(E+E'\cos{\theta})+Q^2] }.
\end{split}
\end{equation}

\subsection{Virtual Photon Absorption Asymmetries} 
\label{sec:compasym}
In section \ref{sec:compton} we gave the spin structure functions in terms of the virtual photon absorption asymmetries, from here on called the spin asymmetries.  We solve equations \ref{eq:virtcomp} for $A_1$ and $A_2$ to get
\begin{equation}
\begin{split}
\label{eq:g1g2}
A_1 &= \nu \frac{MG_1}{W_1} - Q^2\frac{G_2}{W_1}\\
A_2 &= \sqrt{Q^2}\left(\frac{MG_1}{W_1} + \nu\frac{G_2}{W_1}\right).
\end{split}
\end{equation}
From here it is simple to plug in the result of the previous section, equations \ref{eq:ga}, and simplify:
\begin{equation}
\label{eq:a1a2}
\begin{split}
A_1 &=\frac{1}{D'}\left[ A_{180\degrees}\frac{E - E'\cos\theta}{E+E'} + \left( A_{80\degrees} + A_{180\degrees}\cos 80\degrees\right)\frac{E'\sin\theta}{(E+E')\cos\phi \sin 80\degrees}  \right]\\
A_2 &=\frac{1}{D'}\frac{\sqrt{Q^2}}{2E}\left[ A_{180\degrees} + \left( A_{80\degrees} + A_{180\degrees}\cos 80\degrees\right)\frac{E-E'\cos\theta}{E'\sin\theta\cos\phi\sin 80\degrees}  \right]
\end{split}
\end{equation}	

\section{Existing $g_2$ Data}
\label{sec:exist}
We have established $g_2$ as a sort of ugly duckling among the structure functions.  With no simple interpretation in the naive parton model and containing nasty higher twist terms, $g_2$ has the added caveat that it is dominated by the contribution of the transverse target polarization cross sections.  As the experimental complications of a transverse target polarization measurement are myriad, $g_2$ remains scantly measured and poorly understood.

From 1993 to 2003, three experiments at the Stanford Linear Accelerator (SLAC) in Menlo Park, California extracted $g_2$ for the proton and the deuteron using transversely polarized solid targets.  The three experiments, known as E143\cite{e143-1,e143-2,e143-3}, E155\cite{e155} and E155x\cite{e155x}, used the UVa polarized ammonia target and the SLAC polarized electron beam.  SLAC offers a high electron beam energy, but it is in the form of a pulsed beam---instantaneous luminosity is great, but these bursts of high current are intermittent.  E143, E155 and E155x used beam energies of 29 GeV, 38.8 GeV, and 29.1 and 32.3 GeV respectively, achieving $Q^2$ from 0.7 to 20 GeV$^2$.  

\begin{figure}[htb]
  \begin{center}
   \includegraphics[width=4.5in]{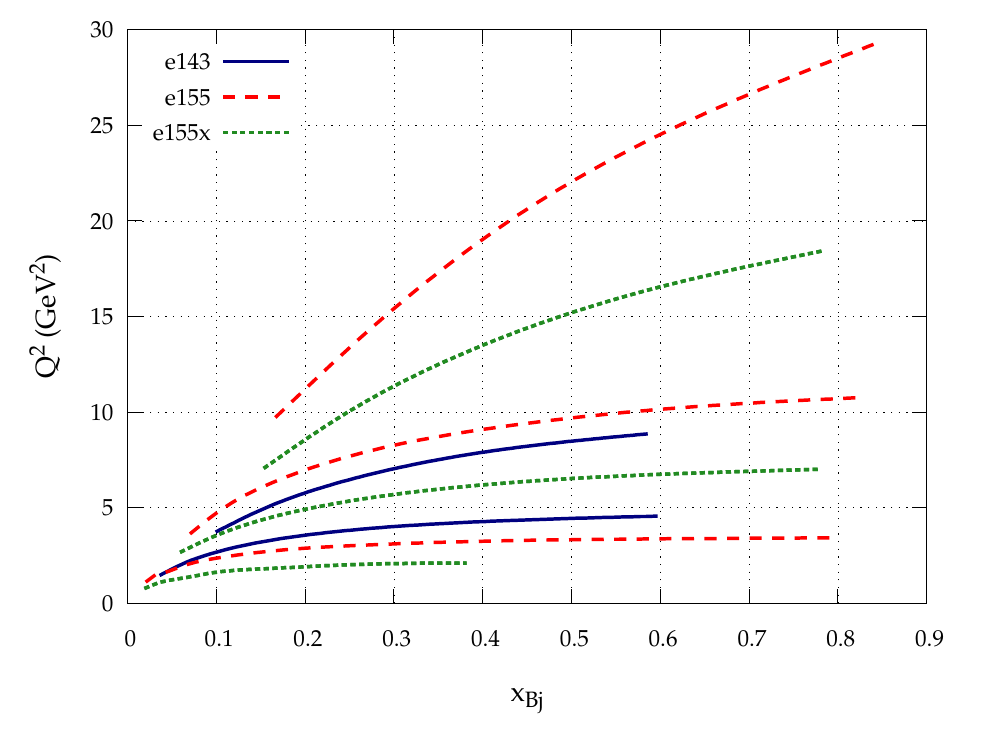}
  \end{center}
  \caption{Kinematics of existing measurements of $g_2$ from SLAC.}
  \label{fig:existkin}
\end{figure}

The kinematics of these three experiments are shown explicitly in figure \ref{fig:existkin}.  Each line represents an angle setting of the spectrometer, as well as beam energy setting in the case of E155x.  The spectrometer takes a small slice around a given $\theta$, which results in swaths of data taken in a line of kinematics as electrons of different final energies are collected.  

\begin{figure}[htb]
  \begin{center}
   \includegraphics[width=4.5in]{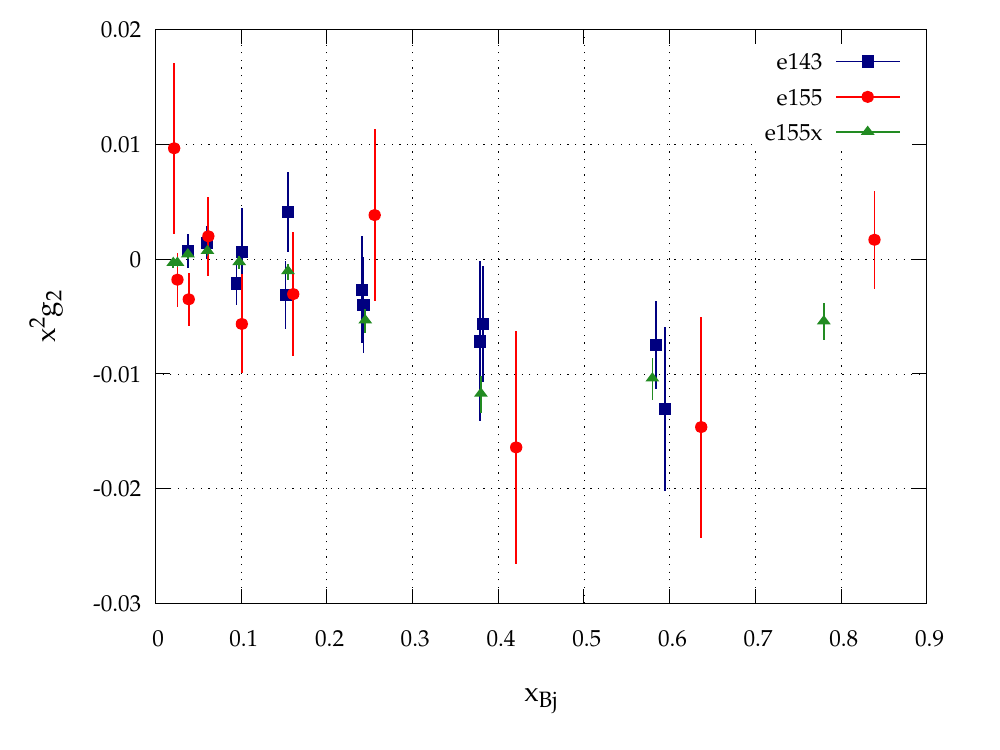}
  \end{center}
  \caption{Existing measurements of $g_2$ from SLAC.}
  \label{fig:existg2}
\end{figure}

Figure \ref{fig:existg2} shows the extracted values of $x^2g_2$, where the kinematic ranges from figure \ref{fig:existkin} have been binned into kinematics points with uncertainty.  We have scaled $g_2$ by $x^2$ to reduce the large variation in values at low $x$. The low $x$ region offers most of the data, but even there any structure away from zero is not convincing.  The paucity of accurate points above $x$ of 0.3 points to the need for more, higher-statistics measurements.

To give some context for these measurements, we turn to the work of the Asymmetry Analysis Collaboration (AAC)\cite{aac}, which publishes parameterizations of the polarized parton distribution functions (\textit{PDFs}) $\Delta q(x)$ discussed in section \ref{sec:partonmodel}.  These PDFs are produced using world data on the spin asymmetry $A_1$, including data from the E143 and E155 experiments.  The lack of transverse data in these computations is notable; the AAC PDFs will not be sensitive to higher twist contributions.

\begin{figure}[htb]
  \begin{center}
   \includegraphics[width=4in]{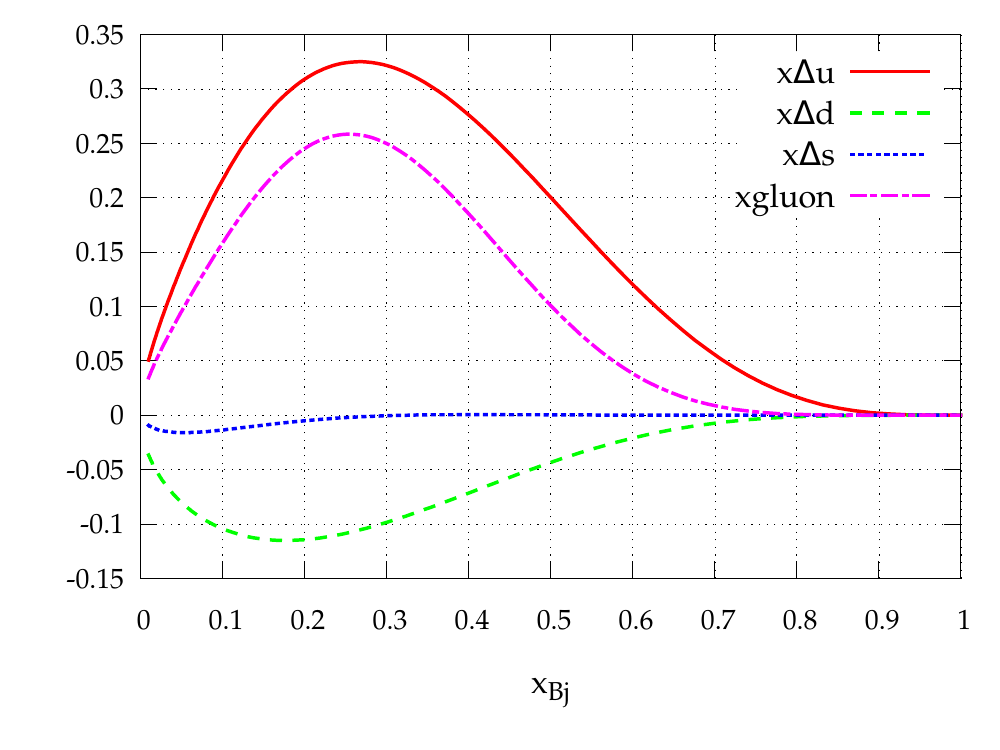}
  \end{center}
  \caption{AAC polarized parton distributions $\Delta q$, scaled by $x$.}
  \label{fig:xppdfs}
\end{figure}

We see the polarized parton distributions, scaled by $x$, as calculated by the AAC in figure \ref{fig:xppdfs}.  Each quark flavor has its own distribution; the anti-quark distributions from the AAC follow that of the strange quark exactly and are not shown.

\begin{figure}[htb]
  \begin{center}
   \includegraphics[width=4in]{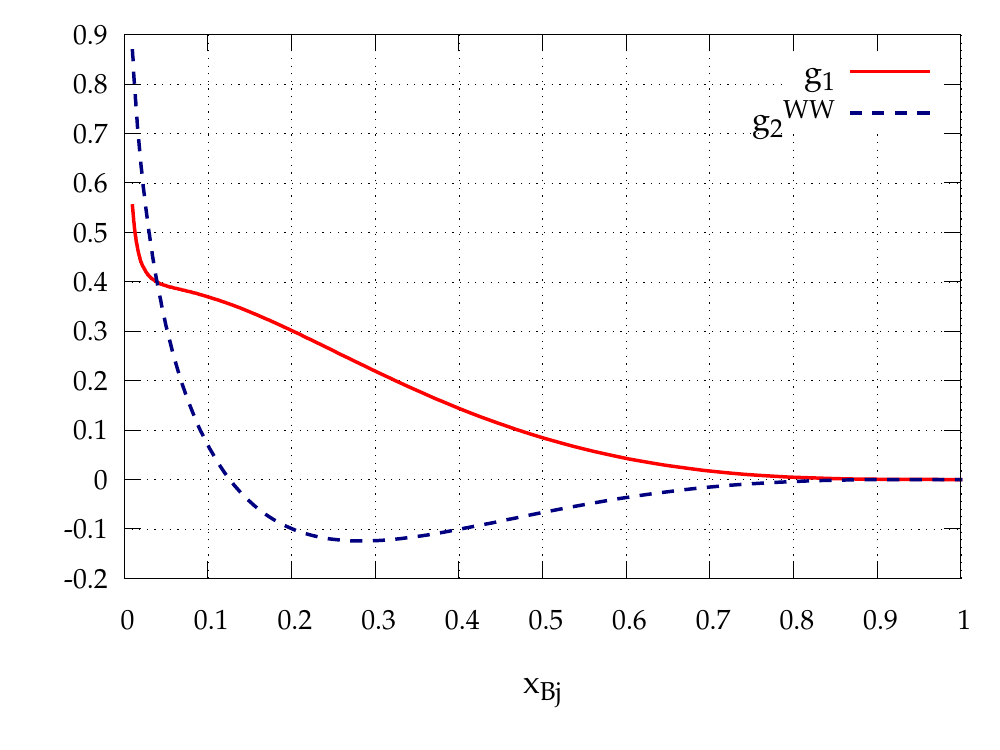}
  \end{center}
  \caption{Spin structure functions $g_1$ and $g_2^{WW}$ computed from AAC polarized parton distributions.}
  \label{fig:g1g2ww}
\end{figure}

Recalling equation \ref{eq:gparton}, we can calculate $g_1$ directly using the polarized parton distribution functions $\Delta q(x)$:
\begin{equation}
g_1(x,Q^2) = \frac{1}{2} \sum_q e_q^2\Delta q(x).
\end{equation}
To relate these ppdfs to $g_2$, we generate $g_2^{WW}$ by integrating over this $g_1$, as shown in equation \ref{eq:ww}
\begin{equation}
g_2^{WW}(x,Q^2)  = \int_0^1 \frac{dy}{y} g_1(y,Q^2) - g_1(x,Q^2).
\end{equation}
The result of these two computations using the AAC PDFs is shown in figure \ref{fig:g1g2ww}.

\begin{figure}[htb]
  \begin{center}
   \includegraphics[width=4.5in]{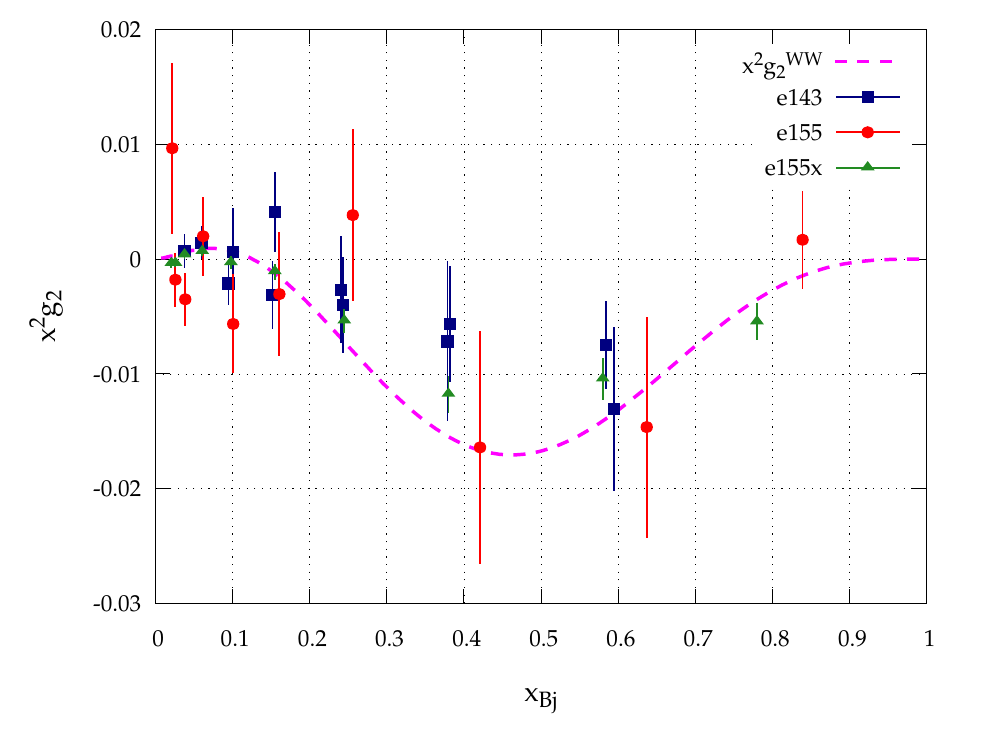}
  \end{center}
  \caption{SLAC data of $g_2$ as a function of $x$, with AAC $g_2^{WW}$.}
  \label{fig:slacg2ww}
\end{figure}

Now we plot this $g_2^{WW}$ with the SLAC data in figure \ref{fig:slacg2ww}, scaling again by $x^2$.  Any statistically significant deviation of the SLAC data points from the $g_2^{WW}$ would indicate higher twist behavior.  Unfortunately, the sparsity and uncertainty in the data currently do not allow for any such conclusions.

\chapter{Description of the Experiment}

\begin{figure}[bh]
  \begin{center}
    \includegraphics[width=4.5in]{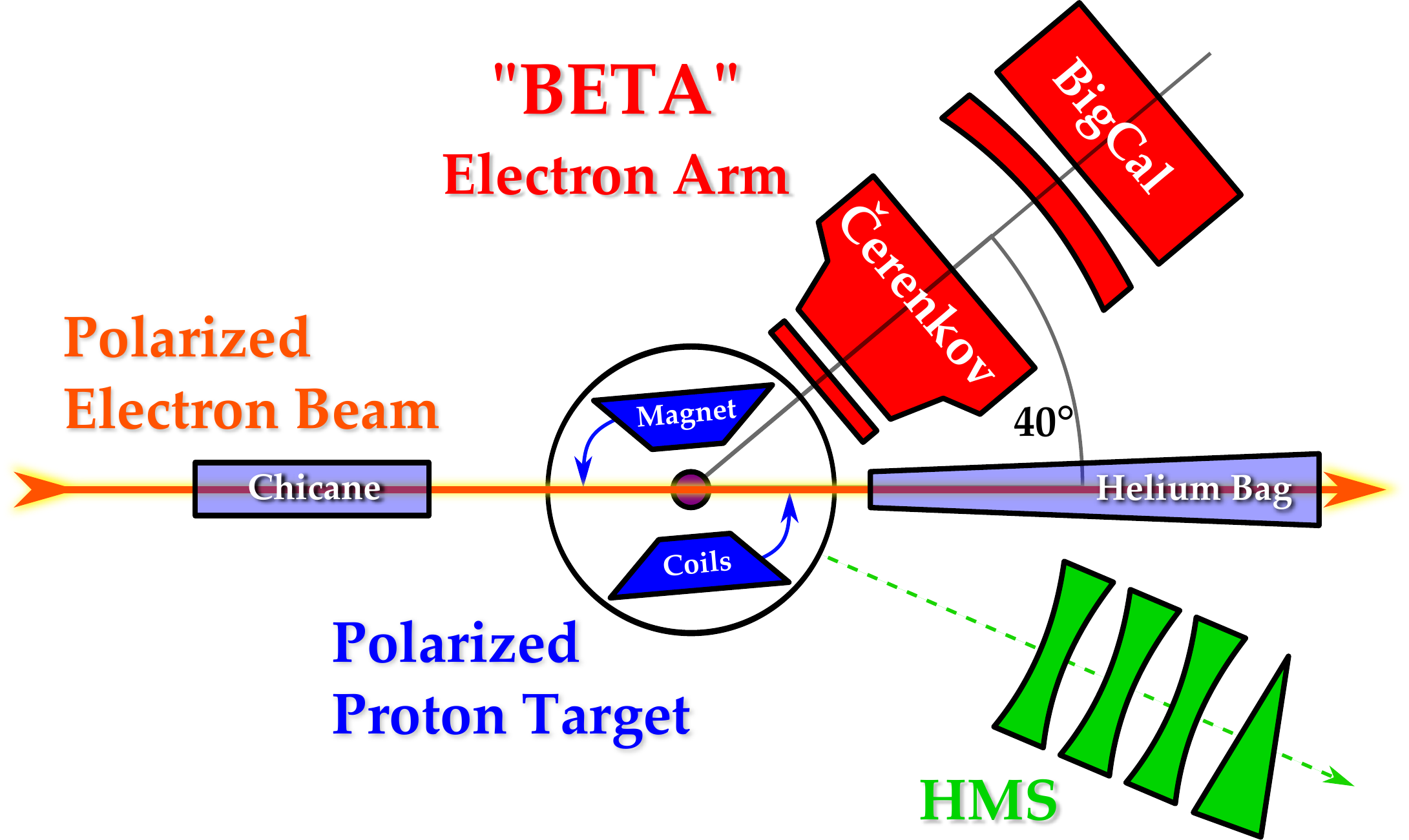}
  \end{center}
  \caption[Schematic overview of SANE.]{Schematic overview of SANE's experimental layout, with a novel electron arm at 40$\degrees$ viewing double polarized electron--proton scattering with the target alignment at 180\degrees\ and 80\degrees\ to the beam.}
  \label{fig:sane-over}
\end{figure}

\begin{figure}[htb]
  \begin{center}
   \includegraphics[width=3.5in]{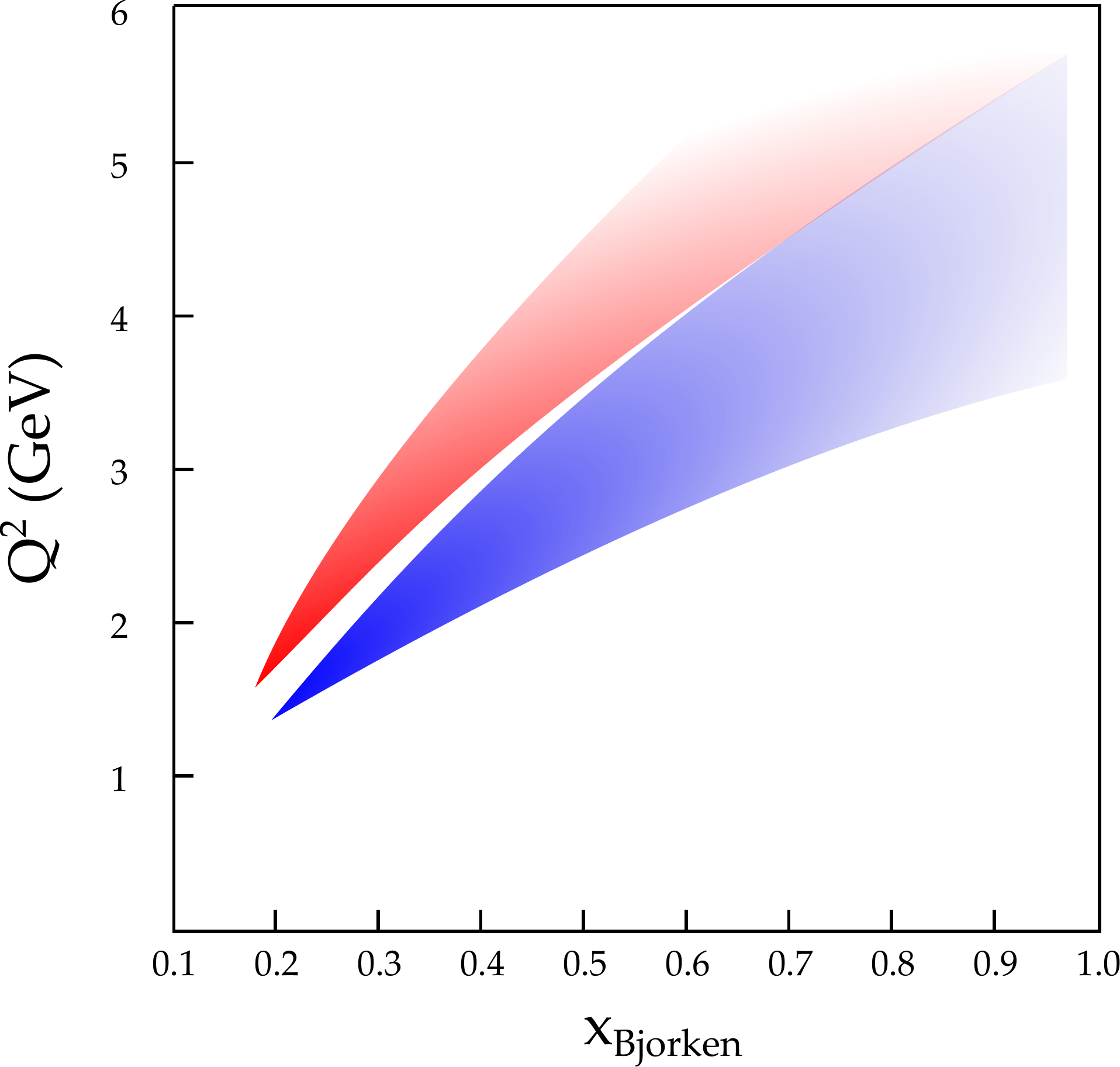}
  \end{center}
  \caption[SANE's experimental kinematics.]{SANE's experimental kinematics.  In red are kinematics achieved at 5.9 GeV beam energy, and in blue are those from 4.7 GeV.  
 }
  \label{fig:kine}
\end{figure}

Experiment E03-007, known as the \textbf{S}pin \textbf{A}symmetries of the \textbf{N}ucleon \textbf{E}xperiment (\textbf{SANE}), took data in Hall C of Jefferson Lab from January to March of 2009.  A telescope array of detectors was used to view the CEBAF polarized electron beam incident on a polarized ammonia ($^{14}$NH$_3$) target, to make an inclusive measurement of spin asymmetry A$_1$ and spin structure function g$_2$ via deep inelastic scattering.  The electron arm sat at 40$\degrees$ to the beam with a solid angle of approximately 0.2 sr to detect scattered electrons at kinematics of $2.5<Q^2<6.5$ GeV$^2$ and $0.3<x_{Bj}<0.8$ using incident electron beam energies of approximately 4.7 and 5.9 GeV.  SANE's kinematics can be seen in figure \ref{fig:kine}.
As shown in section \ref{sec:measure}, to produce effective measurements of both A$_1$ and A$_2$ and the spin structure functions, it was necessary to measure DIS asymmetries in which the target polarization included orthogonal components; for SANE this meant polarization of the target nearly transverse to the incident beam, as well as longitudinal.

This chapter outlines the experimental design of SANE, discussing each subsystem in turn, with the exception of the target, which is described in chapter \ref{sec:target}.  A brief introduction to the CEBAF accelerator begins in section \ref{sec:beam}.  A description of the electron detector package is given in section \ref{sec:beta}, followed by a discussion of the triggers and data acquisition used during the experiment in section \ref{sec:trigger}.  Although the standard Hall C high momentum spectrometer was used during SANE in an auxiliary role to determine effective target thickness, this analysis doesn't include HMS asymmetry data.

	\section{Polarized Electron Beam}
	\label{sec:beam}
	
		\subsection{Accelerator}
\begin{figure}[htb]
  \begin{center}
    \includegraphics[width=4.5in]{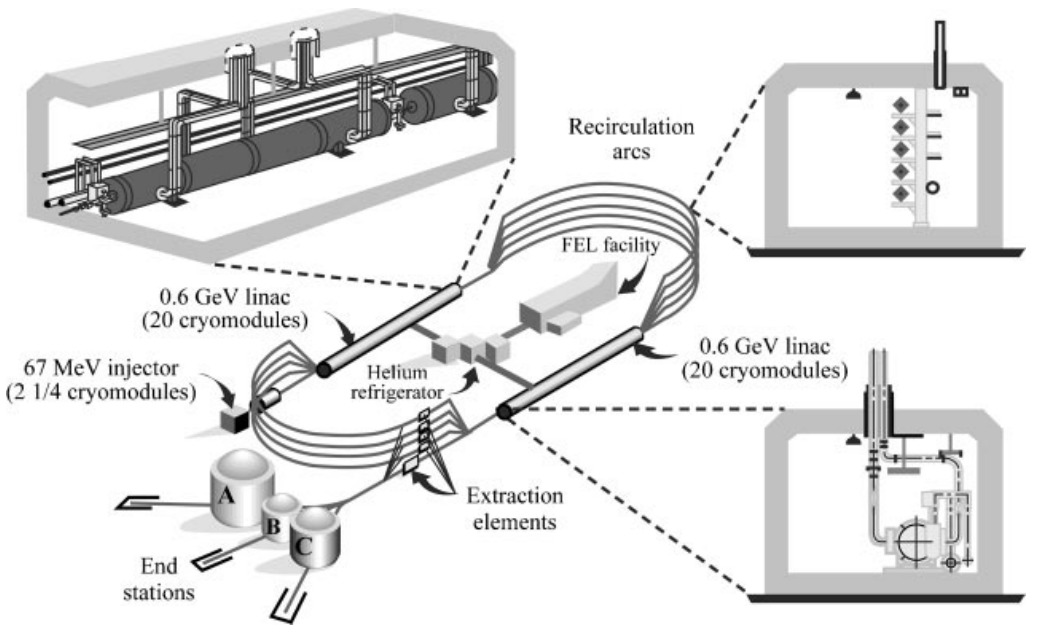}
  \end{center}
  \caption[Schematic overview of JLab's CEBAF accelerator.]{Schematic overview of JLab's CEBAF accelerator. Reproduced from \cite{doi:10.1146/annurev.nucl.51.101701.132327}.}
  \label{fig:cebaf}
\end{figure}

Jefferson Lab's Continuous Electron Beam Accelerator Facility (CEBAF) consists of two, anti-parallel linear accelerators, each capable of approximately 600 MeV of acceleration.  These accelerators are connected in series via 9 recirculating arcs, 5 at the north end and 4 at the south, to form a ``race-track'' allowing up to 5 passes through the linacs and providing a maximum beam energy of around 6 GeV. After extraction, the accelerator can deliver polarized, continuous wave beam at currents up to 200 $\mu$A to be divided among the three experimental halls.  Figure \ref{fig:cebaf}  shows a schematic overview of the accelerator.

			\subsubsection{Polarized Electron Source}
\begin{figure}[hbt]
  \begin{center}
    \includegraphics[width=6in]{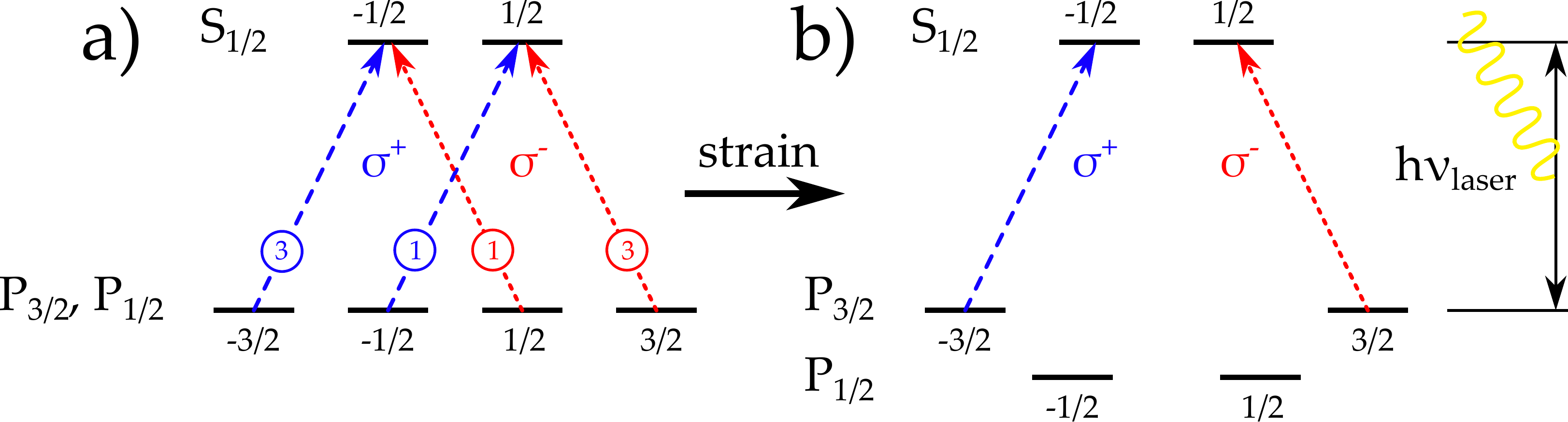}
  \end{center}
  \caption[Energy levels and laser induced transitions for GaAs]{Energy levels and laser induced transitions for unstrained (a) and strained, doped (b) GaAs.  Straining the GaAs breaks the degeneracy of the P state, allowing a theoretical maximum polarization of 100\%.}
  \label{fig:source}
\end{figure}

			CEBAF's polarized electron beam begins with a polarized electron source---electrons excited from a photocathode using circularly polarized light.  The gallium arsenide (GaAs) cathode emits polarized electrons when illuminated by circularly polarized laser light with a frequency that matches the bandgap energy of the material.  Right handed polarized light excites electrons from P$_{-3/2}$ and P$_{-1/2}$ valence band states into S$_{1/2}$ ($-$) and ($+$) conduction band states respectively, and left-handed light takes P$_{3/2}$ and P$_{1/2}$ to S$_{1/2}$ ($+$) and ($-$).  These transitions can be seen in a) of figure \ref{fig:source}, with right-handed circularly polarized light inducing the blue transitions, and left-handed the red.  
			
			In GaAs, the P$_{1/2}$ and P$_{3/2}$ level states are degenerate, so light of the band-gap energy will induce transitions of both P$_{1/2}$ and P$_{3/2}$.  The Clebsch-Gordan coefficients for these processes mean the transition rate is three times higher for the P$_{3/2}$ states, creating a theoretical polarization of 50\%\cite{hernandez-garcia:44}. 
						
			To access higher polarizations, the degeneracy of the P states can be broken by mechanically straining the GaAs.  Jefferson Lab's GaAs cathodes are strained via a phosphorus doping in every other layer of the so-called ``superlattice.''  This strain changes the bandgaps of the P states such that one can be pumped at a time to produce a theoretical maximum of 100\% electron polarization, as seen in b) of figure \ref{fig:source}.

			Three diode lasers provide the circularly polarized light used to illuminate the cathode, one for each experimental hall.  Three bunches at 499 MHz pulses make a train of 1497 MHz, which is equal to the resonant frequency of the RF accelerating cavities in the accelerators.  The circular polarization of the light is controlled by Pockels cells, which use electric field dependent birefringence to shift the phase of the light.  This allows rapid reversal of the polarization of the light and thus the helicity of the electrons, and is used in practice to create pseudo-random 30Hz helicity batches.  A half-wave plate can also be inserted to reverse the helicity to observe any time-dependent systematic effects.
			An excellent overview of polarized particles beams is given in reference \cite{mane}.

			\subsubsection{Acceleration and Delivery}
			
	Electrons from the polarized source are accelerated into the injector by a 100kV electron gun, and the injector provides as much as 67 MeV of additional acceleration as it sends the electrons into the north linear accelerator.  The injector and each linear accelerator consist of 2 1/4 and 20 cryomodules respectively; these cryomodules themselves contain 8 superconducting RF cavities as well as supporting cryogenics and power.  Each cavity provides a nominal acceleration of roughly 28 MeV, giving each linac a nominal acceleration of 570 MeV.  At 5 passes through the race-track, this provides 5.7 GeV beam energy, although accelerator improvements have pushed this to above 5.9 GeV.
			
	The accelerating cavities are superconducting niobium cooled to 2 K, and each is powered by an RF klystron at 1497 KHz.  Electrons ride the crest of the RF waves in the superconducting cavities, building energy while their speed remains very near to the speed of light.  Since the electrons are already relativistic after leaving the injector, they can stay in phase with the RF field in the cavities, and they will remain so even after several linac passes.  In this way the cavities carry as many as five sets of electron beams from each successive pass simultaneously.
	
	Once the beam reaches the end of a linac, a series of dipole magnets sorts the beams according to their energy, routing each to a recirculating arc.  These arcs steer the beam back around to the other linac, with each successive arc using a larger field integral to carry beam of higher momentum around the turn in the race-track.
	
	The beam can be extracted from the racetrack at the beam switching yard, which uses RF separator magnets at 499 MHz to separately extract the three beams after any number of passes to send to each of the three experiments halls \cite{doi:10.1146/annurev.nucl.51.101701.132327}.
		
		\subsection{Standard Hall C Beamline}
	SANE took advantage of the standard beamline equipment installed in Hall C to provide precise data on the energy, position, current and polarization of the beam as it passes through the arc.  The beamline leading from the switching yard into Hall C consists of 8 dipoles, 12 quadrupoles, 8 sextupoles which steer and focus the beam.  In addition to this steering, the beam is rastered to increase its spot size to spread the heat load over a wider area of the target\cite{yan:571}.
			
			\subsubsection{Beam Position}
	The position of the beam within the beam line is unsurprisingly a crucial piece of data to track during experimental running.  In addition to ensuring that the beam's trajectory follows directly to center of the 2.5 cm diameter target cup, the beam position also provides information on the beam energy, as described in the next subsection.
	
	The Beam Position Monitors (BPMs) each consist of a resonant cavity with a resonant frequency equal to that of the accelerator.  Inside the cavity are four antennae: a pair for x and a pair for y position, but rotated 45 degrees from the vertical to avoid synchrotron radiation damage.  An asymmetry of the amplitudes of the signals on opposite antennae is proportional to the distance between the beam and the midpoint of the antennae \cite{gueye}.  The BPMs used for SANE were hand-picked for low current operation, as usual beam current in Hall C is on the order of 100 $\mu$A, not 100 nA.

			\subsubsection{Beam Energy}
		The arc magnets leading the beam into Hall C are used as a spectrometer to allow the measurement of the beam energy as it enters the hall.  Under normal operation, three pairs of high resolution superharps\cite{Yan1995261}, or wire scanners, determine the position and direction of the beam at the entrance, exit and middle of the arc.  Using these measurements of the curvature of the beam over its 34.3$\degrees$ deflection by the dipoles, we can determine the energy of the beam with precise knowledge of the dipole field:
\begin{equation}
E \simeq p = \frac{e}{\theta} \int \vec{B} \cdot \vec{dl}
\end{equation}
with electric charge $e$, arc bend angle $\theta$, and the magnetic field integral over the path of the beam \cite{yan-carlini}. 

	However, beamline infrastructure needed for the polarized target necessitated the removal of a superharp.  Instead, less accurate position data from the beam position monitors, available throughout the experiment, was used.  The average readings of the beam energy measurements, averaged per run for each beam energy and target field configuration are shown in table \ref{tab:beame}.

\begin{table}[htb]
  \begin{center}
\begin{tabular}{llll}
\toprule
Nominal $E$ & Target Field Angle & Average $E$ (MeV) & Standard Deviation\\
\midrule
4.7 GeV & 180$\degrees$    &    4736.7    &    0.9 \\
4.7 GeV & 80$\degrees$   &    4728.5    &    0.8\\
4.7 GeV & 80$\degrees$   &    4729.1   &    0.5\\
5.9 GeV & 180$\degrees$    &    5895.0   &    1.9\\
5.9 GeV & 80$\degrees$    &    5892.1    &    4.9\\
\bottomrule
\end{tabular}
\caption{Table of beam energies averaged per run for each SANE run period.}
  \label{tab:beame}
\end{center}
\end{table}

			\subsubsection{Beam Current}
	Measurement of the beam current entering Hall C is provided by three devices---two resonant cavity Beam Current Monitors (BCMs 1 and 2) and one so-called Unser monitor.  The beam current can be measured by measuring the RF power coupled out of the resonant cavities of the BCMs.  The BCMs are designed to resonate in the transverse magnetic mode (TM$_{010}$) at the same frequency of the accelerator's RF.  Antennae inside the cavities give a voltage signal proportional to the square of the beam current. 
	
	The Unser monitor is a parametric current transformer\cite{unser:266}, which consists of toroidal transformers through which the beam passes, giving an inductive measure of the current.  The stable gain of the Unser makes it the standard against which the BCMs are calibrated.  More information on beam current measurement is available in appendix A of reference \cite{armstrong}.
	

			\subsubsection{Beam Polarization}
			\label{sec:beampol}
	A M{\o}ller polarimeter was used to measure the polarization of the beam at nine points throughout the experiment.  These polarimeters leverage our precise understanding of $\vec{e}+\vec{e} \to e+e$ scattering, whose cross section is well known from QED.  By polarizing an electron target parallel to the beam axis $P^\parallel_t$, we can relate the beam polarization $P^\parallel_b$ to the measured polarized cross section by way of the unpolarized cross section $d\sigma_0/d\Omega = [\alpha(4-\sin^2\theta)/(2m_e\gamma\sin^2\theta)]^2$ for scattering angle $\theta$:	
	\begin{equation}
	\begin{split}
	\frac{d\sigma}{d\Omega} = \frac{d\sigma_0}{d\Omega} \left[ 1+ P^\parallel_t P^\parallel_b A_{zz}(\theta) \right], \\
	\mathrm{for} \,\,\,\, A_{zz}(\theta) = -\sin^2\theta\frac{8-\sin^2\theta}{(4-\sin^2\theta)^2},
	\end{split}
	\end{equation}	
the analyzing power $A_{zz}$.  Forming an asymmetry of the cross sections for beam and target spins parallel and anti-parallel, we have:
	\begin{equation}
	\label{eq:moller}
	\epsilon = \frac{d\sigma^{\uparrow\uparrow} -d\sigma^{\uparrow\downarrow}}{d\sigma^{\uparrow\uparrow} +d\sigma^{\uparrow\downarrow}} = A_{zz}(\theta)P^\parallel_tP^\parallel_b.
	\end{equation}
	
\begin{figure}[htb]		
\begin{center}
    \includegraphics[width=6in]{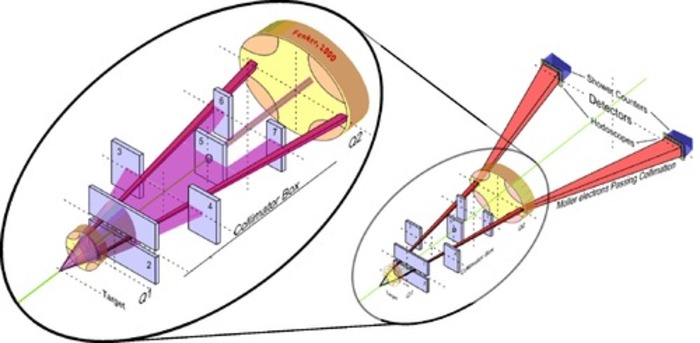}
  \end{center}
  \caption[Diagram of the Hall C M{\o}ller Polarimeter.]{Diagram of the Hall C M{\o}ller Polarimeter by H. Fenker, reproduced with permission from \cite{mollerdiag}.}
  \label{fig:moller}
\end{figure}

To make this asymmetry measurement, an iron film target is polarized by a 4 T superconducting split-coil solenoid.  As the analyzing power is maximized for electrons scattered at 90$\degrees$ in the center of mass frame, pairs of electrons around this angle are detected in coincidence.  This coincidence removes the background from other scattering processes, and a series of movable collimators allows selection of a tight range about 90$\degrees$ in the center of mass frame.  A diagram of the polarimeter is seen in figure \ref{fig:moller}.

After passing through quadrupole magnets and collimators, the electrons are detected by one of two lead-glass shower counters equipped with photomultiplier tubes to create a signal from the \v{C}erenkov shower.  The coincidence counting rate between these two shower counters at different beam helicities is used to produce the asymmetry in equation \ref{eq:moller}.  The large acceptance of these detectors reduces sensitivity to the Levchuk effect due to the orbital motion of electrons in the iron atom\cite{gaskell-moller}. Since the iron film target degrades the beam, polarization measurements cannot occur during data taking, but are performed routinely to monitor the beam polarization.  More information on the Hall C M{\o}ller Polarimeter can be found in references  \cite{Hauger2001382,gaskell-moller}.

\begin{table}[p]
  \begin{center}
\begin{tabular}{lllllll}
\toprule
Date	&Run&HWP	&Wien Angle	&Beam $E$ (MeV)	&QE (\%)	&Polarization (\%)	\\ \midrule
1/25	&71942	&IN	&29.99$\degrees$	&4730.46	&0.1844	&87.79 $\pm$ 1.54	\\
	&71943	&IN	&29.99$\degrees$	&4730.48	&0.1844	&88.21 $\pm$ 0.98	\\
	&71944	&IN	&29.99$\degrees$	&4730.51	&0.1844	&85.13 $\pm$ 0.93	\\
	&71945	&IN	&29.99$\degrees$	&4730.53	&0.1844	&87.71 $\pm$ 0.99	\\
	&71946	&IN	&29.99$\degrees$	&4730.53	&0.1844	&88.24 $\pm$ 1.01	\\
	&71947	&IN	&29.99$\degrees$	&4730.53	&0.1844	&86.76 $\pm$ 0.95	\\
	&71948	&IN	&29.99$\degrees$	&4730.53	&0.1844	&87.33 $\pm$ 1.55	\\
	&71949	&IN	&29.99$\degrees$	&4730.52	&0.1844	&86.58 $\pm$ 0.99	\\
	&71950	&IN	&29.99$\degrees$	&4730.52	&0.1844	&85.38 $\pm$ 0.97	\\
	&71951	&IN	&29.99$\degrees$	&4730.53	&0.1844	&86.71 $\pm$ 0.97	\\\midrule
2/1	&72209	&IN	&29.99$\degrees$	&4729.25	&0.0888	&89.00 $\pm$ 1.02	\\
	&72210	&IN	&29.99$\degrees$	&4729.29	&0.0888	&87.32 $\pm$ 1.10	\\
	&72211	&IN	&29.99$\degrees$	&4729.28	&0.0888	&83.45 $\pm$ 1.04	\\\midrule
2/5	&72300	&IN	&29.99$\degrees$	&4728.23	&0.0708	&87.26 $\pm$ 0.68	\\
	&72301	&IN	&29.99$\degrees$	&4728.27	&0.0708	&85.64 $\pm$ 0.93	\\\midrule
2/11	&72465	&OUT	&29.99$\degrees$	&5892.84	&0.3124	&-61.16 $\pm$ 1.10	\\
	&72466	&OUT	&29.99$\degrees$	&5892.70	&0.3124	&-60.56 $\pm$ 1.11	\\
	&72467	&OUT	&19.99$\degrees$	&5892.81	&0.3124	&-72.83 $\pm$ 1.02	\\
	&72468	&OUT	&19.99$\degrees$	&5892.43	&0.3124	&-72.04 $\pm$ 0.98\\
	&72469	&OUT	&19.99$\degrees$	&5891.65	&0.3124	&-75.35 $\pm$ 0.97	\\
	&72470	&OUT	&22.99$\degrees$	&5891.75	&0.3124	&-71.88 $\pm$ 1.06	\\
	&72471	&OUT	&22.99$\degrees$	&5891.46	&0.3124	&-70.82 $\pm$ 1.06	\\
	&72472	&OUT	&22.99$\degrees$	&5891.08	&0.3124	&-70.64 $\pm$ 2.17	\\\midrule
2/14	&72537	&OUT	&22.99$\degrees$	&5891.24	&0.2790	&-73.36 $\pm$ 1.08	\\
	&72538	&OUT	&22.99$\degrees$	&5891.11	&0.2790	&-73.70 $\pm$ 1.05	\\
	&72539	&OUT	&22.99$\degrees$	&5891.03	&0.2790	&-72.19 $\pm$ 1.83	\\\midrule
2/24	&72767	&OUT	&13.00$\degrees$	&5892.92	&0.0830	&-75.51 $\pm$ 1.08	\\
	&72768	&OUT	&13.00$\degrees$	&5892.85	&0.0830	&-76.90 $\pm$ 1.00	\\\midrule
2/28	&72839	&IN	&29.99$\degrees$	&4728.95	&0.2516	&87.63 $\pm$ 0.96	\\
	&72840	&IN	&29.99$\degrees$	&4728.88	&0.2516	&86.28 $\pm$ 1.08	\\\midrule
3/9	&72965	&OUT 	&-18.00$\degrees$	&5895.58	&0.1635	&-90.22 $\pm$ 1.29 	\\
	&72966	&OUT 	&-18.00$\degrees$	&5894.22	&0.1635	&-86.81 $\pm$ 1.27 	\\\midrule
3/12	&72977	&OUT	&21.19$\degrees$	&4736.33	&0.1789	&65.83 $\pm$ 0.97 	\\
	&72978	&OUT	&21.19$\degrees$	&4736.34	&0.1789	&66.36 $\pm$ 0.99 	\\

\bottomrule
\end{tabular}
\caption{Table of SANE M{\o}ller Runs.}
  \label{tab:moller}
\end{center}
\end{table}

Nine M{\o}ller measurements, shown in table \ref{tab:moller}, were taken during SANE, and were used by SANE collaborator D. Gaskell to create a fit to the salient accelerator data to extrapolate beam polarizations throughout the experiment.  The fit included three degrees of freedom: the magnitude of the polarization at the source $P_{source}$, the degree of imbalance between the north and south linear accelerators, and a global correction from the beam energy $F_{corr}$\cite{gask}.   For Wien angle $\theta_w$, correction for the quantum efficiency of the cathode $F(\epsilon_q)$, and half wave plate status $n_{hwp}$, the expression for beam polarization $P_B$ is
\begin{equation}
\label{eq:pb}
P_B = (-1)^{n_{hwp}} P_{source} F_{corr} F(\epsilon_q) \cos (\theta_w + \varphi_{precession}),
\end{equation}
where $\varphi_{precession}$ is determined by following the spin precession through each bend in the accelerator.  The correction due to the quantum efficiency was based on a fit to GEp-III data.  The spin precession of an electron of mass $m_e$ bent in an angle $\theta$ in a magnetic field while traveling with energy $E$ is
\begin{equation}
\varphi = \frac{(g-2)}{2m_e} E \times \theta
\end{equation}
where $g$ is the electron's gyromagnetic ratio \cite{PhysRevSTAB.7.042802,Montague:1983yi}.

The east and west recirculating arcs are 180$\degrees$ bends, $\theta_{arc}$, and the Hall C arc is a 37.52$\degrees$ bend in the opposite direction, $\theta_{bend}$.  This means as an electron travels from the source to the target, the total spin precession is
\begin{equation}
\varphi_{precession} = \left(\frac{g-2}{2m_e}\right)  \left\{ \sum_{n_{arc}=1}^{2N_p-1}\left[E(n_{arc})  \theta_{arc}\right] -  E_b \theta_{bend}  \right\}
\end{equation}
for $E(n_{arc})$ the energy of the beam upon reaching that arc for that pass (the energy accumulated through each previous linac pass plus the injector energy) and $E_b$ the final beam energy.

The Wien angle is the initial spin angle as determined by a Wien filter at the accelerator's electron source.  This filter rotates the spin relative to the particle's momentum using uniform and orthogonal magnetic and electric fields.  As can be seen in equation \ref{eq:pb}, the Wien angle directly affects the final polarization, but as the bend angles and thus precession into the three experimental halls are different, it's not possible to give all halls maximum polarization for most beam energy settings.  Thus a compromise between halls is made to choose a Wien angle that provides the best polarization possible in the circumstances \cite{Higinbotham:2009ze}.	

\begin{figure}[htb]		
\begin{center}
    \includegraphics[width=5in]{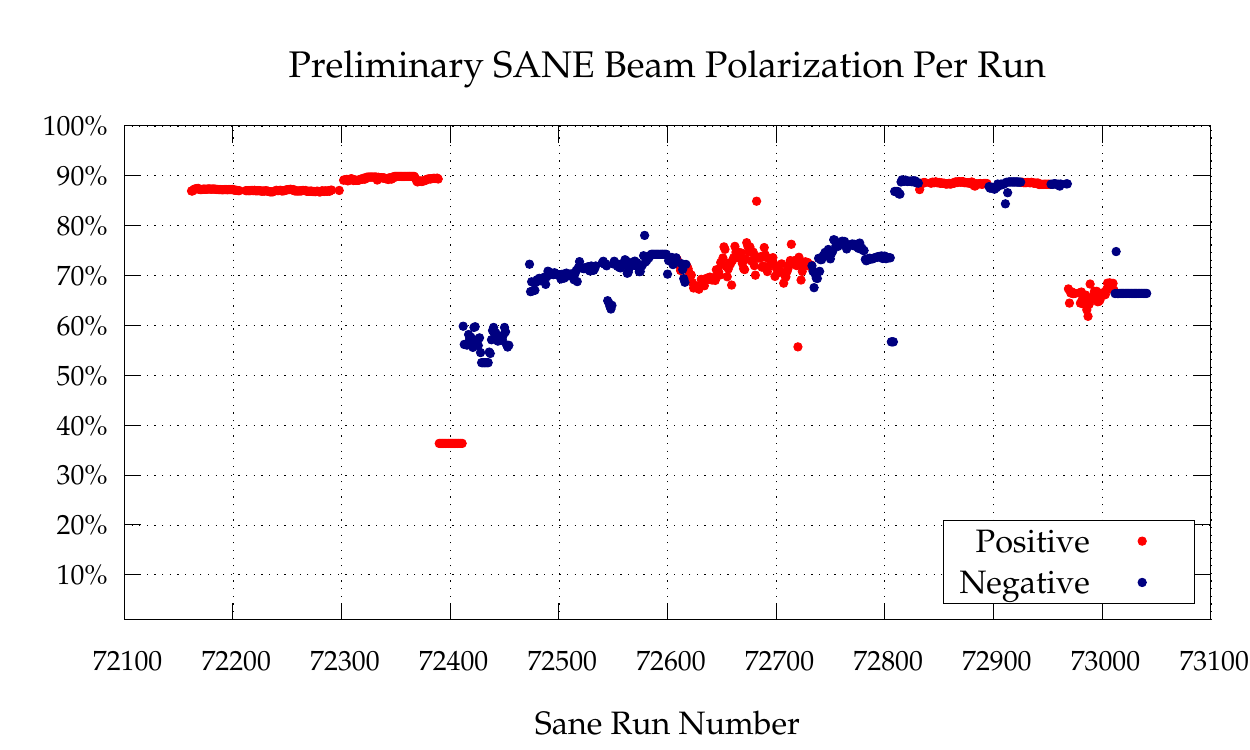}
  \end{center}
  \caption[Electron beam polarization for each SANE experimental run.]{Electron beam polarization for each SANE experimental run.  The polarizations fall in groups depending most strongly on the beam energy of the run.  The 4.7 GeV beam energy setting allows near 90\% polarization throughout, while the 5.9 GeV setting polarizations (in the middle of the experiment) drop significantly as the beam energy increases.  At the end of the experiment, cryomodule failures necessitated 5 accelerator passes to achieve 4.7 GeV, and the polarization suffered.}
  \label{fig:beam_pol}
\end{figure}

Using beam energy, Wien angle, quantum efficiency, and half wave plate status as collected over time by JLab's EPICS system, the beam polarization for each run during SANE was calculated using the above formulation.  The original sane{\_}pol.f code by D. Gaskell was translated into Perl for this purpose.  Figure \ref{fig:beam_pol} shows the beam polarization per run as averaged over charge accumulated on target during SANE.

			\subsubsection{Fast Raster}
			
	Hall C's fast raster system uses two air-core magnets to spread the beam spot from below 100 $\mu$m to $2\times 2$ mm$^2$.  The intense local heating by such a small spot requires the increase of the beam spot to prevent target damage; even the aluminum windows of the target cryostat could be melted by such intense local heat.  The deflection of the beam is achieved by two bedstead ``air-core'' magnets sitting roughly 25m upstream of the target. These magnets are formed by gluing cables together without the use of potting material, and they offer quick response and resistance to eddy effects.
	
\begin{figure}[htb]
  \begin{center}
    \includegraphics[width=2.75in]{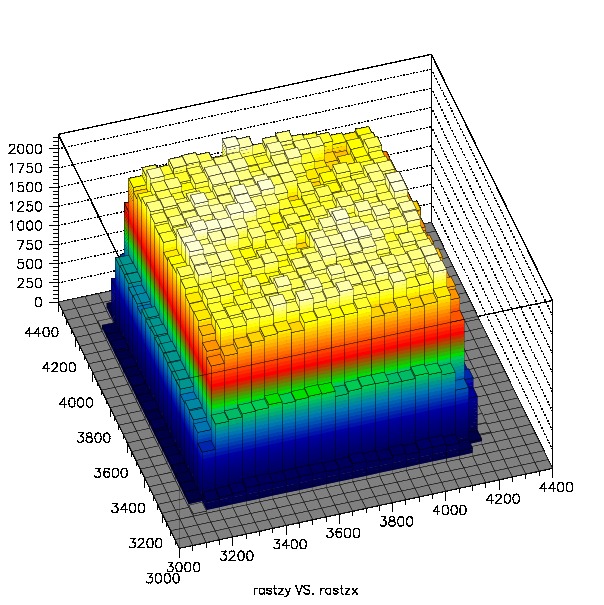}
  \end{center}
  \caption[Fast raster pattern.]{Plot of number of hits in BETA versus the fast raster position for SANE production run 73041, showing the fast raster pattern.}
  \label{fig:fastraster}
\end{figure}

	The magnets are driven by purpose-built power sources implementing bipolar MOSFET switching bridges which are controlled by pulse generators at the desired raster frequency.  To produce a uniform square beam spot, triangle waveforms are used to drive the magnet currents.  Figure \ref{fig:fastraster} visualizes the fast raster via hits during an example run in SANE plotted against the fast raster position at that time.  More information on Hall C's fast raster system can be found in references \cite{Yan199546} and \cite{Yan20051}.
			
		\subsection{SANE Hall C Beamline}
		
	In addition to the standard beamline equipment in Hall C, SANE required extra beamline equipment to accommodate the UVa polarized target.	The fast raster spreads the beam onto a $2\times 2$ mm square;  a slow raster was added to spread the beam evenly over a larger portion of the target material cup.  When the target magnetic field is near perpendicular to the beam, the beam is deflected down, away from the center of the target.  To counteract this, the beam was sent through a chicane which bent it down and then back up at the target.  After the beam passed through the center of the target, it would continue to bend down, missing the beamline, so a helium bag was used to transport the beam to the beam dump.
	
			\subsubsection{Slow Raster}
	The beam spot area after the fast raster is $2\times 2$ mm, but the cups which hold the target material are one inch in diameter.  As radiation dose---the beam's charge deposited in the target over area---damages the polarizability of the ammonia target material (see section \ref{sec:ammoniaexp}), the beam was rastered a second time to spread it evenly over the material.  This second raster was circular, unlike the square fast raster, both to match the cylindrical target cups and pass more easily through the 1.5 inch beam pipe.  Throughout most of the experiment, the slow raster's diameter was 2 cm.	
	
	Three waveform generators were used to drive the slow raster magnets.  The angular velocity of the beam about its undeflected trajectory was kept constant and amplitude modulation was used to uniformly draw the beam through a spiral to form a circle.  For a constant angular velocity radial pitch $dr / d\theta = A$, we assume a much larger azimuthal velocity than radial velocity in the spiral \cite{Fukuda199745}, to obtain $\omega(t) = v_0/r(t)$.  After integrating to determine constant $A$ and combining these two expressions, we have
	\begin{equation}	
	 r\frac{dr}{dt} =  v_0\frac{R}{2\pi N}\, \, \,  \implies  \, \, \,r(t) = \sqrt{\frac{R}{\pi N}v_0t}	
	\end{equation}
for raster radius $R$ and number of revolutions per radius traced $N$.  

\begin{figure}[htb]
  \begin{center}
    \includegraphics[width=2.75in]{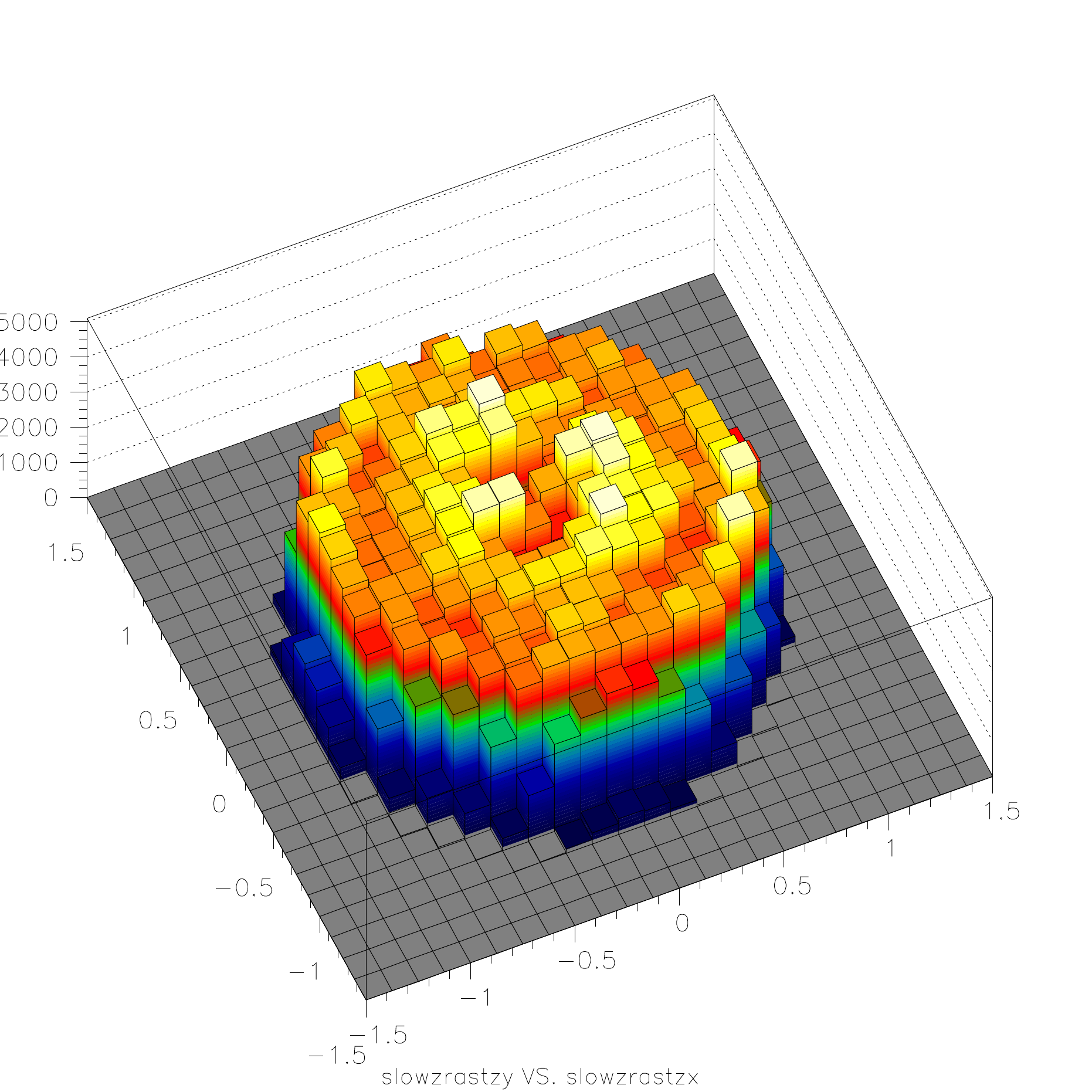}
  \end{center}
  \caption[Slow raster pattern]{Plot of number of hits in BETA versus the slow raster position for SANE production run 73041, showing the slow raster pattern.}
  \label{fig:slowraster}
\end{figure}

	 To create the amplitude modulation to control the radius of the spiral, a Wavetek programmable waveform generator (G1) was used to generate a 30 Hz waveform of the function $t^{1/2}$.  Two other Waveteks (G2 and G3) were used to generate 100Hz sine waves with a 90$\degrees$ phase difference, creating a circle.  The G2 and G3 are phase locked to the clock of the G1, and their amplitudes are controlled by the G1, producing the final spiral raster pattern.  These signals controlled two pulse width modulation amplifiers which drive the x and y slow raster deflection magnets \cite{yan-slowraster}.  In figure \ref{fig:slowraster} we show an example plot of hits versus the slow raster position in x and y for a sample run.
	 
			\subsubsection{Chicane}
\begin{figure}[tbh]
  \begin{center}
    \includegraphics[width=6in]{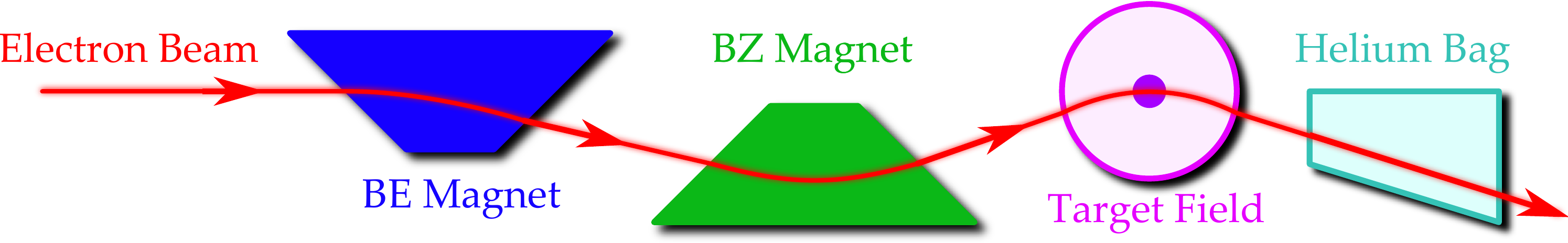}
  \end{center}
  \caption[Diagram of the SANE beamline with perpendicular target field.]{Diagram of the SANE beamline during perpendicular target field running (not to scale).}
  \label{fig:sanebeamline}
\end{figure}

	While the trajectory of the beam is unaffected by the target's 5 T magnetic field when it is coaxial to the coils, SANE required near perpendicular target polarization and thus magnetic field alignment for much of the experiment.  The standard Hall C beam would be deflected down by the target magnetic field in this case, causing the beam to miss the center of the target.  To counteract the bend of the beam before it met the target, a chicane was used, as seen in figure \ref{fig:sanebeamline}.	
		
	The chicane consisted of two dipole magnets, BE and BZ.  BE bent the incoming beam downwards toward the BZ, which in turn bent the beam back up at the target.  These magnets were precisely positioned and tuned to allow the beam to strike the center of the target after being bent by the target magnetic field. 	Table \ref{tab:chicane} shows the positioning, deflection and integrated $\int B \cdot dl$ of the chicane magnets for the two beam energy settings used while the target was in its perpendicular configuration.

\begin{table}[ht]
  \begin{center}
\begin{tabular}{ccccccc}
\toprule

Beam $E$ & BE Bend& BZ  Bend & Target Bend   &BE $B dl$&BZ $B dl$&Target $B dl$\\ \midrule 
 4.7 GeV& 0.878$\degrees$ & 3.637$\degrees$ & 2.759$\degrees$ & 0.513& 1.002  & 1.521 \\ 
 5.9 GeV& 0.704$\degrees$ & 2.918$\degrees$ & 2.214$\degrees$ & 0.513& 1.002  & 1.521  \\
\bottomrule
\end{tabular}
\caption[Table of chicane parameters.]{Table of chicane parameters for 80$\degrees$ field for both beam energy settings.  Integrated $Bdl$ given in Tm.}
  \label{tab:chicane}
\end{center}
\end{table}
			
			\subsubsection{Helium Bag}
			
	The final consideration to be made for the beam while the target field was near perpendicular was transport to the beam dump.  In figure \ref{fig:sanebeamline} the beam can be seen bending down in the target field after passing through the target, which would cause it to miss the standard Hall C beamline to the beam dump. Were the beam to pass through the air in the hall to reach the beam dump, ionization would create unacceptable amounts of harmful by-products such as ozone.
	
	To address the beam transport to the beam dump, an 80-foot-long helium bag was devised.  The helium bag included  0.04 inch aluminum windows at the entrance on an extension piece as well as at the exit to beam dump for both straight-through and bent beam running.  The exit windows were large enough to accept the beam at both 4.7 and 5.9 GeV when bent by the target magnet in perpendicular running to 2.8$\degrees$ and 2.2$\degrees$ nominal beam deflection, respectively.

	\section{Electron Detector Package}
	\label{sec:beta}
\begin{figure}[tbh]
  \begin{center}
    \includegraphics[width=5in]{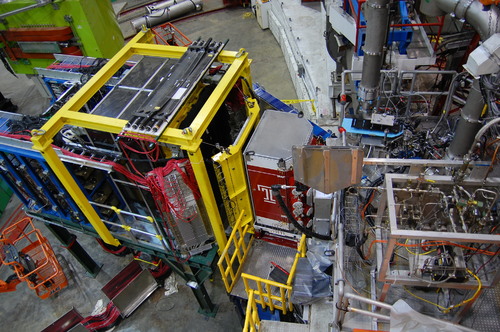}
  \end{center}
  \caption[Photograph of BETA]{Photograph of the experimental hall during SANE.  To the left is BETA, with the calorimeter and its support electronics in blue and yellow, the \v{C}erenkov tank in red, the hodoscope in yellow between them, and the target to the right.}
  \label{fig:sanepic}
\end{figure}

The electron arm of the experiment, known as the Big Electron Telescope Array or \textbf{BETA} and seen in figure \ref{fig:sanepic}, was a non-magnetic detector array designed for large acceptance, high pixelization, high background rejection and low deadtime with adequate energy resolution to observe high $x_{Bj}$ DIS electrons.  BETA was comprised of 4 main systems; a large electromagnetic calorimeter, a threshold \v{Cerenkov} detector, and two tracking hodoscopes.  The drift space between the \v{C}erenkov and calorimeter gave a pointing accuracy to isolate events within the scattering chamber, effecively making it a telescope to view the scattering interaction.  

Together the calorimeter and \v{C}erenkov allowed for effective identification of electrons from the target.  The threshold \v{C}erenkov was used primarily for the differentiation of electrons and photons; a \v{C}erenkov TDC event which matched the timing of a calorimeter event was the primary criteria.    In fact, the calorimeter was capable of differentiating electrons from charged pions on its own.  As the radiation length and physical length of the bars ensured nearly all of an incoming electron's energy was deposited in the calorimeter, a simple energy cut was sufficient to exclude charged pions, which were unlikely to exceed 500 MeV.  The eight mirrors of the \v{C}erenkov enabled the separation of the calorimeter into eight segments, each segment in the ``shadow'' of one mirror, which made it possible to place a geometric cut to ensure an electron event at a given position in the calorimeter was seen on the appropriate \v{C}erenkov mirror.
		
		\subsection{BigCal}
		\label{sec:bigcal}
		
\begin{figure}[tbh]
  \begin{center}
    \includegraphics[width=5in]{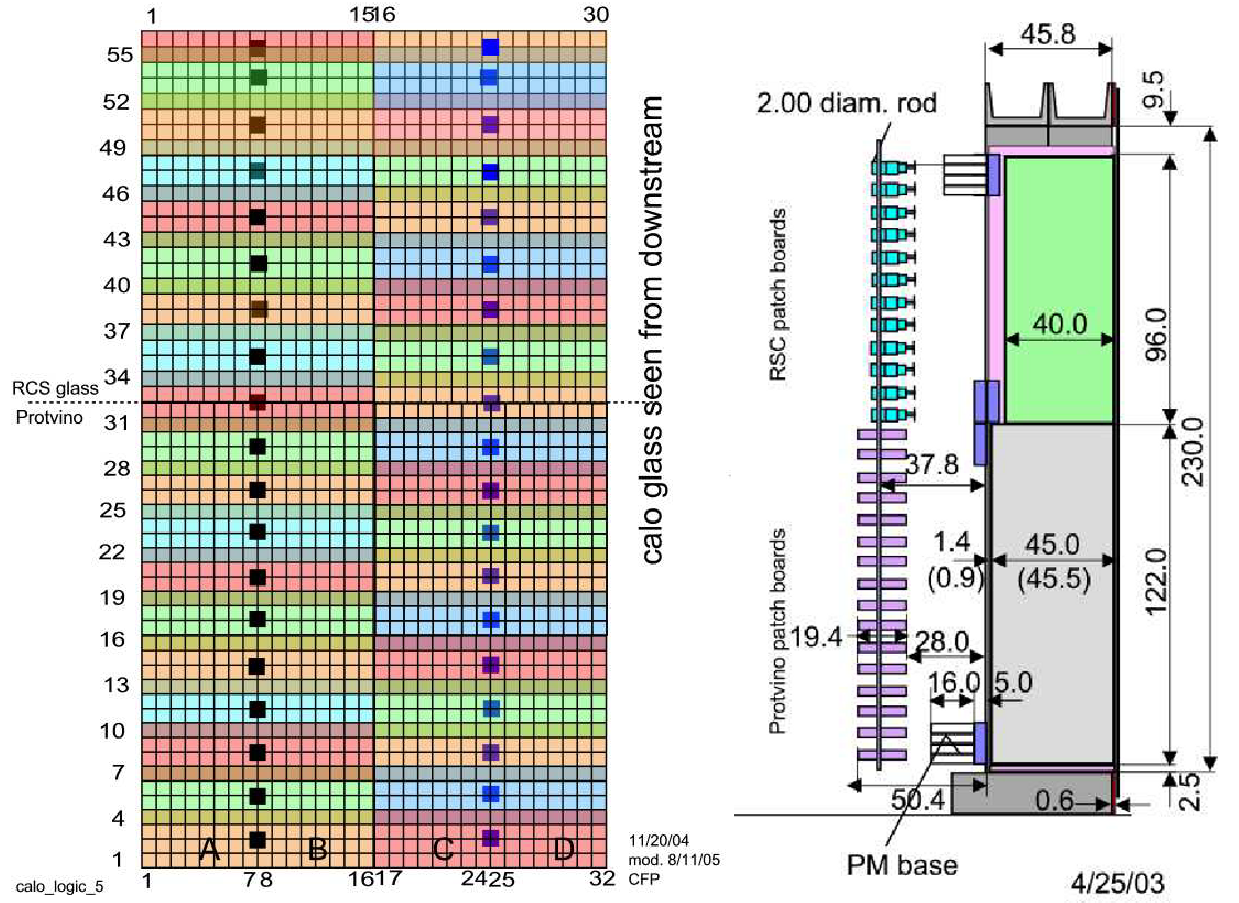}
  \end{center}
  \caption[Diagrams of BigCal]{Left is the face of BigCal, showing 1,744 lead glass blocks, with different colors indicating the groupings of the trigger channels.  Right shows a cutaway view of the calorimeter from the side. Diagrams from reference \cite{bigcal}.}
  \label{fig:bigcal}
\end{figure}

	BETA's electromagnetic calorimeter, nicknamed \textit{BigCal}, consisted of 1,744 TF1-0 lead glass blocks; 1,024 of these were $3.8\times3.8\times45.0$ cm$^3$ blocks contributed by the Institute for High Energy Physics in Protvino, Russia, while the remaining 720 were $4.0\times4.0\times40.0$ cm$^3$ and came from Yerevan Physics Institute, most recently used to study real Compton scattering (RCS) in Hall A.  The calorimeter was assembled by the GEp-III collaboration \cite{PhysRevLett.104.242301,PhysRevLett.106.132501}. The Protvino blocks were stacked $32\times32$ to form the bottom section of BigCal, and the RCS blocks were stacked $30\times24$ on top of these, as seen in figure  
	 \ref{fig:bigcal}.  The assembled calorimeter had an area of roughly $122\times218$ cm$^2$, making a large solid angle of approximately 0.2 sr with the face of the calorimeter placed 3.50 m from the target cell.

\subsubsection{Shower Counters}

While the mechanism of electromagnetic calorimeters is well known, a brief discussion is worthwhile.  When transversing a given material, electrons or positrons of energies above a material's \textit{critical energy} $E_c$ lose energy primarily through bremsstrahlung---``braking radiation'' \cite{leo,pdb}.  Photons emitted via bremsstrahlung from a high energy electron are most likely to produce an electron--positron pair, which will radiate via bremsstrahlung in turn.   This chain of events leads to a ``shower'' of electrons, positrons and photons which continues until the energies of the secondary particles falls below the critical energy, when ionization and excitation of the material take over.  In addition, primary and secondary electrons and positrons move very close to the speed of light, exceeding $c/n$ for the index of refraction of the glass, so that they emit \v{C}erenkov radiation at optical wavelengths, adding to the shower.  This shower can be collected by photomultiplier tubes to obtain a measurement of the energy of the incident particles\cite{fernow}.

We express the characteristic distance particles travel through a given material as a \textit{radiation length}, which is the mean distance over which a high energy electron loses $1/e$ of its energy to bremsstrahlung.  We can also use it to approximately describe the electromagnetic cascade in a material: after traveling 2 radiation lengths, an electron and it's secondaries are likely to have interacted twice, resulting in two electrons, a positron and a photon, for instance\cite{fernow}.  Radiation length $X_0$ is expressed approximately by Fernow in terms of the atomic mass and number of the absorber $A$ and $Z$: $X_0 = 180 \cdot A/Z^2$. A more precise expression is given in the Particle Data Book\cite{pdb}. 
	
\begin{table}[tbh]
  \begin{center}
\begin{tabular}{ll}
\toprule
Index of Refection $n$& 1.6522\\
Density $\rho$ &3.86 g/cm$^3$\\
Radiation Length $X_0$ &2.74 cm\\
Moliere Radius $R_M$ &4.70 cm\\
Critical Energy $E_c$ &15 MeV\\
\bottomrule
\end{tabular}
\caption{Table of TF1-0 lead glass characteristics for calorimetry.}
  \label{tab:leadglass}
\end{center}
\end{table}
		
	The characteristics of the TF1-0 lead glass used in BigCal are shown in table \ref{tab:leadglass}. A high density, index of refraction and transparency, along with a small radiation length make it ideal for calorimetry.  The thickness of the glass was approximately 16 radiation lengths (16.2 for the RCS section, 16.4 for the Protvino section), which will stop electrons of up to 10 GeV.  The Moliere radius of 4.7 cm means that an electron shower will expand into several of the 4 cm or 3.8 cm square bars.

		\subsubsection{BigCal Configuration}
	Each lead-glass bar was wrapped in aluminized mylar to optically isolate it from its neighbors.  The end of each bar is optically coupled to a Russian FEU84-12 stage ``venetian blind'' photomultiplier tube by a 5 mm thick silicon pad, or ``cookie.''  PMT's, cookies and bars were enclosed within a black box, and signal and high-voltage power cables enter the black box by labyrinth openings to keep out external light.

\begin{figure}[tbh]
  \begin{center}
    \includegraphics[width=4.5in]{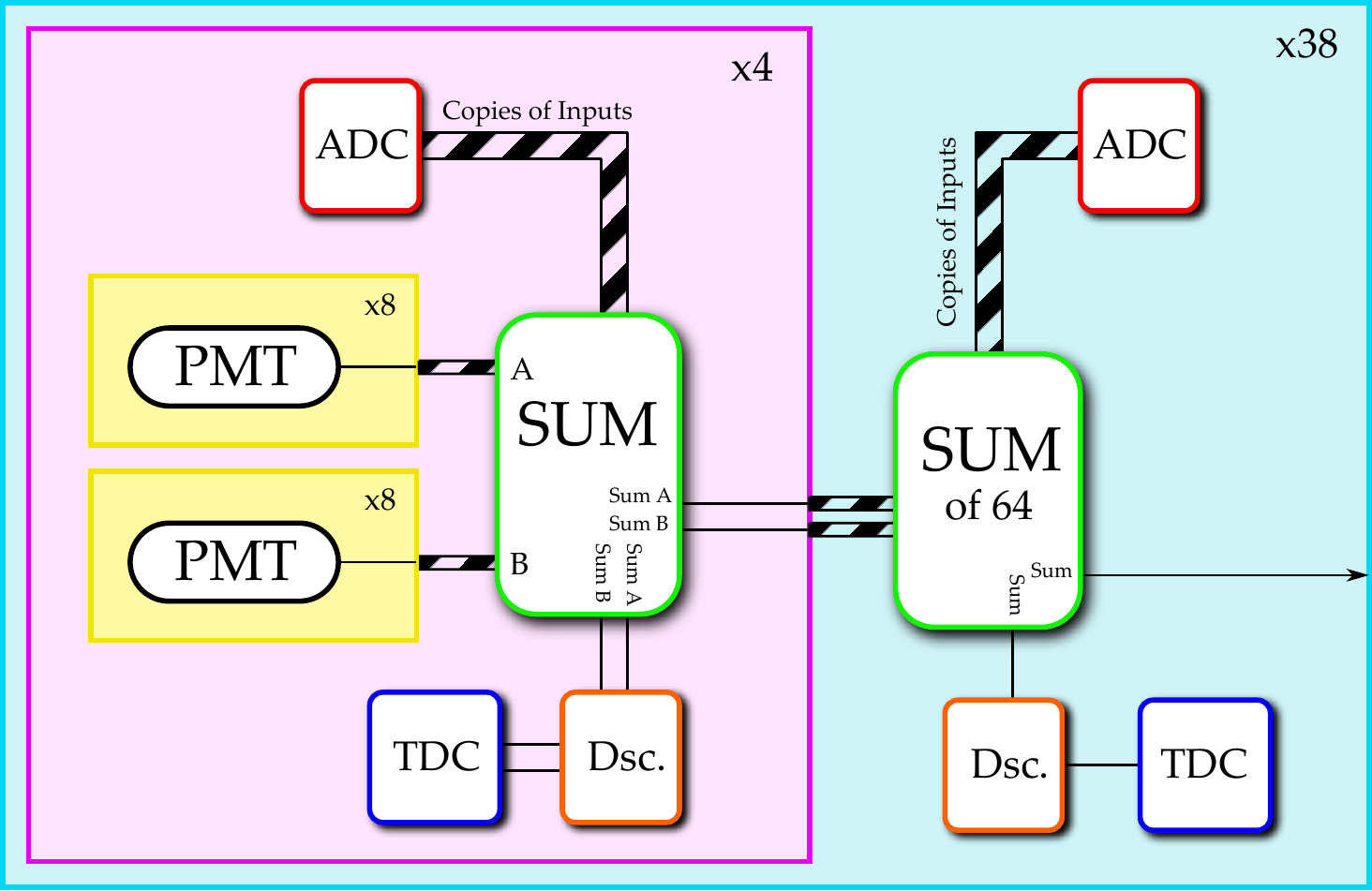}
  \end{center}
  \caption[Schematic of BigCal wiring.]{Schematic of BigCal wiring, showing individual signals, sums of 64 and summing to produce final trigger.}
  \label{fig:bigcalwiring}
\end{figure}

BigCal's photomultiplier signals are taken through several stages of summing and discrimination to produce final ADC and TDC signals for the calorimeter as a whole.  A schematic is shown in figure \ref{fig:bigcalwiring} to accompany this description.  The signals from the photomultipliers are first sent to one of 224 first-level summing modules which each handle 8 signals, amplifying by a factor 4.2 and combining groups of 8 signals to produce a summed output, as well as passing along the individual amplified signals to the ADCs to read out.

The summed outputs from the first-level sums of 8 go to a discriminator and thence to a TDC, for a total of 224 TDCs.  A copy of the first-level summing is sent to the second level summing modules, which sum two groups of 4 such inputs to produce what is now a sum of 64 PMT signals.  The sets of 64 PMTs that go into these sums are illustrated in figure \ref{fig:bigcal}, which shows that each group of 64 is 16 blocks wide by 4 blocks tall.  There are 38 such sums of 64, which allows each set of 64 to overlap with the set above and below it by one row, as seen by mixing of colors in the figure.   The sums of 64 then go to a logical OR to be sent to the trigger supervisor, which will be described in section \ref{sec:trigger}.  

Further description of BigCal wiring is available in references \cite{bigcal,puckbigcal}, and A. Puckett's thesis contains discussion on the background and use of the calorimeter in great detail\cite{puckett}.
		
		\subsection{Gas Cerenkov}
		
	A \v{C}erenkov counter was designed and built for SANE by Temple University to provide electron detection and pion rejection of 1,000:1.  Each of the eight roughly $40\times40$ cm$^2$ mirrors focused \v{C}erenkov photons onto a single, 3 inch, quartz-window, Photonis XP4318B photomultiplier tube. 
	
	A threshold \v{C}erenkov detector leverages the \v{C}erenkov effect of charged particles which ``exceed'' the speed of light in given medium of index of refraction $n$: $v>c/n$.  \v{C}erenkov radiation comes in the form of an electromagnetic shock wave, a conical wavefront following the particle, emitted at angle $	cos(\theta_C) = 1/\beta n(\omega)$. 
	
	 Careful selection of the material based on its index of refraction allows indication of charged particles with speed above a given threshold.  While electrons and pions of similar energy may be collected in the calorimeter, the much heavier pions will not exceed the threshold speed, allowing rejection of the unwanted background.
	 
	Dry N$_2$ gas at near atmospheric pressure was used as a radiator in SANE's \v{C}erenkov tank.	The index of refraction of N$_2$ is approximately 1.000279, which gives a $\beta$ threshold for \v{C}erenkov emission by pions of $\beta_{threshold} = 1/n = 0.999721$, which corresponds to a momentum of 5.9 GeV.  As the highest beam energy used during SANE was 5.9 GeV, pions above that threshold should not occur.
	
	The number of \v{C}erenkov photons per wavelength per unit of length travelled is given by
	\begin{equation}
	\frac{d^2N}{d\lambda dx} = \frac{2\pi z^2 \alpha}{\lambda^2} \left( 1 - \frac{1}{\beta^2 n^2(\lambda)}\right )
	\end{equation}
for a particle of charge $ze$ and index of refraction $n(\lambda)$, which is generally dependent on the wavelength of the particle traveling through the medium\cite{leo}.  For $n=1.000279$, a conservative cutoff of $\lambda = 200$ nm and a radiator thickness of 125 cm, we expect on the order of 20 photoelectrons after considering the photocathode sensitivity.
	
	\begin{figure}[tbh]
  \begin{center}
    \includegraphics[width=6in]{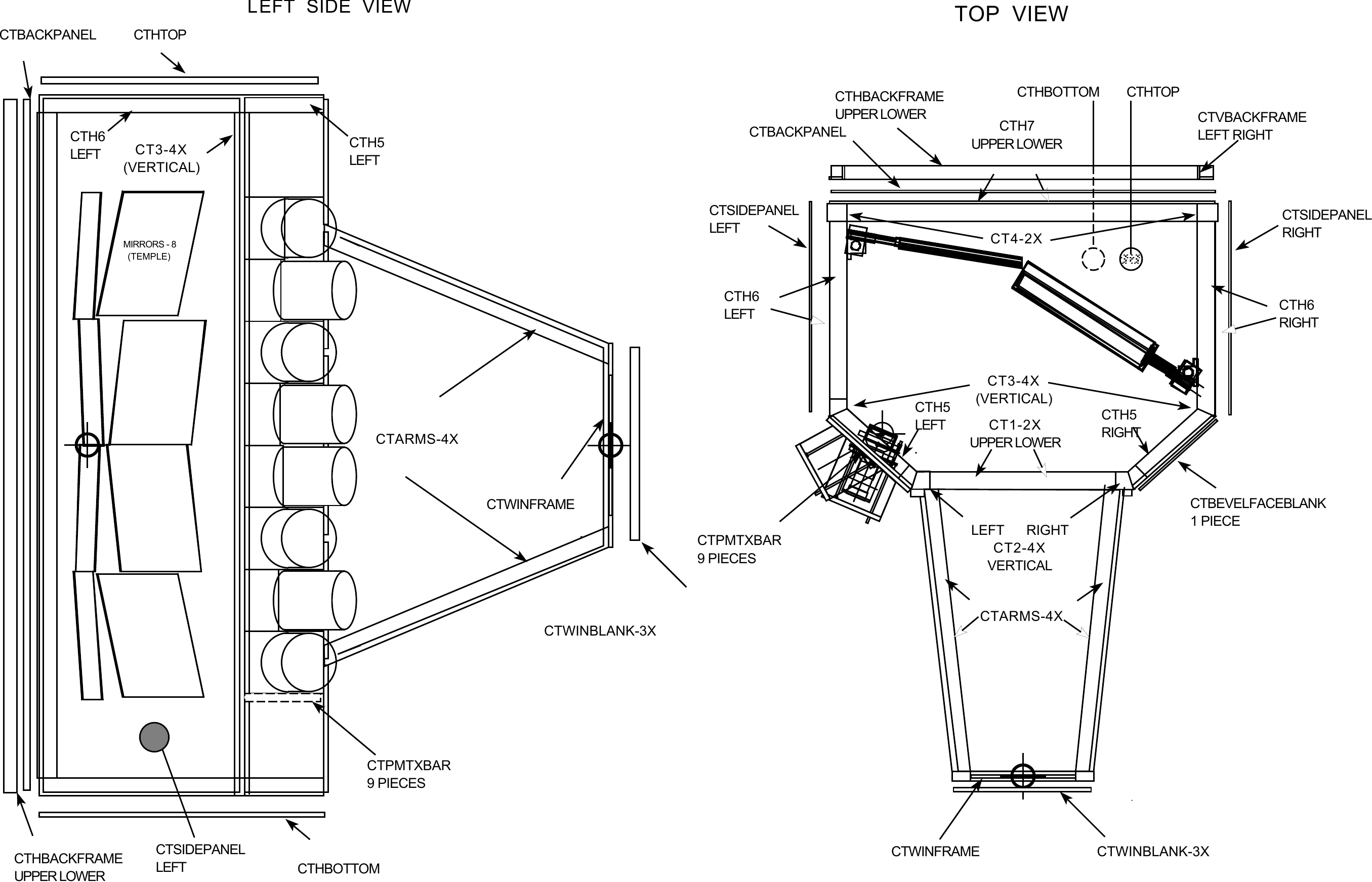}
  \end{center}
  \caption[Drawing of \v{C}erenkov detector]{Drawing of \v{C}erenkov detector design, showing photomultiplier tube and mirror placement.  From Temple University fabrication schematics.}
  \label{fig:cerenkov}
\end{figure}

	The \v{C}erenkov tank's 8 mirrors were designed for point-to-point focusing from the target cell to the photomultiplier photocathodes, and were arranged in two columns.  Four spherical mirrors covered the large scattering angle column and four elliptical mirrors in the small angle column, as seen in figure \ref{fig:cerenkov}.  The mirrors were positioned so that they covered the entire face of BigCal as viewed from the target, with slight overlaps in the mirrors, dividing BigCal into 8 equal sectors.  This allowed geometrical correlation of particle hits in BigCal, providing further background rejection.
	
	The 8 Photonis photomultiplier tubes were positioned on the large angle side of the tank to protect the tubes from both the more intense magnetic field from the target and heavier particle flux from the target and beam line.  Extensive shielding surrounded the tank to decrease background not from the target cell.  While the phototubes were shielded from the target magnet's field with $\mu$-metal, during the near perpendicular magnetic field setting, an additional inch-thick iron plate was positioned between the phototubes and the target as the field affected the performance of the tubes significantly.

		\subsection{Hodoscopes}
		
	Two tracking hodoscopes were included in BETA: one directly in front of the face of BigCal, contributed by Norfolk State University, and the second sandwiched between the \v{C}erenkov and target outer vacuum chamber, contributed by North Carolina A\& T State University.
	
	\subsubsection{Lucite Hodoscope}	
	
	Mounted directly onto the BigCal platform, 80 cm from the face of the calorimeter, the 28 bars of the lucite hodoscope provided background rejection and position data.  The $3.5\times6.0\times80.0$ cm$^3$  bars were curved to a radius of 240 cm, providing normal incidence of particles originating in the target.  The ends of the bars were cut at 45$\degrees$ angles to avoid reflections as the bar met the light guides.  The light guides took the $4.9 \times 60$ cm$^2$ rectangular bar to 4.9 cm circular to optically couple to 2 inch Photonis XP2268 photomultiplier tubes.  
	
	\begin{figure}[tbh]
  \begin{center}
    \includegraphics[width=3.5in]{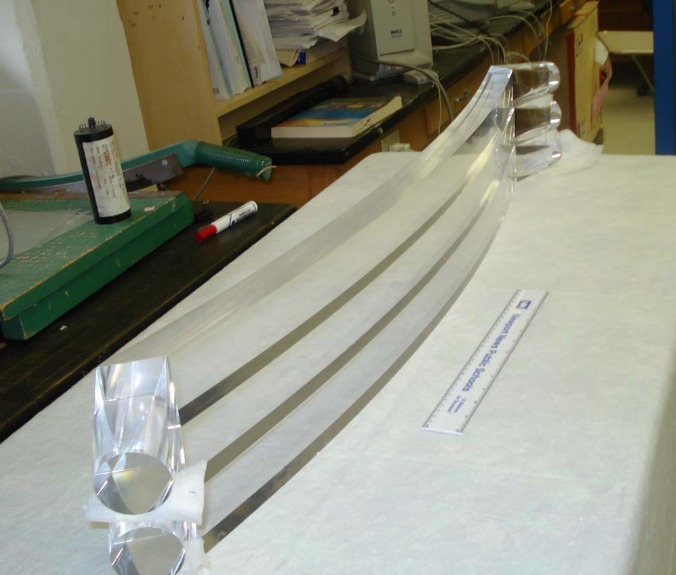}
  \end{center}
  \caption[Photograph of three of the 28 hodoscope bars.]{Photograph of three of the 28 hodoscope bars before preparation and installation, showing the curvature of the bars and 45$\degrees$ coupling to the light guides.}
  \label{fig:hodobar}
\end{figure}

	The lucite hodoscope bars offered an index of refraction of $n=1.49$, allowing \v{C}erenkov radiation from charged particles above $\beta_{threshold}=0.67$.  Charged particles above this threshold create \v{C}erenkov light which totally internally reflects down the length of the bar.  As phototubes collect this light from both ends, the position of the incidence along the bar can be inferred from the time separation of arrival of the signals in the photomultiplier tubes.
	
	The photomultiplier tubes were each shielded from the target magnetic field with 1.5 mm $\mu$-metal, as well as a magnetic shielding box which enclosed each of the 2 set of 28 tubes.  The signals from the tubes were sent to discriminators then TDCs, as well as ADCs, for recording.
		
	\subsubsection{Front Tracker}
	
The front tracker consisted of three planes of $3\times3$ mm$^2$	Bicron BC-408 plastic scintillator bars positioned as close to the target cell as feasible.  It sat just outside the target's outer vacuum chamber, 48 cm from the target cell.  The purpose of this hodoscope was to provide tracking data on particles while they were still under the influence of the target's magnetic field.  Combining this position data with final positions caught in BigCal, the curved trajectory of the particle in the magnetic field should be discernible, allowing the differentiation of positively and negatively charged particles.  This would provide rejection of the positron background which diluted BigCal's yield of DIS electrons.

The active area of the tracker was 40 vertical by 22 horizontal cm and the three tracker planes included a set of 133 vertical scintillator bars, the X plane, and 2 sets of 73 horizontal bars, the Y planes. The two Y planes were offset by half the height of a bar, 1.5 mm, to provide redundant Y information on particles traveling through the tracker.  A Bicron BCF-92MC blue-green wave-length shifting fibers were coupled along the length of each bar of the tracker.  These 2.5 M long fibers acted both to carry the light from the bars to the magnetically shielded PMTs nearly 2 meters away, and to shift the wavelength of the scintillated light in the bars into the most sensitive range of the Hamamatsu H8804 64 channel photomultiplier tubes.


	\section{Triggers and Data Acquisition}
	\label{sec:trigger}
	
	The collection of event data was coordinated by a trigger supervisor (TS), which received trigger information from BigCal, \v{C}erenkov and HMS TDCs.  The trigger supervisor will accept triggers from Readout Controllers (ROCs) if it is not busy reading the previous event.  If a trigger is accepted, a signal is sent to generate gates for ADCs and start signals for TDCs.   The ROCs then readout their data, which is assembled by the event builder on a host server.  From there, the data is copied to long-term tape storage.  
	
	\subsection{Triggers}
	\label{sec:triggers}
	Eight trigger types were defined for SANE in the trigger supervisor, and of these only three are important to this analysis.  BETA1 triggers, defined as trigger type 2, were the result of BigCal hits, while BETA2 triggers, defined as trigger type 4, were the results of the coincidence of BigCal and \v{C}erenkov hits.  A second BigCal only trigger for $\pi^0$ particles, was defined as trigger type 3.	
	
	Prescale factors could be set 
	into the trigger supervisor to allow the reduction of triggers of that type accepted: a prescale of 5 on a given trigger means that only 1 in 5 triggers of that type are used. Prescales are useful for controlling deadtime, the portion of total time when new events are not being accepted as the data acquisition system is busy processing and recording events.

\subsubsection{BETA1 Trigger}
	The BETA1, or type 2, trigger was the result of a hit in BigCal.  As described in section \ref{sec:bigcal}, the 1,744 calorimeter bars and photomultiplier tubes were summed into groups of 8 in a first-level sum, and then 64 in a second-level sum. There were 38 such sums, as each set of 64 (4 rows and 16 columns) included an overlap of one row with the group above and below it.  This overlap addressed efficiency issues that could occur when a hit at a boundary gives half its energy to one and half to another summed set, while not breaking the trigger threshold in either.   
	
	These 38 sums of 64 were sent to one of four sixteen-channel discriminators, dividing BigCal into four quadrants with its own trigger threshold.  The 38 discriminator outputs were routed to a logical fan-in/fan-out unit to perform an OR on the 38 trigger sums.  This meant if any of the sums of 64 exceeded its threshold, a trigger was generated.  Trigger type 2 consisted of this OR of the BigCal PMT signals.
	
\begin{figure}[phtb]
  \begin{center}
    \includegraphics[width=6in]{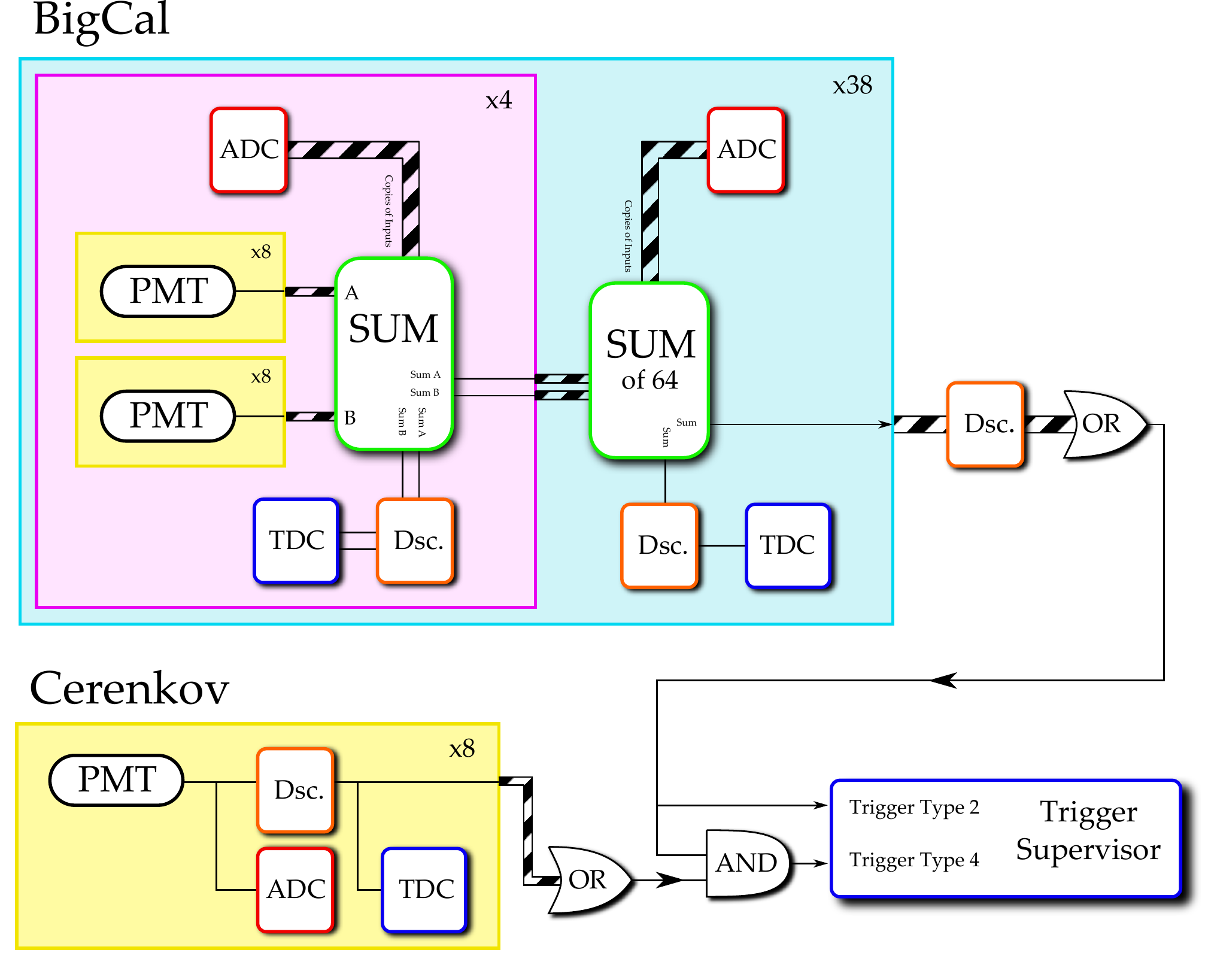}
  \end{center}
  \caption[Diagram of SANE DAQ wiring.]{Diagram of SANE DAQ wiring, from PMTs to trigger supervisor.  Both trigger type 2 and 4 are shown.}
  \label{fig:sanedaq}
\end{figure}
	
\subsubsection{BETA2 Trigger}

	The main BETA trigger, BETA2 or trigger type 4, was the coincidence of a hit in the calorimeter and \v{C}erenkov.  The \v{C}erenkov detector's 8 photomultiplier tubes were discriminated and sent to a logical unit which performed an OR of these signals. The results of the OR of sums of 64 from BigCal, and the OR of the 8 mirror PMTs were then sent to a logical unit to perform an AND to obtain a coincidence of the two systems. Figure \ref{fig:sanedaq} shows the creation of both trigger types 2 and 4.

\subsubsection{$\pi^0$ Trigger}
\label{sec:pizero}
Production of $\pi^0$ particles in the target was used for calibration purposes. The primary branching ratio of the $\pi^0$ is two photons, and while the photons would not be observed in the \v{C}erenkov, they were picked up in the calorimeter.  By knowing the separation angle and energy of both photons from the $\pi^0$, we have a known energy point based on the mass of the pion.  To this end, a trigger was set up to collect $\pi^0$, looking for two hits on BigCal separated vertically.  An AND of sums of 64 caused this trigger to fire.


\subsection{Data Acquisition}

SANE's data acquisition was handled by the CEBAF Online Data Acquisition system, a framework of software and hardware guidelines started by the Jefferson Lab Data Acquisition group as the lab was being constructed.  CODA provided a front-end user interface, as well as a server component which controlled all the data acquisition parameters of a run.

\begin{figure}[htb]
  \begin{center}
    \includegraphics[width=4.5in]{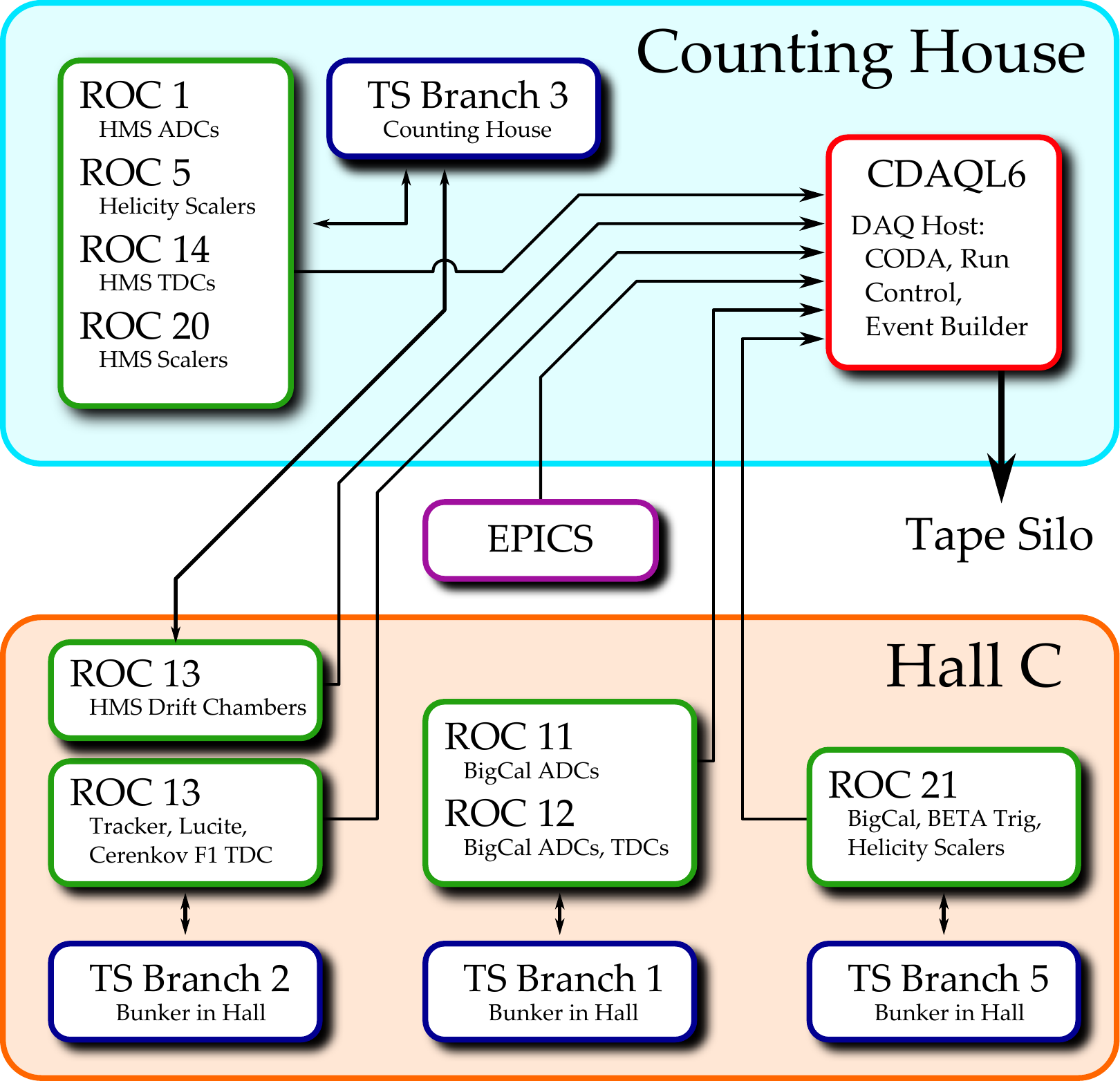}
  \end{center}
  \caption{Diagram of the Hall C DAQ data flow during SANE.}
  \label{fig:hallcdaq}
\end{figure}

The trigger supervisor controlled the readout of data from all sources when a run is in progress. The TS sat in the electronics bunker in Hall C and accepted triggers---as defined in the section \ref{sec:triggers}---via one of four branches in various locations connected to the TS by long branch cables.  The TS can handle 8 ROCs on each of its 4 branches, allowing as many as 32 ROCs to be coordinated.  The layout of the Hall C data acquisition during SANE, including the trigger supervisor branches, the ROCs that reported to each, and which systems reported to each ROC, is shown in figure \ref{fig:hallcdaq}. 
 Further information on the trigger supervisor is available in references \cite{TS1,TS2}.

The trigger supervisor coordinates the readout of data by accepting triggers if the system is not busy, then informing the individual ROCs to record data based on that trigger.  The ROCs are single-board CPUs in each crate which collect data from the ADCs, TDCs or scalers in its crate into banks of memory.  These memory banks become event fragments assembled later on.  When a trigger is accepted, the TS sends a level 1 accept signal to all its branches, forming gate, start and stop signals for the ADCs and TDCs.  Once the ROCs have collected and processed the resulting data, they send an acknowledgement signal (ack), back to the TS which remains in a ``busy'' state until all the ROCs report back.

CODA's \textit{event builder} assembles incoming event fragments from the ROC banks into full physics events.  Four classes of events were used in the Hall C CODA setup.  Upon the start or stop of a run, a \textit{status event} is inserted into the data stream to record salient run parameters.  Experimental triggers from the TS create \textit{physics events} which contain all the data from the ROC banks from that trigger.  Every two seconds a \textit{scaler event} is created to read all the experimental scalers to the data stream and \textit{EPICS events} are inserted every 30 seconds to record slow control data recorded in the EPICS system.

During SANE, the Linux data acquisition machine CDAQL6 hosted the CODA run control and event builder programs.  Event data were written to this machine's hard disk, to later be transferred to tape silos at Jefferson Lab's mass storage system.  Further information on JLab's CODA system can be found in \cite{coda}.


\chapter{Polarized Target}
\label{sec:target}

SANE utilized frozen ammonia ($^{14}\textrm{NH}_3$) as a proton target, polarized via dynamic nuclear polarization in a 5 T magnetic field at around 1 K.  An introduction to the theory and mechanisms behind solid polarized targets begins in section \ref{sec:introtarget}.  Section \ref{sec:material} addresses the materials used in the target, and the means of measuring the polarization via NMR follows in section \ref{sec:nmr}.  A description of the systems and methods used for the target during the experiment is given in section \ref{sec:targetsetup}, and the data analysis and results of the target polarizations achieved during the experiment are discussed in sections \ref{sec:targetanal} and \ref{sec:results}.  

	\section{Dynamic Nuclear Polarization}
	\label{sec:introtarget}

	
The method of dynamic nuclear polarization (DNP) was first developed for metals by Overhauser in 1953\cite{overhauser}, and was applied to solid insulators by others by 1958 \cite{Jeffries-PhysRev.106.164}\cite{abragam}.  In DNP, nucleon polarization is achieved in a high magnetic field by transferring the polarization of free electrons in the medium to the nucleon using a microwave field.  Several mechanisms are known to contribute to the DNP process, and each will be addressed in the following subsections.

			\subsection{Thermal Equilibrium Polarization}
The simplest method to polarize a given material, and the starting point for other mechanisms, is the interaction of the magnetic moment of the particle of interest with an external magnetic field.  Placing the material, assumed at first to be simple collection of non-zero spin particles, in a high magnetic field and cooling it to a low temperature induces polarization as particles tend to align themselves with the field. 

\begin{figure}[htb]
  \begin{center}
    \includegraphics[width=2.5in]{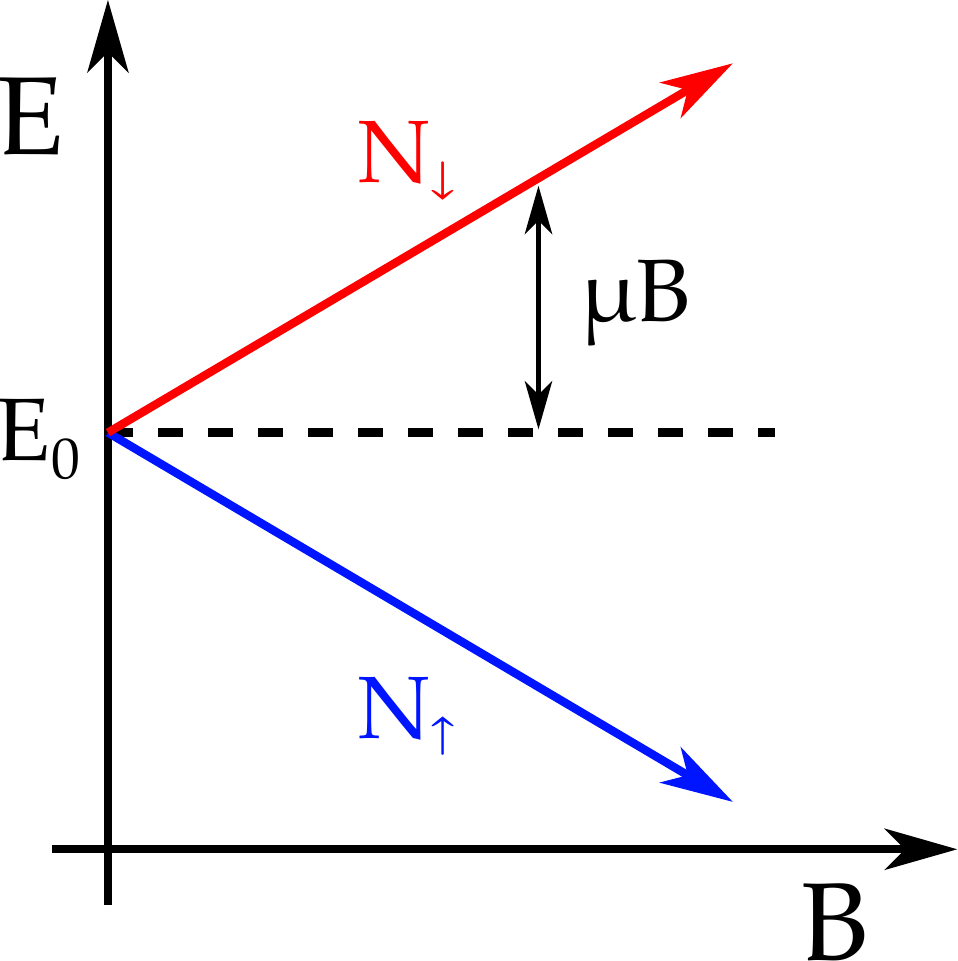}
  \end{center}

  \caption[Zeeman splitting of a spin-$\frac{1}{2}$ particle in magnetic field $B$.]{Zeeman splitting of a  spin-$\frac{1}{2}$ particle with magnetic moment $\mu$, in magnetic field $B$.}
  \label{fig:zeeman}
\end{figure}

A magnetic moment $\vec{\mu}$ in the external field $\vec{B}$ creates a set of $2J + 1$ energy sublevels via the Zeeman interaction (as in figure \ref{fig:zeeman} for spin $\frac{1}{2}$ particles), where $J$ represents the spin of the particle.  With statistical mechanics we can express the relative population of energy sublevels via the Boltzmann law:
\begin{equation}
N_1 = N_2 \cdot exp \left ( \frac{-\Delta E}{k_B T} \right )
\end{equation}
where $N_{1,2}$ are the population numbers of the sublevels, $T$ is the temperature and $k_B$ is the Boltzmann constant.  The energy of the Zeeman interaction is $\vec{\mu}\cdot\vec{B}$; thus for the case of spin-$\frac{1}{2}$ particles, the ratio of aligned to anti-aligned states, where now $N_1$ and $N_2$ become $N_\uparrow$ and $N_\downarrow$, is given as:
\begin{equation}
\label{eq:boltz}
\frac{N_\uparrow}{N_\downarrow} = exp \left ( \frac{2\mu B}{k_B T} \right ).
\end{equation}

The \textit{vector polarization} of the material, $P$, is a measure of the particle's spin alignment in the magnetic field.  Again considering the case of a spin-$\frac{1}{2}$ ensemble of particles, the vector polarization is given as:
\begin{equation}
P = \frac{N_\uparrow-N_\downarrow}{N_\uparrow+N_\downarrow}.
\end{equation}
This can be combined with equation \ref{eq:boltz} to give the polarization when the system is at thermal equilibrium:
\begin{equation}
P_{TE} = \frac{e^{\frac{\mu B}{kT}} - e^{\frac{-\mu B}{kT}}}{e^{\frac{\mu B}{kT}} + e^{\frac{-\mu B}{kT}}} = \tanh \left (\frac{\mu B}{kT} \right).
\label{eq:PTE}
\end{equation}

	Using equation \ref{eq:PTE} we find that the electron polarization in a 2.5T magnetic field and at 1K, for example, is approximately 92\%.  However, the magnetic moment of the proton is much smaller than that of the electron ($\mu_{e} \approx 660\mu_p$), which results in a far lower proton polarization of 0.25\% at 2.5T and 1K \cite{crabb97}.  As magnetic fields far beyond 2.5T and temperatures far below 1K are difficult to achieve, other mechanisms must be pursued to create high proton polarizations.

			\subsection{Solid-State Effect}
\begin{figure}[htb]
  \begin{center}
    \includegraphics[width=4in]{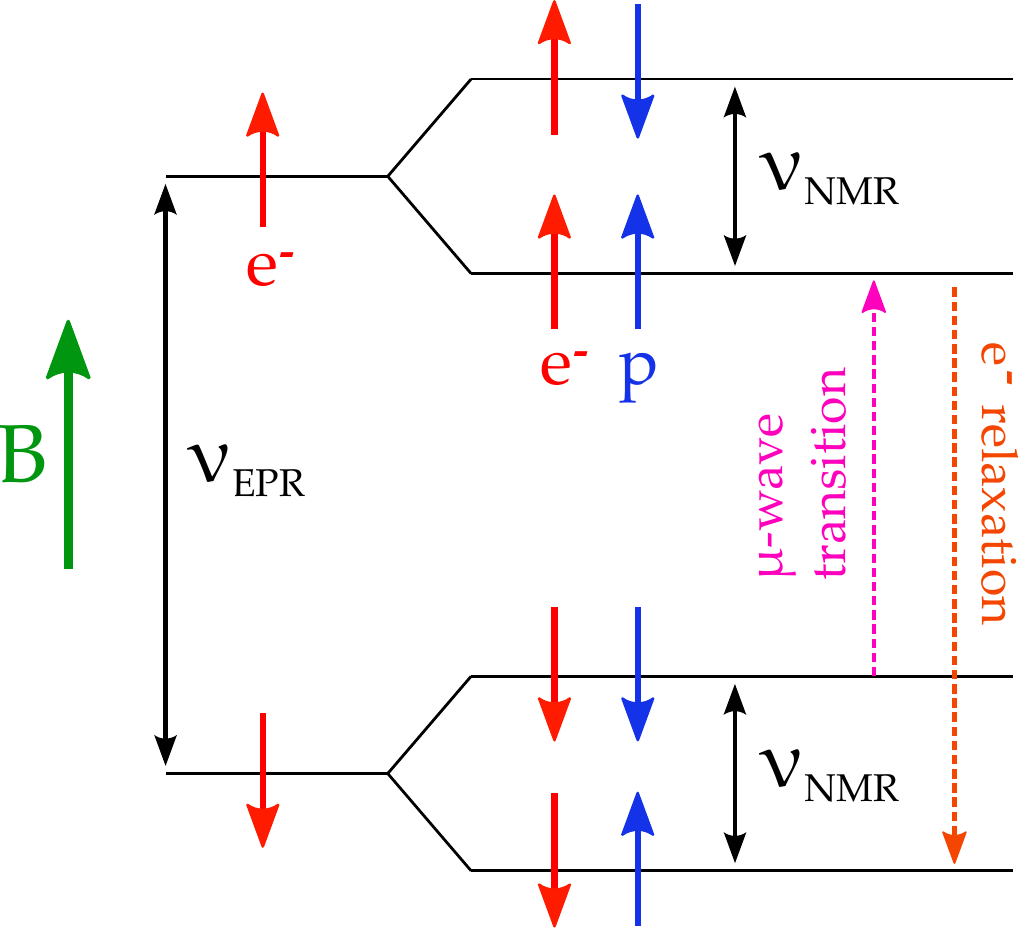}
  \end{center}

  \caption[Schematic of the solid-state effect.]{Schematic of the solid-state effect, showing the microwave driven transition to positively polarize protons.  Based on figure from \cite{crabb97}.}
  \label{fig:solid}
\end{figure}

	The solid-state effect is the simplest view of the DNP process in which microwaves are introduced to the thermal equilibrium polarization method.  In a target material with a suitable number of unpaired electron spins, hyper-fine splitting from the spin-spin interaction of the proton and electron in the magnetic field gives four discrete energy levels corresponding to the 4 permutations of aligned and anti-aligned spins, as in figure \ref{fig:solid}.  The Hamiltonian of such a system, which includes the spin-spin interaction term $H_{ss}$, is seen in equation \ref{eq:hamsolid}.  By applying an RF-field at the correct frequency, the coupled electron-proton spin system can be driven to preferentially fill the desired proton spin state.  
	
\begin{equation}
\label{eq:hamsolid}
H = \vec{\mu}_e \cdot \vec{B}\, +\, \vec{\mu}_p \cdot \vec{B} + H_{ss}
\end{equation}	
	
Electron spins are flipped by applying microwaves at the EPR (electron paramagnetic resonance) frequency, $\nu_{\textrm{EPR}}$, which corresponds to the Zeeman energy of the electron's magnetic moment in the given $B$ field---again $\vec{\mu}_{e}\cdot\vec{B}$.  Likewise the proton spins can be flipped by microwaves at the NMR (nuclear magnetic resonance) frequency, $\nu_{\textrm{NMR}}$, corresponding to the proton Zeeman energy $\vec{\mu}_p\cdot\vec{B}$.

Although simple dipole selection rules ($\Delta m_j = \pm 1$) forbid the simultaneous flipping of both spins, the spin--spin interaction term $H_{ss}$ creates mixing through which we access the previously forbidden transitions. The spins of the electron and proton then can be simultaneously flipped by applying microwaves of frequency higher or lower than $\nu_{\textrm{EPR}}$ by $\nu_{\textrm{NMR}}$.  Thus the transition $e_\downarrow p_\downarrow \rightarrow e_\uparrow p_\uparrow$ can be induced with microwaves at $\nu_\mu = \nu_{\textrm{EPR}} - \nu_{\textrm{NMR}}$.  The electron will tend to relax into the lowest energy state, $e_\uparrow p_\uparrow \rightarrow e_\downarrow p_\uparrow$, allowing it to be used to polarize another proton and making possible a continual driving of protons into positive polarization.  In the same manner, aligned protons can be anti-aligned ($e_\downarrow p_\uparrow \rightarrow e_\uparrow p_\downarrow$) using microwaves at $\nu_\mu = \nu_{\textrm{EPR}} + \nu_{\textrm{NMR}}$.  In this way both positive and negative proton polarizations can be achieved with the same magnetic field by altering the microwave frequency.

It is the relaxation times of the proton and electron at a given temperature which allow the polarization to continue to grow.  At 1K, the proton relaxes on the order of tens of minutes, whereas the electron's relaxation time is on the order of milliseconds (see section \ref{sec:relax} for a further treatment of spin relaxation).  The quick relaxation of the electron means it can be used to polarize a different proton.  This creates a rate of polarization higher than the rate of depolarization due to proton relaxation and allows polarization to be constantly built and maintained by microwaves.  This time development of the system can be reduced to linear rate equations \cite{abragam1983}.

The rate equations for such a case can be seen in equation \ref{eq:rateeq}, with electron polarization $P_S$ and proton polarization $P_I$. Here $N_S$ and $T_S$ represent the number of free electrons and their relaxation time, while $N_I$ and $T_I$ are the number of protons and their relaxation time.  $V$ is the probability of a proton and electron flipping due to the microwaves per unit time.  The superscript $L$ denotes thermal equilibrium polarizations, when no microwaves are driving the transitions.  As mentioned in the previous section, the thermal equilibrium polarization of the electrons, $P_S^L$, is much larger than that of the protons, $P_I^L$ \cite{Goertz2002403}.
 
\begin{equation}
\label{eq:rateeq}
\begin{split}
\frac{dP_S}{dt} &= -V(P_S - P_I) + \frac{1}{T_S}(P_S^L - P_S)\\
\frac{dP_I}{dt} &= \frac{N_S}{N_I}V(P_S - P_I) - \frac{1}{T_S}(P_I - P_I^L)
\end{split}
\end{equation}

Setting these equations equal to zero leads to the maximum polarizations:  

\begin{equation}
\label{eq:maxpol}
\begin{split}
P_S &= \frac{P_I^L \frac{N_IT_S}{N_ST_I} + P_S^L(\frac{N_I}{N_ST_IV} +1)}{\frac{N_I}{N_ST_IV}+\frac{N_IT_S}{N_ST_I} +1}\\
P_I &= \frac{P_I^L (\frac{N_I}{N_ST_IV}+\frac{N_IT_S}{N_ST_I}) + P_S^L}{\frac{N_I}{N_ST_IV}+\frac{N_IT_S}{N_ST_I} +1}.
\end{split}
\end{equation}

The condition of highest polarization occurs in the limit when $\frac{N_IT_S}{N_ST_I} \ll 1$, i.e. when the total electron relaxation rate, $N_S/T_S$, is much greater than that of the protons, $N_I/T_I$.  In this limit, the polarizations of the electron and proton systems are approximated by:
\begin{equation}
\label{eq:maxes}
\begin{split}
P_S &\approx P_S^L\\
P_I &= \frac{P_I^L\frac{N_I}{N_ST_IV}+P_S^L}{\frac{N_I}{N_ST_IV}+1}.
\end{split}
\end{equation}
If the induced transitions of the electron spins, $N_SV$, are much faster than the proton relaxation, as in $(N_ST_IV)/N_I \gg 1$, the upper limit of proton polarization reaches its theoretical maximum, the thermal equilibrium polarization of the electrons:
\begin{equation}\label{eq:pmax}
P_I \le P_S^L.
\end{equation}

			\subsection{Equal Spin Temperature Theory}
\begin{figure}[ht!]
  \begin{center}
    \includegraphics[width=5.2in]{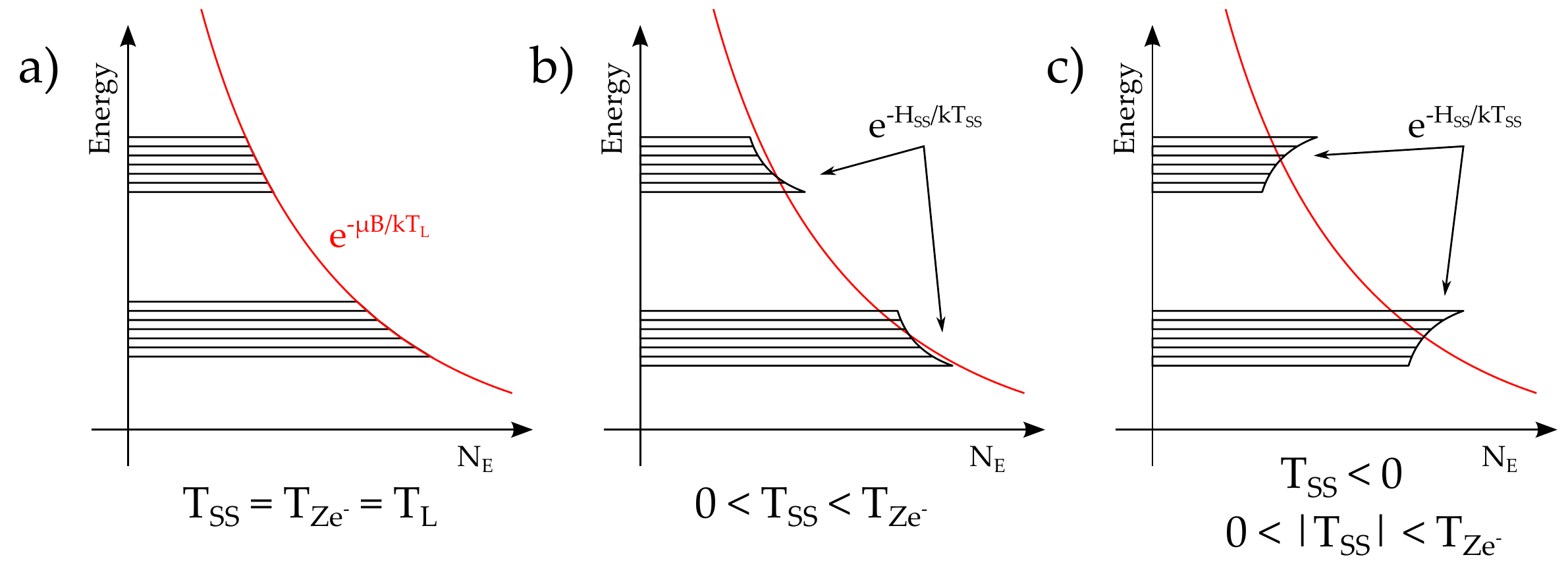}
  \end{center}
  \caption[Diagram of equal spin temperature theory]{Diagram of equal spin temperature theory, showing the system at thermal equilibrium (a), ``cooling'' the spin system to negatively polarize (b), and ``heating'' it to positively polarize (c).}
  \label{fig:est}
\end{figure}

	In most target materials in use today, the theory of equal spin temperature more accurately describes the DNP process\cite{borghini}.  The solid effect provides no facility to address the dipolar interactions between electrons which occur in materials with high electron concentrations.  While these electron spin interactions are weak in comparison to the Zeeman interaction with the external field, they create a band of quasi-continuous energy states, illustrated in figure \ref{fig:est}, which cannot be ignored in a realistic model\cite{crabb97}.  Borghini called this spin-spin mechanism the ``DONKEY effect,'' meaning ``dynamic orientation of nuclei by cooling electron interactions'' \cite{Borghini:1968gd}.
	
	The spin-spin interaction between electrons introduces a separate energy reservoir which is dependent on the Zeeman and lattice energies only through relaxation processes \cite{goldman1970}.  We describe the populations of these bands of energy states using a Boltzmann distribution with temperatures $T_{SS}$ of the electron spin-spin interaction reservoir and $T_{Ze}$ the electron Zeeman energy.  Likewise, the proton spin system is represented by a Zeeman reservoir, $T_{Zp}$.  Thermal equilibrium, seen in a) of figure \ref{fig:est}, occurs when $T_{SS}$ and $T_{Ze}$ are equal to the temperature of the lattice, $T_L$.   The Zeeman temperatures determine the population of the two bands of spin states, much like in the solid-effect, while the spin-spin temperature gives the distribution within the state.

In dynamic nuclear polarization, microwaves are used to change the spin-spin temperature $T_{SS}$, which in turn interacts with the proton Zeeman system $T_{Zp}$\cite{abragam1982}.  Microwaves of frequency slightly greater or less than that corresponding to the electron's Zeeman energy, $\nu_e$, are applied to the target material.  For microwave frequency $\nu_e+\delta$, energy $h(\nu_e+\delta)$ is absorbed, $h\nu_e$ by the electron Zeeman system and $h\delta$ by the spin-spin system.  For $\delta>0$, the spin-spin system absorbs energy, heating $T_{SS}$; $\delta<0$ causes the system to emit the energy, cooling $T_{SS}$.  This cooling can result in a negative spin temperature $T_{SS}$, which corresponds to a negative polarization.  Figure b) in \ref{fig:est} shows the cooling\footnote{Although we have referred to this ``cooling,'' the Zeeman energy is still increased at higher negative temperatures and thus negative polarizations; cooling is rightly defined by the reduction of the absolute value of the temperature\cite{Borghini:1968gd}.} of the spin-spin system and c) its heating.  

Thermal mixing between the proton Zeeman system and the spin-spin system results in heating or cooling of $T_{Zp}$ and thus polarization of protons.   $T_{Zp}$ is cooled (or heated) by a double spin flip of electron spins and an accompanying single flip of the protons\cite{pierce}.  The energy of the electron Zeeman spin system stays the same, but the proton system emits or absorbs $h\nu_p$, its Zeeman energy, to or from the electron spin-spin system.  Via this process $T_{Zp}$ will come into thermal equilibrium with the spin-spin system, $T_{SS}$.  The continued polarization of the proton system results from the maintenance of $T_{SS}$ by the microwaves.  The proton polarization is then given by the Boltzmann distribution of $T_{Zp}$:
	
\begin{equation}
 P = \frac{e^{\frac{\mu B}{kT_{Zp}}} - e^{\frac{-\mu B}{kT_{Zp}}}}{e^{\frac{\mu B}{kT_{Zp}}} + e^{\frac{-\mu B}{kT_{Zp}}}}
\end{equation}

			\subsection{Overhauser and Cross Effects}

		The \textit{Overhauser effect} is the primary DNP mechanism in metals\cite{overhauser}, the  first realized polarized targets.  Electrons in metals follow Fermi statistics, and the polarizing mechanism proceeds via the saturation of the material with microwaves at the electron frequency $\omega_e$, with $\hbar \omega_e = E_F^+ - E_F^-$ from the Fermi energies of the electrons in the metal.  This saturation destroys the electron polarization, which drives the nuclear polarization\cite{abragam1983}.  The relative nucleon spin population is then:
\begin{equation}
\frac{N_+}{N_-} = exp \left( \frac{\hbar(\omega_e-\omega_n)}{kT} \right).
\end{equation}		
		
		In the \textit{cross effect}, nuclear polarization originates from cross-relaxation transitions between electron spins coupled with the application of microwaves to saturate electron polarizations for electrons with resonance frequencies around the microwave frequency\cite{borghini}.  For two electronic spin packets which differ by a given frequency, a spin flip-flop can occur when a nuclear spin flips along with them to conserve energy.  By saturating one of the packets with microwaves at its Larmor frequency, the nuclear spins will preferentially fill one spin state, driving nuclear polarization\cite{abragam1982}.

			\subsection{Spin Diffusion}
		These processes all depend on nuclei in close proximity to a free electron to provide coupling.  We count on \textit{spin diffusion} to carry nucleon polarization away from the free electron impurity sites.  An excess of nuclear magnetization surrounding a polarizing free electron will tend to decay by a diffusion process mediated by the successive flip-flops of dipole-dipole interacting pairs of nucleons\cite{abragam1982}.  Since these flip-flops are energy conserving, they are very frequent, on the order of $10^4$ times per second \cite{Borghini:1968gd}, allowing the nuclear polarization to travel quickly throughout the material.
			
			\subsection{Spin Relaxation}
			\label{sec:relax}
		We have shown that the difference between the electron and proton relaxation times is crucial to continued proton polarization, so a brief explanation of the mechanism is warranted.  \textit{Spin-lattice relaxation} is the mediating process providing thermal contact between the lattice and spins, and it leads to the relaxation times which can vary so greatly---from $10^{-3}$ seconds to $10^6$ seconds depending on the case\cite{Borghini:1968gd}.  In general, the difference between the proton and electron lattice relaxation times is due to the coupling strength of each with the lattice, which in turn depends on their magnetic moments.
		
		The occurrence of a depolarizing ``flip'' without a corresponding ``flop'' to conserve energy, requires Larmor energy $\hbar \omega_S$ to come from or go to the lattice.  In the case of electron relaxation, this energy is absorbed or emitted in the form of a phonon to the lattice.  In DNP, free electrons are in the form of paramagnetic impurities of the ``Kramers'' type\cite{abragoldman}.  For Kramers ions, the relaxation rate is
		\begin{equation}
		\frac{1}{T_{1e}} = \eta \left(\frac{\omega}{v}\right)^5 \coth \left(\frac{\hbar \omega_S}{2kT}\right)\frac{\hbar}{\rho}
		\end{equation}
for velocity of sound in the crystal $v$, density $\rho$, and $\eta$, a dimensionless structure coefficient.  Typical values of these parameters result in $1/T_{1e} \approx 1000$, on the order of milliseconds.

		Nucleon relaxation is less likely to proceed in this way as a nucleon is weakly coupled to the lattice.  They instead relax through coupling with the electron spin-spin system via the same transitions we employ to polarize the nucleons with microwaves.  As these transitions are ``forbidden,'' they are far less likely---between $10^4$ and $10^6$  times smaller than electron relaxations\cite{Borghini:1968gd}.  Once a nucleon has flip-flopped with the electron via the forbidden transition, the relaxation can proceed to the lattice via electron relaxation.
				
		\section{Target Material}
		\label{sec:material}		
		The main considerations when choosing a target material to polarize via DNP are its maximum achievable polarization (as well as the rate at which it is achieved), its resistance to radiation damage caused by an experimental beam, and the prevalence of polarizable nucleons of interest in the material.  This presence of available nucleons for scattering, in our case protons, is quantified by the material's \textit{dilution factor}---the ratio of free, polarizable protons to total nucleons in the material.  The running time of the experiment $t$ depends on the luminosity $\mathscr{L}$ \footnote{Luminosity is the product of beam current and areal density of target particles in the target.}, the polarization achieved, and this dilution factor, $f$\cite{Meyer200412}.  To measure within a chosen accuracy $\Delta A$ of the measured asymmetry $A$, we expect
		\begin{equation}
		\label{eq:time}
		t \approx \frac{1}{f P^2 \mathscr{L} \Delta A^2}. 
		\end{equation}
The careful choice of target material with high maximum polarization and high dilution factor is thus crucial to achieving an accurate measurement.

Throughout section \ref{sec:introtarget}, we assumed a target material with a sufficient number of free electrons to provide coupling to the nuclear spins.  These free electrons take the form of paramagnetic radicals which must be introduced to a given material by doping.  The first successful DNP material was hydrated Lanthanum Magnesium Nitrate (La$_2$Mg$_3$(NO$_3$)$_{12}$ $\cdot$ 24H$_2$O), known as LMN, which was chemically doped with neodymium \cite{PhysRevLett.9.268} and could achieve above 70\% proton polarization.  Unfortunately, LMN proved a poor target, as its resistance to radiation damage was poor and its dilution factor was small.   From 1965 to 1971 Borghini, Mango, Sheffler and others at CERN tested over 200 materials in over 500 mixing ratios in a great search for new target materials\cite{masaike}, which led to chemically doped alcohols: butanol (with porphyrexide) and diol (with Cr$^{5+}$).  Today chemical radicals such as EHBA, a synthesized chromium radical, and TEMPO, a stable nitroxyl radical, are commonly used as dopants.

Niinikoski was the first to obtain substantial polarizations with ammonia doped with paramagnetic centers via irradiation\cite{Niinikoski1979141},  which produces radicals in the material via an ionizing particle beam.  Generally this is performed before an experiment in a smaller electron or proton accelerator facility, but it can be done using experimental beam as well, as long as the energy of the beam is ionizing.  Typical irradiation doses produce on the order of $10^{19}$ spins/ml, either via a ``warm dose'' between 80--90 K at a smaller accelerator, or via a ``cold dose'' in the experimental beam and cryostat at 1 K.  

Today irradiation doping is commonly used to create DNP target materials from ammonia ($^{14}$NH$_3$ and $^{15}$NH$_3$) and lithium hydride ($^{7}$LiH and $^6$LiH), as well as their deuterated counterparts  ($^{14}$ND$_3$, $^{15}$ND$_3$ and $^{7}$LiD).  In addition, the alcohols butanol and pentanol offer attractive deuterated forms.  Deuterated materials offer polarized deuterons in place of polarized protons to allow spin structure measurements on the neutron.

			\subsection{Ammonia as Target Material}
The material used during SANE, irradiation doped ammonia ($^{14}\textrm{NH}_3$), presents many attractive qualities.  After doping, ammonia can polarize to a high degree ($>$90\% at 1 K and 5 T \cite{crabb97}) quickly ($<$30 minutes), as opposed to the hours-long polarization cycle of lithium hydride.  It offers good radiation damage resistance---an order of magnitude better than chemically doped butanol---and a dilution factor of roughly 17.6\%.  
Ammonia containing a nitrogen isotope, $^{15}$NH$_3$, is used in scattering experiments which trade the higher dilution factor of $^{14}$NH$_3$ for a nitrogen atom with paired neutrons which do not polarize.

The desired paramagnetic centers in ammonia for use in DNP are atomic hydrogen and N$\overset{\bullet}{\textrm{H}}_2$ produced by the ionization of the NH$_3$ molecule.  At ``warm'' irradiations above 77 K, only N$\overset{\bullet}{\textrm{H}}_2$ radicals are produced, whereas atomic hydrogen is made below 4 K during a ``cold dose''\cite{materials}.   The production of radicals by irradiation give the normally colorless frozen ammonia beads a deep purple hue.  When the material is kept at 77 K in liquid nitrogen, these radicals can remain in the material for months to years\cite{Meyer198365}, though the color will fade to a pale violet.

			\subsection{Material Preparation}
			\label{sec:prep}
Ammonia is a gas at room temperature, in which state it is hazardous to breathe or to expose to the eyes.  As its melting point is 195.5 K, it is generally handled and stored under liquid nitrogen (77 K)\cite{Meyer200412}.  To produce ammonia beads usable as target material, ammonia is flowed into a sealed aluminum cylinder in a bath of liquid nitrogen (hereafter LN$_2$).  The ammonia freezes into a solid slug, which can be crushed through a series of mesh screens to form irregular beads of approximately the desired size (2 mm).  The size and shape of the beads must be a compromise between the need for cooling in experimental beam, and the desire for a high packing fraction of material.

Once the beads have been produced, the irradiation is performed at an electron accelerator to knock out protons from the NH$_3$ to form N$\overset{\bullet}{\textrm{H}}_2$ paramagnetic centers.  The material used during the SANE experiment was irradiated at the MIRF\footnote{MIRF: medical industrial radiation facility} at NIST in Gaithersburg, Maryland.  Early experimentation with irradiations of ammonia under a bath of LN$_2$ resulted in unexpected explosions, so the irradiations were performed under a liquid argon bath (LAr$_2$).  The main danger of this method is the production of radioactive chlorine gas.

\begin{figure}[hb!]
  \begin{center}
    \includegraphics[width=5.5cm]{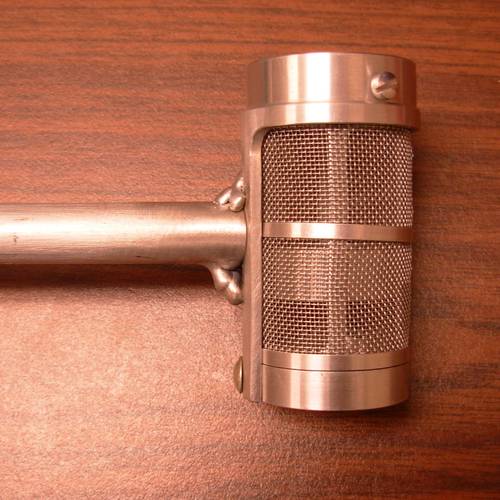}
  \end{center}
  \caption{Irradiation basket used during target material irradiation at NIST.}
  \label{fig:basket}
\end{figure}

Using the MIRF electron beam, electrons of 19 MeV struck the material under the 87 K LAr$_2$ bath at a beam current of between 10 to 15 $\mu A$.  The material was suspended in the bath in the aluminum mesh cup of an irradiation insert, seen in figure \ref{fig:basket}.   The mesh cylinder is 2.5 cm in diameter and 6.6 cm long. A mesh piston within allows variable volume, and the aluminum mesh lid locks by rotation for easy removal under liquid nitrogen.  After approximately 30 minutes of beam, the insert is rotated 180\degrees \,to allow even irradiation throughout the material sample.  This process is continued until a dose of approximately 10$^{17}$ e$^-$/cm$^2$ is achieved.

			\subsection{Performance in an Experimental Setting}
			\label{sec:ammoniaexp}
	The polarization performance of irradiated ammonia in an experimental beam follows three basic stages.  First, beam heating produces an immediate effect.  Second, excess radicals produced by the radiation dose of the beam cause a longer term decay in polarization.  While this polarization decay can be recovered by anneals, the third stage is the rapid increase of these decay rates after repeated anneals which indicate the end of the material's useful life.		
			
				\subsubsection{Beam Heating}
The first effect of the beam is an immediate reduction in DNP efficiency, and thus polarization, due to heating.  We recall from section \ref{sec:introtarget} that the maximum polarization of a material is limited by the thermal equilibrium polarization of the electron spins, which in turn depends on the material's temperature.  The experimental beam produces a heat load which cannot be entirely absorbed by the cryogenic systems maintaining the material temperature, which will be covered in section \ref{sec:cryogen}.	

In SANE, the CEBAF electron beam at around 100 nA produces roughly 500 mW of heat while passing through the target.  This heat load will generally produce a reduction in polarization of approximately 5\%.  This polarization reduction is easily visible on any graph of polarization, as beam trips result in the removal of the beam heat load.  The loss of beam allows the polarization to climb, but it will fall again once the beam returns.  This process generally produces many small spikes in a graph of polarization over time.

				\subsubsection{Radiation Damage}
The next effect of the experimental beam is an exponential decay of polarization due to radiation damage of the material.  As radiation dose from the experimental beam builds on a given material sample, further paramagnetic centers are created.  As more free electrons permeate the material, the careful balance between the electron and proton relaxation rates is upset.  More paramagnetic centers allow more relaxation paths through the forbidden transitions, and more relaxation paths leads to a higher proton relaxation rate which reduces DNP efficiency.  This reduction of efficiency with accumulated dose causes an exponential decay of the polarization over time.

\begin{figure}[hbt]
  \begin{center}
    \includegraphics[width=4in]{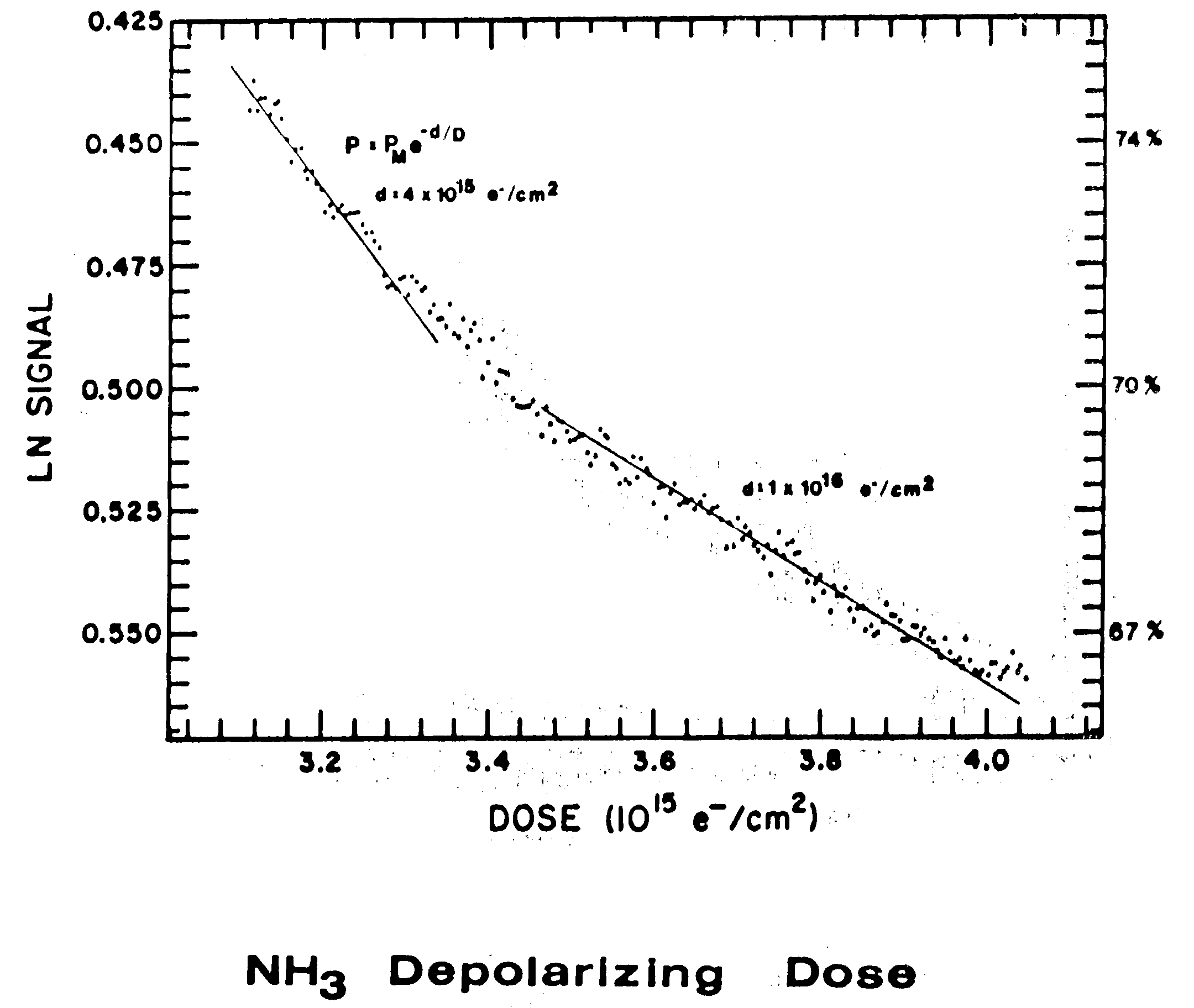}
  \end{center}
  \caption[Polarization decay in ammonia versus radiation dose at SLAC.]{Polarization decay in ammonia versus radiation dose in experimental beam at SLAC.  Two exponential decays are present with decay constants $d = 4 \cdot 10^{15}e^-/cm^2$ and $d = 1 \cdot 10^{16} e^-/cm^2$. From reference \cite{seely}.}
  \label{fig:seely}
\end{figure}
The decay of polarization with accumulated radiation dose can be approximated as two exponential decays, seen in data from SLAC in 1982 shown in figure \ref{fig:seely}.  These distinct exponential decays can be attributed to different radicals being produced in the beam.  In ammonia, the production of more atomic hydrogen and  N$\overset{\bullet}{\textrm{H}}_2$ radicals, the same as those used during DNP, creates more sites for polarization, but also relaxation.  Once an optimal number of these ``good'' radicals are created, the excess act to decrease DNP efficiency.  But other types of radicals can be produced which have slightly different EPR resonance frequencies than atomic hydrogen and  N$\overset{\bullet}{\textrm{H}}_2$.  These ``bad'' radicals cannot aid DNP using microwaves of the same frequency, and thus serve only to allow relaxation and depolarization.

				\subsubsection{Anneals}
The decay of polarization due to radiation damage will continue until the measurement time for a given accuracy, as mentioned in equation \ref{eq:time}, is unacceptable.  In $^{14}$NH$_3$, polarization can drop from above 80\% to 60\% after a dose of approximately 2 to 4 Pe$^-$/cm$^2$.\footnote{Hereafter Pe$^-$ indicates 10$^{15}$ electrons.}  In an experiment using a 7 nA beam current such as those in Hall B, this dose is reached in around 110 hours; for experiments using 100 nA beam current like SANE, this can occur in about 8 hours.  Fortunately, the process of \textit{annealing} allows the recombination of paramagnetic centers to restore polarization.  To anneal, the target material is moved out of the beam and the polarizing microwave radiation and is heated to between 70-100 K for between 10 and 60 minutes.  The increased temperature induces recombination of radicals, but care must be taken to avoid removing too many of the paramagnetic centers used in DNP.  Atomic hydrogen centers recombine at lower temperatures than  N$\overset{\bullet}{\textrm{H}}_2$, so these are the first to be removed\cite{demarco}.

In the Univ. of Virginia target used during SANE, anneals are accomplished using a coiled heater wire at the bottom of the target insert.  Current through the wire heats helium in the target space, which rises to convectively heat the material.  To provide rough control over the temperature in the material, the current in the wire is controlled by a PID loop which monitors the temperature in thermistors on the material cups.
				
				\subsubsection{End of Life}

\begin{figure}[hbt!]
  \begin{center}
    \includegraphics[width=3.5in]{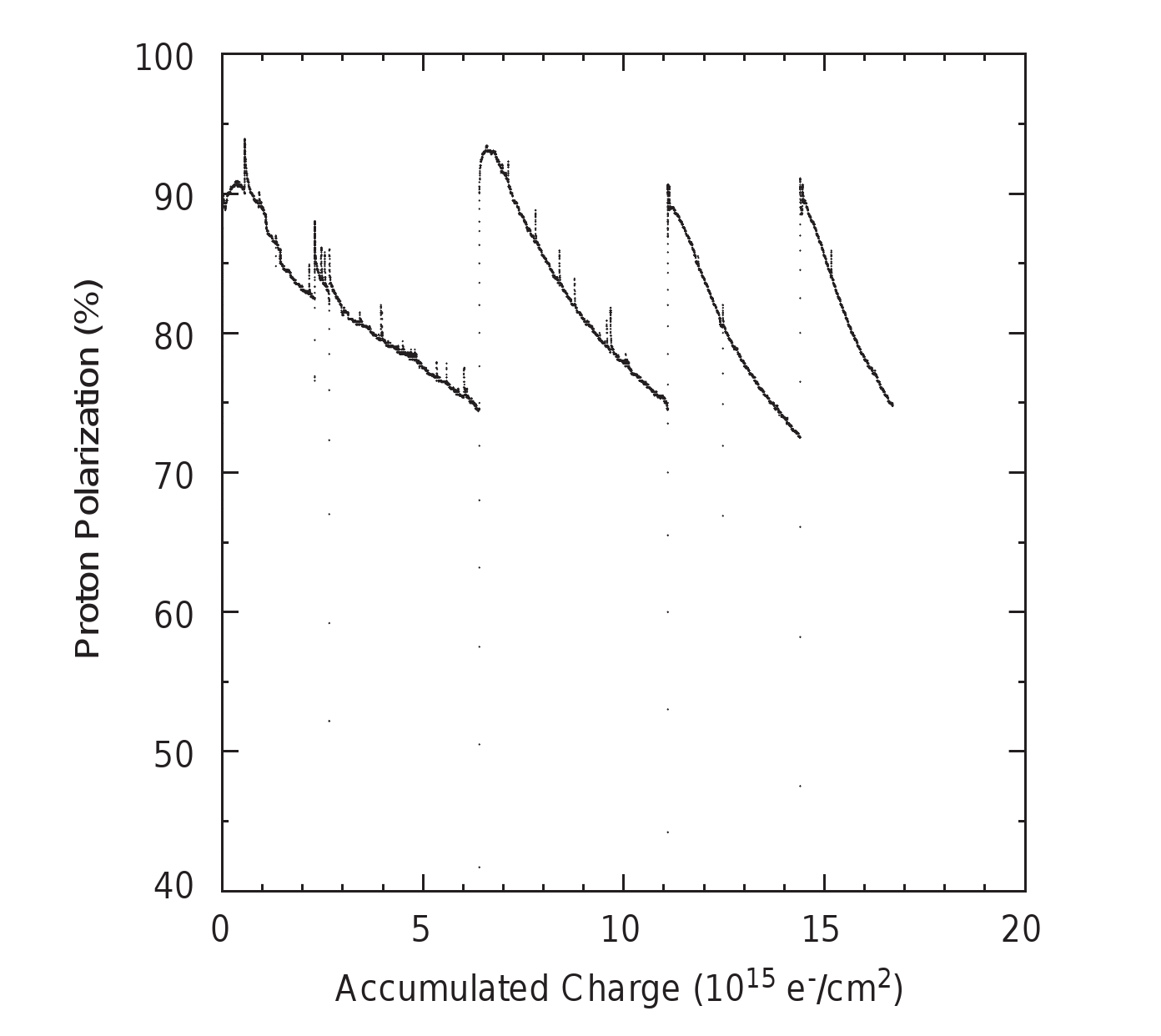}
  \end{center}
  \caption[Polarization decay with radiation dose from SLAC]{Polarization decay with radiation dose from SLAC E155, showing increase in decay rate after successive anneals. From reference \cite{McKee200460}.}
  \label{fig:mckee}
\end{figure}

While anneals allow polarization recovery for a given target material, the material sample still has a limited lifetime of total accumulated radiation dose.  After successive cycles of irradiation dose in the experimental beam followed by polarization recovery via anneals, the rate at which the polarization decays due to radiation will increase.	This material exhaustion is seen in figure \ref{fig:mckee} as the charge accumulated increases\cite{McKee200460}.  

The eventual exhaustion of the material is thought to be due to the creation of different, ``bad'' radicals.  Radicals such as hydrazine, N$_2\overset{\bullet}{\textrm{H}}_4$, can be formed from radicals produced in the beam, such as  N$\overset{\bullet}{\textrm{H}}_2$, when the material is heated during an anneal.  These hydrazine radicals recombine at higher temperatures than N$\overset{\bullet}{\textrm{H}}_2$ and atomic H, so once they are produced it is impractical to remove them without the loss of the wanted, ``good'' centers from the material\cite{demarco}.

		\section{Nuclear Magnetic Resonance Measurements}
		\label{sec:nmr}
Knowledge of the degree of target polarization is crucial to running and analysis of a double spin asymmetry measurement.  The same Zeeman splitting due to a particle's spin in the target magnetic field, which is leveraged in DNP, can also be used to query the polarization of spins in the material using a small, variable magnetic field from a coil embedded in the material.  This secondary field induces spin flips of the nuclei, and by observing the energy absorbed or emitted, a proportional measure of the polarization can be made.  In this section we outline the theory behind the resonance measurements used for DNP, as well as the method of measurement.		
					
			\subsection{NMR Theory}
As mentioned in section \ref{sec:introtarget}, a particle of spin $I$ placed in a magnetic field $\vec{B}$ results in $2I+1$ energy levels due to Zeeman splitting.  The separation between these levels is $\hbar\omega_L = \vec{\mu} \cdot \vec{B}/I = g\mu_nB$ for particle g-factor $g$,  particle Larmor frequency $\omega_L$ and magnetic moment $\mu$.  An RF field at the Larmor frequency of the particle can cause a flip of spin as it absorbs or emits energy interacting with the field.  The system's response to this RF radiation is its \textit{magnetic susceptibility}, $\chi(\omega)$, a function of RF frequency $\omega$.  When the RF in the coil creates a time-varying magnetic field perpendicular to the static target field, the magnetic susceptibility can be expressed as a difference of a dispersive term $\chi^\prime(\omega)$ and absorptive term $\chi^{\prime\prime}(\omega)$\cite{crabb97}:
\begin{equation}
\chi(\omega) = \chi^\prime(\omega) -  i\chi^{\prime\prime}(\omega).
\end{equation}

The absorptive portion of the magnetic susceptibility can be integrated over frequency to give a proportional measure of the polarization\cite{Goldman1975393}:
\begin{equation}
P = K \int_0^\infty{ \chi^{\prime\prime}(\omega) d\omega}. 
\end{equation}
for a constant $K$ containing information on spin species, spin density, gyromagnetic ratio and other NMR system quantities.  Generally $\chi^{\prime\prime}(\omega)$ is only non-zero in a small frequency range centered around the particle's Larmor frequency, so the integral only need be performed in this smaller range.

In the lab, this absorptive signal can be observed using an inductor, called an \textit{NMR coil} hereafter, embedded in or surrounding the target material sample, which creates a field with a component perpendicular to the target magnetic field.  The coupling between the spins in the material and the coil's magnetic field creates an inductance
\begin{equation}
L_C(\omega) = L_0 [1 + 4\pi\eta\chi(\omega)]
\end{equation}
with coil inductance $L_0$ for unpolarized material and  coil filling factor $\eta$, a function of the coupling of the coil to the material\cite{Court1993433}.  The coil's impedance is then
\begin{equation}
\begin{split}
Z_C &= r_C + i\omega L_C(\omega)\\
	&= r_C + i\omega L_0 \bigl[ 1 + 4\pi\eta\chi^\prime(\omega) - i4\pi\eta\chi^{\prime\prime}(\omega)\bigr]\\
	&= r_C + 4\pi\omega L_0\eta\chi^{\prime\prime}(\omega) + i\Bigl[\omega L_0\bigl(1+4\pi\eta\chi^\prime(\omega)\bigr)\Bigr]
\end{split}
\end{equation}
for coil resistance $r_C$.  A measurement which isolates the real part of this expression will give the absorptive component of the magnetic susceptibility and thus a proportional measure of the polarization.

			\subsection{Q-Meter Measurement}
			\label{sec:qmeter}
To integrate the real portion of the NMR coil's impedance over frequency, we create a series LCR circuit using a capacitor $C$, a damping resistor $R$, and the coil's inductance $L_C$.  By choosing the capacitance $C$ such that the circuit's resonant frequency is exactly the proton's Larmor frequency, $\omega_0 = \sqrt{L_0C}$, the power dissipated or generated in the circuit can be observed versus frequency using a \textit{Q-meter}.  In general, a Q-meter measures the quality factor of a given circuit, and Q-meters designed expressly for polarization measurements were developed at the University of Liverpool\cite{Court1993433}.

\begin{figure}[htb!]
  \begin{center}
    \includegraphics[width=4in]{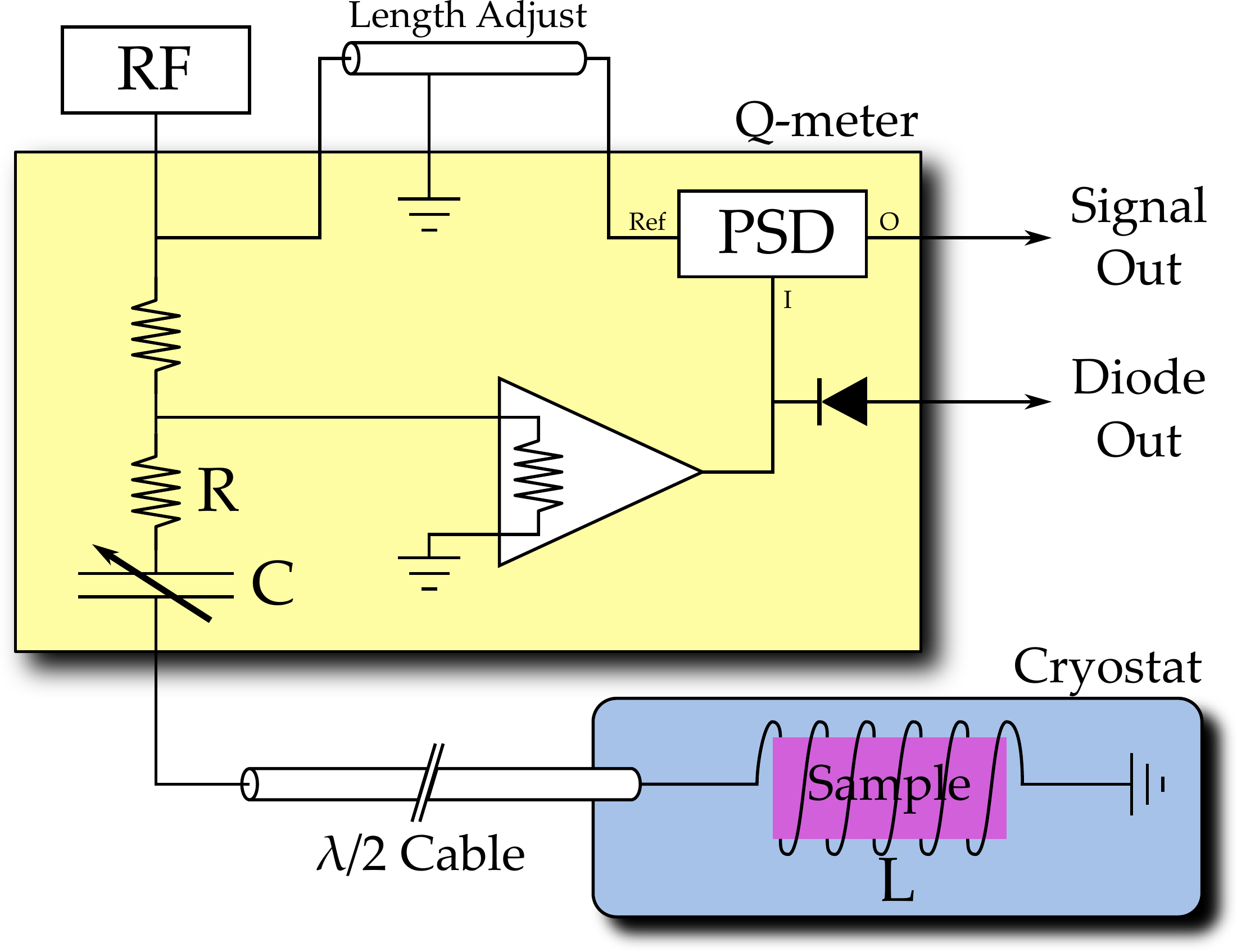}
  \end{center}
  \caption[Diagram of Q Meter circuit]{Diagram of Q Meter circuit showing RF generator, Phase Sensitive Detector (PSD), and LCR component with target material inside inductor coil.}
  \label{fig:NMR-circ}
\end{figure}

A schematic of such a Q-meter, like the one used during SANE, can be seen in figure \ref{fig:NMR-circ}.  After impedance matching, an RF generator drives the circuit, providing AC voltage which sweeps in frequency through the LCR circuit's resonant frequency.  This circuit is tuned to the Larmor frequency of the intended particle, which for a proton in a 5 T magnetic field is 213 MHz.  Current through the NMR coil drives the impedance signal, so it is crucial that it be independent of frequency.  This is achieved by using a high impedance amplifier to connect to the phase matching portion of the circuit.

The output of the amplifier is split to be sent to a full-wave diode detector for diagnostic output and to a \textit{phase-sensitive detector} (PSD), which is a \textit{balanced ring modulator}, (BRM) in the case of the Liverpool Q-meter.  This device accepts an input and a reference signal and outputs the input multiplied by the input's phase, relative to the reference signal.  The input signal comes from the amplified LCR circuit and the reference from the RF generator.  To measure only the real part of the input signal and thus the real part of LCR impedance, the reference signal must be adjusted so there is zero phase difference between it and the input signal.  This is accomplished by simply adjusting the length of the \textit{phase cable} which carries the reference signal.  The output of the BRM is then a proportional measure of the absorptive term, $\chi^{\prime\prime}$, and is amplified for data collection.

When the RF generator sweeps through the frequencies surrounding the circuit's resonant frequency, a \textit{Q-curve} is created.  This Q-curve is produced by the output of the BRM plotted against the RF frequency used as reference, and it represents the impedance of the circuit versus frequency.  With no polarization, the Q-curve is a parabola with its maximum at the resonant frequency of the circuit; this is the \textit{background signal} which depends on many quantities of the system.  For positive polarizations in the target material, impedance is increased around the Larmor frequency of the particle as spins absorb energy from the RF to flip from aligned to anti-aligned.  For negative polarizations, impedance is decreased around the Larmor frequency as spins emit energy while flipping from anti-aligned to aligned.  Integrating the dip or peak in the signal due to this absorption or emission by first subtracting out the background signal gives a proportional measure of the material's polarization.

During the course of the experiment, the target material must be subjected to extreme cold and high radiation, which can damage the electrical components.  During SANE this was addressed by locating the electronics outside the cryostat which held the target and thus NMR coil.  A semi-rigid cable provides connection between the electronics and the NMR coil, but a long transmission cable is susceptible to frequency dependent reflections which would critically degrade the signal.  To avoid these reflections, a standing wave can be created in the transmission cable by choosing a length of cable that is an integer multiple of the half-wavelength of the resonant frequency.  This cable is thus called a $\lambda/2$ cable.  However, we must remember we are sweeping in frequency, which will result in distortion as the frequency departs from the resonant frequency\cite{mckeethesis}.

\subsubsection{Thermal Equilibrium Calibration}
\label{sec:te}
Once we integrate the Q-curve, we have a proportional measure of the polarization which must be calibrated to give the true polarization.  To find this constant of proportionality, called a \textit{calibration constant}, we must measure the area of the Q-curve at a point of known polarization.  Fortunately, we recall from section \ref{sec:introtarget} that the polarization at thermal equilibrium is a known quantity: $P = \tanh \left (\frac{\mu B}{kT} \right)$.  By allowing the system to relax into thermal equilibrium at a given temperature, we know the polarization and can form a calibration constant using the Q-curve area which will allow us to correct Q-curve areas with polarizations that have been enhanced by DNP:
\begin{equation}
\frac{P_{\textrm{Enh}}}{P_{\textrm{TE}}} = \frac{A_{\textrm{Enh}}}{A_{\textrm{TE}}}\frac{G_{\textrm{TE}}}{G_{\textrm{Enh}}}
\end{equation}
for TE and enhanced (Enh) polarizations $P$, signal areas $A$, and amplifier gains used for each measurement $G$.

	\section{Target Setup and Equipment}
	\label{sec:targetsetup}
\begin{figure}[htb]
  \begin{center}
    \includegraphics[width=4.5in]{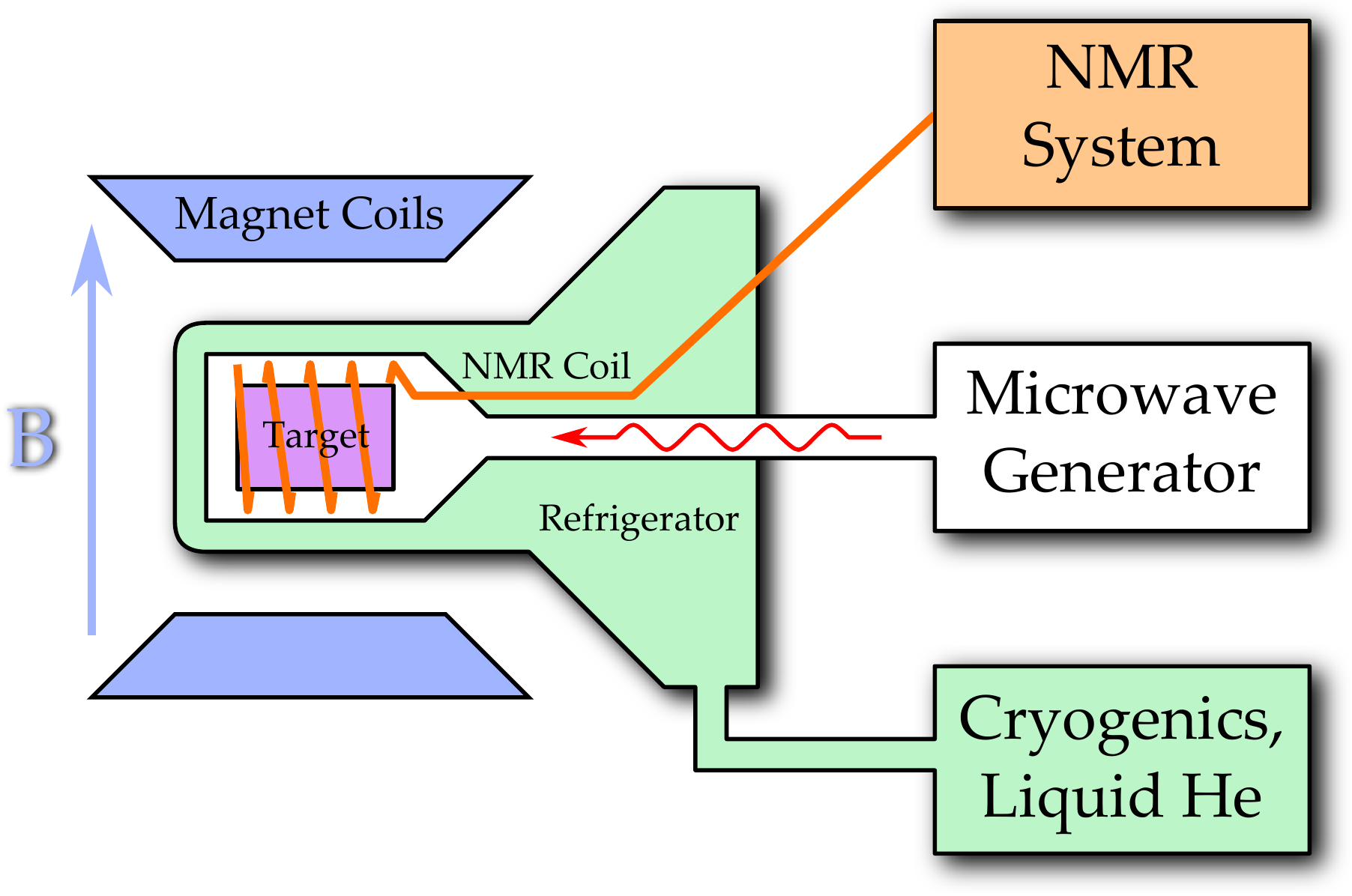}
  \end{center}
  \caption[Schematic overview of systems required for DNP.]{Schematic overview of the systems required for dynamic nuclear polarization.}
  \label{fig:DNP-over}
\end{figure}

	This section describes the application of the preceding techniques in the operation of the University of Virginia polarized target used during SANE.  An overview of the necessary systems is seen in figure \ref{fig:DNP-over}.  A superconducting Helmholtz pair magnet provided the 5 T field in a target region kept at 1 K by a liquid helium evaporation refrigerator.  Microwaves were provided by an Extended Interaction Oscillator (EIO), and the data acquisition electronics, including the NMR system, provided an online approximation of target polarization and recorded operating conditions.
	
\subsection{Cryogenics}
	\label{sec:cryogen}
\begin{figure}[]
  \begin{center}
    \includegraphics[width=6.0in]{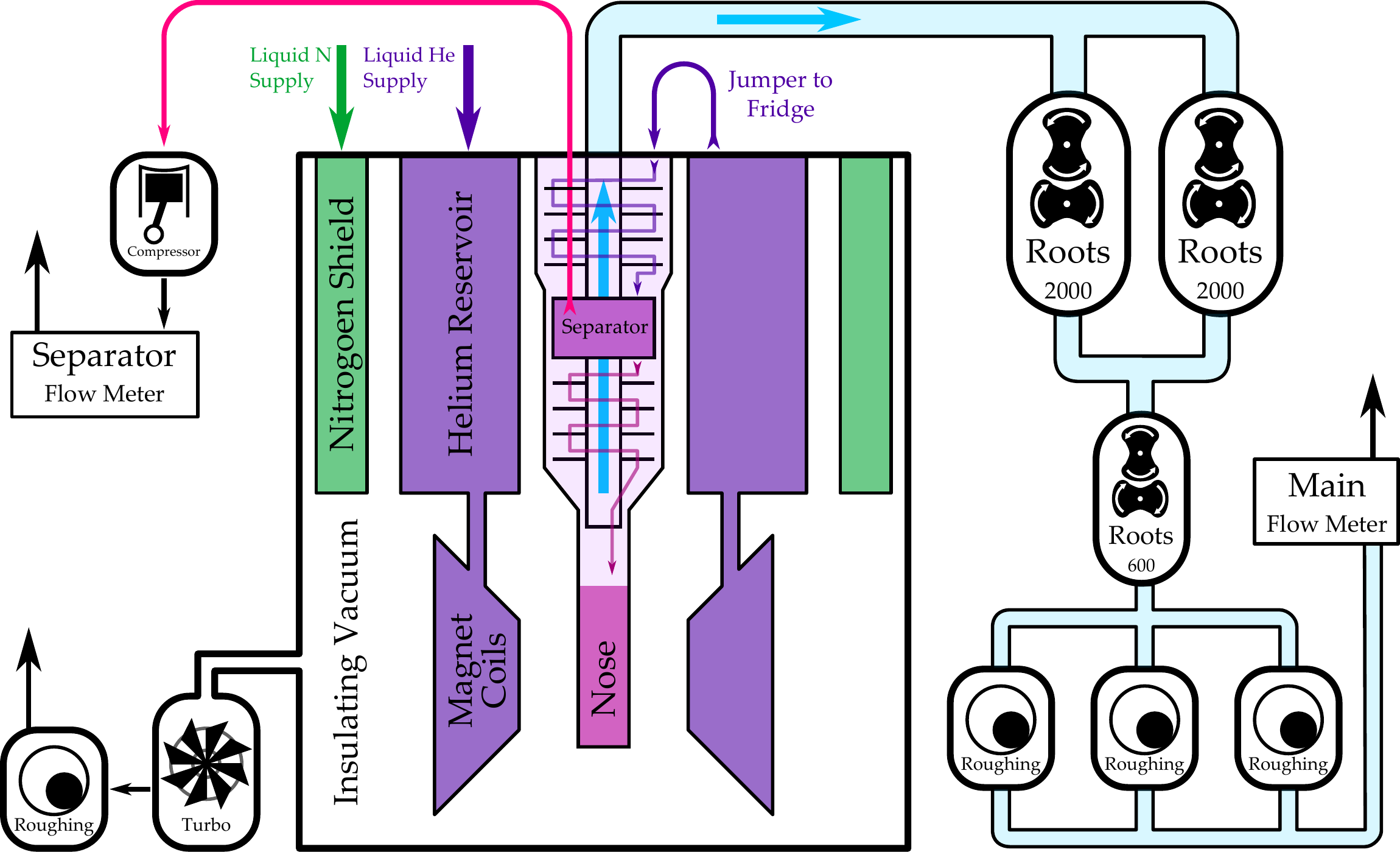}
  \end{center}
  \caption[Schematic overview of the cryogenic systems used for DNP.]{Schematic overview of the cryogenic systems used for DNP in the Univ. of Virginia test lab.  The experimental setup in Hall C was effectively the same, although the stray magnetic field meant a diffusion pump was used instead of a turbo, and Roots pumps of different capacities were used.}
  \label{fig:pumps}
\end{figure}

The efficiency of the DNP process depends strongly on extreme low temperatures.  An insulated cryostat is used in the Univ. of Virginia target which contains a $^4$He evaporation refrigerator fed by the magnet's liquid helium reservoir and insulated by a liquid nitrogen shield.  The system in use at UVa is seen in figure \ref{fig:pumps}. 

The target material is enclosed in an insert which extends into the \textit{nose} of the refrigerator, where a bath of 1 K liquid helium provides cooling.  The nose is supplied and cooled by the refrigerator above it.  Liquid helium is drawn from the magnet helium reservoir through an insulated jumper and flows through baffles which cool the liquid.  The helium then reaches the \textit{separator}, which serves as a reservoir of cooled helium to supply the nose.  Pumping on the separator draws the helium from the magnet reservoir into the separator, and the flow from this pump is monitored and recorded.  From the separator, helium can flow down into the nose through two valves.  The first, the \textit{run valve}, leads to piping which is thermally coupled to many layers of heat exchangers which serve to cool the helium further.  The second, the \textit{bypass valve}, leads directly down into the nose and is used to start the evaporation process.

The helium bath in the nose is pumped to low pressure to allow helium evaporation to cool the refrigerator.  As the helium evaporates, it is pumped up and out of the refrigerator, passing the many layers of heat exchangers and baffles to convectively cool them.  High capacity pumps are required to maintain low pressure in the nose; three Roots blowers were used in series and backed by rotary vane pumps to deal with the high flow rate.  This final flow rate out of the refrigerator through the pumps is also monitored and recorded.

The cryostat is insulated by a vacuum space which was pumped to approximately 10$^{-7}$ torr by a diffusion pump.  To prevent black body radiative heat loss to the 300 K outer walls of the cryostat, a reservoir of liquid nitrogen acts as a 77 K heat shield. 
	
\subsection{Target Insert}

\begin{figure}[htb]
  \begin{center}
    \includegraphics[width=1.5in, angle=-90]{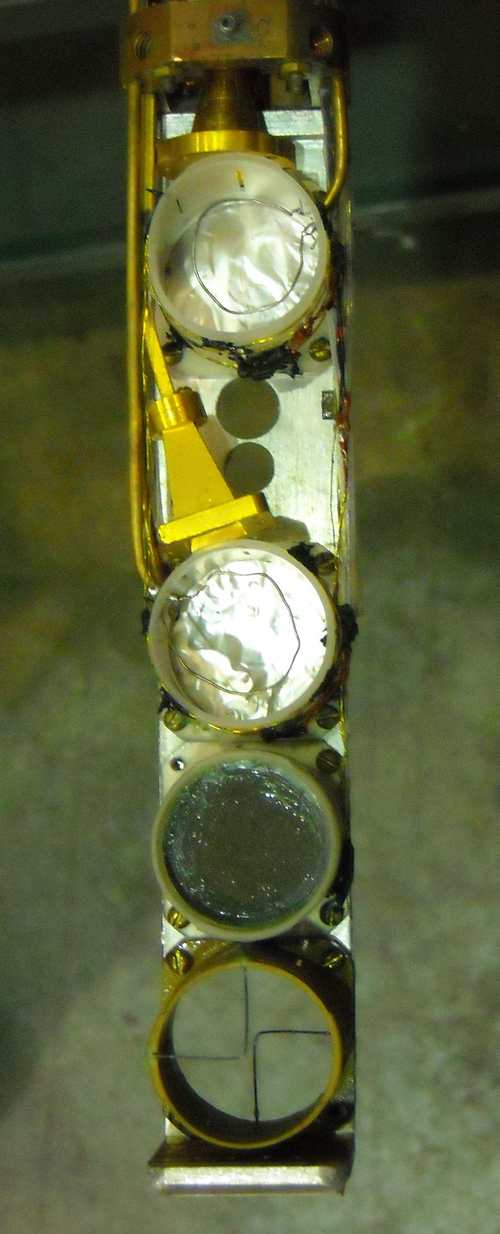}
  \end{center}
  \caption{Photograph showing the bottom of a SANE target insert.}
  \label{fig:insert}
\end{figure}

The target material was suspended in the magnet's uniform field region in the refrigerator's nose by the \textit{target insert}.  The insert is roughly 1.5 m long and provides storage for two target material samples in 
 2.5 cm diameter Kel-F target cups at the bottom, shown in figure \ref{fig:insert}.  In addition to two target cavities, there are spaces for a carbon disk and tungsten wire cross-hairs.  The inserts carry semi-rigid cable down to the NMR coils inside the target cavities, and microwave guides extend down to horns trained on each of the target material cup.  As anneals of the material require precise temperature data, the insert is equipped with thermocouples, platinum resistors and carbon-glass resistors in crucial locations.  Heater wire runs to the bottom of the insert to provide the heat needed to perform anneals, and the entire insert was raised and lowered by a mechanized lift to position the correct target cup in the beam.


\subsection{Microwaves}

The microwaves needed to drive the polarization enhancement in DNP were supplied by an Extend Interaction Oscillator (EIO) by CPI Canada.  The EIO itself sat directly above the target during the experiment, coupled to either target material cups by a switching junction and over-sized, CuNi wave-guides which terminated in horns to broadcast microwaves evenly over the cups.  The Varian microwave power supply sat in the shielded area of the Short Orbit Spectrometer hut, an unused Hall C spectrometer, and a remote control module was used to control the power supply from the counting house.

To measure EIO frequency and power output, a small portion of the microwaves were directed into an EIP frequency counter and an HP power meter.  An additional check of microwave power is available by monitoring the flow rate caused by helium boiling off due to the heat; the standard operating power for the EIO tube was less than 1 W.
	
\subsection{Magnet}
\label{sec:magnet}
The 5 T magnetic field was provided by a NbTi, split-pair, superconducting magnet built by Oxford Instruments in 1991. It provides a $10^{-4}$ field uniformity in a $3\times3\times3$ cm$^3$ volume, and this uniform field region may be tuned using shim coils.  The magnet is a veteran of experiments E143, E155, and E155x at SLAC, and GEN98, GEN01 and RSS at JLab.   The open geometry of the coils allows experimental beam to be directed at the target at both parallel and perpendicular to the magnetic field; this crucial quality allowed the measurement of perpendicular asymmetries providing access to g$_2$.  Directly above the superconducting coils is an 85 L reservoir of liquid helium and a ``donut'' of support electronics.  The donut also contains diodes which trip over a given voltage to dissipate the magnet's current into resistors to prevent damage to the coils in the event of a quench.

The magnet is powered by an Oxford IPS-120 power supply which provides careful regulation of current and voltage as the magnet is ``ramped'' up to the full 77 A to provide 5 T.  Voltage in the coils while the current in changed, or ``ramped,'' is related to the resistance, the change in current and the magnet's inductance: 
\begin{equation}
 V_C \approx  L_C\frac{dI}{dt}+I_LR_L.
\end{equation}
The resistance term is due to the non-superconducting leads which connect the power supply to the coils.  High voltage can cause arcing and damage of the coils; the quench protection is designed to trip above 7 V.  

The leads of the power supply are connected to the superconducting coils by a superconducting switch.  They attach to the coils on either side of a section of superconducting wire which is thermally coupled to a heater wire.  When current is driven through the heater wire, the heat causes the region of superconductor between the leads to become resistive.  For a sufficiently high resistance in the switch, the coils are effectively connected in series with the power supply.  When the heater is turned off, the switch again becomes superconducting; the coils are now persistent and the current from the power supply can be ramped down without affecting the magnet's current.  The magnet should stay in persistent mode, with no external power, for a very long time; the fractional current loss rate is on the order of $10^{-10}$ per day.

		\subsubsection{Magnet Failure \& Repair}
During SANE, the quench protection circuitry failed, and a quench caused damage to a superconducting joint.  Quenches are not uncommon occurrences as the magnet is ``trained'' to operate at its intended current after a long hiatus.  The first SANE quench occurred in the JLab EEL testing building before the experiment in June of 2008.  A rupture in the refrigerator nose allowed helium into the cryostat's insulating vacuum.  The vacuum loss allowed heat transfer to the magnet which led the coils to become resistive in a quench.  The second SANE quench occurred on October 31st and was possibly caused by ramping down the magnet power supply before the superconducting switch was completely cooled.  A problem in the GPIB communications buffer which remotely controls the power supply may have led to starting the ramp before the prescribed 30 second wait which allows the switch to fully cool.  The third quench occurred a few days later, after the magnet had been up at 5 T taking experimental beam on a CH$_2$ target for two days.  In this case, a target operator attempted to ramp the magnet down at too high a rate.  Although ramping limits are set into the firmware of the power supply, the magnet quenched as the rate increased from 1.5 A/m to 2.0 A/m at 60 A.

Tests of the magnet after this quench showed resistive elements in the superconductor, necessitating repair.  Inspecting the support electronics showed damage to a superconducting joint and wires connecting this to the quench protection circuit.  The quench protection circuit consists of pairs of diodes, with opposing bias, for each coil to shunt current into resistors. In this case one of the diodes had failed.  Although the other diodes worked properly, the current from this diode traveled to the neighboring diode, which caused the damage to the wires and joint.  Repairs were performed by Hall C technician J. Buffet and Oxford Instruments specialist P. Brodie.  The offending protection diode, a MBRP30045CT, was out of production and was replaced with an equivalent, MBRP40045CT.

After repair, the magnet operated at 5 T, but at compromised efficiency.  The first effect was a loss in persistence; the magnet current decayed at the rate of approximately 0.05\% per day.  This necessitated the current to be lifted occasionally to keep it at 5 T; the drift of the magnet current was visible in the shift of Larmor frequency of the protons as observed in the NMR signal.  The second effect was overall fragility of the magnet system, observed in frequent quenches throughout the experiment.  This fragility may have been from a combination of effects, including the rotation to perpendicular field, the introduction of iron shielding close to the magnet, beam heating and slight geometrical shift causing field instability.  The third effect was seen in voltage jumps and miniature quenches during ramping, which were later attributed to a failed shim coil heater switch.  If the shim coils were not connected to their power supply to dissipate current induced by the inductively couple main coils, these shim coils could quench, dumping induced current back into the main coils.

A thorough discussion of the magnet failure and repair is given in appendix \ref{sec:magapp} and reference \cite{maxwell}.

	\subsection{Data Acquisition}
The target data acquisition was centered around a purpose-built computer running LabView on Windows XP.  Two identical machines were built to provide a redundant backup, to guard against an eventuality which later occurred in the form of a hard drive failure of the DAQ machine in the experimental hall.  The computers ran Intel Core2 Duo processors, with 2 GB of RAM, and included 4 PCI card components to allow them to interface with the target equipment.  The GPIB card, a National Instruments PCI-GPIB, was used for communications with the RF generator, magnet power supply, and several temperature monitors; addresses for the two GPIB interfaces used for the target are shown in table \ref{tab:gpib}.  The MIO card, a National Instruments PC-MIO016XE-10, was used in combination with a BNC breakout box, a National Instruments BNC-2090, to serve as an ADC and DAC for the NMR system.  The DIO card, a National Instruments PC-DIO-96,  was used to control a switch box which provided the TTL signals required to control the Q-meter and gain selection.  The final card, a Measurement Computing DAS card, measured resistances and voltages on less time sensitive quantities---the ``slow controls'' monitor.

\begin{table}[bht]
  \begin{center}
\begin{tabular}{lcl}
\toprule
&Address & Device \\
\cmidrule(r){2-3}
\multirow{6}{*}{GPIB0 (GPIB-ENET)} & 2 & Keithley Voltmeter for Fridge Temps\\
 & 8 & $^4$He Manometer \\ 
 & 12 & Lakeshore Voltmeter for Insert Temps \\
 & 13 & Lakeshore Gaussmeter \\
 & 15 & MKS 670 $^3$He Manometer \\
 & 25 & Oxford IPS-120 Magnet Power Supply   \\ \cmidrule{2-3}
\multirow{2}{*}{GPIB1 (PCIe-GPIB)} & 3 & EIP Frequency Counter \\
 & 28 & Rohde \& Schwarz RF Generator \\ \bottomrule
\end{tabular}
\caption{Table of GPIB interfaces and addresses used for target data acquisition.}
  \label{tab:gpib}
\end{center}
\end{table}

Data acquisition equipment was located in three places in the hall while the experiment ran.  The defunct SOS hut provided radiation shielding and housed the data acquisition computer, microwave power supply, and BNC breakout box.  Long lines carried signals from these components out of the hut through the spectrometer's nose, where incident particles would enter the SOS when it was in operation, to the upper and lower platforms.  The ``upper platform'' surrounded the top of the target cryostat, and held the Q-meters, RF generator, microwave generator, and an oscilloscope for monitoring the Q-curve while tuning.  The ``lower platform'' held a table with a remote LCD monitor, mouse and keyboard controlling the data acquisition computer, as well as the magnet power supply and numerous cryogenic controls and monitors.

The online data acquisition software, originally written by P. McKee, is in National Instrument's LabView language, and it consists of many modules designed for individual tasks, which then intercommunicate via TCP messaging.  Individual modules include PDP, user control and display for most target tasks; SMC, superconducting magnet control; QCA, NMR curve acquisition; OLA, online signal analysis; EDL and EDR, logger and retriever for JLab's EPICS interface; SCM, slow controls monitor; TEB, target event builder; TMC, the target vertical position control; and TCL, local target data logger.
	
	\subsubsection{NMR Control and DAQ}
	\label{sec:nmrdaq}
\begin{figure}[ht!]
  \begin{center}
    \includegraphics[width=6.0in]{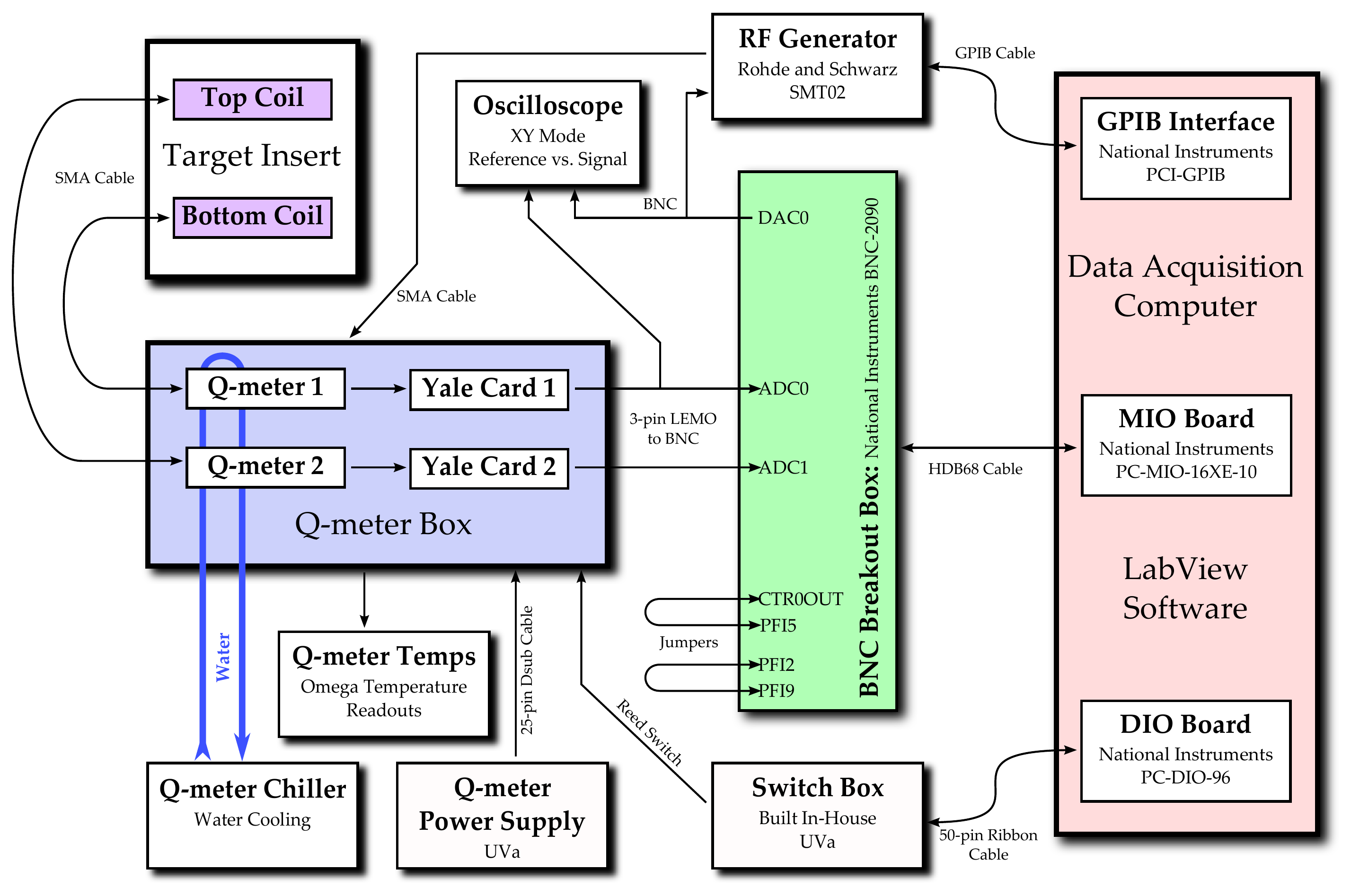}
  \end{center}
  \caption[Schematic overview of the DAQ systems used in the NMR system.]{Schematic overview of the data acquisition systems used in the NMR system to collect polarization and system status data during SANE.}
  \label{fig:daq}
\end{figure}

Shown in figure \ref{fig:daq} is a diagram of the components of the NMR control and data acquisition electronics.  Upon initialization, the LabView code communicated with the Rohde \& Schwarz RF generator (R{\&}S), setting the center frequency to be the Larmor frequency of the desired particle; in the case of SANE this was always 213 MHz for protons.  When the LabView software was switched into run mode by a user, the R{\&}S is set into ``RF On'' mode to broadcast RF following an external modulation signal.  This external modulation is provided by the MIO card via the BNC breakout box, which is in turn controlled by the QCA LabView module.  QCA generates a triangle wave of 1 kHz to provide the sweep through frequency.  The R{\&}S uses this external modulation signal to sweep linearly from 400 kHz below to 400 kHz above the proton's Larmor frequency.
	
The RF signal from the R{\&}S is connected to the NMR coil embedded within the material by $\lambda/2$ semi-rigid cable with a teflon dielectric.  The operation of the Q-meter system is discussed in section \ref{sec:qmeter}.  The resultant NMR signal from the Q-meter is amplified by approximately 1, 20 or 50 times by a Yale amplification card, before being sent to the ADC unit of the BNC breakout box.  From here the MIO card digitizes the signal.  The system performs a predetermined number of frequency sweeps, generally 500, averaging the signals to reduce noise.  The resultant averaged signal is then sent to the OLA module for fitting and integration.

\begin{figure}[hb!]
\begin{center}
    \includegraphics[width=6in]{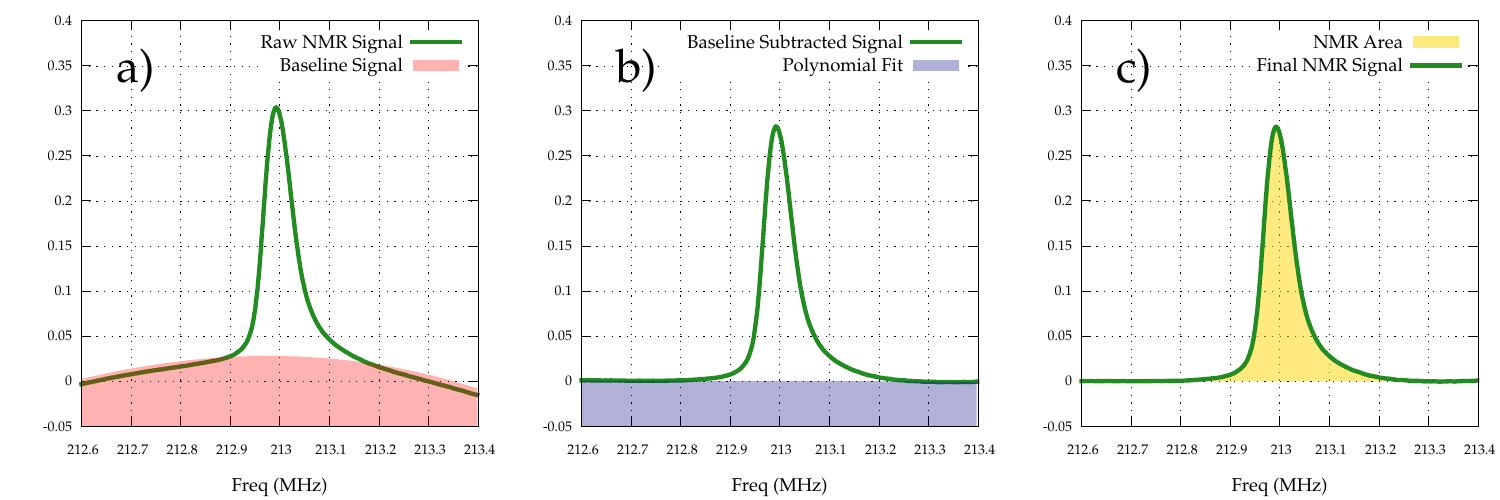}
    \end{center}
  \caption{Steps of NMR Signal Analysis, see text.}
  \label{fig:curves}
\end{figure}

The OLA module performs the signal integration as illustrated in figure \ref{fig:curves}, an example NMR signal from a negatively enhanced polarization on March 8th, 2009.  A background signal is first subtracted from the signal, as in a) of figure \ref{fig:curves}; this background is the Q-curve signal with the polarization signal removed, usually by moving the magnetic field and thus Larmor frequency, out of the range of the frequency sweeps.  The background signal is dependent on many target variables, such as small temperature shifts in and around the target electronics, so this subtracted signal generally still contains some ``background'' as the true background signal shifts.  To remove any residual background and isolate the area of the signal which is due to the polarization of the material, a polynomial fit is performed on the wings of the signal, as in b) of \ref{fig:curves}.  After subtracting this polynomial fit, the final signal can be integrated to produce the NMR area in arbitrary units, as in c) of \ref{fig:curves}. 

	\subsubsection{Cryogenic Data}
\begin{figure}[hb!]
\begin{center}
    \includegraphics[width=3in]{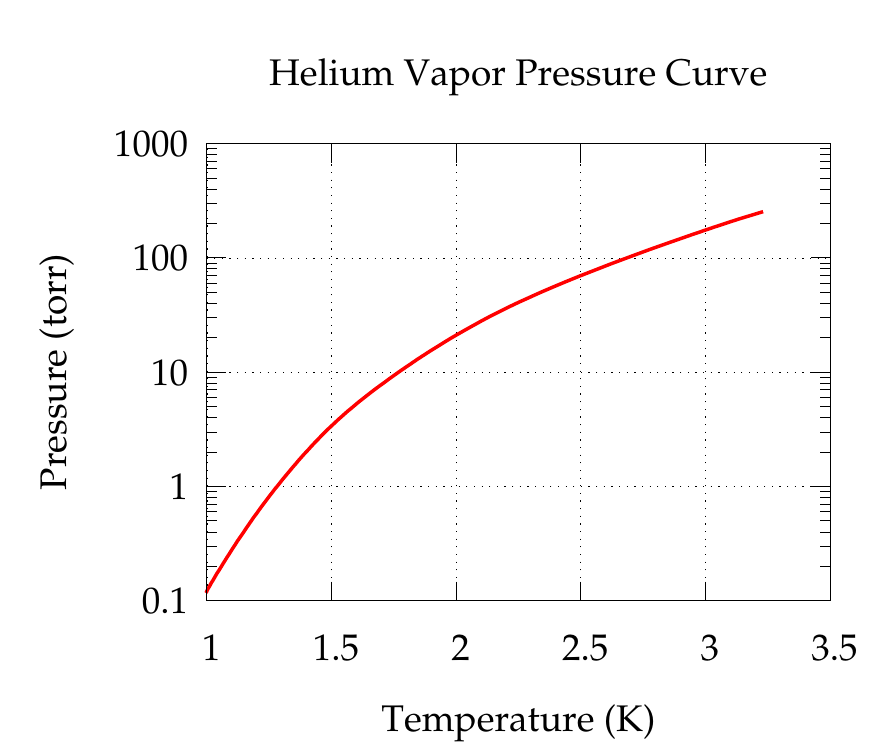}
    \end{center}
  \caption{Helium Vapor Pressure Curve.}
  \label{fig:helium}
\end{figure}

Accurate temperature and pressure information is crucial to target operation, particularly during the sensitive thermal equilibrium measurements when the temperature is used to calibrate the enhanced polarization measurements.	The primary measurement of pressure and temperature in the target nose was taken with a $^4$He manometer.  This manometer, which consists of a tube which extends from the refrigerator up to a measurement head outside the cryostat, gives the pressure in the refrigerator in torr.  When liquid helium is in the nose, the vapor pressure of helium can be used to calculate the temperature, as in figure \ref{fig:helium} reproduced from data from the Royal Society of London, 1941\cite{helium}.

As a check of the $^4$He manometer, a $^3$He manometer was used for much of the experiment, measured with the MKS 670.  The refrigerator included Allan-Bradley thermistors in various locations throughout, allowing observation as evaporative cooling is begun.  A level probe in the separator was not operational throughout the experiment, as was a level probe in the nose.  Despite an attempt to include a level probe on the target insert, the liquid level in the nose was measured only by temperatures on the insert for much of the experiment.

During anneals, temperature measurements are particularly crucial.  The temperature in each target cup was measured by a 100 $\Omega$ platinum resistor, 1000 $\Omega$ RuO$_2$ chip resistor in balanced resistance bridge, and gold/chromel thermocouples.  These were measured by an Oxford ITC-4, Lakeshore voltmeter and Keithley voltmeter.  

Unfortunately, it was difficult to control the temperature of both cups during an anneal.  In normal operation, helium is heated as current is passed through heater wire at the bottom of the insert.  As the terminus of the refrigerator's helium line is usually below the insert, the upward flow of heat is unimpaired.  During SANE, a leak in the separator of the refrigerator necessitated its replacement with a spare from Charlottesville.  To do this replacement without further delays, it was necessary to shorten the copper tube which carries liquid helium from the heat exchangers to the bottom of the nose.  This meant that during SANE, helium was blown at the target cups from above.  This change meant the bottom target cup, which was closer to the heater and further from the helium tube, could have a temperature as much as 20 K higher.

Most other cryogenic quantities, such as cryogenic liquid levels and pump flow levels, were measured by the JLab target group's equipment to be received into the UVa target data stream through EPICS\footnote{Experimental Physics Industrial Control System\cite{epics}}.
	
	\subsubsection{Data Storage and EPICS Reporting}
All target data from the UVa system was stored in proprietary LabView data files.  New data files were started after each anneal, and copies of the data files were simultaneously created locally and remotely on JLab's group disk.  A daily backup of local data was also performed to a USB hard drive using a timed Perl script.	

To ensure the polarization data was included in the experiment's overall data stream, upon completion of each set of sweeps salient quantities were sent to JLab's EPICS server.  In addition, crucial target quantities which were measured by JLab target group equipment were imported through EPICS to save in the UVa target data files.  A list of quantities sent and received through EPICS and their variable names are shown in table \ref{tab:epics}.
	
\begin{table}
  \begin{center}
\begin{tabular}{lll}
\toprule
& Variable & Quantity Stored \\
\cmidrule{2-3}
\multirow{11}{*}{Sent} & RF\_Freq & RF Frequency as sent to RF Generator \\
 & uWave\_Freq & Microwave Frequency as measured by EIP\\
 & Polarization & Online Polarization from NMR Area and CC \\ 
 & VPT\_3He & $^3$He Manometer Pressure \\
 & VPT\_4He & $^4$He Manometer Pressure \\
 & NMR\_Area & Area of Polysubtracted NMR Signal \\
 & Magnet\_Current & Current reported from Magnet PS \\
 & PT\_Encoder & Target Lift Encoder Value \\
 & PT\_Position & Target Lift Encoder Position Integer \\ 
 & Event\_Num & Target DAQ Event Number, based on Unixtime  \\ 
 & LabViewTime & Target Computer Timestamp for Event \\ 
\cmidrule{2-3}
\multirow{8}{*}{Received} & LL91111 & Liquid Helium Level in Magnet \\
& LL91112 & Liquid Helium Level in Nose \\
& LL91101 & Liquid Helium Level in Buffer \\
& LL91110 & Liquid Nitrogen Level in Shield \\
& FI91127 & Separator Flow Measurement \\
& FI91148 & Main Flow Measurement \\ 
& PI91131 & Insulating Vacuum Pressure \\
& ISD3H001G0AAD3 & Magnet Witness Field \\\bottomrule
\end{tabular}
\caption{Table of EPICS variables sent and received in SANE's target DAQ.}
  \label{tab:epics}
\end{center}
\end{table}

\section{Target Data Analysis}	
	\label{sec:targetanal}
This section outlines the analysis steps undertaken to produce final, ``offline'' polarizations, and lays out the results of these steps.  The majority of the analysis code was written by the author in Perl, with a crucial piece in the form of a C program originally written by P. McKee and updated by the author to access data saved in proprietary, binary LabView data files.  This section takes much from an internal, SANE technical note by the author \cite{maxwellpol}.  Although a good estimate of the target polarization is produced during the running of the experiment, the ``online polarization,'' it is necessary to reintegrate the NMR signals after the fact to produce ``offline polarizations'' which include corrections for Yale Card gain estimates and offline thermal equilibrium measurements, among others.

	\subsection{Yale Card Gains}
The first correction to be made in the offline analysis is the inclusive of the true Yale gains.  In the online calculation of polarization, the gain settings are taken to be ideal---exactly x1, x20 or x50.  In actuality, the gain factor of each Yale card is different.  The true Yale card gains are shown in table \ref{tab:gain}.  A thermal equilibrium measurement is generally taken at gain x50, as the signal is small.  In the case of gain card 9, the multiplication factor of the signal was 57.8273.  An enhanced signal is recorded at x1, a multiplication factor of 1.14927.  In an online polarization, the the ratio between the same area at x50 and x1 is taken to be 50.  To be accurate, it should be $57.8273/1.14927 = 50.3165$; thus this is an effect which can cause nearly 1\% change from the online to offline polarizations.
\begin{table}[htb]
  \begin{center}
\begin{tabular}{cccccl}
\toprule
Slot & Yale Card & Gain x1 & x20 & x50 & SANE Channel\\ \midrule
1 & 9 & 1.14927 & 23.4362 & 57.8273 & Top Proton \\
4 & 23 & 1.15129 & 22.2085 & 58.8023 & Bottom Protom \\ \bottomrule
	\end{tabular}
\caption{Table of Yale Card Gains Settings used during SANE.}
  \label{tab:gain}
\end{center}
\end{table}
	
		\subsection{Thermal Equilibrium Measurements}
		\label{sec:temeas}
As discussed in section \ref{sec:te}, the proportional measure of polarization measured in the area of the NMR signal is calibrated to an actual polarization using a thermal equilibrium measurement.  In practice, this involves carefully regulating the pressure and thus temperature in the refrigerator to keep the value as constant as possible.  The system must come into total thermal equilibrium before $P = \tanh \left (\frac{\mu B}{kT} \right)$ holds true.  The rate at which thermal equilibrium is reached is dependent on the temperature itself, so thermal equilibrium measurements (TEs) are taken at a higher temperature---around 1.6 K---to speed up the process.  At this temperature, a TE can still take longer than 2 hours.  Fortunately, both target material cups can be brought to equilibrium at once as long as they both remain in the liquid helium bath.

While the material comes into equilibrium, its NMR signal is recorded to give an indication of the extent it has come into equilibrium.  The signal area will change with the temperature; once the signal area is constant, thermal equilibrium has been reached.  After the experiment, each TE measurement is examined to ensure all the data points used to calculate the thermal equilibrium calibration constant for that measurement are within the period of constant temperature.
	
All successful thermal equilibrium measurements taken during SANE are shown in table \ref{tab:te}.  Table includes both online and offline calibration constants from the measurement, as well as the starting run that this calibration constant was used in the experiment.	
\begin{sidewaystable}
 \begin{center}
\begin{tabular}{lllllcc}
\toprule
Date		& Online Top CC	& Offline Top CC & Online Bot CC & Offline Bot CC 	& Top Start Run &	Bot Start Run\\ \midrule
2/1/2009	&	-3.032553 & -2.423104 &	-2.923663 & -2.335818 &	72162 &	72164	\\
2/8/2009	& Bad Coil	& NA & -2.372758	& -2.04475 & 72428	& 72378	 \\
2/11/2009	& -2.576022	&-2.297514& -2.634438	& -2.208803& 72428	& 72417	\\
2/14/2009	& -2.39848	&-1.852405& -2.634438	& -2.039955 & 72428	& 72417	\\
2/18/2009	& -2.348974	& -2.124947& -2.350482	& -2.000138& 72428	& 72417 \\
2/20/2009	& -2.775968	&-2.359297 & -2.273682	& -2.045262& 72669	& 72657	\\
2/25/2009	& -2.605546	&-2.204987 & -2.255362 &-2.045262& 	72669	& 72657	\\
2/26/2009	& -2.867905	&-2.455009 & -2.844311	&-2.605065 & 72837	& 72824	\\
3/2/2009	& -2.745752	&-2.361584 & -2.846276	&-2.562661 & 72837	& 72824	\\
3/6/2009	& -4.797562 &-4.652708 & 	-5.411936	&-4.203433 & 72913	& 72929 \\
3/9/2009	& -4.624355 &-3.982140  & 	-4.971622	&-4.104187 & 72913	& 72929	\\
3/10/2009	& -2.374633	&-2.035103 & -2.385056	&-1.956892 & 72986	& 72984	\\ \bottomrule

\end{tabular}
\caption[Table of online thermal equilibrium measurements.]{Table of online thermal equilibrium measurements during SANE, including their corresponding run ranges.}
  \label{tab:te}
\end{center}
\end{sidewaystable}

\begin{figure}[htb!]
\begin{center}
    \includegraphics[width=5in]{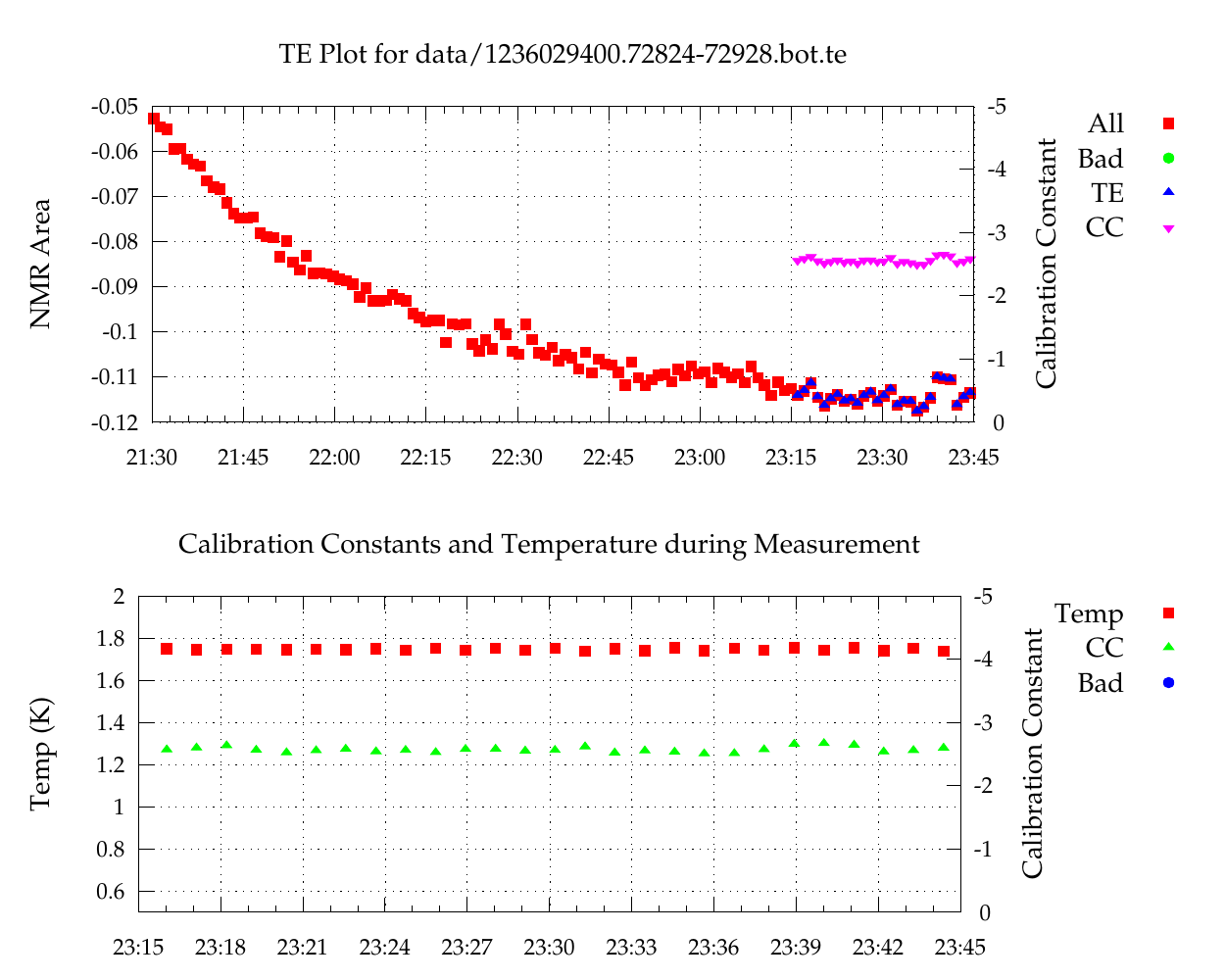}
    \end{center}
  \caption[Example thermal equilibrium measurement plot.]{Example thermal equilibrium measurement plot, showing TE decay, selected points, their corresponding calibration constants and temperatures.}
  \label{fig:tecurve}
\end{figure}

To ensure the quality of the points chosen to be included in each TE measurement, a plot such as figure \ref{fig:tecurve} is created for each measurement.  The title for the plot indicates that this was the TE that began at 1236029400 unixtime, to be applied to the material that was used in the bottom target cup during runs 72824 to 72928.   Ideally, these plots show the decay curve as the material reaches thermal equilibrium, shown in the figure in red.  The points which have been chosen to be included in the measurement should be constant in NMR area, shown in blue.  Appendix \ref{sec:teapp} contains all such plots for the experiment. 

Since thermal equilibrium measurements are the key to calculating true polarizations, as many TE measurements are taken per material sample as time allows.  One calibration constant (CC) for the material is then formed by averaging the calibrations, weighted by the statistical error in the measurement.  

		\subsection{NMR Signal Integration}
Once offline calibration constants and Yale card gains are determined, each NMR signal, technically an average of around 500 such sweeps in frequency, is re-analyzed and 	integrated.  This step is done using P. McKee's ``polcalc'' C program, which decodes the LabView data files, performs summing, and outputs data in plain text files.  The signal subtraction and integration is identical in process to that described in section \ref{sec:nmrdaq}.

		\subsubsection{Correction Due to Field Drift}
As discussed in section \ref{sec:magnet}, the current in the target magnet current decayed at a rate of about  0.05\% per day, causing a decrease in the magnetic field.  Although small, this decay caused the Larmor frequency of the proton to shift from 213 MHz at 5 T to 212.8 MHz at 4.995 T in just two days; as the sweep range is 800 kHz wide, a 200 kHz shift can be problematic.  The only effective change that this required was in the polynomial fit performed to the signal wings which subtract any residual background.

In online polarization calculation, the signal ``wings'' are assumed to be the channels 5-90 and 410-495 of the 500 channel signal.  Offline, using polcalc, the default is channels 3-125 and 375-497.  If the polarization peak of the signal encroaches upon these wings, the polyfit is no longer valid and creates an erroneous poly-subtracted signal and thus NMR area.

Generally, the magnet field drift necessitated the reconnection of the magnet power supply every 2 to 3 days to re-establish 5 T.  Unfortunately, the magnet's instability could result in a quench and hours of lost time each time this was attempted.  This meant occasionally the polarization peak would drift well into the wings of the polynomial fit before the magnet current was corrected.  During the first run of offline analysis, the difference between the  wing channel selection of the online and offline analyses was apparent as the polarization appeared to plummet in the offline compared to the online.  
\begin{figure}[htb!]
\begin{center}
    \includegraphics[width=6in]{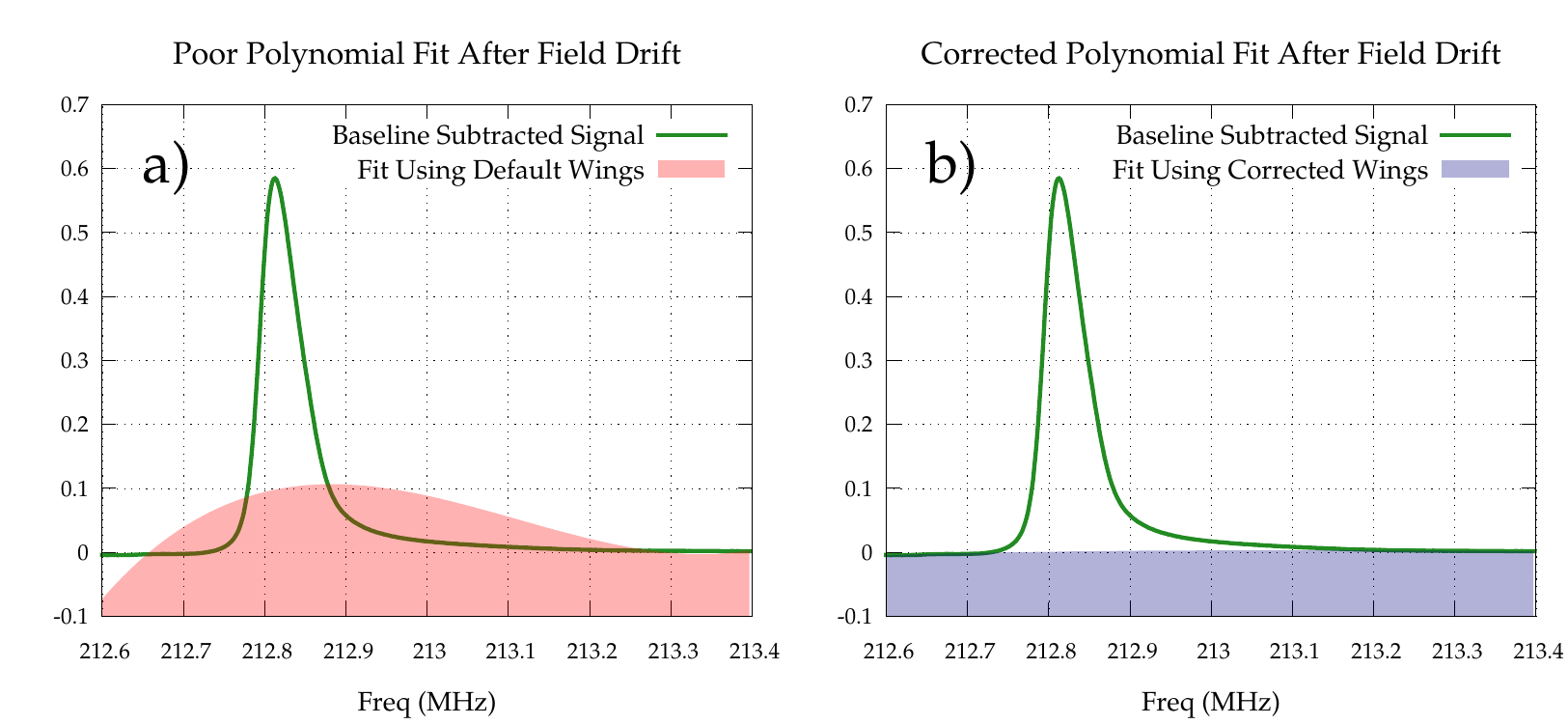}
    \end{center}
  \caption[Correcting NMR signal fit.]{Effect of magnetic field drift on polynomial fit to background and corrected polynomial fit.}
  \label{fig:drift}
\end{figure}

Upon further investigation, it was discovered that at several points during SANE, the standard signal wing definition would not produce a realistic background to the signal.  In an example NMR signal shown in a) of figure \ref{fig:drift}, the polynomial fit to the wings of the signal is erroneous, and will result in an inaccurate polarization measurement.  To remedy this problem, several different procedures were investigated, including a variable wing definition that began a given number of channels away from the peak, and fitting the peak itself to integrate directly.  The simplest and most effective method was to shrink the size of the left-hand wing included in the fit only in the case when the peak encroached the sides of the signal, thereby ensuring only the background portion of the signal was included in the fit.  In addition, the smaller wing definitions used in the online analysis were applied to the offline code.  The result of these changes can be seen in b) of figure \ref{fig:drift}.

\begin{figure}[htb!]
\begin{center}
    \includegraphics[width=6in]{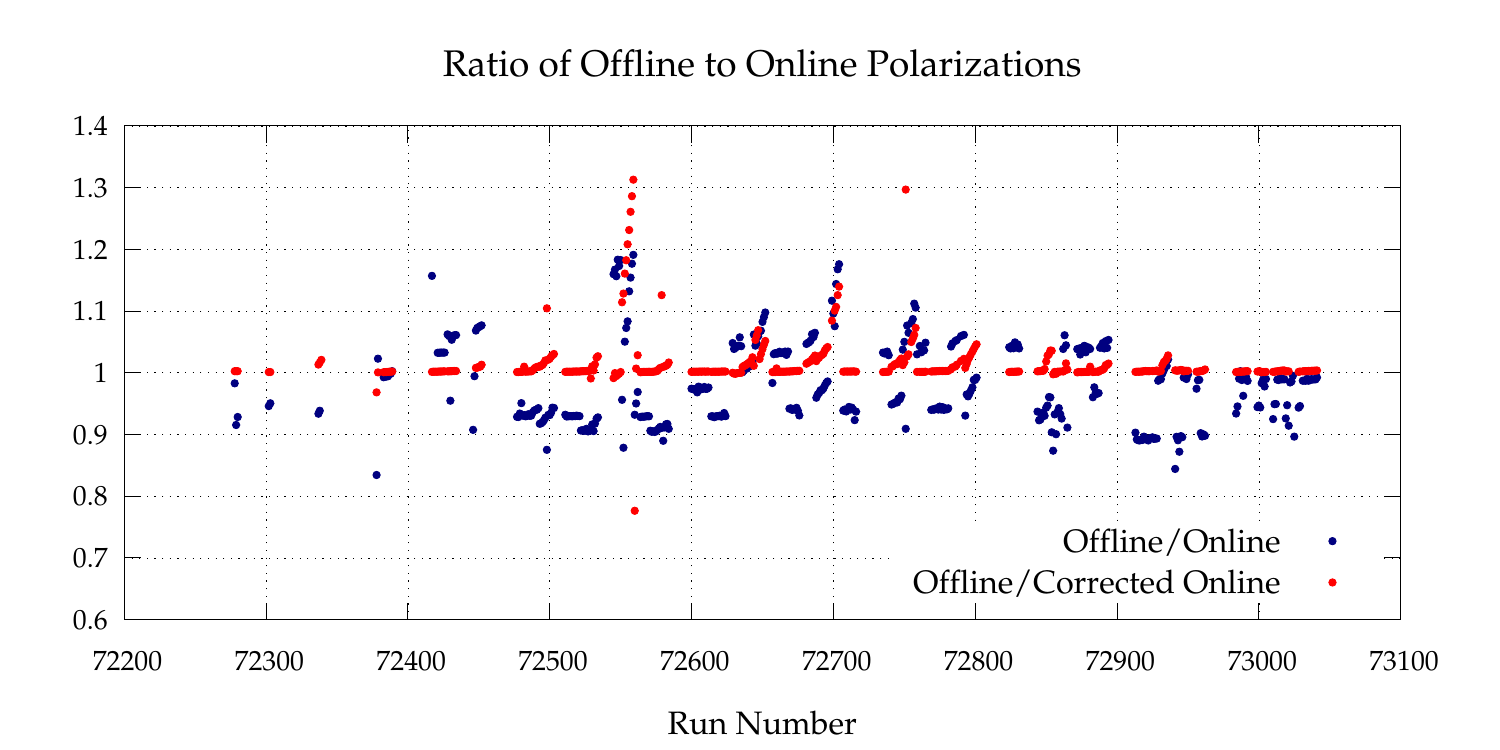}
    \end{center}
  \caption{Ratio of offline to online polarizations during SANE.}
  \label{fig:onoff}
\end{figure}

Shown in figure \ref{fig:onoff} are the overall changes between the online polarizations collected while SANE was running and the offline polarizations produced in the author's analysis afterwards.  In blue are data points representing the ration of online to offline polarizations.  The contributions to any difference from unity in this ratio are due to the gain correction, offline TE calibration constants and magnet field drift correction.  In red are ratios of offline to corrected online polarizations, where corrected online polarizations use online NMR signal areas, with offline calibration constants and Yale card gains applied.  Thus the red points isolate the effects of the magnetic field drift correction; this correction generally raised the absolute final polarization.  The ratio is seen to increase quickly as the polarization peak encroaches upon the signal wing area, but drops back to unity after the field is restored to 5 T.

		\subsection{Polarization per Run}
While the target polarization is measured over time, the change in the polarization is typically within the error of the polarization measurement over the course of a given experimental run.  In addition, as this is an asymmetry measurement to be extracted as a function of kinematics, any polarization change over the course of the run will average out.  It is sufficient then, to apply the charge averaged polarization over a run to all events in the run.		
		
		To create a file of all salient target data, including offline polarizations, over time for each SANE run, Perl scripts were used to call polcalc.  Run start and stop times were determined from ``EPICS scaler files,'' plain text files listing many experimental quantities over time written out by the replay.  To determine any offset in time from the EPICS server and the target DAQ computer, a script was written to compare online polarization values written by the target DAQ to EPICS (and thus having an EPICS timestamp) against the online polarization values written to disk with a LabView timestamp.  This offset was found to be 59 minutes and 40 seconds, until daylight savings time on March 9th, when it became 1 hour, 59 minutes and 40 seconds.  

Next a Perl script examines all events written to target data files to produce a database of which files contain data for which time frame.  Using the corrected run start and stop time, and a run list of which target cup was used for each run, polcalc is passed the correct baseline files and event file to extract and integrate all target data during the run.

To create one polarization value to apply to each run, the polarizations measured over time are charge averaged.  A Perl script reads an EPICS datafile with all measurements from the beam current monitors (BCMs), which are recorded every two seconds.  Taking into account the time offset between EPICS and LabView, the currents are averaged over the time between target events, about 30 seconds, and this averaged value for current is taken to be the current for the time period when that polarization is valid.  Finally, the currents are used to charge average the polarizations and calculate the charge accumulated on target for that run.  These charge averaged polarizations are the end product of the target analysis, to be used to produce corrected asymmetries per run.  Charge averaged polarizations for all SANE runs are discussed further in section \ref{sec:offpols}.

\section{Polarized Target Results}	
\label{sec:results}	
Despite near catastrophic failures of both the target superconducting magnet and cryogenic refrigerator, over 300 hours of beam were taken on polarized ammonia during the course of SANE.  Eleven different ammonia target loads were used, requiring 7 material changes, 23 thermal equilibrium measurements and 26 anneals.  This section discusses the target performance and results of this portion of the analysis.

		\subsection{Thermal Equilibrium Results}
\begin{figure}[htb!]
\begin{center}
    \includegraphics[width=5in]{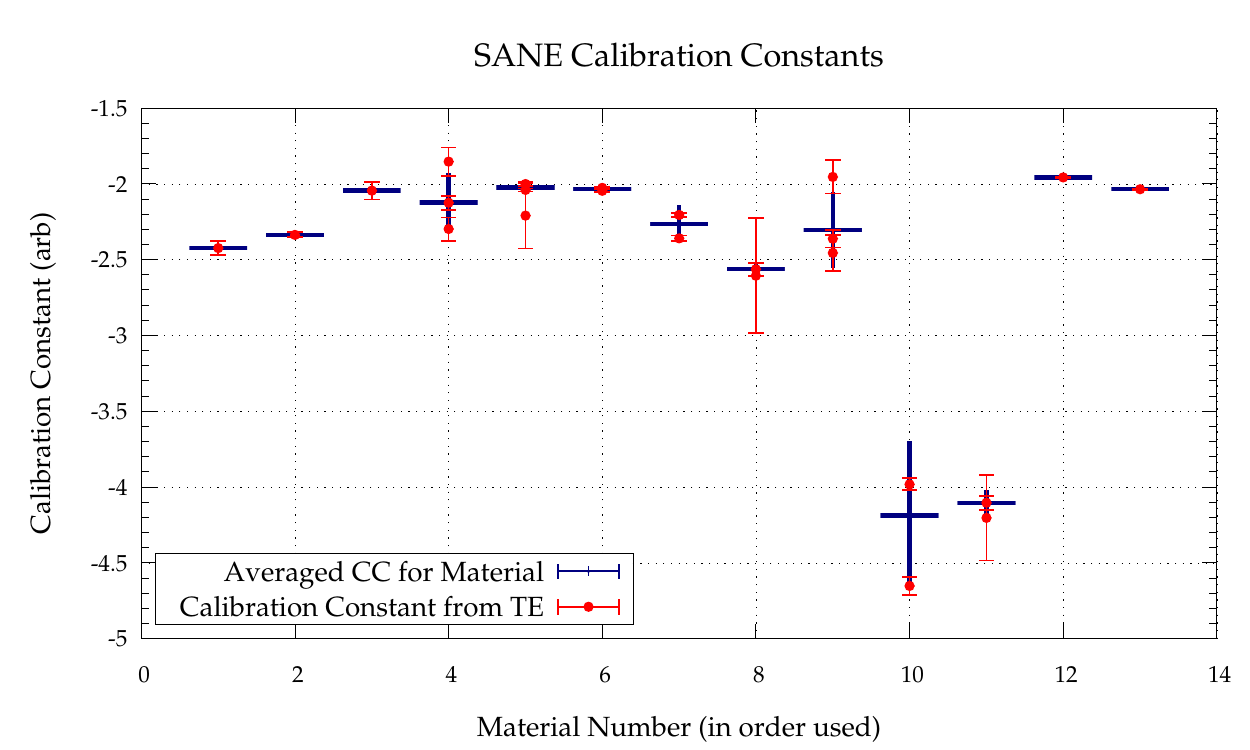}
    \end{center}
  \caption[Calibration constants used during SANE.]{Calibration constants and their averages for each target material sample used during SANE.}
  \label{fig:te}
\end{figure}

In figure \ref{fig:te}, all ``13'' target loads\footnote{Eleven loads were used, with two of these used for both target field configurations, thus requiring 2 separate calibration constants for each.} used during the experiment are shown with all calibration constants on the material, as well as the averaged calibration constant used for all polarizations from that material.  Error bars of the individual CCs are statistical, and are due to the number and distribution of the area and pressure measurements that went into the CC calculation.  The error bars on the averaged CCs are the weighted standard deviations of the CCs that went into the calculation of the average.  The final error on the polarization will be determined using this scatter of the CCs around their average.  In theory, the CCs of any given material sample should not change, and thus any large difference between CCs indicates uncertainty.  Materials with only one CC will use an average of the errors of other materials.

The most obvious feature in the figure \ref{fig:te} is the sudden drop for the calibration constants on materials 10 and 11.  These two materials have calibration constants nearly a factor a two larger than the others, but this is not unexpected.  These two ``materials'' are actually the same material samples as 8 and 9; the samples were not fully exhausted by radiation damage when the time came to rotate the magnet to switch the direction of the field from perpendicular to the beam to parallel.  To save time, materials 8 and 9, from the top and bottom target cups in the insert, were removed from the cryostat and stored in liquid nitrogen until the rotation was complete.  Then the insert was returned to the cryostat without disturbing the ammonia samples; they were dubbed materials 10 and 11 in their new role.  While this saved precious time, it meant that the NMR coils---which were oriented vertically in the target cup---had their axial component, and thus their induced magnetic field, near parallel to the target magnetic field.  As mentioned in section \ref{sec:nmr}, the magnetic susceptibility is probed by a time-varying field \textit{perpendicular} to the static target magnetic field.  Since this time-varying field was near parallel for these materials, the NMR signal was much fainter.  Hence the larger calibration constant.

		\subsection{Material Performance}
The ammonia samples used during the course of the experiment are listed in table \ref{tab:materials}, which shows each sample's averaged calibration constant as well as its total charge accumulated in the beam.  A sample lasted on average about 11 Pe$^-$/cm$^2$, although this average is skewed to the low end by a few materials removed early due to poor performance.  We reiterate that 8 and 11, as well as 9 and 10, are the same material sample used at two different magnetic field configurations.  The charge accumulated here was calculated using BCM1.
	
\begin{table}[htb]
  \begin{center}
\begin{tabular}{lllcl}
\toprule
Sample & Position & Run Range & Calibration Constant & Charge (Pe$^-$/cm$^2$)\\\midrule
1  & Top & 72162--72427 & -2.945048 & 3.8\\
2  & Bottom & 72164--72377& -3.015994 & 4.4\\
3  & Bottom & 72378--72416& -2.044750 & 2.0\\
4  & Bottom & 72417--72656&-2.122256 & 19.7\\
5  & Top & 72428--72668& -2.023154 & 22.9\\  
6  & Bottom & 72657--72823& -2.032478 & 12.7 \\
7  & Top & 72669--72836& -2.263753 & 16.4 \\ 
 \multirow{2}{*}{8 \& 11}  & Bottom & 72824--72928& -2.563189 &\multirow{2}{*}{11.3} \\
  & Bottom & 72929--72983& -4.106710 & \\  
 \multirow{2}{*}{9 \& 10}   & Top & 72837--72912& -2.303744 & \multirow{2}{*}{12.5}\\
  & Top & 72913--72985& -4.187268 &  \\
12  & Bottom & 72984--73029& -1.956892 & 5.5\\
13  & Top & 72986--73014& -2.035103 & 11.0\\
 \bottomrule
	\end{tabular}
\caption[Table of ammonia samples used during SANE.]{Table of ammonia samples used during SANE, showing run range and position, as well as calibration constant and total charge accumulated on the material.}
  \label{tab:materials}
\end{center}
\end{table}

			\subsubsection{Material Lifetime}
			\label{sec:life}
An example material lifetime is shown in figure \ref{fig:life}\footnote{Plots representing all material samples available in appendix \ref{sec:applife}.}.  Here negative target polarization are red points and positive as blue, and vertical gold lines represent the anneal of the material. 
\begin{figure}[htb!]
\begin{center}
    \includegraphics[width=5in]{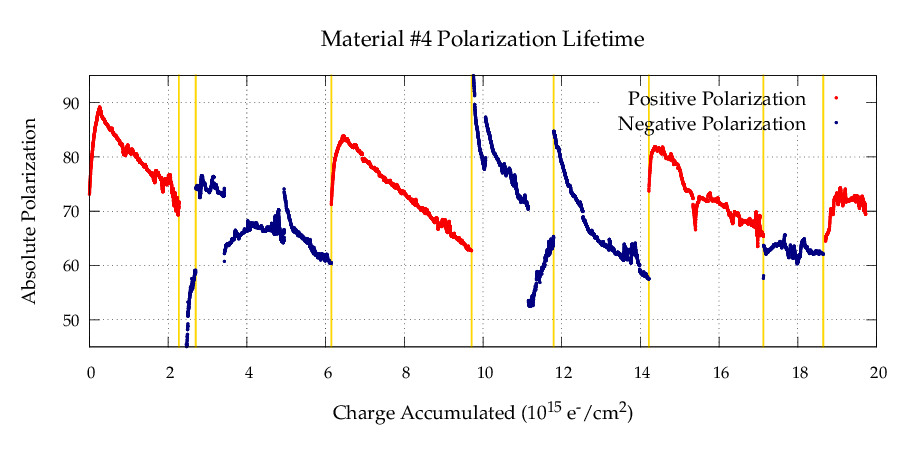}
    \end{center}
  \caption[Example material lifetime in total charge accumulated.]{Example material lifetime in total charge accumulated, showing anneals of the material as vertical gold bars.}
  \label{fig:life}
\end{figure} 

At charge accumulations of 0, 6 and 14 Pe$^-$/cm$^2$ the polarization is seen to rise under the influence of experimental beam.  The polarization is built up in these instances as too few paramagnetic radicals were in place within the sample; in the case at 0 Pe$^-$/cm$^2$ this indicates under-irradiation at NIST, and in the other two cases it indicates an over-anneal of the material.	

At around 3 and 11 Pe$^-$/cm$^2$, there are spontaneous drops in polarization.  These spots are due to loss of liquid level in the refrigerator nose.  After the replacement of the original refrigerator with a spare, no level probe was available to observe the liquid level in the nose.  This meant the only indicators of liquid level were temperature measurements from the target insert itself.  A careful balance is crucial, too little flow from the separator to the nose via the run valve allows the helium to boil off in the beam, but too much allows the refrigerator to overfill, potentially freezing O-rings and spoiling the vacuum.  This balance is kept using the flow indicators from the separator and main fridge, but is a technique that requires a level of expertise that was not available at all times during the experiment.  Thus, loss of level in the nose, which allowed heating of the material and loss of polarization, was not an uncommon occurrence.

Positive leaps in polarization when there is no anneal, such as at 10 Pe$^-$/cm$^2$ are due to a beam trip.  As the beam's heat load is removed, the polarization will recover due to increased DNP efficiency, and in the case of the jump at 10 Pe$^-$/cm$^2$, a long break in the beam allowed extra time to build up polarization that is rarely available when beam is ready to be brought into the hall.

			\subsubsection{Optimal Microwave Frequency}
\begin{figure}[hbt!]
\begin{center}
    \includegraphics[width=4.5in]{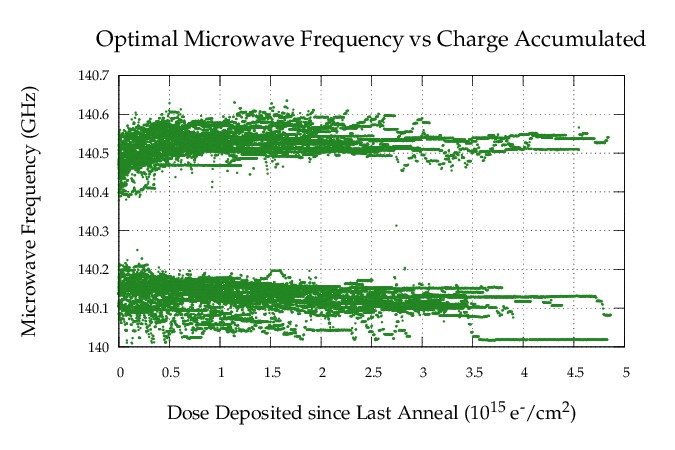}
    \end{center}
  \caption[Plot of the change in microwave frequency with accumulated charge.]{Plot showing the change in microwave frequency with accumulated charge from beam.  The upper grouping shows frequencies used to polarize negatively, the lower, positively.}
  \label{fig:micro}
\end{figure}

Figure \ref{fig:micro} shows the ``optimal'' polarizing microwave frequency versus charge accumulated since the last anneal.  The plot shows the microwave frequency data for the entire experiment, with the accumulated dose being reset to zero after the anneal.  The upper grouping of points, around 140.5 GHz, represents frequencies used to polarize negatively (we recall $h(\nu_{e} + \nu_p)$ from section \ref{sec:introtarget}), and the grouping around 140.1 GHz was used to polarize positively ($h(\nu_{e} - \nu_p)$).  

We have used the word ``optimal'' in quotes to indicate that these frequencies are perhaps not ideal.  These are the frequencies used to polarize the target during SANE, as chosen by target operators to maintain the polarization over time.  As can be seen in the plot, the frequency which produces the highest polarization changes as dose from the beam accumulates, due to the production of more and varied paramagnetic radicals.  As target operators are fallible, and generally not target experts, the frequencies chosen are in most cases actually sub-optimal.  For instance, a horizontal line of points, such as the one visible at around 140.46 GHz, represents a long period where the polarizing microwave frequency was not changed at all.
			
The figure shows that the polarizing frequencies used to generate positive and negative polarizations tend to drift apart as dose accumulates.  However, this drift is not precisely the same for both polarities.  In the case of negative polarities, the best polarizing frequency rises swiftly with the first dose, but this increase slows, forming a curve approaching about 140.5 GHz.  In the positive case, the drop in frequency is much more gradual, nearly forming a linear slope from 140.2 to 140.1 GHz over 4 Pe$^-$/cm$^2$.  Appraising target operators of this most crucial part of their task is an important part of operator training.

			\subsubsection{Anneals}
\begin{figure}[htb!]
\begin{center}
    \includegraphics[width=6in]{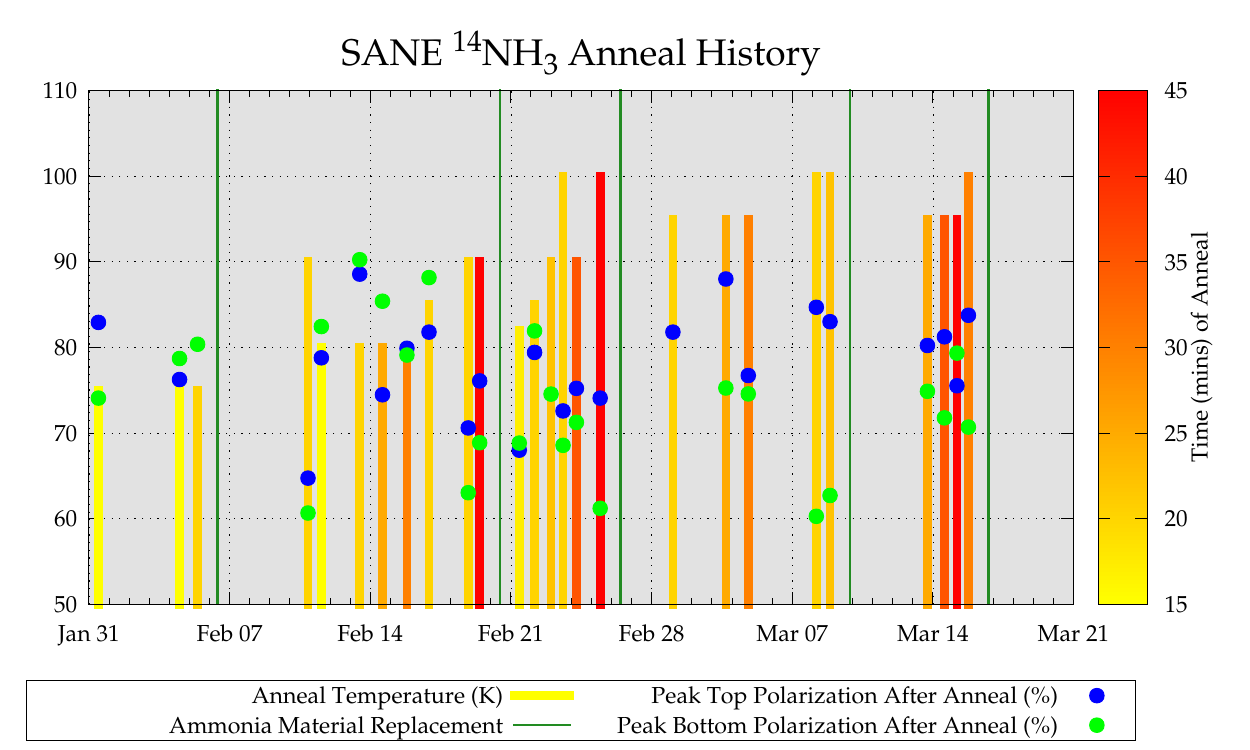}
    \end{center}
  \caption{Plot detailing pertinent data from all anneals performed during SANE.}
  \label{fig:anneals}
\end{figure}

Figure \ref{fig:anneals} shows a plot detailing each anneal performed during SANE.  Green vertical lines represent replacement of material, and the height of the red to yellow bars give the temperature at which the top material was held.  The length of time this temperature was maintained is represented in the color of the bar, with yellow being the shortest and red being the longest.  Also shown for each anneal are blue and green dots which give the peak polarization of the top and bottom cup material samples after the anneal was performed.	
			
As has been mentioned, the replacement cryogenic refrigerator made it difficult to maintain similar temperature in both the bottom and top material cups.  In the new configuration, liquid helium fell from above the target cups, while the heater wire boiled helium to heat the cups from the bottom.  This led to the top material being as much as 20 degrees hotter than the bottom.  The difficulty in performing consistent anneals adversely affected the peak target polarizations during SANE; the anneals just after March 7 are an example of relatively short anneals in which the top peak polarizations are much higher than the bottom.  In this case, the top was properly annealed, but the bottom did not have enough paramagnetic centers removed.  As time went on, target experts tried to err on the side over over-annealing material, as removing too many centers results in a slight delay as centers are built back up in the beam, but removing too few means that another anneal must be performed.
			
			\subsubsection{Unexpected Effects}
\begin{figure}[htb!]
\begin{center}
    \subfigure{\includegraphics[width=2.2in]{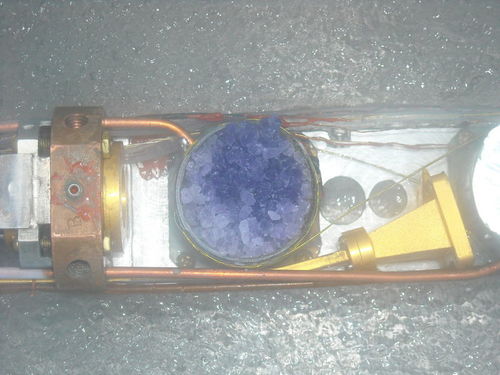}}
    \subfigure{\includegraphics[width=2.2in]{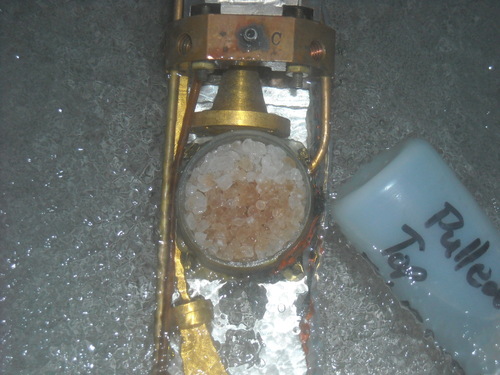}}
    \end{center}
  \caption{Photograph of materials 7 (left) and 5 (right) upon removal from use.}
  \label{fig:color}
\end{figure}

The first unexpected observation in ammonia samples used during the experiment was a brown discoloration was observed in some of the material samples, but not others.  Generally, ammonia turns a deep purple under radiation dose, as seen in the photo on the left of figure \ref{fig:color}.  This photo shows the beam spot was a bit to the right and bottom of the target cup, as this is where the hue is deepest.  However, several materials used during SANE developed a brown or dun hue, as seen in the right photo of the figure.  There was no apparent correlation between amount of dose received and this coloration, and some samples showed both dun and violet coloration at different areas within the cup.  This discoloration is currently not fully understood.
			
The second unexpected effect was persistent radioactivity in two of the materials used during the experiment.  The radioactivity of materials has typically fallen to safe levels within a week of leaving the beam, but two samples in particular remained ``hot'' longer than 2 weeks.  They were not the samples with the highest dose accumulated, so the cause of their continued radioactivity is not fully understood.

These samples were take to JLab's Radiation Control Group where gamma spectrum analysis was performed.  These two materials showed strong emission peaks as 477.7 keV, which corresponds to Be$^7$ decay.  The cross section of $^{14}$N$(\gamma,X)^7$Be is 0.12 mb \cite{RP}, so the presence of beryllium should not have been an entire surprise.  Investigation of other ammonia target materials, including some used in Hall B at the same time, confirmed the presence of Be$^7$ in them as well.  It is not clear why some samples developed more than others, possibilities include the loss of helium in the nose for some samples and not others, and the differences in anneal conditions between samples.

As previous electron scattering experiments by our group at SLAC and JLab used $^{15}$NH$_3$, SANE was our first use of  $^{14}$NH$_3$ at such high doses and thus our first opportunity to observe these behaviors.

		\subsection{Offline Polarizations}
		\label{sec:offpols}
\begin{figure}[htb!]
\begin{center}
    \includegraphics[width=6in]{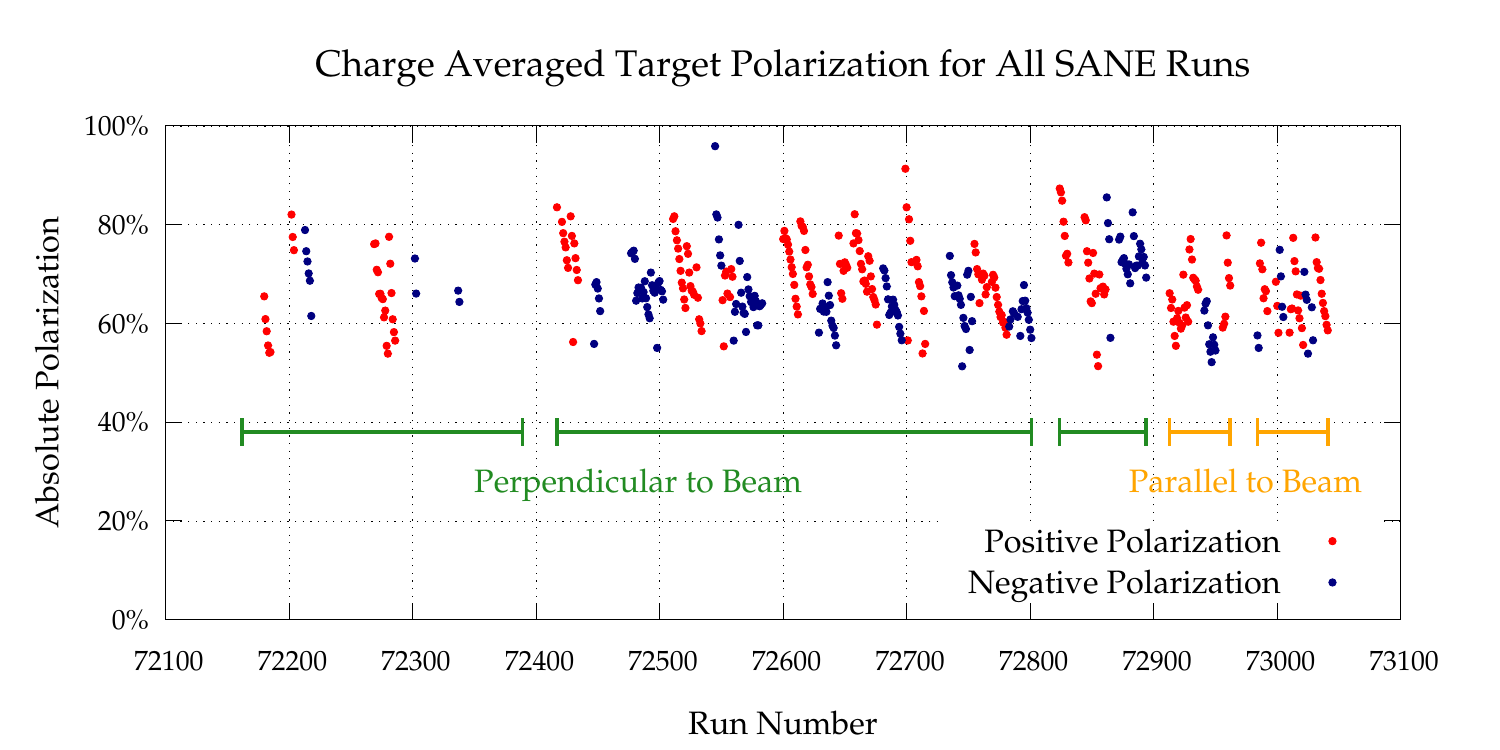}
    \end{center}
  \caption[Offline target polarizations for all SANE runs.]{Offline target polarizations for all SANE runs, showing run ranges for perpendicular and parallel magnetic field configurations.}
  \label{fig:pols}
\end{figure}

Shown in figure \ref{fig:pols} are the final results of the target analysis in the form of a plot of charge averaged polarizations for each run of SANE.  Red points represent positive polarizations and blue points, negative.  Horizontal bars of green and gold represent the different run periods of SANE, separated by magnetic field orientation and beam energy setting.  This data is summarized in table \ref{tab:pols}.  At 68\%, the charge averaged absolute polarization fell short of the anticipated polarization quoted in SANE's proposal of 75\%.  However, considering the unforeseeable difficulties in the operating conditions during the run, a near 70\% average polarization should be seen as a success.
		
\begin{table}[htb]
  \begin{center}
\begin{tabular}{lcc}
\toprule
B Field Orientation & Beam Energy Setting & Absolute Polarization\\\midrule
\multirow{2}{*}{Perpendicular} & 5.9 GeV & 69\%\\
 & 4.7 GeV& 66\%\\
 \multirow{2}{*}{Parallel} & 5.9 GeV&66\% \\
 & 4.7 GeV& 68\%\\
 \midrule
 \multicolumn{2}{c}{Entire Experiment} & 68\% \\ \bottomrule
	\end{tabular}
\caption[Table of absolute, charge-averaged, offline polarizations.]{Table of absolute, charge-averaged, offline polarizations per run setting during SANE.}
  \label{tab:pols}
\end{center}
\end{table}


\chapter{Data Analysis}
\label{sec:anal}

As laid out in section \ref{sec:measure}, the spin structure functions $g_1$ and $g_2$ can be determined using measured electron--proton scattering asymmetries with orthogonal target polarization components.  For $g_1$, longitudinal target polarization dominates, but no completely model-independent measurements can be made without transverse polarization.  Transverse polarization dominates in $g_2$, and offers a gateway to higher--twist physics.  Although SANE's electron arm acceptance would be unfavorably blocked by running the UVa polarized target at exactly transverse to the beam, the equations given in section \ref{sec:measure} account for the near-transverse target polarization angle of 80\degrees.

The measured asymmetries from the BETA detector package take the following form:
\begin{equation}
A = \frac{1}{fP_BP_T}\frac{N_+-N_-}{N_++N_-}
\end{equation}
where $N_+$ and $N_-$ are yields of electrons from positive and negative beam helicities.  By making partitioning cuts on the kinematic properties of the hits, which are then put in corresponding \textit{bins}, we can form the yields ---and thus the asymmetry $A$--- as a function of kinematics.  The factors $f$, $P_B$ and $P_T$ correspond to the necessary dilution factor, beam polarization and target polarization corrections to the asymmetry.  The dilution factor is discussed in section \ref{sec:dilution}, and the beam and target polarizations were covered in sections \ref{sec:offpols} and \ref{sec:beampol} respectively.

This chapter details the production of asymmetries and spin structure functions from the data collected during SANE.  This analysis uses the data from the calorimeter and \v{C}erenkov detector, and their calibration is discussed first.  Next the identification and reconstruction of scattered electrons of interest is covered.  Finally, the production of corrected experimental asymmetries, virtual Compton asymmetries and spin structure functions is detailed.

\section{Calibration}
	\subsection{Calorimeter}
	\label{sec:calcalib}
The energy deposited by incident particles into each lead--glass block of the calorimeter results in an ADC signal from the corresponding photomultiplier tube, but these ADC signals must be calibrated to provide accurate results.  The first step towards calibration is done in hardware, adjusting the high--voltage power supplied to the phototubes so that each ADC channel corresponds to roughly 1 MeV.  This rough calibration was performed using cosmic ray events before the experiment began to approximately equilibrate the signals, followed by pion events  to assign 1 MeV channel width during commissioning.  More precise calibration requires the analysis of ADC signals of known energy. The large number of $\pi^0$ events coming from the target throughout the experiment offer just such reference signals, and effectively allow the gain to be monitored in time without interrupting data-taking.  

Neutral pions produced in the target decay very rapidly; the primary decay mode is to two photons with a 98.82\% probability and a mean lifetime of $8.4\times10^{-17}$ seconds\cite{pdb}.  Even traveling near the speed of light, the pions have decayed to photons before exiting the target.  By measuring the angle of separation of the photons, we have a relation which gives us the relative energy of the two photons $E_1$ and $E_2$ in terms of the pion mass and photon separation angle $\alpha$:
\begin{equation}
\label{eq:pi0}
m^2 = 4E_1E_2\sin^2\alpha/2
\end{equation}

As mentioned in section \ref{sec:pizero}, $\pi^0$ events, consisting of two vertically separated clusters in BigCal, were collected under trigger type 3.   To turn the collection of ADC signals from all the calorimeter blocks from each event into useful data, we must reconstruct the \textit{clusters} of hits in blocks which correspond to the shower of one incident particle.

		\subsubsection{Clustering}	
The shower of an incident particle, discussed in section \ref{sec:bigcal}, can deposit the particle's energy in several blocks surrounding the point where it enters the calorimeter.  To determine the energy and position of the incident particle, we must discern which blocks were involved by building clusters.

\begin{figure}[htb]
  \begin{center}
   \includegraphics[width=4.5in]{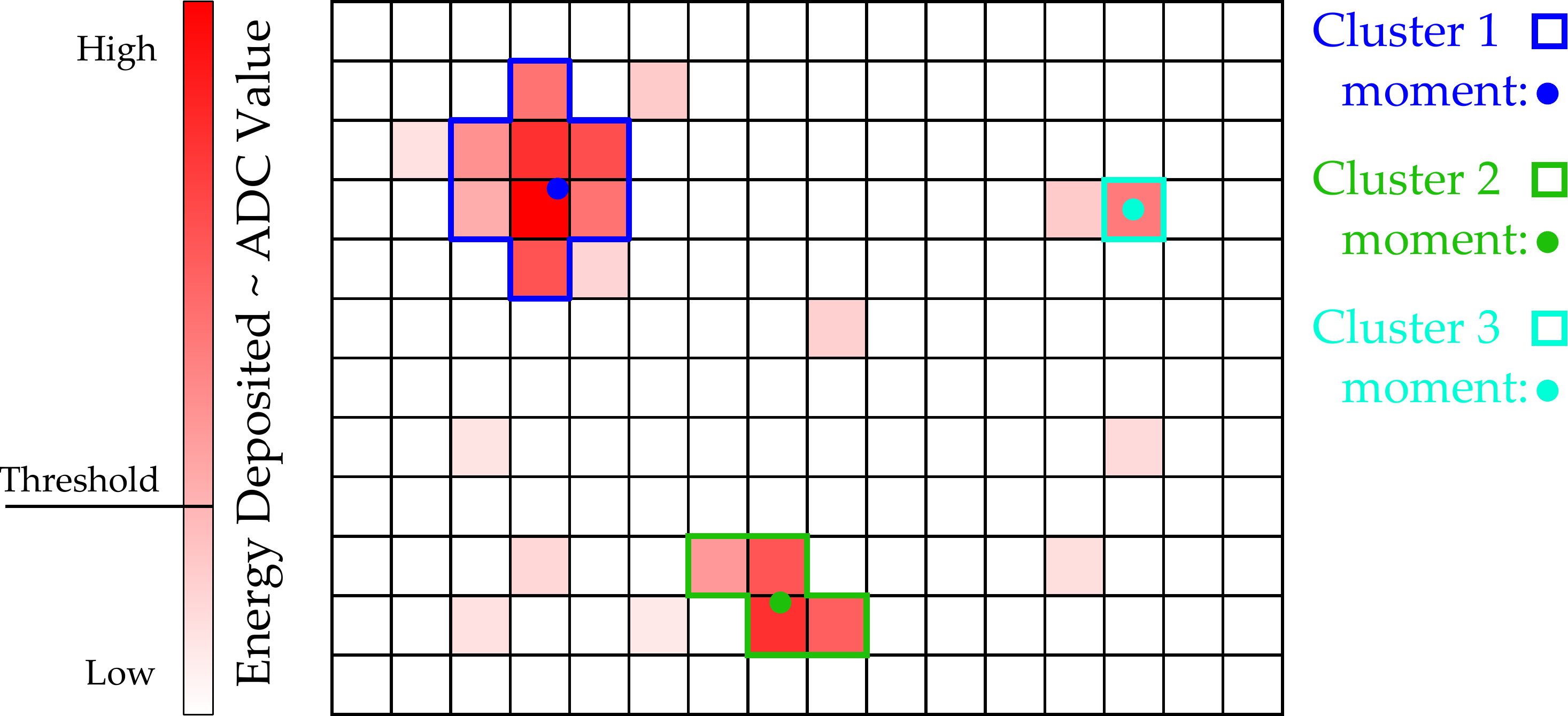}
  \end{center}
  \caption[Diagram showing an example of clustering.]{Diagram showing an example of clustering for a hypothetical set of calorimeter blocks, including the energy--averaged cluster moments.}
  \label{fig:cluster}
\end{figure}

The first step in cluster reconstruction is identifying the highest ADC values from BigCal phototubes during a given event.  These local maxima will be used as seeds to build clusters.  Starting with the highest energy block, blocks which physically neighbor the seed block and whose energy exceeds a given threshold, nominally 10 MeV, are added to the cluster.  Next, these newly added blocks are considered, so that their neighbors which exceed the threshold are also added.  This process continues to grow the cluster by adding blocks which are not already spoken for until a maximum number of 25 blocks in a $5\times5$ grid is reached.  Once a cluster is completed, the next highest maxima in BigCal is found and a cluster is grown from it, until all the blocks exceeding a given threshold are used.  Figure \ref{fig:cluster} illustrates clustering for a hypothetical set of blocks.

Once our clusters are built we can assign them an energy $E_c$ based on the ADC value $A_i$ of each constituent blocks $i$ and a calibration constant for that block $c_i$: 
\begin{equation}
E_c = \sum_{i}c_iA_i.
\end{equation}
The constants $c_i$ are the end goal of the calibration.  We start out with rough values for $c_i$ assuming each ADC channel corresponds to 1 MeV.

The position which the incident particle entered the calorimeter can be approximated using the positions  of the blocks in the cluster and their deposited energies.  By performing an energy--weighted average of the block positions in the cluster we produce the average position, or \textit{moment}
\begin{equation}
\begin{split}
\langle x\rangle &= \sum_i \frac{c_iA_i}{E_c}(x_i - x_{seed})\\
\langle y\rangle &= \sum_i \frac{c_iA_i}{E_c}(y_i - y_{seed})
\end{split}
\end{equation}
of the cluster, for individual block coordinate on the BigCal face $(x_i,y_i)$ and coordinate of the seed block $(x_{seed},y_{seed})$.  The cluster position on the face of BigCal is then taken to be $(x_{seed}+\langle x\rangle,y_{seed}+\langle y\rangle)$.

		\subsubsection{$\pi^0$ Mass Reconstruction}
Once clusters are built from the hits in the calorimeter, we can begin to adjust the calibration constants $c_i$ using our $\pi^0$ signals.  For $\pi^0$ trigger events, discussed in section \ref{sec:pizero}, we should have two vertically separated clusters of energy $E_1$ and $E_2$ with moments $(x_1,y_1)$ and $(x_2,y_2)$.  To remove electron--positron pairs which were produced after passing through the \v{C}erenkov and might be mistaken for $\pi^0$ events, we can add a cut to choose clusters between 20 and 80 cm apart.  Clusters closer than 20cm run the risk of ambiguity with large clusters; the cluster limit of 5 $\times$ 5 results in a square of 20 $\times$ 20 cm for the larger calorimeter blocks.  Further cuts ensure the photons arrived in the same time window.  
 We then calculate the invariant mass of the supposed pion according to equation \ref{eq:pi0}:
\begin{equation}
\label{eq:sep}
m_\textrm{inv}^2 = 2E_1E_2(1-\cos\alpha), \quad
\textrm{for}\ \cos\alpha = \frac{x_1x_2 + y_1y_2 + z^2}{(x_1^2+y_1^2+z^2)(x_2^2+y_2^2+z^2)}
\end{equation}
the angle between the trajectories from the target to each cluster moment and $z$ the distance of the calorimeter face from the target.  

\begin{figure}[htb]
  \begin{center}
   \includegraphics[width=3in]{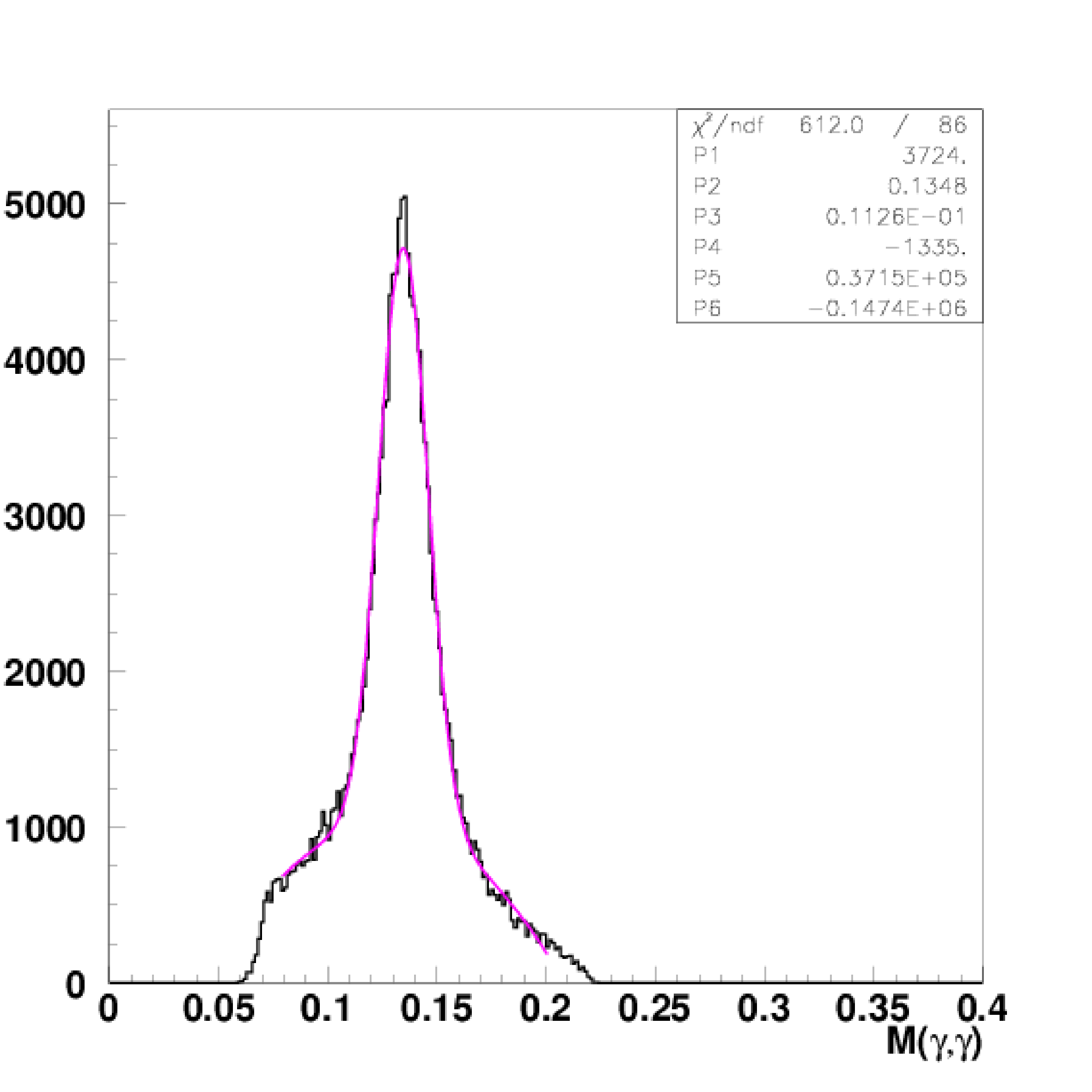}
  \end{center}
  \caption[Plot of neutral pion mass reconstruction.]{Plot of neutral pion mass reconstruction after block calibration.  The energy resolution of this peak is directly proportional to the energy resolution of the clusters in the calorimeter.  Plot by H. Baghdasaryan from reference \cite{doe10}.}
  \label{fig:mpi}
\end{figure}

To calibrate a block, we form a histogram of the invariant mass results for all the clusters which include the block.  By dividing this invariant mass by the known $\pi^0$ mass $m_{\pi^0}= 134.9$ MeV, this histogram should show a distribution which is peaked above or below unity.  Dividing the calibration constant $c_i$ by the peak value of this distribution and squaring gives the new calibration constant to be applied to that block.  Once new constants are produced in this way for all the blocks, we start again, forming new histograms to fit.  By iterating in this manner many times, our invariant mass peaks for all the blocks should converge about one and our constants $c_i$ are achieved.  Figure \ref{fig:mpi} shows the $\pi^0$ mass reconstruction after calibration for a subset of calorimeter blocks.

		\subsubsection{Angle Correction: Neural Network}
\begin{figure}[htb]
  \begin{center}
   \includegraphics[width=4in]{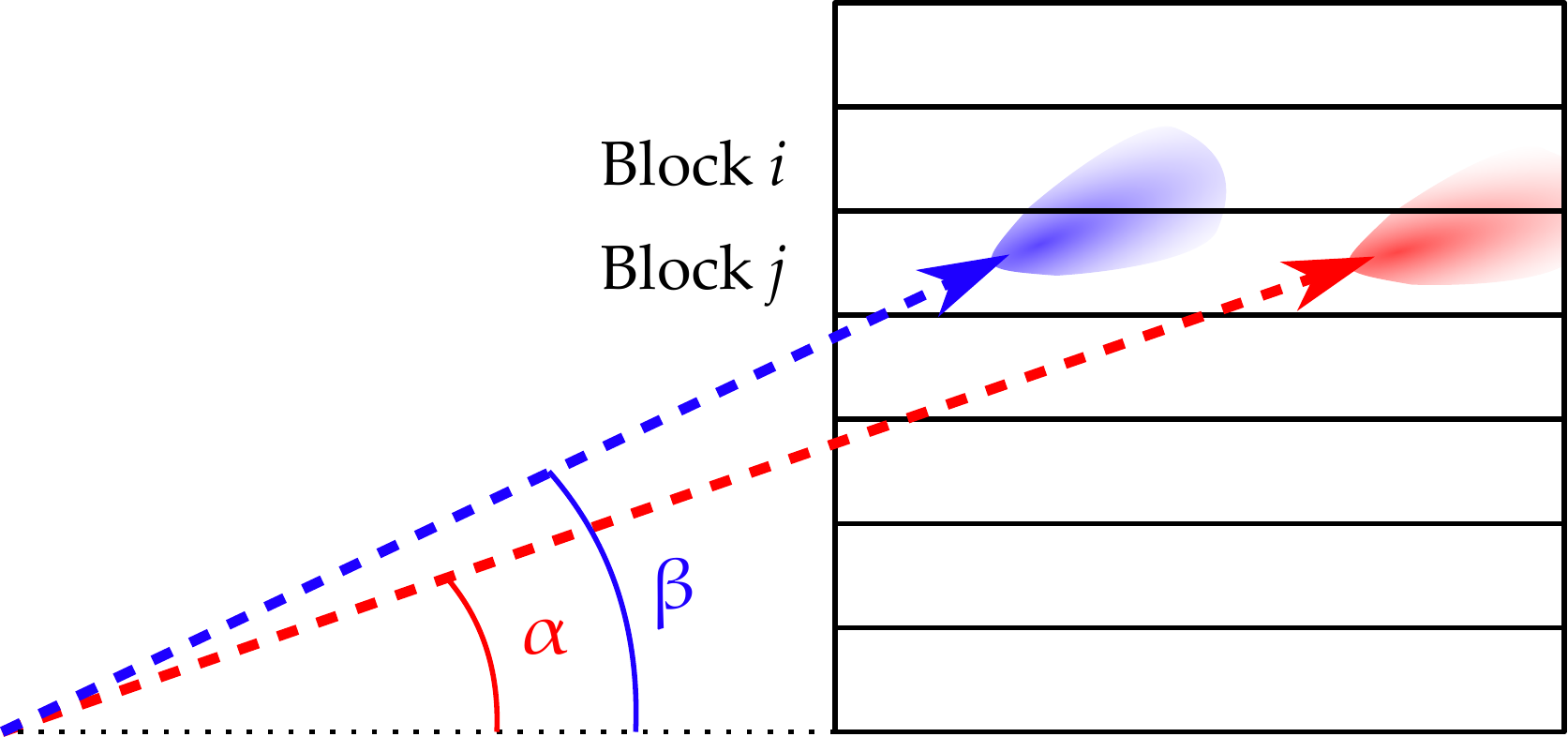}
  \end{center}
  \caption[Diagram showing the need for angle correction based on shower depth.]{Cross-sectional diagram showing the need for angle correction based on shower depth; here blocks $i$ and $j$ fire equally for both the red and blue incident particle trajectories, despite different incident angles $\alpha$ and $\beta$.}
  \label{fig:angrec}
\end{figure}

	Although the calibration constants for each of the phototubes have been obtained, we have shown that the process depends on our ability to correctly reconstruct the angle of separation between the secondary photons of the $\pi^0$.  In equation \ref{eq:sep} we took the distance from the target $z$ to simply be the position of the calorimeter face.  However, the depth of the inception of the shower can vary with the energy of the incident particle.  For particles arriving farther from the center of the calorimeter, which thereby have more oblique angles, the shower depth has an increasing effect upon the resolved cluster moment, an idea illustrated in figure \ref{fig:angrec}.	

To correct our separation angles for this shower depth effect, we turn to a neural network.  In a neural network, a set of inputs is transformed to a set of outputs via a number of sigmoidal activation functions weighted to reproduce specific results.  This process is designed to mimic the function of the brain, where neural nodes fire in varying strengths to produce output.  These computational automata have been used in the past to process calorimeter data \cite{nnet}, and in SANE they were used to aid in cluster position reconstruction.  While a full introduction to neural networks is beyond the scope of this document, a thorough treatment can be found in reference \cite{neuralnet}.

\begin{figure}[htb]
  \begin{center}
   \includegraphics[width=3.5in]{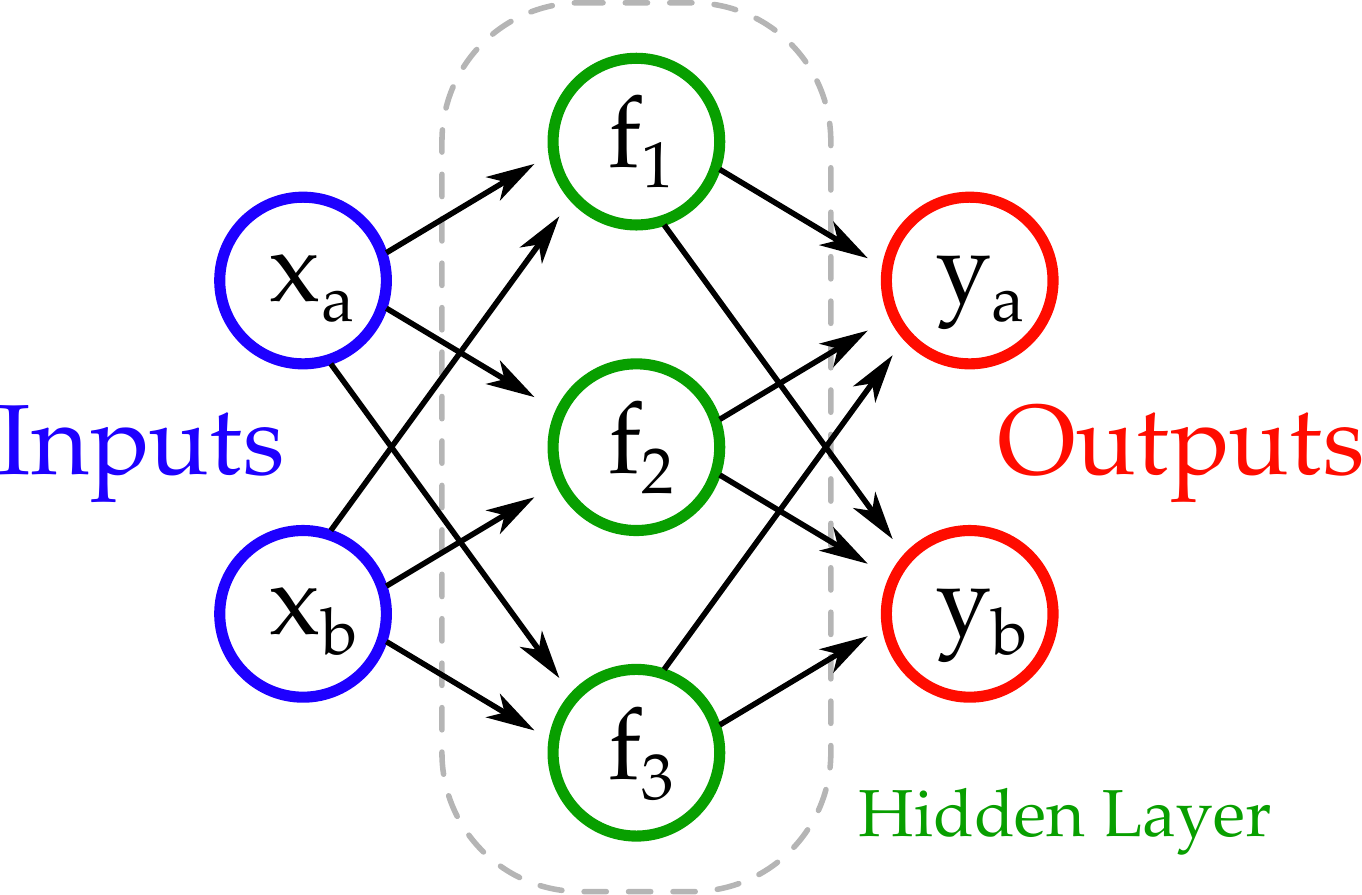}
  \end{center}
  \caption[Neural Network Diagram.]{An example neural network with 2 inputs, 2 outputs and 3 internal nodes.}
  \label{fig:nnet}
\end{figure}

An example neural network is shown in figure \ref{fig:nnet}.  Each arrow connecting the nodes represents a multiplicative weighting $w_{ij}$ which controls how the inputs $x_a$ and $x_b$ become outputs  $y_a$ and $y_b$ via functions in the ``hidden-layer,'' $f_1(x)$, $f_2(x)$ and $f_3(x)$.  Here I will label the weights by the subscripts of the nodes they connect; for instance, $w_{a1}$ connects $x_a$ to $f_1$ and $w_{2b}$ connects $f_2$ to $y_b$. There is one additional weight $w_{0i}$ for each function which takes into account biasing.  The output $y_a$ is then, for example: 
\begin{equation}
y_a = \sum_{i=1,3} w_{ia}f_i(w_{ai}x_a + w_{bi}x_b + w_{0i}).
\end{equation}

The effectiveness of the neural network comes from the training process. The weights which control the output are created by training the network with a system of inputs and outputs which the network will emulate in producing its results.  We produce these weights $w_{ij}$ by iterating the calculation of the outputs and changing the weights until the error between the calculated output and training example output is minimized.

\begin{figure}[htb]
  \begin{center}
   \includegraphics[width=3.5in]{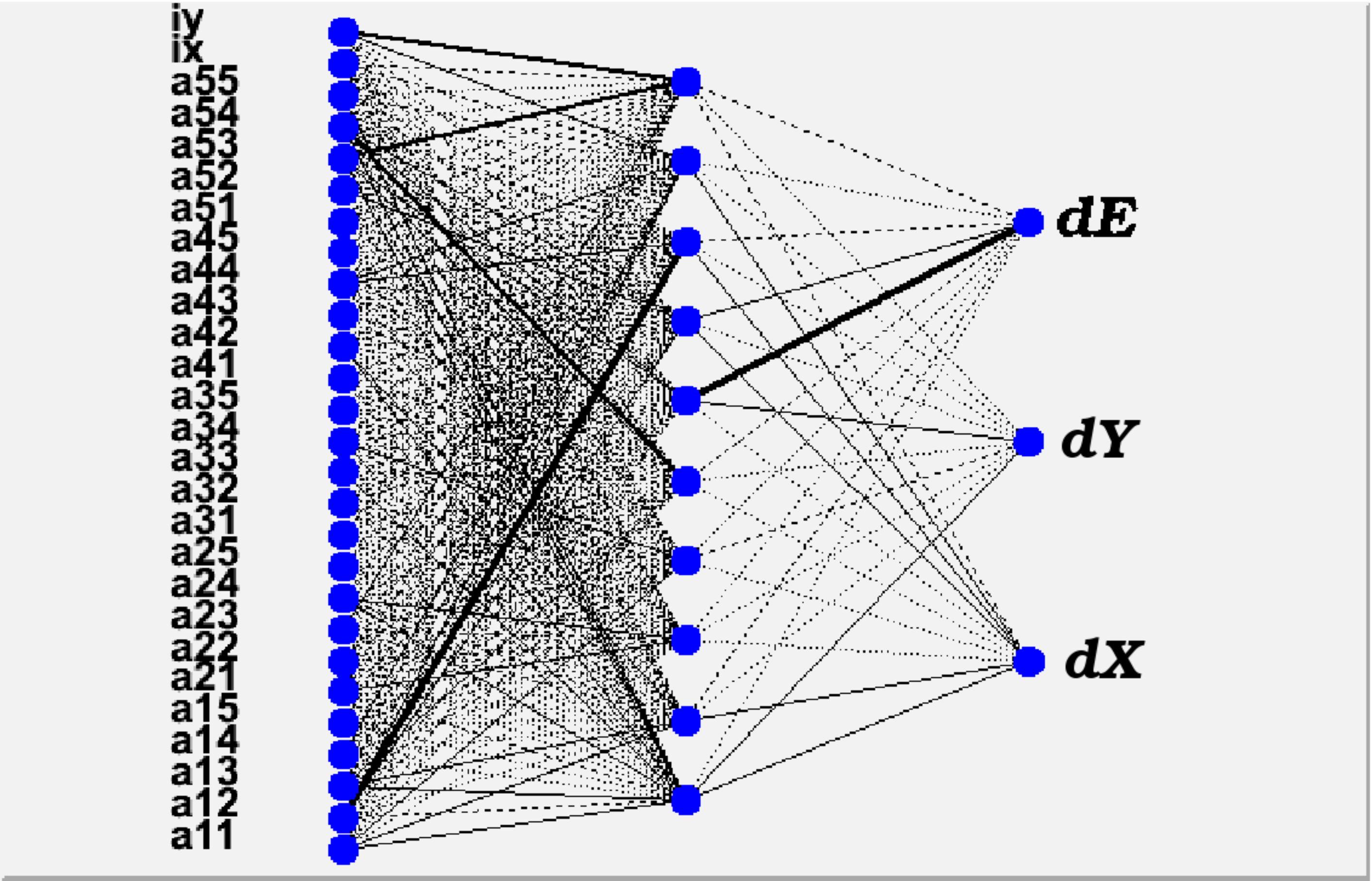}
  \end{center}
  \caption[Neural network diagram for SANE.]{A diagram of SANE's neural network, with 27 inputs and 3 outputs.  Here the thickness of the connecting lines represents the weight of that node connection.  Diagram by H. Baghdasaryan from reference \cite{doe10}.}
  \label{fig:nn}
\end{figure}

SANE's neural network, seen in figure \ref{fig:nn}, was based on ROOT's Multilayer Perceptron class\cite{rootnn} and had 27 inputs: two for the x--y coordinates of the seed block, and 25 for the energy  deposited in each of the up to 25 blocks which could be included in the cluster.  The three outputs from the neural network were the x--y coordinate of the new cluster center and the cluster energy.  As the neural network provided the x--y position of the cluster, the cluster moments mentioned in the previous section were no longer needed.

The crucial training of SANE's neural network was performed using a Monte Carlo simulation of the electron detector package, magnetic field and target written in GEANT3.  Events thrown from the target in the simulation have known energies and trajectories, and the response of the calorimeter for each event was simulated.  More than 20 million simulated events, both electrons and photons, were thrown to train the neural network, each event refining the weight factors.   The calibration effort was lead by SANE collaborator H. Baghdasaryan.

\begin{figure}[htb]
  \begin{center}
   \includegraphics[width=6in]{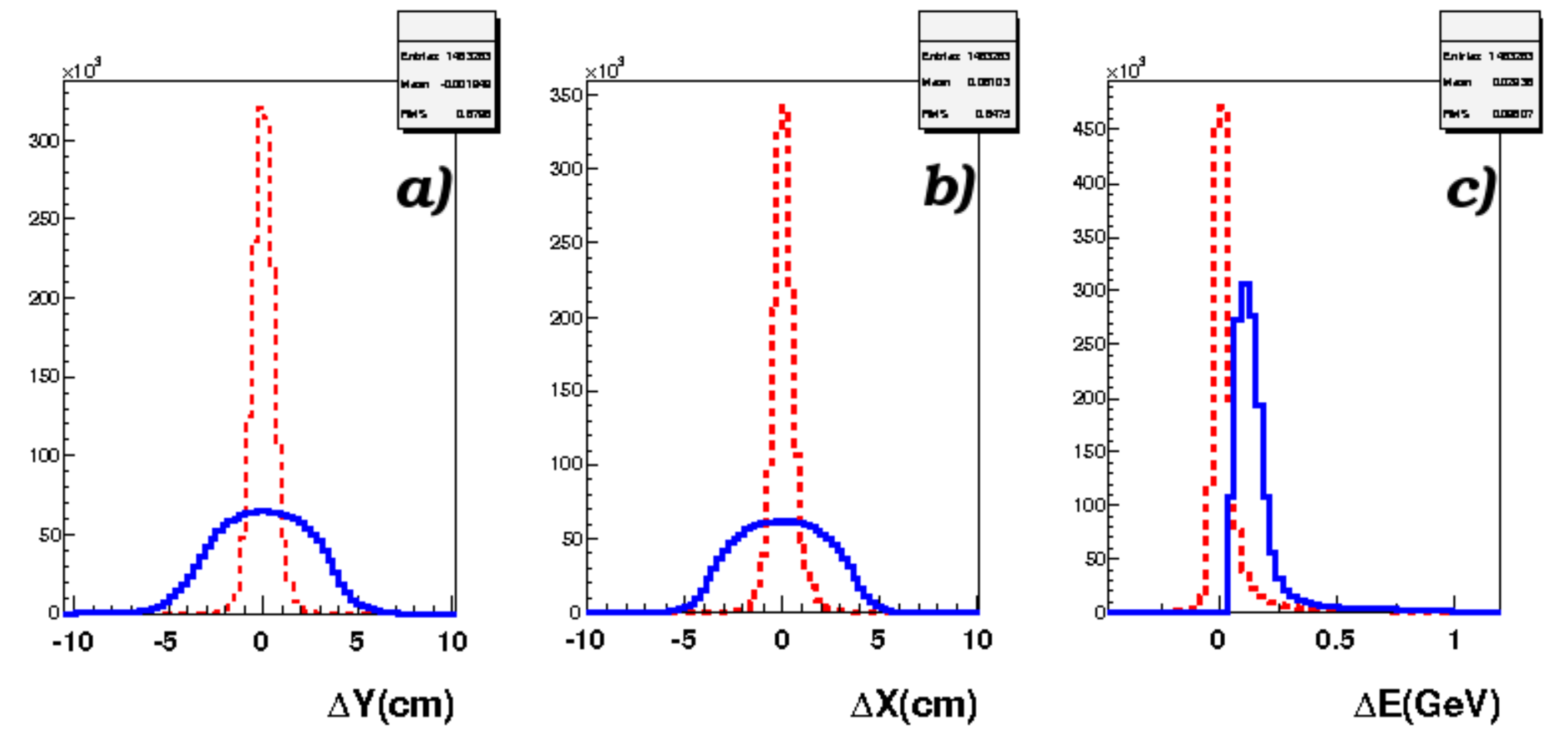}
  \end{center}
  \caption[Neural network vs. moment approach.]{A plot showing the difference between thrown and reconstructed events in the simulation.  The blue lines represent the moment method and the red, dashed line is the neural network approach.  Plot by H. Baghdasaryan from reference \cite{doe10}.}
  \label{fig:nnvm}
\end{figure}

The effectiveness of the neural network reconstruction is illustrated in figure \ref{fig:nnvm}, compared against the results obtained by the moments method alone.  The plot shows the difference between the generated event and reconstructed quantity for the $x$ and $y$  position in the calorimeter (a, b) and energy (c).  This gives a good impression of the accuracy of the method, and the much tighter peaks of the neural network approach, in dashed red, plainly result in better resolution than those of the moment method in blue.  

The energy resolution achieved after the neural network calibration was shown by collaborator J. Mulholland to be $0.096/\sqrt{E'}$ during the $80\degrees$ target field  and $0.107/\sqrt{E'}$ during the parallel.  These values were determined by measuring neutral pion mass reconstruction peak widths as a function of $E'$, such as that seen in figure \ref{fig:mpi}. His method is discussed further in reference \cite{jmul}.

	
	\subsection{\v{C}erenkov Detector}
	For this analysis, only the TDC values of an electron event were needed.  The TDC value for an event, which was triggered by a threshold on a photomultiplier ADC, was sufficient to tag an event as charged or not.

		\subsubsection{\v{C}erenkov Time-Walk}
\begin{figure}[htb]
  \begin{center}
   \includegraphics[width=2.5in]{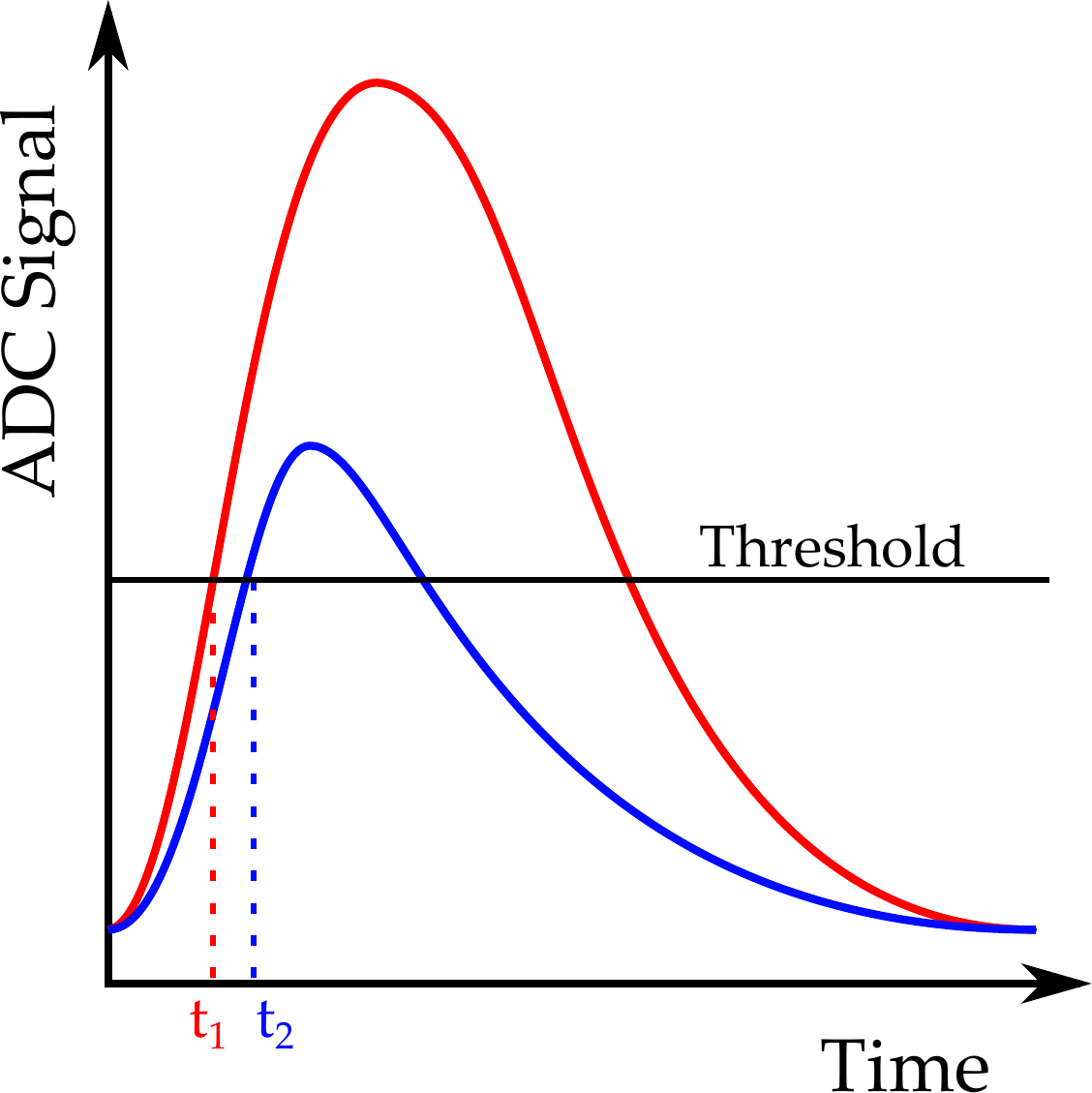}
   \includegraphics[width=2.7in]{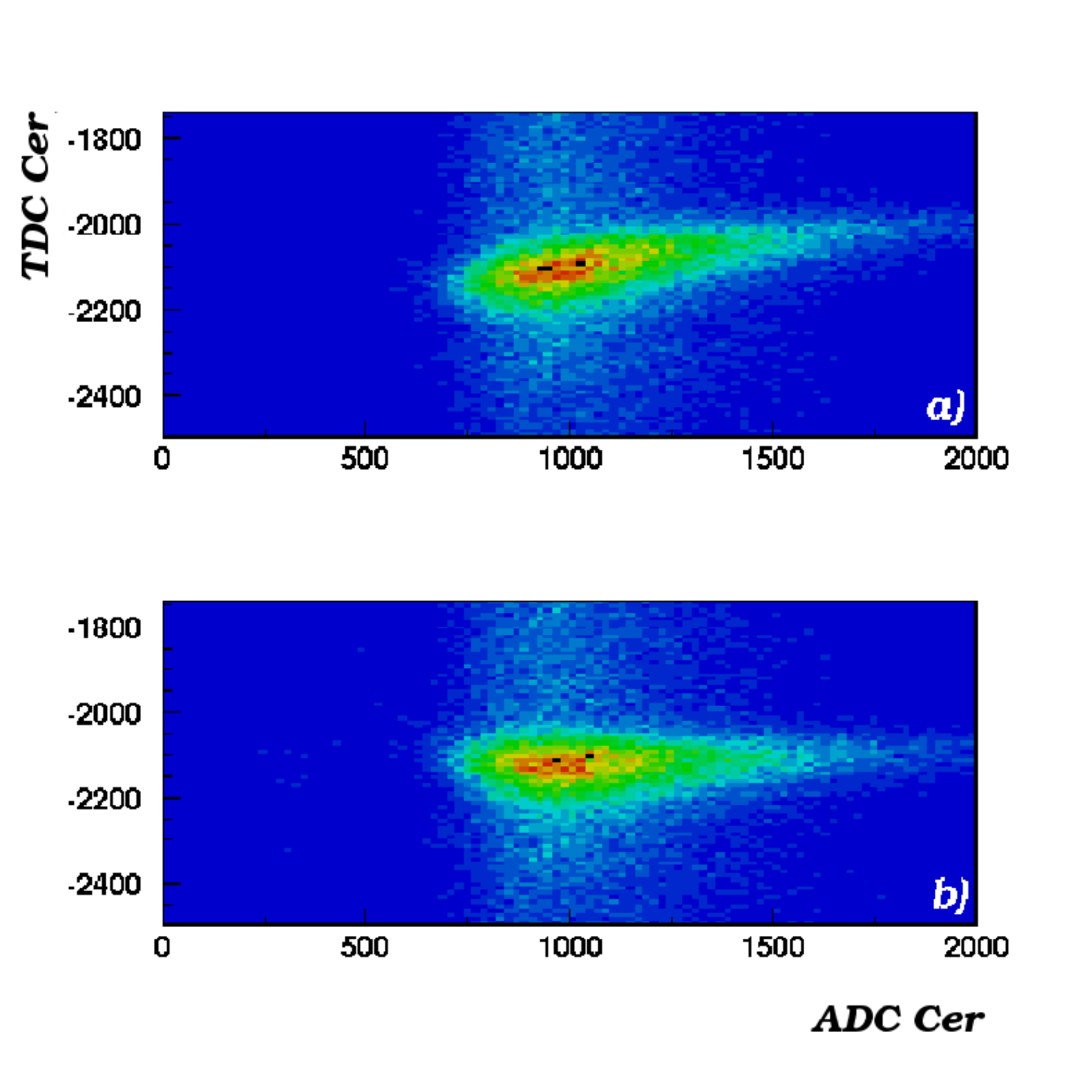}
  \end{center}
  \caption[Time walk plots.]{The time-walk effect illustrated with theoretical ADC signals is shown to the left, where the stronger, red peak passes the threshold before the weaker blue peak, even though they should arrive at the same time.  To the right\cite{doe10} is an example of the \v{C}erenkov TDC versus ADC values, showing the uncorrected (a) and time-walk corrected signals (b).}
  \label{fig:timewalk}
\end{figure}

		A \textit{time-walk} is a shift in the trigger time based on the peak height  of an ADC signal where a discriminator triggers on a threshold of the ADC signal from a photomultiplier tube.  This correction is necessary because a large signal will reach the threshold sooner than a weaker signal would, as seen to the left of figure \ref{fig:timewalk}.  To correct for this shift in time, a $y = c_1 + c_2/x$ fit was made to the scatter-plot of the \v{C}erenkov TDC vs. ADC signals, allowing the adjustment of the TDC signals so that this trend has no slope, as seen in the right of figure \ref{fig:timewalk}.		  
		
\section{Event Reconstruction and Selection}
	With properly calibrated event data from our electron detector array in hand, we can begin to determine the physical characteristics of the events in order to build our asymmetries.  As we produce yields of events for our asymmetries, we combine a set of cuts to maximize the number of good electron events in our sample.

	\subsection{Event Physics Reconstruction}
After calibration, we have a set of events, each of which consists of ADC and TDC values from our various detectors.  To move forward in the analysis, these detector signals must be reconstructed into the path of an electron of energy and trajectory which must be determined.

We can make use of the procedures established in section \ref{sec:calcalib} for the clustering of hits on the calorimeter and the cluster correction using our neural network.  By passing the neural network the seed block position and energies per cluster block, we are returned an energy and position for each hit in the calorimeter.  These data lead us to the three quantities of interest for each event: the final electron energy $E'$ and the electron scattering angles $\phi$ and $\theta$.

\begin{figure}[htb]
  \begin{center}
   \includegraphics[width=4in]{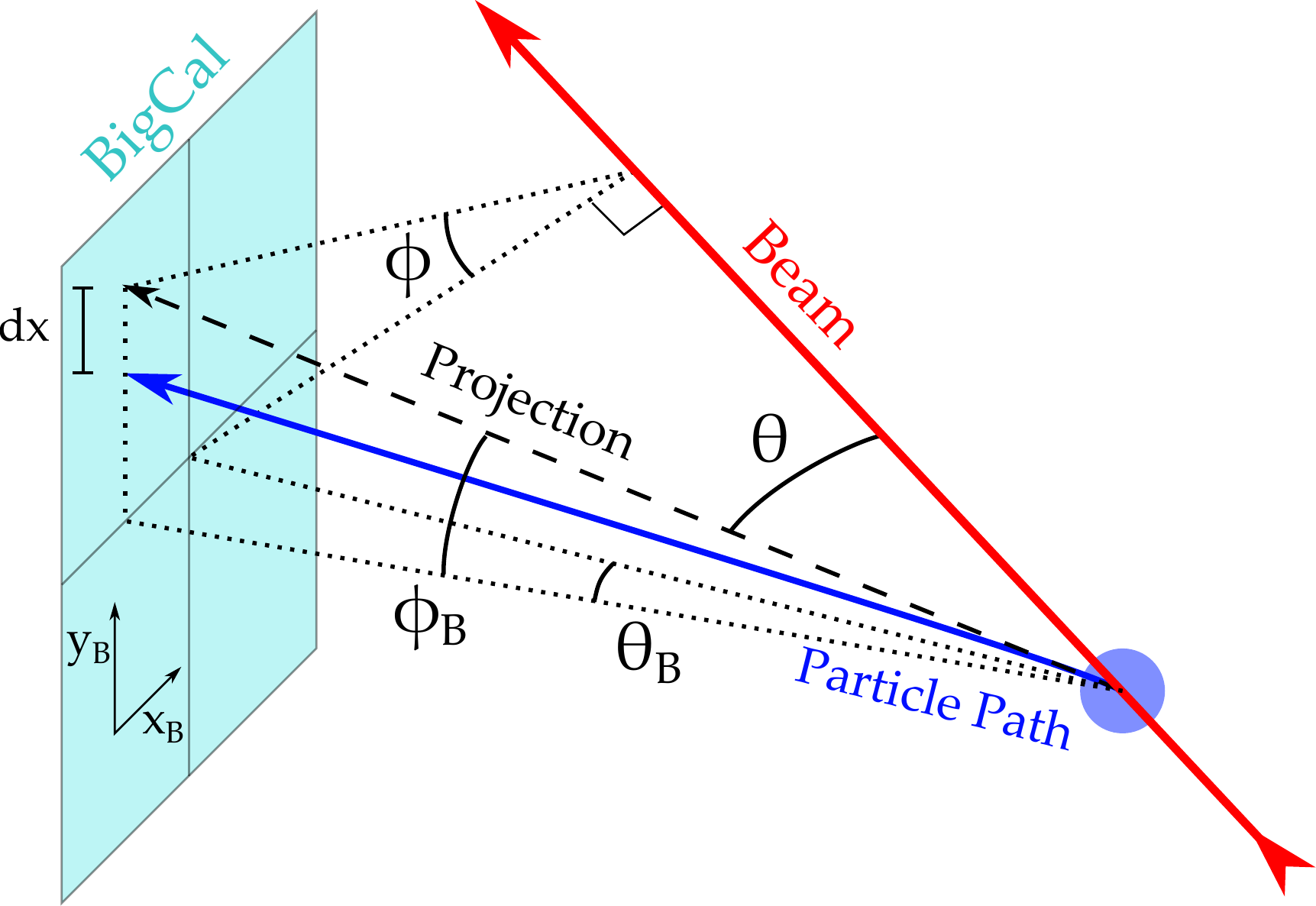}
  \end{center}
  \caption[Diagram of BETA and Physics angles.]{Diagram of BETA and Physics angles, showing the bent particle path and its straight-line projection from the target.}
  \label{fig:betacoor}
\end{figure}

With the $x$ and $y$ position of a cluster on the face of BigCal, it's trivial to form scattering angles $\phi_B$ and $\theta_B$ in BigCal coordinates.  Figure \ref{fig:betacoor} shows these BigCal scattering angles, as well as the physics scattering angles $\phi$ and $\theta$, which are related by
\begin{equation}
\begin{split}
\cos\theta &= \cos(\theta_{\textrm{BETA}} - \theta_B)\cos\phi_B, \\
\tan \phi &= \frac{\tan\phi_B}{\sin(\theta_{\textrm{BETA}} - \theta_B)},
\end{split}
\end{equation}
for BigCal central angle from the beam $\theta_{\textrm{BETA}} = 40\degrees$.	
	
	\subsubsection{Target Field Deflection}
As we calculate the scattering angles of the electron from detector data, it is critical we take into account the deflection due to the magnetic field of the target.  For a magnetic field $B$, we express the angular deflection of an electron, charge $e$ and momentum $p$ as $\Delta\phi = e/p \int B dl$, where $\Delta\phi = \phi_B-\phi_r$ results in the $dx$ of figure \ref{fig:betacoor}. This $\phi_r$ is the observed scattering angle of the cluster.  The computation of $\Delta\phi$ is accomplished via a 15 parameter fit: 
\begin{equation}
\begin{split}
\phi_B = \phi_r &+ (a_1 + a_2\theta_r+ a_3\phi_r+ a_4\theta^2_r+ a_5\phi^2_r + a_6\theta_r\phi_r)\\
&\times (a_7 + a_8/E_r + a_9/E^2_r)\\ &\times (a_{10} + a_{11}s_x + a_{12}s^2_x) \times (a_{13} + a_{14}s_y + a_{15}s^2_y),
\end{split}
\end{equation}
for cluster energy $E_r$ and slow raster position ($s_x$,$s_y$).  Likewise, we can parametrize any slight deviation in $\theta$:
\begin{equation}
\begin{split}
\theta_B = \theta_r &+ (b_1 + b_2\theta_r+ b_3\phi_r+ b_4\theta^2_r+ b_5\phi^2_r + b_6\theta_r\phi_r)\\
&\times (a.b_7 + b_8/E_r + b_9/E^2_r)\\ &\times (b_{10} + b_{11}s_x + b_{12}s^2_x) \times (b_{13} + b_{14}s_y + b_{15}s^2_y).
\end{split}
\end{equation}
An accurate field map of the target magnetic field, as provided by the manufacturer, is implemented into the GEANT3 Monte Carlo.  By throwing a set of electron events in the simulation, these 15 parameters could be determined for both $\phi$ and $\theta$, allowing the calculation of the field deflection for any case.

	\subsection{Kinematic Binning}
We build electron yields as a function of the kinematic variables $Q^2$ and then $W$ or $x$ by placing cuts on events to exclude events outside a given range from a kinematic value.  These ranges are known as \textit{bins}; each bin contains all the events around the central kinematic value of that bin, or \textit{abscissa}, plus or minus half the distance to the next abscissa.  We choose our kinematic binning to accurately describe any changes in the yield over the kinematics, but the smallest we can meaningfully pick depends on the resolution of our detectors. 

We begin by selecting broad $Q^2$ cuts, separating the data into four roughly equal bins in $Q^2$ to catch any large scale $Q^2$ evolution in our results.  For each of these four bins, we then form tight bins in $x$ or $W$. With a fit to the energy resolution of our calorimeter as a function of energy of the form
\begin{equation}
\delta E '(E') = \frac{C_0}{\sqrt{E'}} + C_1,
\end{equation}
with fit constants $C_0$ and $C_1$, we can produce set of $E'$ bins such that
\begin{equation}
E'_{i+1} = E'_i + \delta E '(E'_i),
\end{equation}
with upper and lower bounds
\begin{equation}
\begin{split}
E'_{lo} &= E'_i - \delta E '(E'_i)/2\\
E'_{hi} &= E'_i + \delta E '(E'_i)/2.
\end{split}
\end{equation}
With this set of bins, we can use the relations of section \ref{sec:variables} to calculate $x$ and $W$ bins corresponding to those in $E'$.

In order to make plots which are easier to read, we also create tables of combined bins.  Although the resolution in $x$, for example, will increase as $x$ decreases, our acceptance also decreases as $x$ falls below around 0.4.  With the dropping statistics at lower $x$, higher resolution only offers many bins with a large statistical error.  To avoid this situation, we combine bins within a given $x$ range of each other to create less cluttered plots.

	\subsection{Event Criteria}
To ensure we minimize background and select electron events only from our process of interest, we apply several criteria to the events which will be included in the helicity yields $N_+$ and $N_-$.  These cuts are enumerated here. 
\begin{enumerate}

\item \textit{Trigger Type}: Only trigger type 4 events are included; these are events with a hit in both the calorimeter and the \v{C}erenkov, as discussed in section \ref{sec:trigger}. 

\item \textit{Single Cluster}: Only events with a single cluster on BigCal were included.  While this cut may not be necessary, and may be removed in the future, it was intended to exclude electron--positron pairs.

\item \textit{\v{C}erenkov Hash}: This cut ensured the \v{C}erenkov hit was pertinent to the event in the calorimeter.  The \textit{\v{C}erenkov hash} flag is greater than zero if a good \v{C}erenkov hit occurred in the correct time frame, and this hit matched a geometrical cut with the calorimeter.  The geometrical cut simply ensures that a hit in any given \v{C}erenkov mirror ended up in the sector on the face of the calorimeter which corresponds to that mirror's projection from the target.

\item \textit{Cluster Energy}: A cut on the cluster energy was placed to exclude charged pion events which are unlikely to be found above 500 MeV; by accepting only cluster energies above 1,300 MeV, we increase the purity of our electron yields. 

\item \textit{Cluster Position}: The edges of the face of BigCal are difficult to calibrate accurately; without blocks surrounding them, the blocks on the sides include partial clusters.  To avoid this area, we add a cut to exclude events which arrive on the edges of the calorimeter.  

\item \textit{Beam Current}: A cut was placed on the beam current to remove events which occurred during beam trips.  Only events occurring when the beam current is over 60 nA are included.
	
\end{enumerate}

	\subsection{Run Selection}
	
The data taken over the course of the experiment was broken up into over 500 experimental runs, each as much as one hour long.  In this first pass of analysis, any run which is suspected of having undesirable traits is rejected, to be added with closer inspection later.  To this end, a list of good runs was compiled to standardize run selection for the experiment.  The criteria for these run evaluations were designed to avoid runs with end--of--run errors, unacceptably low livetimes, asymmetries which were statistical outliers, or those labelled by operators as suspect.

\section{Asymmetry Production}

To generate experimental asymmetries we first produce electron yields in kinematic bins.  To accomplish this, software loops through all the events of an experimental run, keeping two sums of the events that satisfy the event criteria.  These two sums correspond to events which match the criteria of positive and negative beam helicity (the yields $N_+$ and $N_-$).  In addition to these sums, the kinematic quantities which describe the selected event, $\phi$, $\theta$ and $E'$, are averaged along the way.  By including a kinematic binning cut, we collect positive and negative yields for each kinematic bin, so that our measured asymmetry is, for example:
\begin{equation}
A_{\mathrm{measured}}(x;Q^2) = \frac{N_+(x;Q^2)-N_-(x;Q^2)}{N_+(x;Q^2)+N_-(x;Q^2)}
\end{equation}
for Bjorken $x$, or likewise for the invariant mass $W$.  The additional parameter $Q^2$ is included via a broad cut to allow the observation of any $Q^2$ evolution of the quantities of interest.

This measured asymmetry must go through several corrections before it is properly a physics asymmetry.  The dilution factor, and beam and target polarizations, serve to scale the measured asymmetry:
\begin{equation}
A_{\mathrm{physics}} = \frac{1}{fP_BP_T}A_{\mathrm{measured}},
\end{equation}
although corrections are also needed for charge asymmetry, livetime asymmetry, nitrogen polarization, and radiative effects.  This section will discuss these many corrections before outlining the procedure behind the asymmetry generation.

	\subsection{Charge Normalization}
Although the psuedo-random nature of the helicity flops of the beam polarization provides nearly the same number of electrons to the target for each helicity, unequal numbers of electrons can introduce a false asymmetry into our results.  \textit{Charge normalization} is performed to account for this effect.  For the charge accumulated on target, $C_+$ and $C_-$, and positive and negative helicity yields, $N_+$ and $N_-$, our charge normalized asymmetry is
	\begin{equation}
A_{\mathrm{CN}} = \frac{\dfrac{N_+}{C+}-\dfrac{N_-}{C-}}{\dfrac{N_+}{C+}+\dfrac{N_-}{C-}} = \frac{N_+Q-N_-}{N_+Q+N_-},
\end{equation}
for	 $Q=C_-/C_+$.  These $C_+$ and $C_-$ values were taken from the EPICS data stream helicity scalers for each run.  

\begin{figure}[htb]
  \begin{center}
   \includegraphics[width=2.42in]{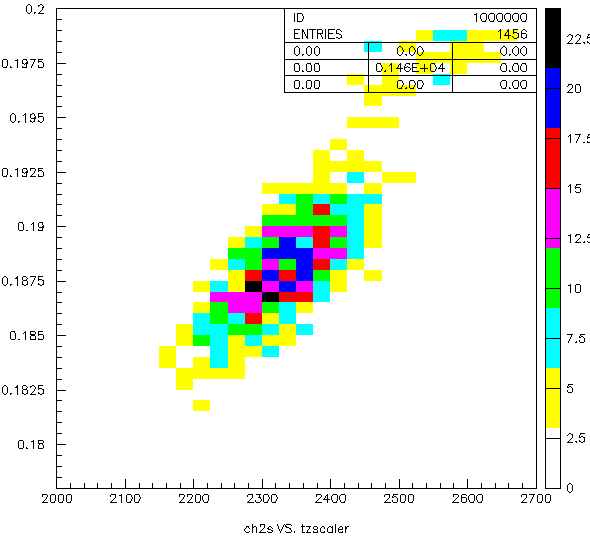}
   \includegraphics[width=2.4in]{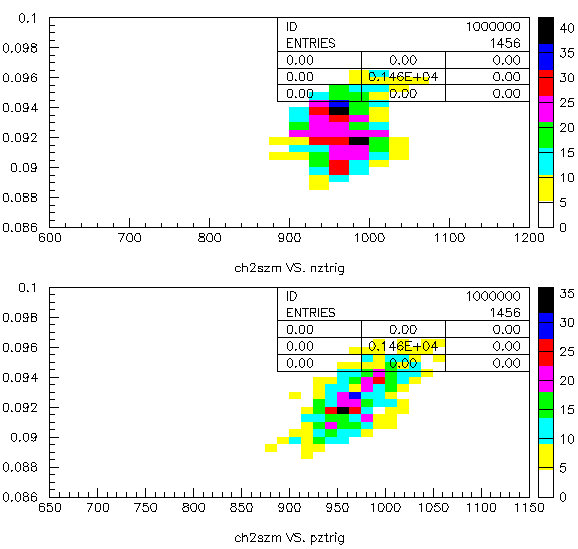}
  \end{center}
  \caption[Charge Helicity Swap.]{Plots showing the correlation between the charge per 2 seconds and trigger scalers for an example event.  The left plot show the charge vs. total scalers, giving the expected, roughly linear correlation.  At right are the negative helicity charge vs. the two helicity scalers, where this linear correlation can only be matched if the negative helicity charge is plotted against the ``positive'' helicity triggers, bottom right. Plot by H. Kang.}
  \label{fig:cnswap}
\end{figure}

It was discovered that in order to produce consistent results, the helicities for these scalers should be swapped.  This strange development was likely due to a swapped or mislabeled cable on the scalers during the experiment.  Figure \ref{fig:cnswap} shows evidence for the swap in the form of charge scalers vs. helicity triggers for an example run.

	\subsection{Livetime Correction}
As the trigger supervisor accepts a trigger to record an event, triggers that arrive while the data acquisition is busy are lost.  We account for this lost time, known as \textit{deadtime}, with a livetime correction to the asymmetry.	Reference \cite{leo} covers deadtime in some detail.

The simplest way to produce a livetime correction to the asymmetry is to directly calculate the computer deadtime for each helicity using the ratio of total accepted trigger events to the total event triggers as recorded by scalers.  Unfortunately, the positive helicity trigger scaler information was lost.  

Two methods were utilized and compared to produce a suitable livetime correction from the data we have; both methods resulted in very small corrections to the asymmetries.  The first approximation makes use of our knowledge of the true physics and background rates.  If we assume that the background events have a small asymmetry $A_B << A$ for physics asymmetry A, we can estimate the physics asymmetry in terms of the measured asymmetry $A_m$, livetime and ratio of physics rate to background rate $f = R/B$ \cite{orlt}:
\begin{equation}
\label{eq:lt}
A = A_m \frac{2+f}{2+fL}.
\end{equation}
If we further assume that the livetimes for the positive and negative helicities are approximately equal $L \approx L_+ \approx L_-$, so that the measured rates $R_m$ are dominated by the background, we have
\begin{equation}
R_m = \frac{R + B}{1 + D_t(R + B)}
\end{equation}
for measured system dead time $D_t$, the ratio of the total input to accepted trigger scalers.  Then the livetime to use in equation \ref{eq:lt} is
\begin{equation}
L = \frac{R_{m}}{R+B} = \frac{1}{1 + D_t(R + B)}.
\end{equation}	

The second method estimates the total positive helicity triggers by assuming the correlation of the accepted positive helicity triggers to the total positive helicity triggers is the same of that of the negative helicity triggers to their total.  By fitting the linear correlation of the negative helicity total scalers and accepted scalers that we have, we can determine the total positive triggers using the accepted positive triggers.  This method, as applied by collaborators H. Bahgdasaryan and H. Kang, gave more reliable results, and was used in this analysis.


	\subsection{Dilution Factor Correction}
	\label{sec:dilution}
We aim to study the spin structure of the proton, but our $^{14}$NH$_3$ material is obviously not a pure proton target.  Scattering from unpolarized material in the target dilutes the e--p scattering asymmetry, requiring the correction of a \textit{dilution factor}.  As we mentioned in section \ref{sec:material}, the dilution factor is simply the ratio of electron rates from the free, polarizable protons to the total rates from all nucleons in the target material.  This ratio is kinematics dependent, depending on the cross sections of the constituents of ammonia.  We thus apply the dilution as a function of invariant mass $W$. 

For $^{14}$NH$_3$, the dilution factor takes the form
\begin{equation}
\label{eq:dilution}
f = \frac{N_1\sigma_1}{N_{14}\sigma_{14} + N_1\sigma_1 + \sum N_A\sigma_A},
\end{equation}
where $N_A$ are the numbers of scattering nuclei of mass number $A$ per unit area in the target, and $\sigma_A$ are the radiated, polarized e--p cross sections and are functions of invariant mass $W$\cite{dfpf}.  The sum in this expression covers everything in the target cell that is not ammonia, such as helium and aluminum.

The numbers of scattering nuclei $N_A$ are computed in terms of Avogadro's number $N_0$, the atomic weight $M_A$, the partial density $\rho_A$ and effective target thickness $z_Ap_f$:
\begin{equation}
N_A = \frac{N_0 \rho_A z_A p_f}{M_A} [1/\textrm{cm}^2].
\end{equation}
This expression's volumetric component comes only from the target thickness due to the target cell's cylindrical shape.  The effective thickness can be computed in terms of the target cup's length and the \textit{packing fraction} $p_f$.
	
		\subsubsection{Packing Fraction}
To move forward with the calculation of the dilution factor of each target material load, we now need a measure of the packing fraction, the proportion of the target material to the liquid helium in which it is immersed.  While we endeavor to fill the target cup completely, differences in the load amount, and the size and shape of the target beads change the packing fraction from load to load.  By comparing the yields from each target load to those using a carbon disc target of known thickness, we can estimate the packing fraction throughout the experiment. 

\begin{figure}[htb]
  \begin{center}
   \includegraphics[width=3in]{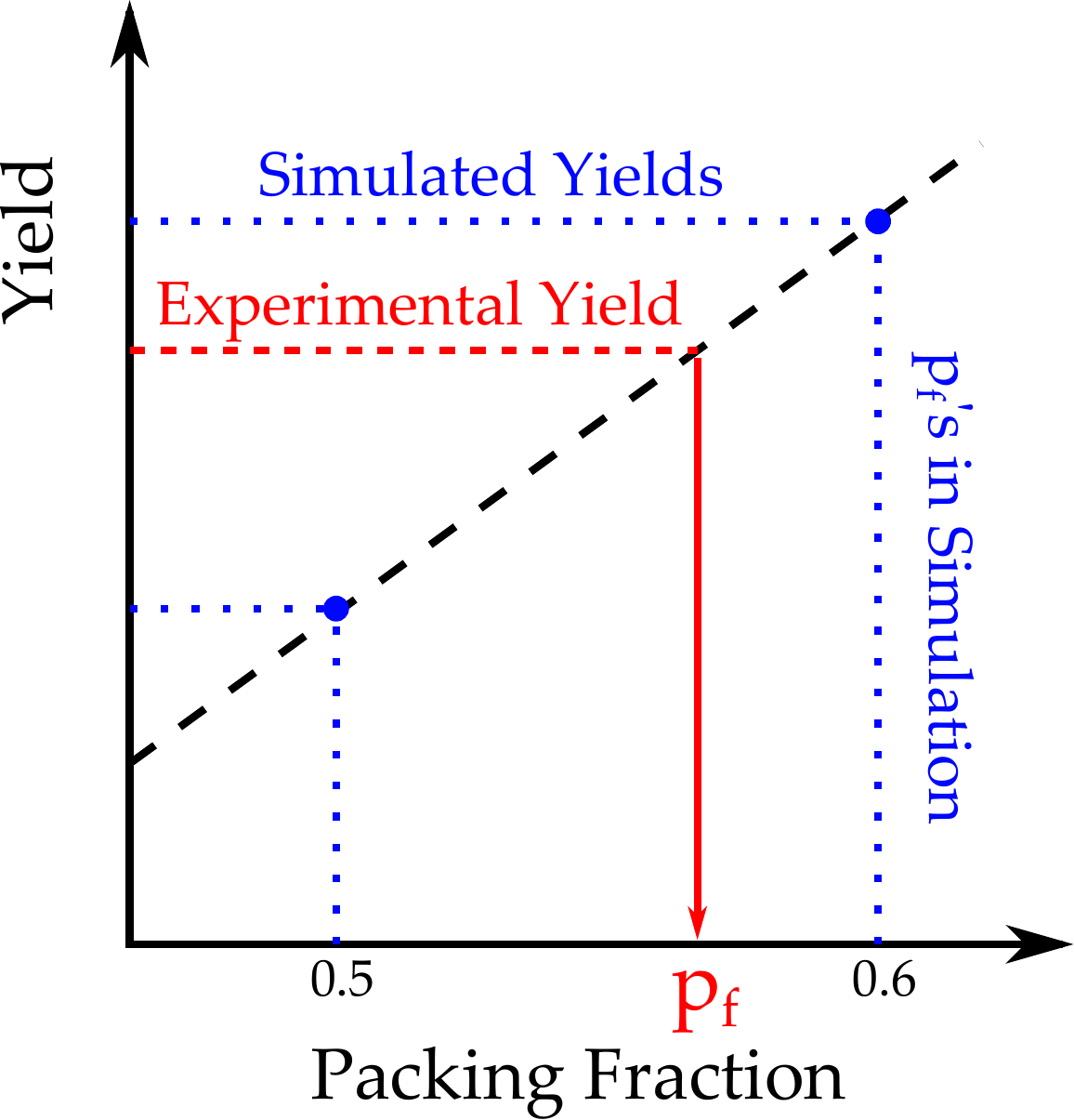}
  \end{center}
  \caption[Packing fraction interpolation.]{An illustration of the interpolation between simulated yields of packing fraction 0.5 and 0.6 to obtain the packing fraction from the experimental yield.}
  \label{fig:pf}
\end{figure}

To provide carbon data with which to compare our ammonia data, runs using a carbon disc were taken at many times during the experiment.  The electron yield from the target will be a linear function of the packing fraction:
\begin{equation}
Y = mp_f+b
\end{equation}
where $m$ and $b$ depend on the beam current, acceptance, partial densities and cross sections.  The linear form allows us to find the packing fraction of a given load by interpolating between two known reference points on the line.  These two points can come from a Monte Carlo simulation which accurately represents the acceptance of the detectors and the cross sections of the target materials involved.  A crucial consideration is the production of a scaling factor to bring the Monte Carlo yields into agreement with the carbon data.  The packing fraction is then a simple linear interpolation between the Monte Carlo yields with a target of packing fraction 0.5 and another of packing fraction 0.6, as seen in figure \ref{fig:pf}.

Table \ref{tab:pf} shows the packing fractions for each target material load used during SANE.  These packing fractions were the work of SANE collaborators H. Kang and N. Kalantarians, using experimental yields from the High Momentum Spectrometer, part of Hall C's standard equipment. 

\begin{table}[ht]
  \begin{center}
\begin{tabular}{lccllcc}
\toprule
Run Range		& Label Run  &$E_{beam}$  & Field & Target Cup  & $p_f$ (\%)  & $p_f$ error (\%)\\
\midrule
72213 - 72233 & 72213 & 4.7 GeV & Perp & Top & 70.1 & 5.16\\ 
72244 - 72256 & 72247 & 4.7 GeV & Perp & Bottom & 68.2 & 5.12\\ 
72271 - 72280 & 72278 & 4.7 GeV & Perp & Top & 49.2 & 4.19\\ 
72281 - 72286 & 72281 & 4.7 GeV & Perp & Bottom & 57.9 & 4.59\\ 
72378 - 72379 & 72379 & 4.7 GeV & Perp & Bottom & 70.1 & 5.16\\ 
72383 - 72416 & 72385 & 4.7 GeV & Perp & Bottom & 72.3 & 5.97\\ 
72657 - 72782 & 72658 & 5.9 GeV & Perp & Bottom & 64.4 & 5.30\\ 
72669 - 72792 & 72672 & 5.9 GeV & Perp & Top & 62.0 & 4.94\\ 
72783 - 72823 & 72790 & 5.9 GeV & Perp & Bottom & 60.2 & 4.98\\ 
72793 - 72836 & 72795 & 5.9 GeV & Perp & Top & 56.9 & 4.81\\ 
72824 - 72928 & 72828 & 4.7 GeV & Perp & Bottom & 62.6 & 4.50\\ 
72929 - 72983 & 72957 & 5.9 GeV & Para & Bottom & 60.6 & 4.68\\ 
72837 - 72985 & 72959 & 5.9 GeV & Para & Top & 59.7 & 4.38\\ 
72984 - 72985 & 72984 & 4.7 GeV & Para & Bottom & 73.7 & 4.86\\ 
72986 - 73018 & 72991 & 4.7 GeV & Para & Top & 68.0 & 4.08\\ 
72986 - 73041 & 73014 & 4.7 GeV & Para & Top & 56.6 & 4.17\\ 
73019 - 73041 & 73019 & 4.7 GeV & Para & Bottom & 58.9 & 4.45\\ 
\bottomrule
\end{tabular}
\caption{Table of packing fractions for all SANE target samples.}
  \label{tab:pf}
\end{center}
\end{table}

		\subsubsection{Dilution Factor Production}	
Once the packing fractions for each target material load have been obtained, the dilution factor can be produced using the GEANT3 Monte Carlo simulation of the target and electron detector package.	 The simulation throws electron events weighted by the partial densities of the constituents of the target, as dictated by the packing fraction.  As the simulation takes into account cross section models as well as the detector acceptance, it allows the creation of realistic, kinematics-dependent dilution factors.

Three million simulated events were thrown for each packing fraction, and each event's properties stored in an ``ntuple'' data file.  The quantity of interest was the weighting factor of the events which arrived as good events in the simulated calorimeter as a function of kinematics --- in this case invariant mass $W$.  By taking the ratio of the weighting factor for events originating from a polarized proton to that of any event, $w_{\textrm{proton}}/w_{\textrm{all}}$ we have a measure of the dilution factor for that packing fraction.  These ratios were then binned in the same kinematic bins chosen for the asymmetries, allowing direct application of the dilution factors.

\begin{figure}[htb]
  \begin{center}
   \includegraphics[width=3.7in]{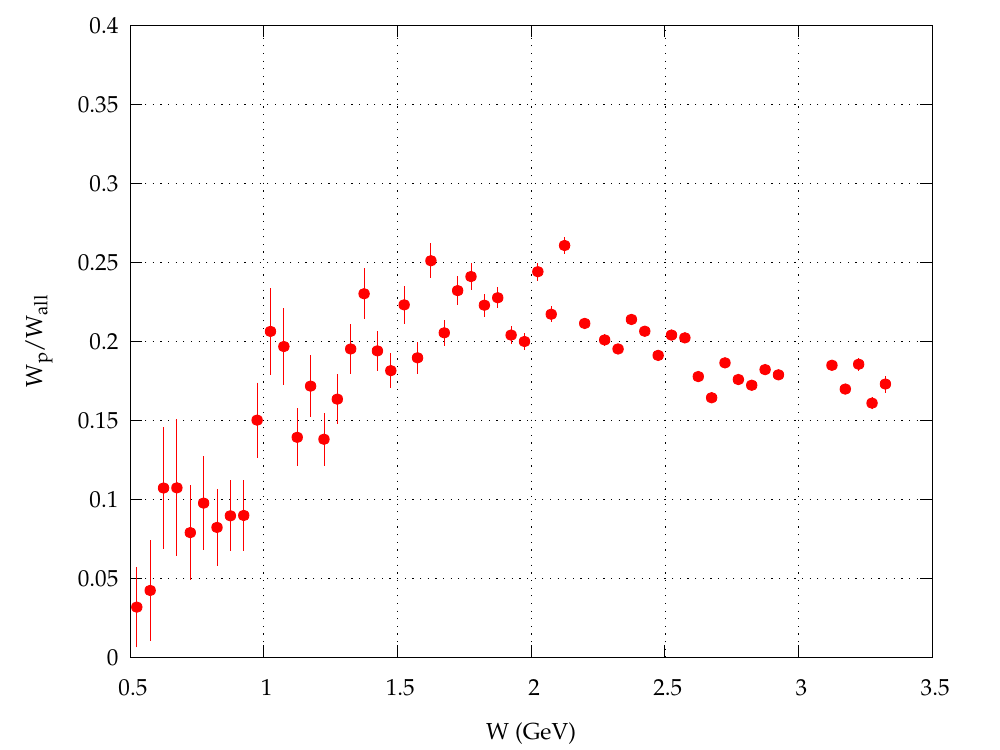}
  \end{center}
  \caption{Example dilution factor for a packing fraction of 0.606.}
  \label{fig:df}
\end{figure}

Once produced, this dilution factor is subject to radiative corrections, which were performed in a similar manner to that described in section \ref{sec:radcor} by SANE collaborator N. Kalantarians.  Figure \ref{fig:df} shows an example preliminary dilution factor for a packing fraction of 0.606.

As of this writing, these dilution factors are still under production by our collaborators.  To serve as a placeholder for the dilution factors, estimations were formulated for in this analysis following equation \ref{eq:dilution} and using cross sections of the neutron and proton from models of $F_2$.  

	\subsection{Nitrogen Correction}
While we take into account scattering from material which is not a free, polarizable protons with the dilution factor, the astute might guess that any polarization of the nitrogen atom in the ammonia target material might necessitate a further correction.  Electrons which scatter off a polarized nitrogen atom will indeed contribute to the polarized asymmetry, and this contribution is calculable.   However, while nitrogen provides a third of the polarizable nucleons, it only polarizes to approximately one sixth of the polarization of the hydrogen in the ammonia.  Since, in nitrogen, a nucleon spin is aligned anti-parallel to the spin of the nucleus one third of the time\cite{orpolcorr}, we find an approximate maximum polarization of anti-parallel nitrogen nucleons of $1/3\times1/3\times1/6 \approx 2\%$.  Even given an accuracy of these estimates of as poor as 10\%, the contribution is small enough that this value is adequate for our needs.

	\subsection{Elastic Radiative Corrections}
	\label{sec:radcor}
	
\begin{figure}[htb]
  \begin{center}
   \includegraphics[width=3.7in]{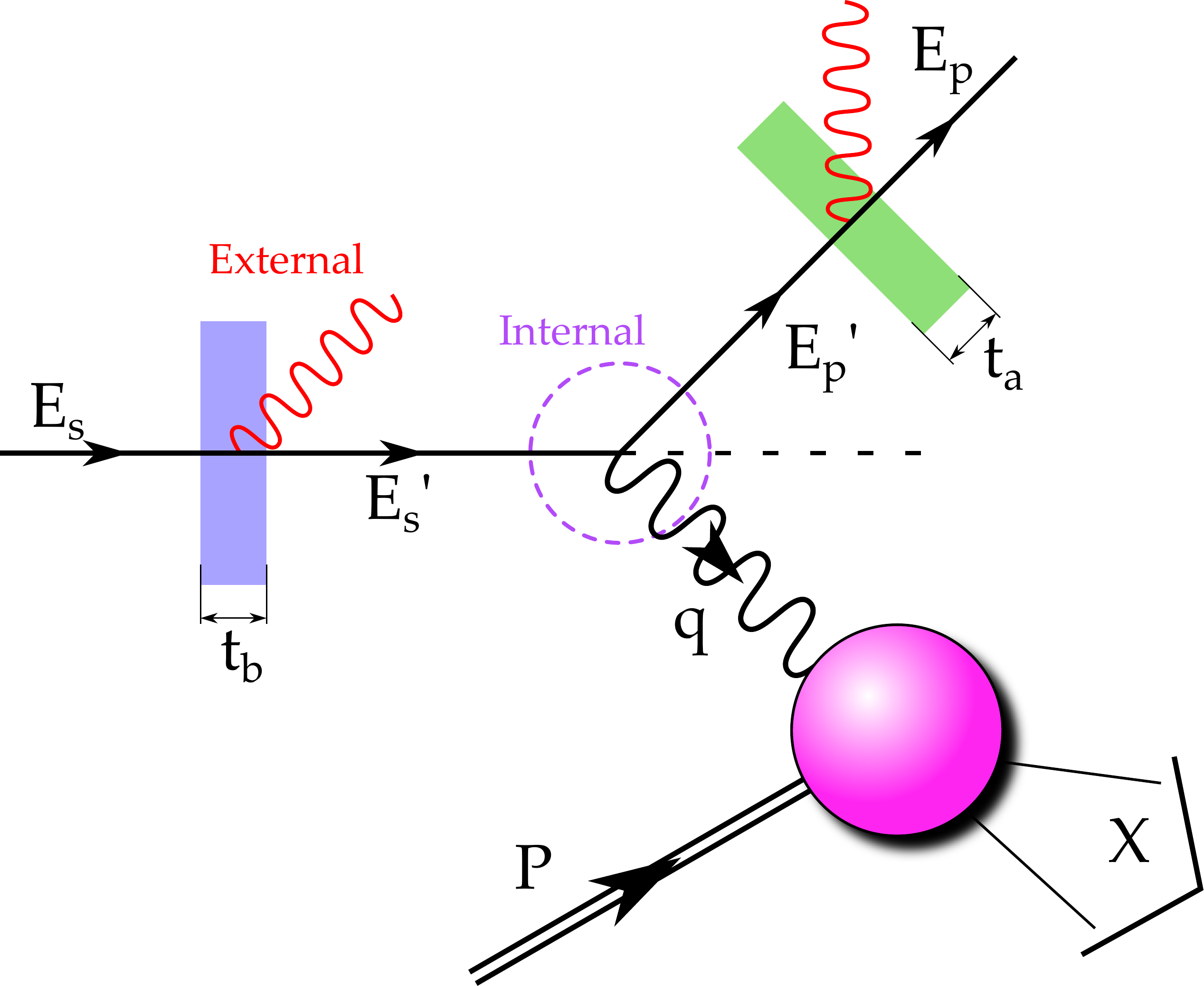}
  \end{center}
  \caption{Diagram showing the mechanisms making radiative corrections necessary.}
  \label{fig:radcor}
\end{figure}

Complicating our calculation of the asymmetries, the scattering electron is subject to energy losses from radiative effects.  As the beam passes through any material before reaching the scattering process of interest, such as the thin aluminum target windows and even the target material itself, an electron can radiate a photon at a probability related to the \textit{radiation length} of the material it passes through.  Likewise, after the scattering of interest, it can radiate a photon before reaching the detectors.  These \textit{external radiative effects} change our incident beam energy and reconstructed final electron energy, $E_s$ and $E_p$, to the actual energies of the e--p interation, $E_s'$ and $E_p'$, as seen in figure \ref{fig:radcor}.  In addition, radiative processes contribute higher-order Feynman diagrams with radiative loops to the process; these so called \textit{internal radiative effects} also require attention.  
	
The procedure for producing radiative corrections to experimental asymmetries is discussed further in appendix \ref{sec:radapp}.  Full radiative corrections require an accurate model of the cross sections involved; as the time-frame for producing such corrections is beyond the scope allowed in this research, only the contribution to the radiative corrections from the elastic tail is considered here.   As the elastic peak is well understood, the so-called \textit{radiative tail} produced by the elastic peak radiating into the deep inelastic region, as discussed in great detail in references \cite{motsai}\cite{stein}, is a more straight-forward contribution to calculate.

Software adapted from several sources by collaborator K. Slifer for the RSS experiment was used for the calculation of SANE's radiative tail.  This code allows the creation of the elastic radiative contribution for polarized targets with field orientations parallel and perpendicular to the incident beam.  We leave specifics of this process to the appendix, but it is worthwhile to address the adaptation of this code  necessary for SANE.

Since the code was written to handle only parallel and perpendicular target field orientations, SANE's 80\degrees\ field setting required some alterations.  The production of an elastic tail requires polarized elastic cross sections to radiate, and by calculating these cross sections at 80\degrees\ the desired results could be achieved.  As the code borrows from the MASCARAD routines by Afanasev, Akushevich and Merenkov\cite{afanasev}, we turned to the advice of A. Afanasev to ensure our 80\degrees\ calculation was correct\cite{afan}.

Following reference \cite{afanasev}, the polarized portion of the cross section is produced via the target polarization four-vector $\eta$.  The longitudinal $\eta_L$ and transverse $\eta_T$ polarization four-vectors are 
\begin{equation}
\begin{split}
\eta_L &= \frac{1}{\sqrt{\lambda_s}}\left( k - \frac{S}{M}p\right),\\
\eta_T &= \frac{1}{\sqrt{\lambda_s\lambda}} [ (-SX + 2M^2Q^2 + 4m^2M^2)k\\
 &\quad + \lambda_s k' - (SQ^2 + 2m^2S_x)p],
\end{split}
\end{equation}
following the momentum notation of chapter \ref{sec:intro}, and with $S=2kp$, $\lambda_s=S^2 - 4m^2M^2$,$\lambda = SXQ^2 -m^2\lambda_q-M^2Q^4$, $\lambda_q = (S-X)^2 + 4M^2Q^2)$ and $X=S-Q^2$.  We can combine $\eta_L$ and $\eta_T$ as orthogonal basis vectors to produce the polarization four-vector $\eta_S$ for any target field angle $\theta_S$:
\begin{equation}
\eta_{S} = \eta_L\cos\theta_S  + \eta_T\sin\theta_S.
\end{equation}
This generalization was inserted into the MASCARAD section of the code, allowing SANE's $\theta_S = 80\degrees$ case to be addressed.
 
The radiative tail correction was applied directly to the asymmetry using a multiplicative and an additive factor\cite{tailsub}.  We define our polarized and unpolarized cross sections, $\Delta$ and $\Sigma$, and asymmetry $A=\Delta/\Sigma$.  These total quantities are separable into inelastic and elastic contributions, as with the asymmetry $A_T = A_{in} + A_{el}$.  To find the inelastic portion of our asymmetry then, we have
\begin{equation}
\begin{split}
A_{in} &= \frac{\Delta_{in}}{\Sigma_{in}}\\
&= \frac{\Delta_T - \Delta_{el}}{\Sigma_{in}}\\
&= \frac{\Sigma_T A_T - \Delta_{el}}{\Sigma_{in}}\\
&= \frac{1}{f_{RC}}A_T - A_{RC},
\end{split}
\end{equation}
with $f_{RC} = \Sigma_{in}/\Sigma_T$ and $A_{RC} = \Delta_{el}/\Sigma_{in}$.  The unpolarized cross sections $\Sigma_{in}$ and $\Sigma_T$ are calculable from existing data, i.e. $F_1$ and $F_2$.  The polarized elastic cross section $\Delta_{el}$ is the result of our elastic radiative tail calculations.
	
\begin{figure}[htb]
  \begin{center}
   \includegraphics[width=3.7in]{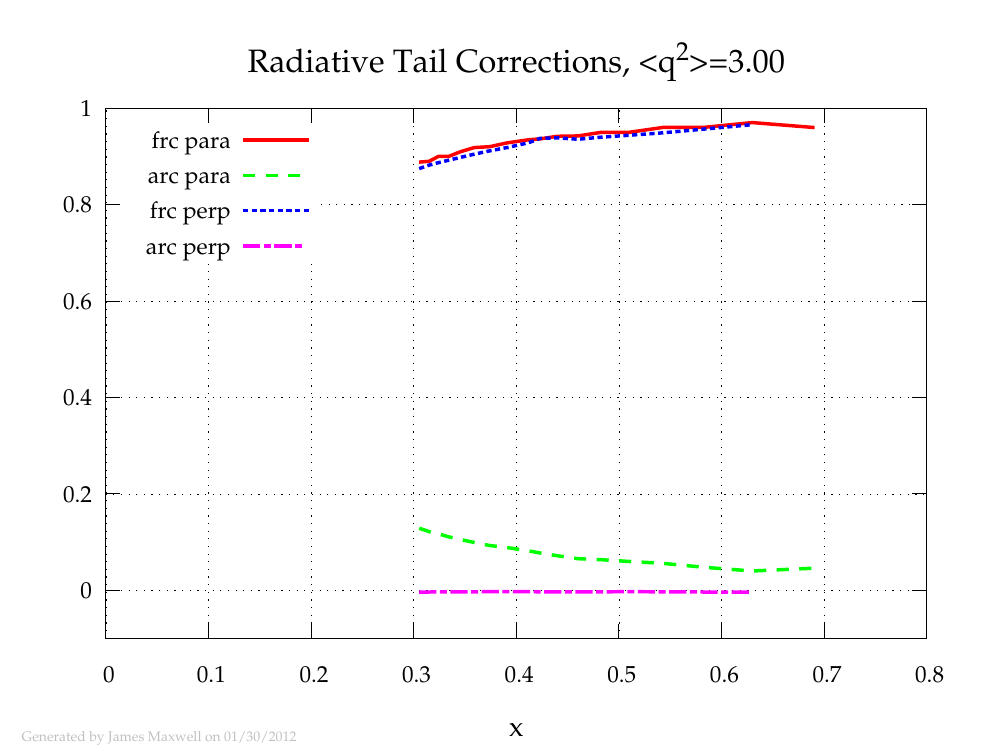}
  \end{center}
  \caption[Example of radiative tail correction factors]{Example of radiative tail correction factors for SANE kinematics, at $<Q^2> = 3$ GeV$^2$.}
  \label{fig:radcorex}
\end{figure}
Figure \ref{fig:radcorex} shows these correction factors $f_{RC}$ and $A_{RC}$ for SANE's  kinematic bins around $Q^2$ of 3  GeV$^2$. These values are averages of all radiative tail corrections applied by the bin per run level, in kinematic bins corresponding to those in the final results.

	\subsection{Physics Asymmetry}
After the sundry corrections above have been taken into account, our physics asymmetry is now
\begin{equation}
A_{\mathrm{physics}}(x,W;Q^2) =  \frac{1}{fP_BP_Tf_{RC}}\frac{Q_rN_+(x,W;Q^2)/L_p-N_-(x,W;Q^2)/L_n}{Q_rN_+(x,W;Q^2)/L_p+N_-(x,W;Q^2)/L_n} + A_{RC},
\end{equation}
where our yields $N_\pm$ are binned in $Q^2$ and are functions of either $x$ or $W$ bins.  Here $Q_r$ is the charge normalization ratio $Q_-/Q_+$, and $L_p$ and $L_n$ are the positive and negative helicity livetimes. This is accompanied by the following expression for the statistical error in the measurement:
\begin{equation}
\sigma_A = \frac{2}{fP_bP_Tf_{RC}}\sqrt{\frac{Q_r^2N_+N_-(N_++N_-)/L_p^2L_n^2}{(Q_rN_+/L_p +N_-/L_n)^4}}.
\end{equation}
A basic treatment of statistical error in the context of particle physics is found in reference \cite{lyons}.

Each run of the experiment is analyzed individually, with a Fortran subroutine looping through each event of the run for every kinematic cut to identify events which pass our selection process of each helicity.  In addition to adding a selected event to the helicity yield, several kinematic properties of the event are averaged into average quantities for each cut.  Thus, each run has a set of kinematic cuts, and each cut has helicity yields and this set of averaged physics quantities: beam energy $E$, scattering energy change $\nu$, final electron energy $E'$, scattering angles $\theta$ and $\phi$, invariant mass $W$, and Bjorken $x$.  The averaged quantities from this list will be used in the extraction of our polarization observables.

Once a table of cuts with yields and averaged physics quantities is produced for each experimental run, the runs must then be combined by target field orientation to create asymmetries $A_{\parallel}(x,W;Q^2)$ and $A_{80\degrees}(x,W;Q^2)$.  First, the target and beam polarizations, livetime, and charge normalization corrections are applied to each table; all the events of a given run share the same corrections.  The dilution factor and radiative corrections are functions of kinematics, and are applied to sets of runs via their bins of $x$ or $W$.  With tables of corrected asymmetries $A_i(x,W;Q^2)$ generated for each run $i$, the runs are combined by averaging each bin, weighted by the statistical error $\sigma_{A_i}$:
\begin{equation}
A_{\parallel,80\degrees}(x,W;Q^2) = \frac{\displaystyle{\sum_{i}A_i/\sigma_{A_i}^2}}{\displaystyle{\sum_{i}1/\sigma_{A_i}^2}}  \quad\quad \textrm{for}\quad i\in (\parallel, 80\degrees\ \textrm{runs}).
\end{equation}


\section{Extraction of Polarization Observables}
With physics asymmetries as a function of kinematics at hand, we can continue with the extraction of the quantities of interest.  We have simple expressions for the spin structure functions in terms of the virtual Compton asymmetries $A_1$ and $A_2$ from section \ref{sec:compasym}, so we will first tackle the extraction of $A_1$ and $A_2$.

	\subsection{Virtual Compton Asymmetries}
	\label{sec:analasym}
Returning to the results of chapter \ref{sec:intro}, we recall equations \ref{eq:a1a2}:	
\begin{equation}
\begin{split}
\label{eq:a1a22}
A_1 &=\frac{1}{D'}\left[ A_{180\degrees}\frac{E - E'\cos\theta}{E+E'} + \left( A_{80\degrees} + A_{180\degrees}\cos 80\degrees\right)\frac{E'\sin\theta}{(E+E')\cos\phi \sin 80\degrees}  \right]\\
A_2 &=\frac{1}{D'}\frac{\sqrt{Q^2}}{2E}\left[ A_{180\degrees} + \left( A_{80\degrees} + A_{180\degrees}\cos 80\degrees\right)\frac{E-E'\cos\theta}{E'\sin\theta\cos\phi\sin 80\degrees}  \right],
\end{split}
\end{equation}
where $D' = (1-\epsilon)/(1+\epsilon R)$ and $\epsilon = 1/(1+2(1+\nu^2/Q^2)\tan^2(\theta/2))$.
	
All the pieces of these two puzzles lie in our data tables of $A_{180\degrees}$ and $A_{80\degrees}$, with the notable exception of $R$.  We recall from section \ref{sec:compasym} that $R$ is the ratio of longitudinal to transverse Compton cross sections, but $R$ can also be expressed in terms of the unpolarized structure functions $F_1$ and $F_2$\cite{Anselmino19951}:
\begin{equation}
R = \frac{F_2}{F_1}\frac{M}{\nu}\left(1 + \frac{\nu^2}{Q^2}\right) -1
\end{equation}
As $F_1$ and $F_2$ are well understood, we can rely on parametrizations of existing measurements to provide $F_1$, $F_2$ and $R$.  In this analysis, the Bosted--Christy data parametrizations \cite{Bosted,Christy} were used to provide these quantities; the Fortran subroutine F1F209.f was called for each $Q^2$ and $W$ bin.

In order to combine our measured $A_{180\degrees}(x,W;Q^2)$ and $A_{80\degrees}(x,W;Q^2)$ with equations \ref{eq:a1a22}, we need to produce the other physics quantities which make up the equations: $E$, $E'$, $\nu$, $\theta$, $\phi$ and $Q^2$.  These quantities were generated per cut as a statistics weighted average over the runs.  Now we must average these averages from the two target polarization orientations, again weighting by statistical errors.  For example, for the averaged values of $E_{i,j}'$ for $\parallel$ runs $i$ and $80\degrees$ runs $j$, we have a weighted average of $E'$ for each bin:
\begin{equation}
\begin{split}
E' &= \left(\frac{\displaystyle{\sum_i E'_i/\sigma_{A_i}^2}}{\displaystyle{\sum_i 1/\sigma_{A_i}^2}}+\frac{\displaystyle{\sum_j E'_j/\sigma_{A_j}^2}}{\displaystyle{\sum_j 1/\sigma_{A_j}^2}}\right)  \left(\frac{1}{\displaystyle{\sum_i 1/\sigma_{A_i}^2}}+\frac{1}{\displaystyle{\sum_j 1/\sigma_{A_j}^2}}\right)^{-1}\\
&\quad \textrm{for}\ i \in\ \parallel \textrm{runs}, j \in 80\degrees \textrm{runs}.
\end{split}
\end{equation} 
All the physics quantities of interest are averaged in this manner, so that each kinematic bin has its own values. These physics quantities are also used to calculate six parameters which are used to simplify the computation of $A_1$ and $A_2$, such that
\begin{equation}
\begin{split}
A_1(x,W;Q^2) &= \frac{1}{D'} \left( A A_{80\degrees} + B' A_{180\degrees} + B A_{180\degrees}\right),\\
A_2(x,W;Q^2) &= \frac{1}{D'} \left(C A_{80\degrees} + D' A_{180\degrees} + D A_{180\degrees}\right),
\end{split}
\end{equation}
where the parameters are
\begin{equation}
\begin{split}
A(x,W;Q^2) & = \frac{E'\sin\theta}{(E+E')\cos\phi\sin 80\degrees}\\
B'(x,W;Q^2) & = A(x,W;Q^2)\cos 80\degrees\\
C(x,W;Q^2) & = -\sqrt{Q^2}\frac{E-E'\cos\theta}{2EE'\sin\theta\cos\phi\sin 80\degrees}\\
D'(x,W;Q^2) & = C(x,W;Q^2)\cos 80\degrees
\end{split}
\end{equation}
from the 80$\degrees$ runs and 
\begin{equation}
\begin{split}
B(x,W;Q^2) & = \frac{E-E'\cos\theta}{E+E'}\\
D(x,W;Q^2) & =  -\sqrt{Q^2}\frac{1}{2E}
\end{split}
\end{equation}
from the 180$\degrees$ runs, and with the term $D'$ unfortunately used twice for different quantities due to our notation.  Like the other physics quantities, these parameter are calculated for each event, then averaged by bin and run, and averaged again to form a value for each bin.  The two primed factors, $B'$ and $D'$ are just constants times the other 80 degree parameters, and are a results of the mixing of parallel and perpendicular asymmetries in the near perpendicular running. 

	\subsection{Spin Structure Functions}

The extraction of the spin structure functions $g_1$ and $g_2$ proceeds simply from the Compton asymmetries if we rewrite equations \ref{eq:g1g2} so that
\begin{equation}
\begin{split}
g_1 &= \frac{F_1}{1+\gamma^2}(A_1 + \gamma A_2)\\
g_2 &= \frac{F_1}{1+\gamma^2}(A_2/\gamma - A_1)
\end{split}
\end{equation}
for $\gamma = \sqrt{4x^2M^2/Q^2}$.

\chapter{Results}

We present here the results of our SANE analysis, culminating in the extracted spin structure function $g_2$ and a preliminary test of the Burkhardt--Cottingham sum rule.  Following the results, we will consider the implications of this work for our understanding of nucleon spin structure, and discuss what remains to finalize these data. 

Over 9 million events passed the event selection and are included in this analysis---nearly 5 million for the parallel target field setting, and over 4.5 million for the 80$\degrees$ setting.  These events fell between $Q^2$ of 1.5 and 6.5 GeV$^2$, with an $x$ range above 0.25.   All the plots of this chapter will show statistical error bars, with the spin asymmetries and structure functions also showing an estimated systematic error band.

\section{Experimental Physics Asymmetries}
\begin{figure}[phtb]
  \begin{center}
   \includegraphics[width=5in]{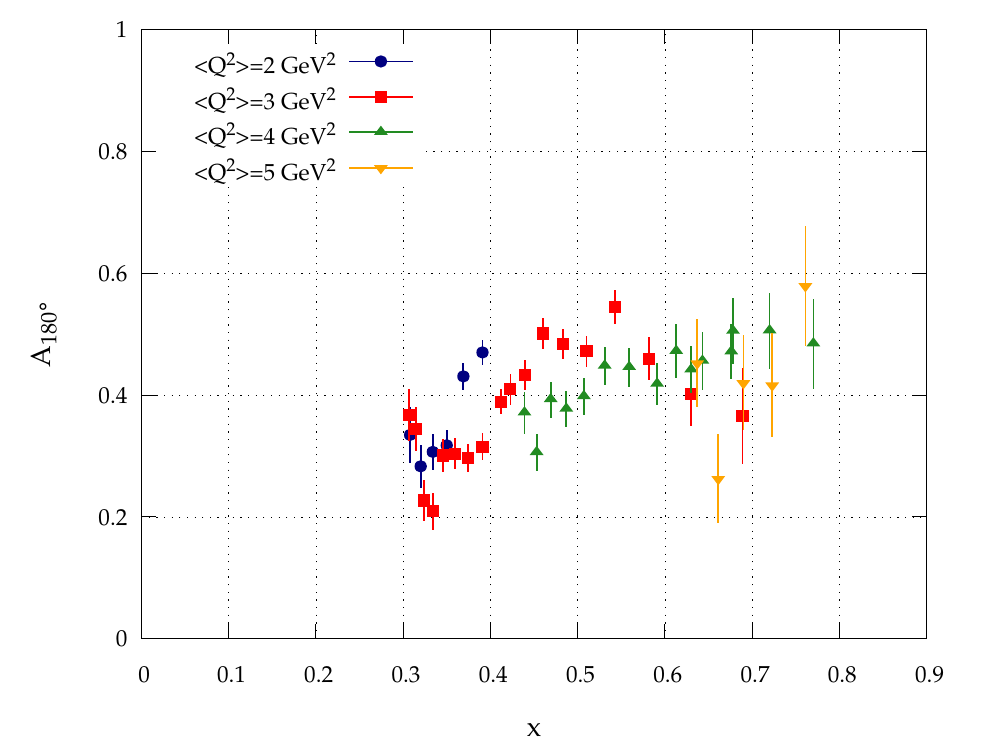}
   \includegraphics[width=5in]{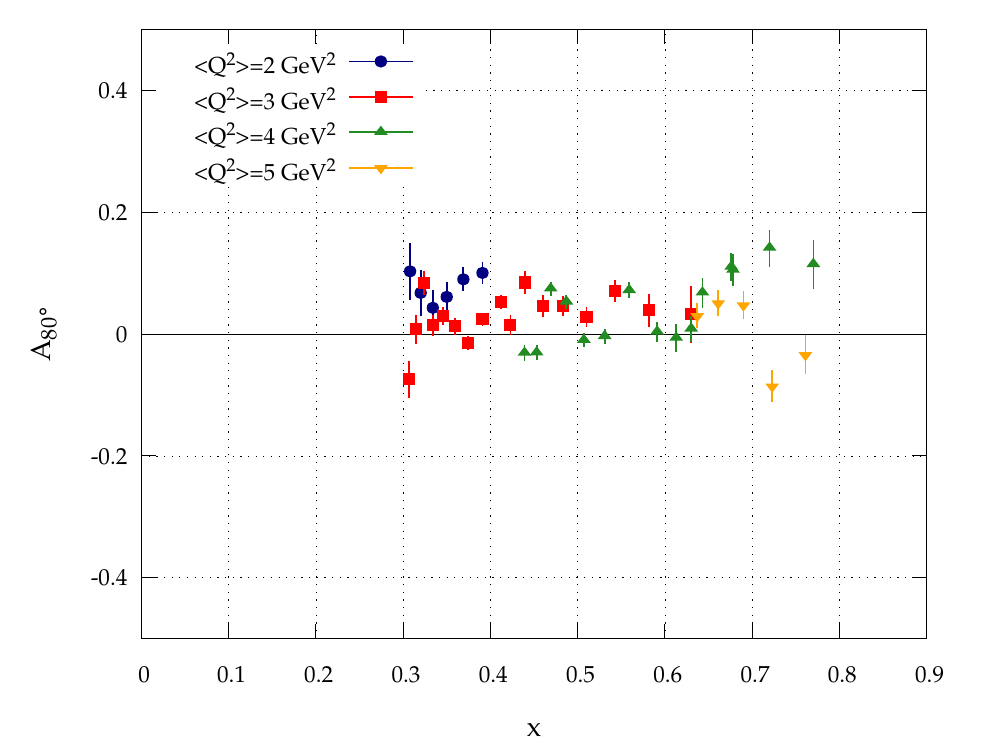}
  \end{center}
  \caption[Experimental physics asymmetries as a function of $x$.]{Experimental physics asymmetries as a function of $x$ for the 180$\degrees$ (top) and 80$\degrees$ (bottom) target field orientation settings in four $Q^2$ bins, calculated with corrections as described in chapter \ref{sec:anal}.}
  \label{fig:physasym}
\end{figure}
Figure \ref{fig:physasym} shows the experimental physics asymmetries, binned in $Q^2$ and $x$ and averaged by run for the two target field configurations, parallel and 80$\degrees$.  As expected, the parallel asymmetries are much larger; the near perpendicular values, while small, are consistently non-zero and positive.  The $Q^2$ behavior is not always consistent across bins, as we will discuss in section \ref{sec:q2dep}.  However, the results are much as expected, and lead us on to the more interesting quantities.

\section{Spin Asymmetries  $A_1$ and $A_2$}
\begin{figure}[htbp]
  \begin{center}
   \includegraphics[width=5.0in]{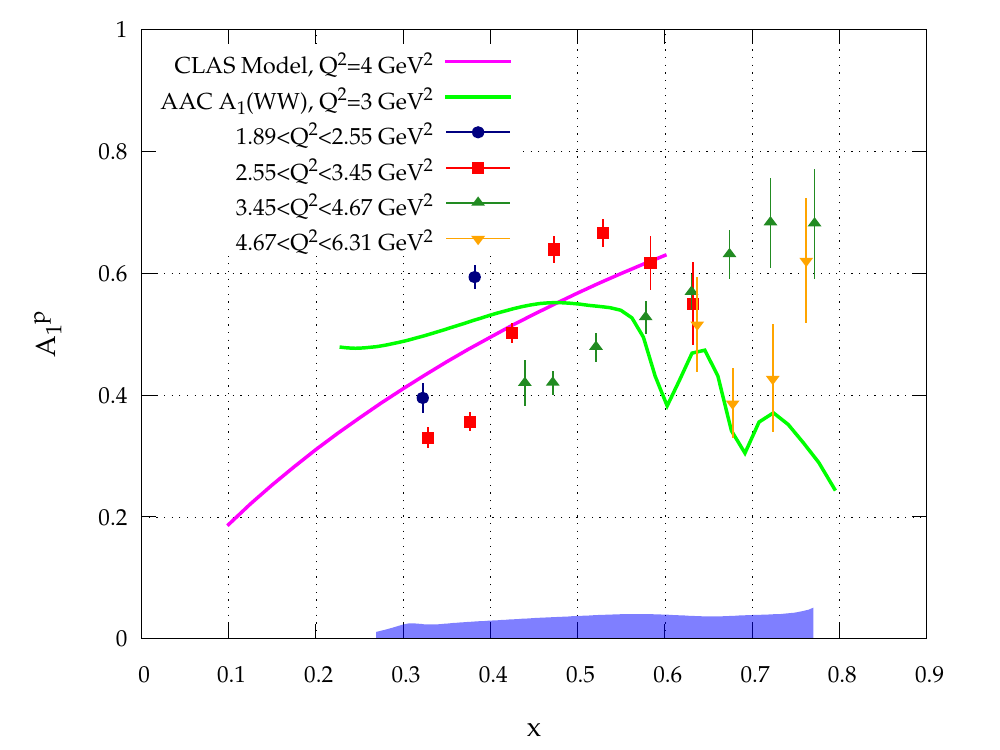}
   \includegraphics[width=5.0in]{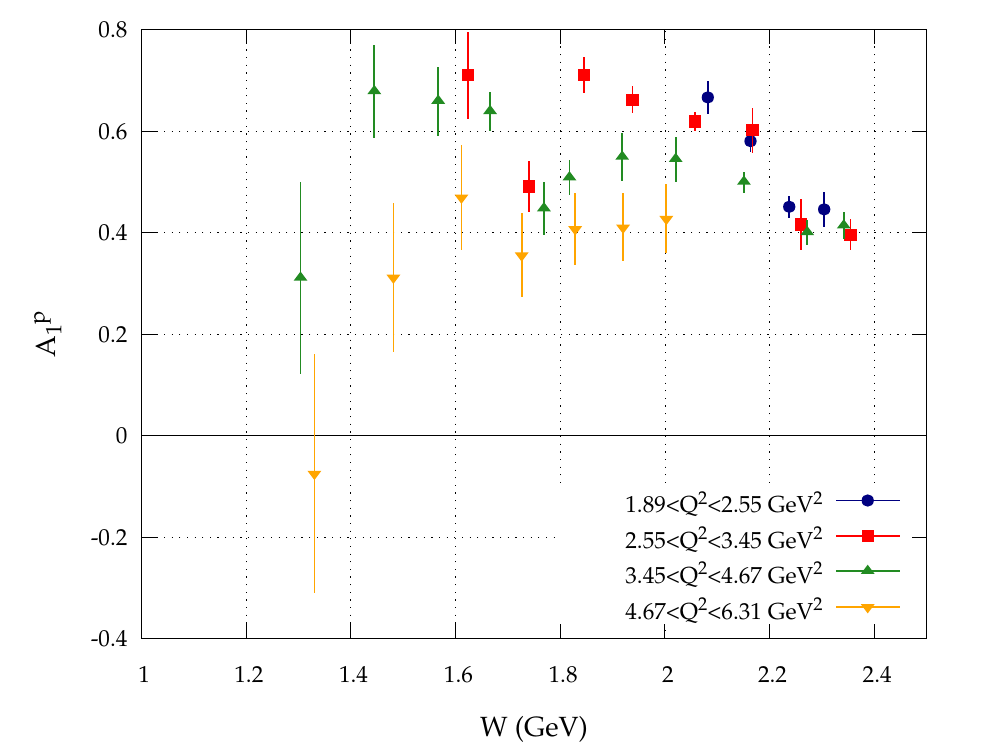}
  \end{center}
  \caption[Spin asymmetry $A_1$ as a function of $x$ and $W$.]{Spin asymmetry $A_1$ as a function of $x$ and $W$ for various $Q^2$.  The blue bar represents the estimated systematic error.}
  \label{fig:a1}
\end{figure}

\begin{figure}[htbp]
  \begin{center}
   \includegraphics[width=5.2in]{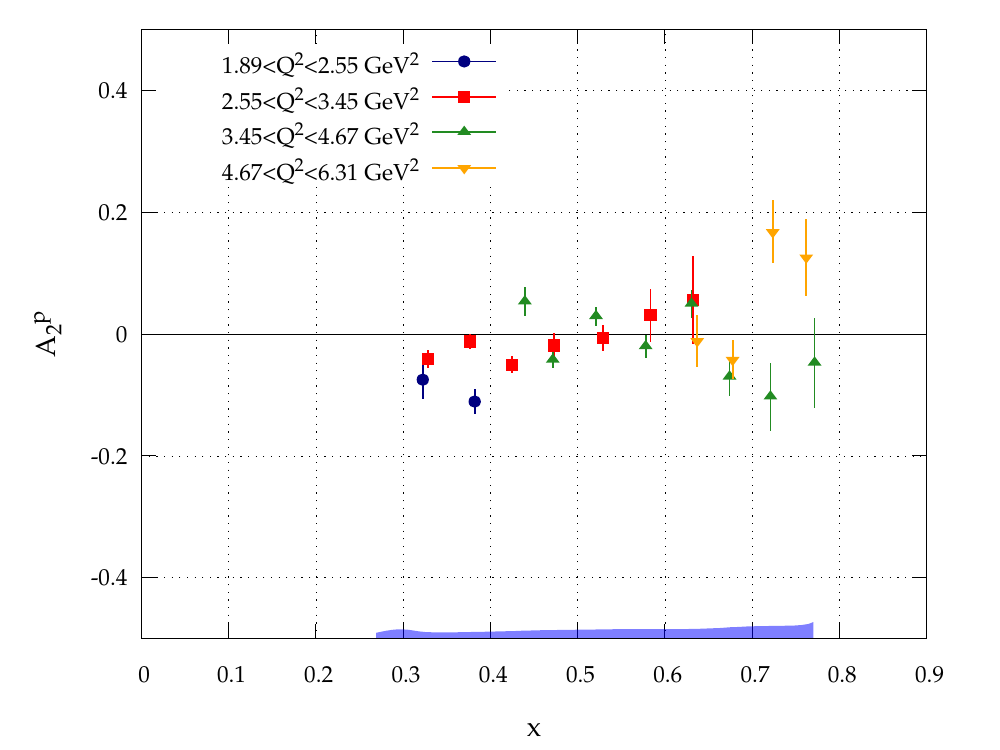}
  \end{center}
  \caption[Spin asymmetry $A_2$ as a function of $x$.]{Spin asymmetry $A_2$ as a function of $x$ for various $Q^2$. The blue bar represents the estimated systematic error.}
  \label{fig:a2}
\end{figure}

Combining the asymmetries for the two field angle settings as prescribed in section \ref{sec:analasym}, we have the spin, or virtual Compton, asymmetries shown in figures \ref{fig:a1} and \ref{fig:a2}.  We present the both asymmetries binned in $x$ using combined binning to reduce the number of data points. $A_1$ also is given as a function of $W$ at the bottom of figure \ref{fig:a1}.  Plotted with $A_1$ versus $x$ is a CLAS collaboration model of $A_1$ using fits to world data, as well as leading twist $A_1$ calculated from AAC polarized parton distributions \cite{aac} discussed in section \ref{sec:exist} and $F_1$ from the Bosted--Christy parameterizations \cite{Bosted,Christy}.

Figure \ref{fig:a1} shows $A_1$ from this analysis following the model calculations to a degree, albeit not within statistical errors.  
Some disparity between the results in different $Q^2$ bins is visible as the results of similar $x$ fall outside their statistical errors;  we discuss the apparent $Q^2$ dependence is greater detail in the coming subsection.

A $W$ dependence in $A_1$ which is visible in the deep inelastic region above $W$ of 2 GeV in the bottom figure \ref{fig:a1} represents a surprising feature.  $A_1$ drops from above 0.6 to 0.4 in a narrow range of $W$ from 2.0 to 2.4 GeV, with general agreement between the three $Q^2$ bin results.   We conjecture that the pair symmetric background correction currently under investigation by our collaborators may shed light on this issue, as it will affect this kinematic region.

The $A_2$ results in figure \ref{fig:a2} are for the most part difficult to statistically differentiate from zero. 
Although the $Q^2$ dependence we see in $A_1$ is present, it is not as pronounced as in $A_1$.  The dip at low $x$ in particular may be an artifact of pair symmetric background.

\subsection{$Q^2$ Dependence}
\label{sec:q2dep}

The $Q^2$ disparity at similar $x$ which we have seen in the experimental and spin asymmetries, particularly in the upper plot of figure \ref{fig:a1}, will be echoed in the spin structure functions as well.  One expects to see little $Q^2$ dependence in these asymmetries versus $x$, so it is worth discussing briefly.

\begin{figure}[htbp]
  \begin{center}
   \includegraphics[width=5.1in]{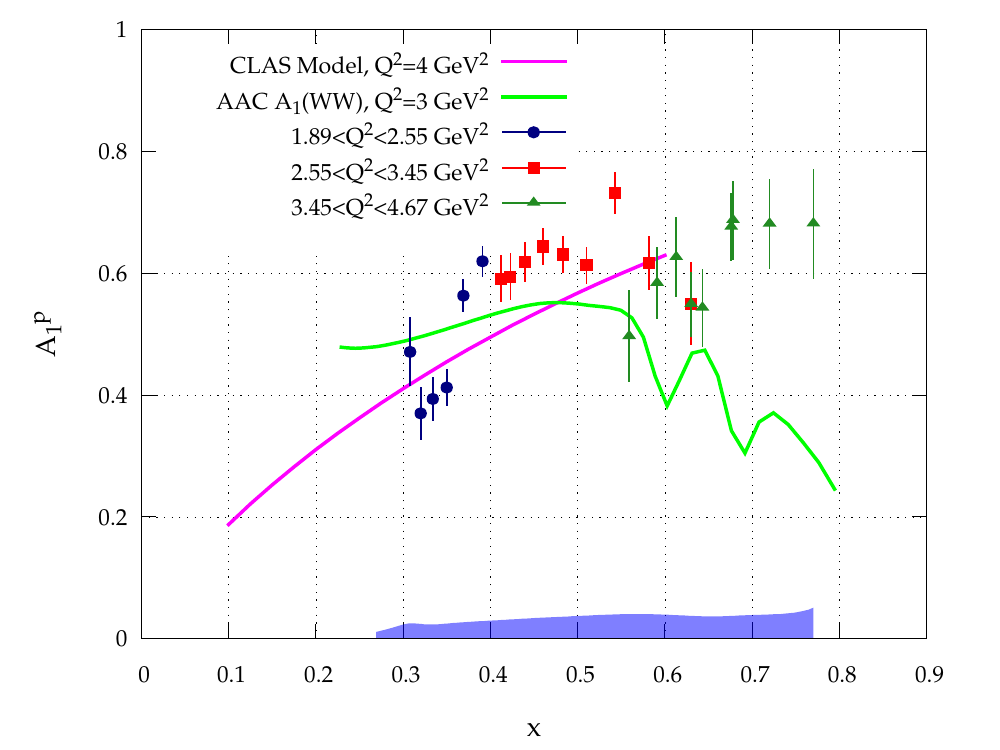}
   \includegraphics[width=5.1in]{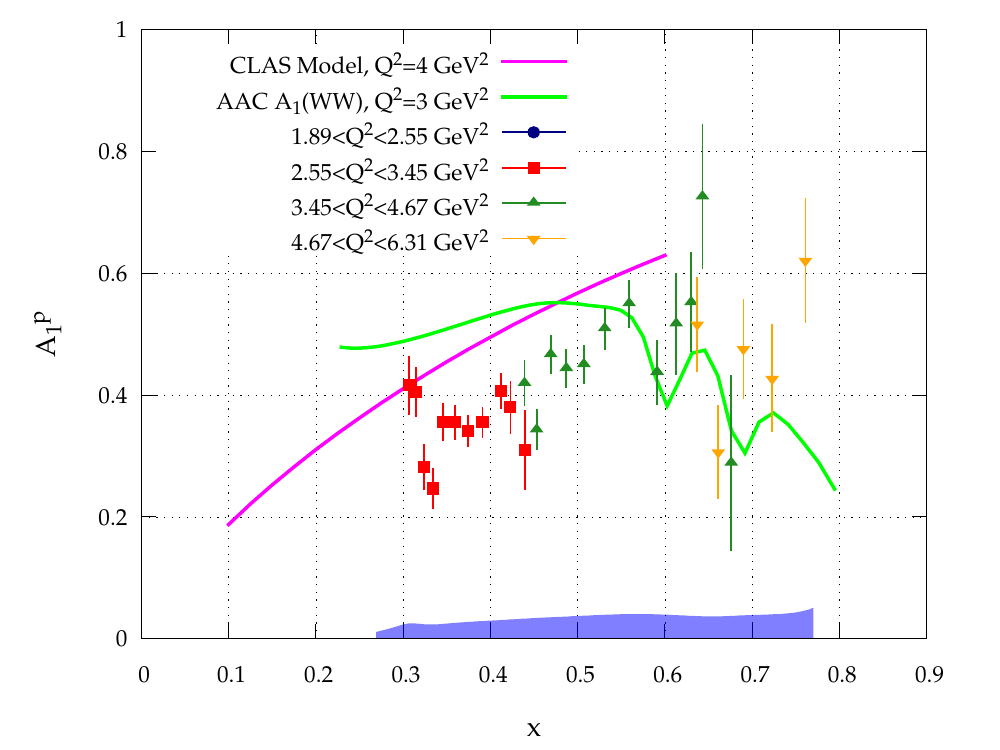}
  \end{center}
  \caption[$A_1$ vs. $x$ for each beam energy setting]{$A_1$ vs $x$ for the 4.7 GeV (top) and 5.9 GeV (bottom) beam energy setting, showing the discrepancy in $A_1$ at the same $x$ between the two cases.}
  \label{fig:a1_energies}
\end{figure}

The discrepancy is quite stark when splitting our data into the portions originating from 5.9 GeV beam energy and 4.7 GeV beam energy, as seen in figure \ref{fig:a1_energies}.  The figure shows $A_1$ as a function of $x$ for the two beam energy settings, here without recombined $x$ bins.  We can see the root of the $Q^2$ behavior in this plot; $A_1$ for each data set is continuous, but the 5.9 GeV set is lower than the 4.7, particularly in the region around 0.4 to 0.5 in $x$.  We can understand the $Q^2$ dependency of $A_1$ for the combined energies plot in this light: as each $Q^2$ set increases in $x$, it moves from a region where the 5.9 GeV data dominates and $A_1$ is lower, to the region where the 4.7 GeV data dominates.

\begin{figure}[htbp]
  \begin{center}
   \includegraphics[width=5.0in]{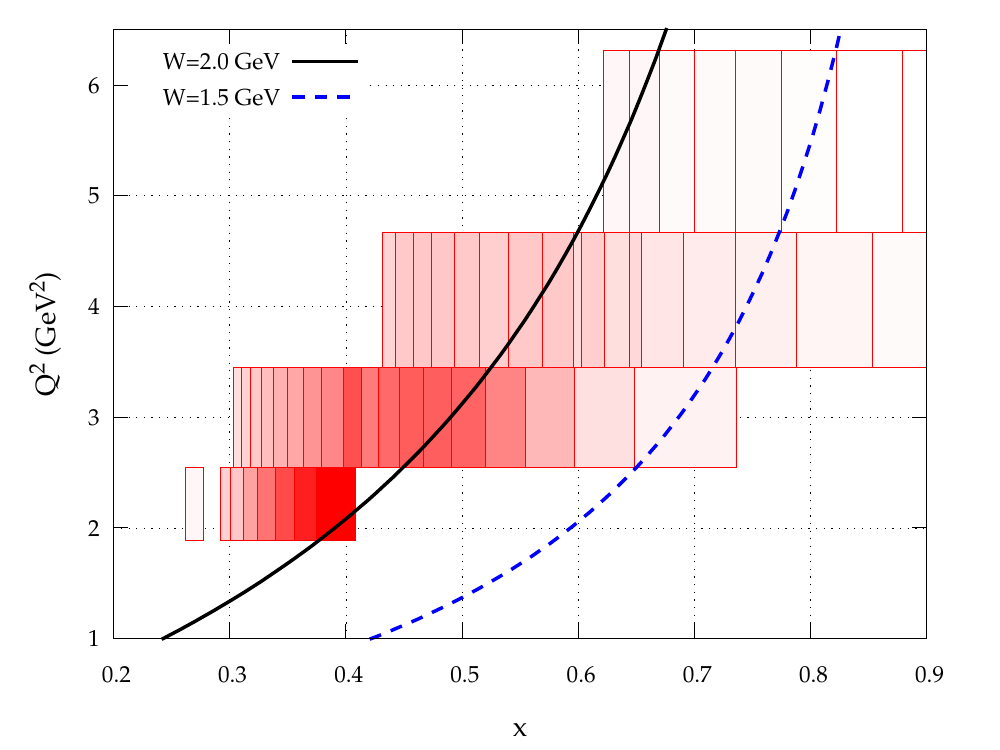}
   \includegraphics[width=5.0in]{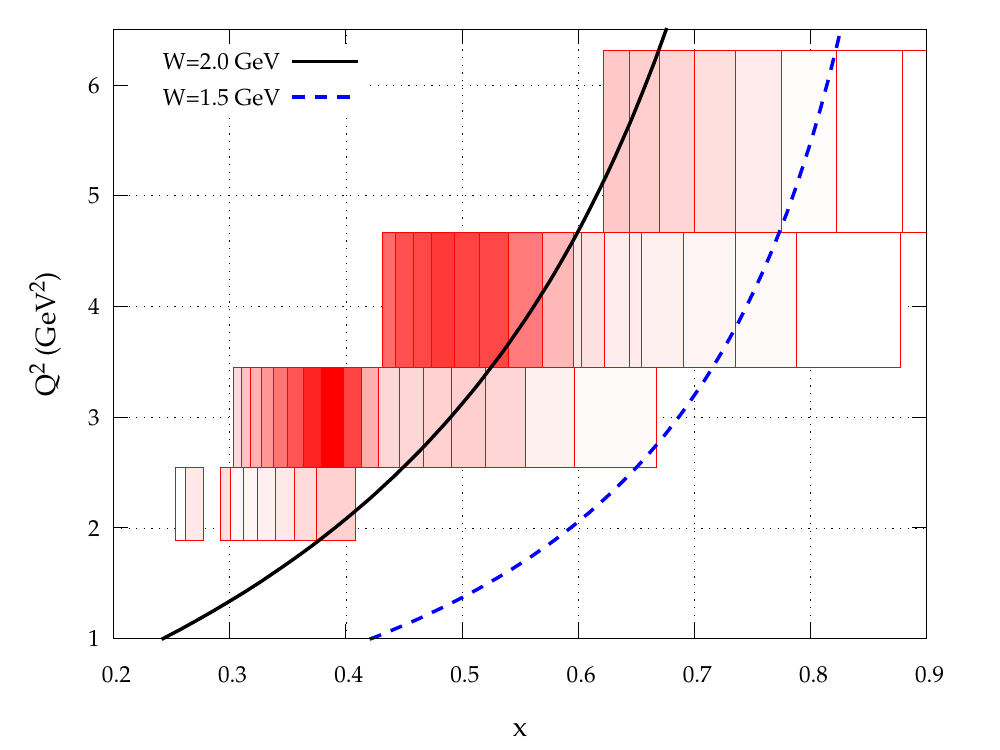}
  \end{center}
  \caption[Kinematic bins and their populations]{Kinematic bins and their populations for parallel (top) and near perpendicular (bottom) target field.  A deeper shade indicates more events.  Constant $W$ of 1.5 and 2 GeV are plotted to illustrate the change in W across the bins.}
  \label{fig:kines}
\end{figure}

Our choice of bins is certainly exaggerating the $Q^2$ disparity seen in the plots versus $x$.  If we consider figure \ref{fig:kines}, we see each kinematic bin and its population, as well as lines of constant $W$ of 1.5 and 2 GeV.  What we notice is that our choice of larger $Q^2$ bins, which makes each bin tall and narrow, also means the inclusion of a broad range in $W$ in each point.  Thus, the $Q^2$ variance in the plots versus $x$ are reflecting not necessarily $Q^2$ dependence, but $W$ dependence.  

\begin{figure}[htbp]
  \begin{center}
   \includegraphics[width=5.0in]{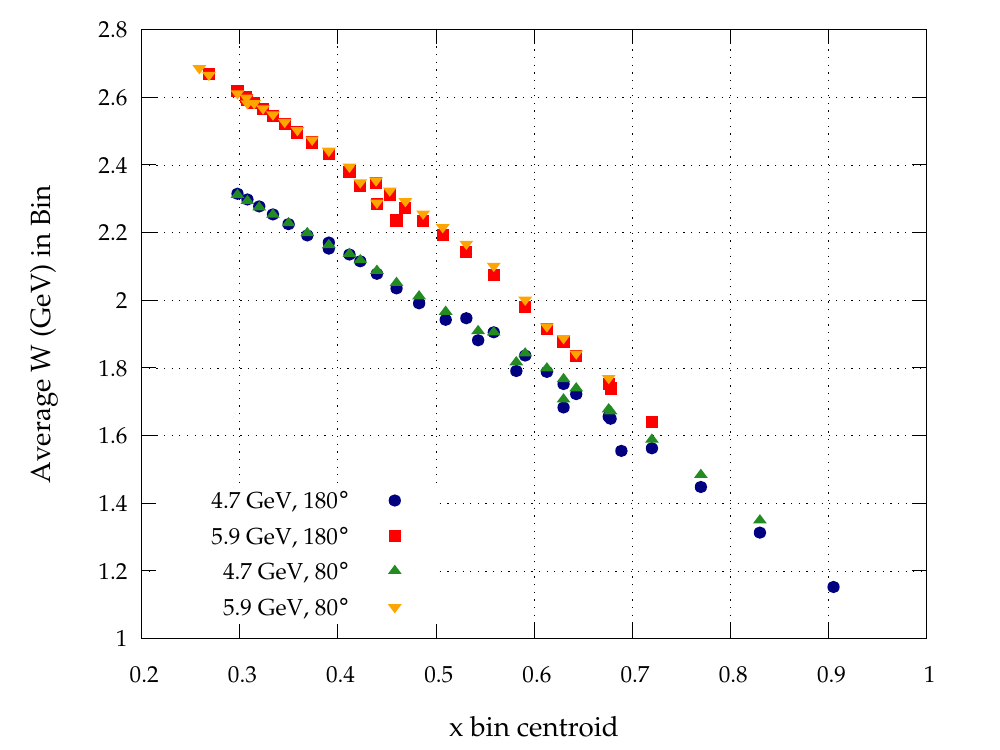}
  \end{center}
  \caption[Average $W$ per bin vs $x$ bin centroid]{Average $W$ per $x$ bin vs $x$ bin centroid for each beam energy setting and target field orientation, covering all $Q^2$ bins.  The gap between the 4.7 and 5.9 GeV data is of interest, and is a consequence of our binning.  Here errorbars in y are small enough to be ignored, and are removed for clarity.}
  \label{fig:w_v_x}
\end{figure}

The real origin of this behavior appears to be the $W$ dependence we are seeing in $A_1$.  We show the difference in average $W$ for each $x$ bin for the two beam energies clearly in figure \ref{fig:w_v_x}.  The gap between the 4.7 and 5.9 GeV $W$ averages at the same $x$ is one piece of the puzzle.  The other piece is the $W$ dependence in $A_1$ discussed in the previous section.  A close look at the deep inelastic region, $W>2$ at the bottom of figure \ref{fig:a1}, shows a dramatic drop of nearly 50\% in $A_1$ in about 0.5 GeV of $W$ above the resonance region for all $Q^2$ bins.  Thus our $x$ bins, which are sampling a range of $W$, reflect the $W$ dependence in the deep inelastic as we plot our different $Q^2$ bins. 

\section{Spin Structure Functions $g_1$ and $g_2$}

\begin{figure}[phtb]
  \begin{center}
   \includegraphics[width=5.0in]{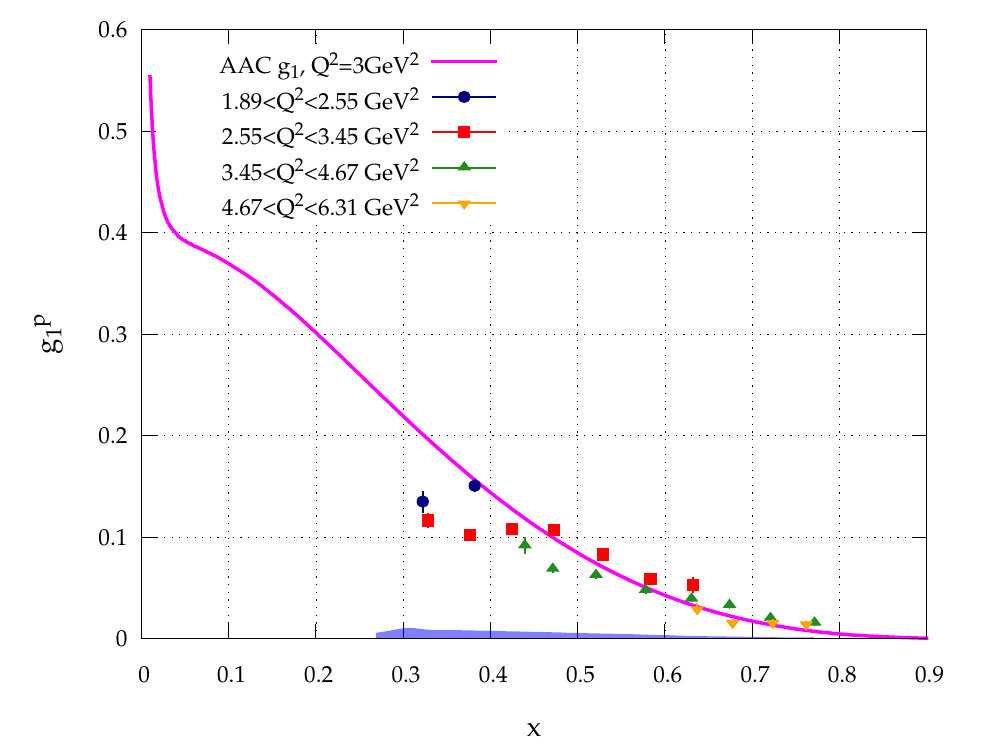}
   \includegraphics[width=5.0in]{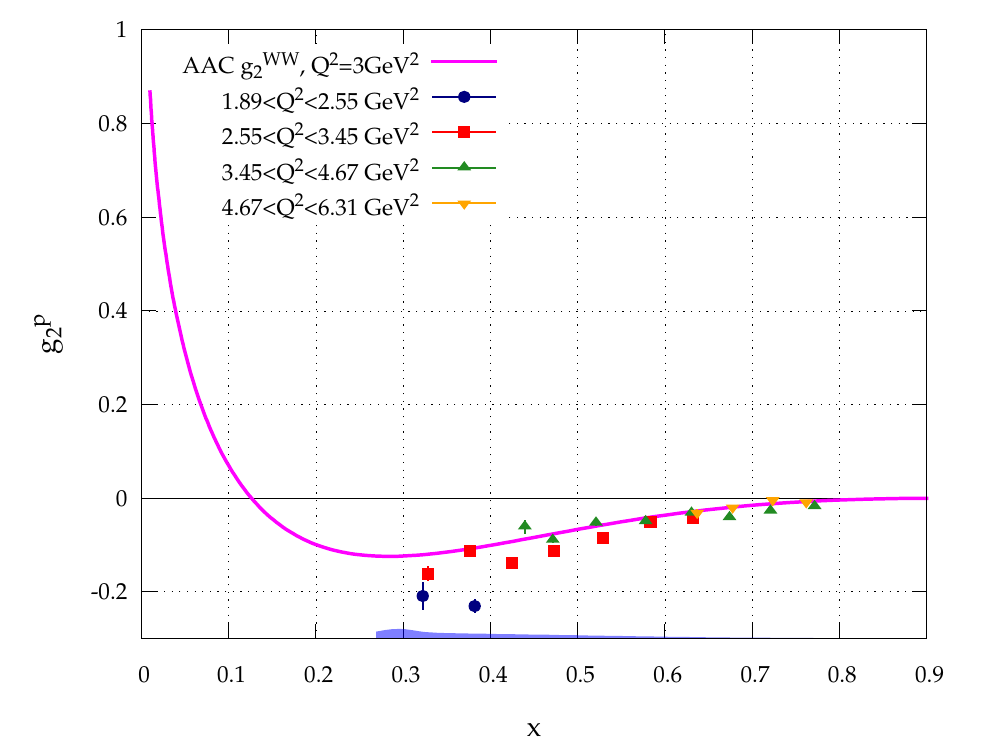}
  \end{center}
  \caption[Spin structure functions $g_1$ and $g_2$ as a function of $x$.]{Spin structure functions $g_1$ and $g_2$ as a function of $x$ for various $Q^2$.  Shown in pink are $g_1$ and $g_2^{WW}$ from AAC polarized parton distribution functions.}
  \label{fig:g1}
\end{figure}

\begin{figure}[phtb]
  \begin{center}
   \includegraphics[width=5.0in]{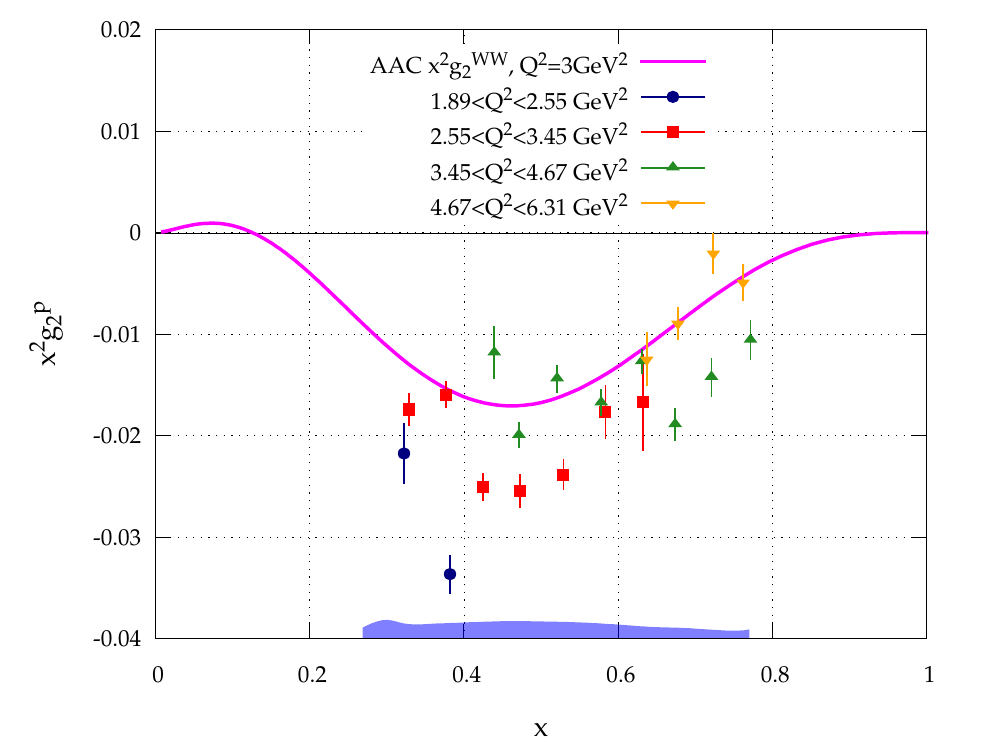}
   \includegraphics[width=5.0in]{figures/existing_g2_g2ww.pdf}
  \end{center}
  \caption[Spin structure function $g_2$ as a function of $x$.]{Spin structure function $g_2$ as a function of $x$ for various $Q^2$, scaled by $x^2$, with AAC $g_2^{WW}$.  Top shows SANE data, bottom shows SLAC data.}
  \label{fig:g2}
\end{figure}

The structure function results are shown in figures \ref{fig:g1} and \ref{fig:g2}.  Looking at $g_1$, we see that the AAC model calculation follows our data quite closely.  Differences between the $Q^2$ bins are again apparent, particularly around  $x$ of 0.4.  In both structure functions, the statistical and systematic error appear quite small as the spin asymmetry errors are scaled and combined into the structure functions.

The bottom of figure \ref{fig:g1} and figure \ref{fig:g2} show the crux of this analysis, the spin structure function $g_2$ as a function of $x$.  As in figure \ref{fig:slacg2ww}, reproduced in this chapter as the bottom of figure \ref{fig:g2}, we show $g_2$ multiplied by $x^2$. The AAC model for $g_2^{WW}$ is shown for easy comparison.  It is immediately apparent that these data fall below the AAC leading twist model of the structure function, particularly at low $x$.  
Any significant deviation of $g_2$ from $g_2^{WW}$ could imply a significant higher twist contribution, however, and these may prove to be exciting results should these artifacts remain after the final corrections are taken into account.

What is also immediately apparent, when comparing SANE data with the SLAC data of the bottom figure  \ref{fig:g2}, is the vast improvement of both kinematic coverage and statistical significance of the world's $g_2$ data.    


\section{First Moment of $g_2$}

With our structure function result achieved pending a few final corrections, we can begin to assess its impact through the first moment of $g_2$. We recall the Burkhardt--Cottingham Sum Rule from section \ref{sec:bcsum}, which asserts this moment should be zero:
\begin{equation}
\Gamma_2(Q^2) = \int_0^1  g_2(x,Q^2) dx= 0.
\end{equation}

Although our result does not cover the entirety of the $x$ range from 0 to 1, by splitting the integral we can draw some conditional conclusions via comparison with $g_2^{WW}$.  In this analysis we follow the work of the RSS collaboration\cite{rss-lett} closely. The full integral contains contributions from leading--twist ($\Gamma_2^{WW}$), elastic scattering ($\Gamma_2^{el}$), and higher--twist ($\bar{\Gamma}_2$),  where the overbar indicates a quantity with the leading twist portion removed.  The higher-twist moment $\bar{\Gamma}_2$ we can separate into the portion from the region in $x$ we have measured ($\bar{\Gamma}_2^m$ for $x>x_0$), and that which have not ($\bar{\Gamma}_2^u$ for $x<x_0$). Since by definition $\Gamma_2^{WW} = 0$, we now have:
\begin{equation}
\Gamma_2 = \Gamma_2^{WW} +\bar{\Gamma}_2 + \Gamma_2^{el} =  \bar{\Gamma}_2^u +  \bar{\Gamma}_2^m + \Gamma_2^{el}.
\end{equation}
By defining
\begin{equation}
\Delta\bar{\Gamma}_2 \equiv \Gamma_2 - \bar{\Gamma}_2^u ,
\end{equation}
we create a quantity which depends only on measured data $\bar{\Gamma}_2^m$ and $\Gamma_2^{el}$.  If we see a significantly nonzero $\Delta\bar{\Gamma}_2$, it would mean either a higher twist contribution in the  unmeasured region $x < x_0$, or that the B--C sum rule does not hold.  

The calculation of  $\Delta\bar{\Gamma}_2$ requires the elastic contribution to the moment, which we can calculate from the elastic form factors \cite{sliferthesis}:
\begin{equation}
\Gamma_2^{el}(Q^2) = \frac{\tau}{2}G_M(Q^2)\frac{G_E(Q^2)-G_M(Q^2)}{1+\tau}.
\end{equation}
We have produced $\Gamma_2^{el}$ for our $Q^2$ bins using the form factor parameterizations of J. Arrington \textit{et al}\cite{arrington}.  

Integrating our $g_2$ result in $x$ produces $\Gamma_2^m$; to calculate $\bar{\Gamma}_2^m$ with leading twist removed, we integrate the difference of the leading twist $g_2^{WW}$ and our $g_2$:
\begin{equation}
\bar{\Gamma}_2^m(Q^2) = \int_{x_0}^1(g_2^{WW}(x,Q^2) - g_2(x,Q^2))dx.
\end{equation}
Here we again compute $g_2^{WW}$ using the polarized parton distribution functions of the AAC\cite{aac}.

Table \ref{tab:bc} shows our results for this B--C sum rule test from the 5.9 GeV beam energy data, along with the results of the RSS experiment.  While the elastic contributions to the moment drops with higher $Q^2$, our measured, higher--twist moment also drops. A conservative estimate of the uncertainties puts them on the same order as the RSS result of $\Delta\bar{\Gamma}_2$.  With this error in mind, our $\Delta\bar{\Gamma}_2$ for the proton is consistent with zero, as in the result of RSS.  In the 5.9 GeV beam energy case, any deviation of $g_2$ from $g_2^{WW}$ is averaged out in the integral, and no deviation from leading twist behavior is seen.  The B--C sum rule appears to remain intact.

\begin{table}[bth]
  \begin{center}
\begin{tabular}{lcccccc}
\toprule
Experiment	& $Q^2$(GeV$^2$)& $\Gamma_2^{el}$  & $\Gamma_2^{m}$ &  $\bar{\Gamma}_2^{m}$ & $\Delta\bar{\Gamma}_2$ & $\Delta\bar{\Gamma}_2$ error\\
\midrule
RSS & 1.28 & -0.0132 & -0.0138 & 0.0126 & -0.0006 & 0.0022\\
\cmidrule{2-7}
\multirow{2}{*}{SANE} & 3.00 & -0.0021 & -0.0339 & 0.0015 & -0.0006 & ...\\
  & 4.06 & -0.0009 & -0.0164 & 0.0015 & 0.0005& ...\\
 \bottomrule
\end{tabular}
\caption[Table of first $g_2$ moment results]{Table of first $g_2$ moment results from SANE 5.9 GeV data and RSS\cite{rss-lett}. We expect error on $\Delta\bar{\Gamma}_2$ on the same order as the RSS result.}
  \label{tab:bc}
\end{center}
\end{table}

\section{Systematic Uncertainty}

A thorough study of SANE systematic error is pending, and is predicated on the completion of the final corrections.  We can however make estimates of these uncertainties, which we have included as an errorband in our result plots, and now enumerate in table \ref{tab:systematics}. 
\begin{table}[htb]
  \begin{center}
\begin{tabular}{lcc}
\toprule
Source	& Estimated Systematic Error \\
\midrule
Target Polarization &  5.0\%\\
Beam Polarization &  1.5\%\\
Dilution Factor &   4.5\%\\
Nitrogen Correction &    0.4\%\\
Radiative Corrections &   1.5\%\\
Kinematic Uncertainty &   4.5\%\\
Background &   1.8\%\\
$R$, $F_1$ &   1.3\%\\
\bottomrule
\end{tabular}
\caption{Table of estimated relative systematic uncertainties on $g_2$ for $E=5.9$ GeV and $x=0.6$.}
  \label{tab:systematics}
\end{center}
\end{table}

The systematic error in the target polarization should be the largest individual contribution to the uncertainty.  The target polarization error is based largely on the accuracy of the calibration constants used to produce the polarization from NMR area; the standard deviation of the individual calibration constants for a given material sample about the mean gives a measure of the systematic error.  The estimate error of the table is the simple average relative error of the calibration constants; a more accurate value will come with a charge average of these errors.  With SANE's calibration constants in mind, we expect that the final systematic error may be slightly greater than that anticipated in the experiment's proposal. 

The error in the beam polarization measurement comes from a global error on the M{\o}ller measurements, plus error due to the fit to these measurements.  The global error is about 0.9\%, and the fit will add 0.5\% or more.

The dilution factor's uncertainty is based on statistical error in the measurement of the packing fraction, as well as error from the simulation.  The dilution factor error will change in kinematics, but should be roughly 4.5\%.  This figure incorporates roughly half of the error in the packing fractions, which are about 5.0\% at this point.  The nitrogen correction should add another 0.4\%.

Uncertainty in the computation of our kinematic properties, $E'$, $\theta$ and $\phi$, add to the error based on the position and energy resolution.  As the position and energy resolution are functions of kinematics, this error contribution will be higher at higher $x$.  Our plots have included an estimation of this systematic error contribution as it changes in kinematics, based on the energy resolution.  The dilution of our events by background will given a further systematic error of at least 1.8\%.

Finally, the parametrization we use to provide $F_1$, $F_2$ and $R$ for the spin asymmetry calculations will contribute additional error, on the order of 1.3\%.  After combining all of these contributions in quadrature, our systematic error should be around 10\% relative.

\section{Remaining Tasks}

These data are quite mature, reflecting most of the corrections necessary for the final results.  However, several corrections remain: final dilutions factors, polarized radiative corrections and pair symmetric background corrections may each serve to alter the results somewhat.

The estimated dilution factors used in these results, calculated using cross section models and the packing fractions of each target material load, are good approximations of the proper dilution factors. 
The finalized dilution factors, discussed in section \ref{sec:dilution}, are under active production by our collaborators.  The final values may differ from our estimates by as much as 20\% in the resonances, but less than 5\% beyond $W$ of 2.

While we have accounted for the effects of the radiative tail, final, polarized radiative corrections are also necessary.  The radiative peak represents a lions share of the events creeping into our kinematic range, but the polarized radiative corrections reflecting all kinematic regimes remain.  This correction may change the central values by as much as 10\%.

Finally, a significant pair symmetric background may be diluting our asymmetries, particularly at low $E'$.  Photons from neutral pion decay can convert to electron--positron pairs before reaching the \v{C}erenkov.  If these events pass the energy threshold, both electron and positron could be accepted in the calorimeter as good electron events from primary scattering.  A correction to this effect takes the form
\begin{equation}
A = \frac{A_m-f_bA_b}{1-f_b},
\end{equation}
for the measured asymmetry $A_m$, the ratio of background to total measured events $f_b$, and the background asymmetry of these events $A_b$.  This correction is being actively pursued by our collaborators, and may change our values as much as 40\% at low $x$, while being negligible above $x$ of 0.6.




\section{Conclusion}

The Spin Asymmetries of the Nucleon Experiment has produced valuable double polarization measurements of the proton's spin structure at $x$ above 0.25 and $Q^2$ from 1.5 to 6.5 GeV$^2$.  We have presented spin structure functions $g_1$ and $g_2$, and virtual Compton asymmetries $A_1$ and $A_2$ in this region, as calculated from these data. With the inclusion of both parallel and rare near perpendicular target orientation asymmetries, these calculations avoid the model dependence required by purely parallel datasets.  We have also shown a preliminary test of the Burkhardt--Cottingham sum rule in a scantly measured kinematic regime.

These data offer a look at spin structure function $g_2$ with unprecedented accuracy.   By both expanding the kinematic scope of existing measurements and contributing vastly to their statistical significance, SANE represents an important advancement in the understanding of nucleon spin structure and an exciting expansion of the frontiers of nuclear physics. 



\appendix

\chapter{TE Calibrations}
\label{sec:teapp}

The 24 offline thermal equilibrium measurements used to produce the target polarization calibration constants are crucial to the accuracy of the experiment, and thus they are worthwhile to illustrate in full in this appendix.  Each measurement has two plots, one upper, one lower, and six such pairs are given per page.   Section \ref{sec:temeas} discusses these measurements further.

The upper plots give the context of the measurement, showing the decay of the NMR area in red squares as thermal equilibrium is approached.  The points chosen for the measurement, which are ideally at a time of constant NMR area and target cell temperature, are given as up--pointing blue triangles.  The calibration constants which correspond to these chosen points are down--pointing pink triangles.

The lower plots focus in on the points chosen.  The red squares here are the target temperature for each point, with temperatures which lie outside two standard deviations from their mean being excluded as the ``bad'' blue diamonds.  Errant readings of the He$_4$ manometer created several bad points throughout the run. 

Two types of TE measurements are recognizable in the upper plots.  The first shows a decay of the NMR area over time, end with the TE measurement.  The second shows a relatively flat, and often more erratic looking  due to the automatic scaling, set of data which represents the second TE in the set of two measurements, top and bottom.  While the top target cup is reaching thermal equilibrium, the second is submersed in liquid helium and also equilibrating.  Thus, after a top cup TE measurement was taken, the target insert was immediately moved to the bottom cup to take a measurement without the long wait and lost beam time which would be necessary otherwise.

\begin{figure}[p]		
\begin{center}
 \includegraphics[width=2.9in]{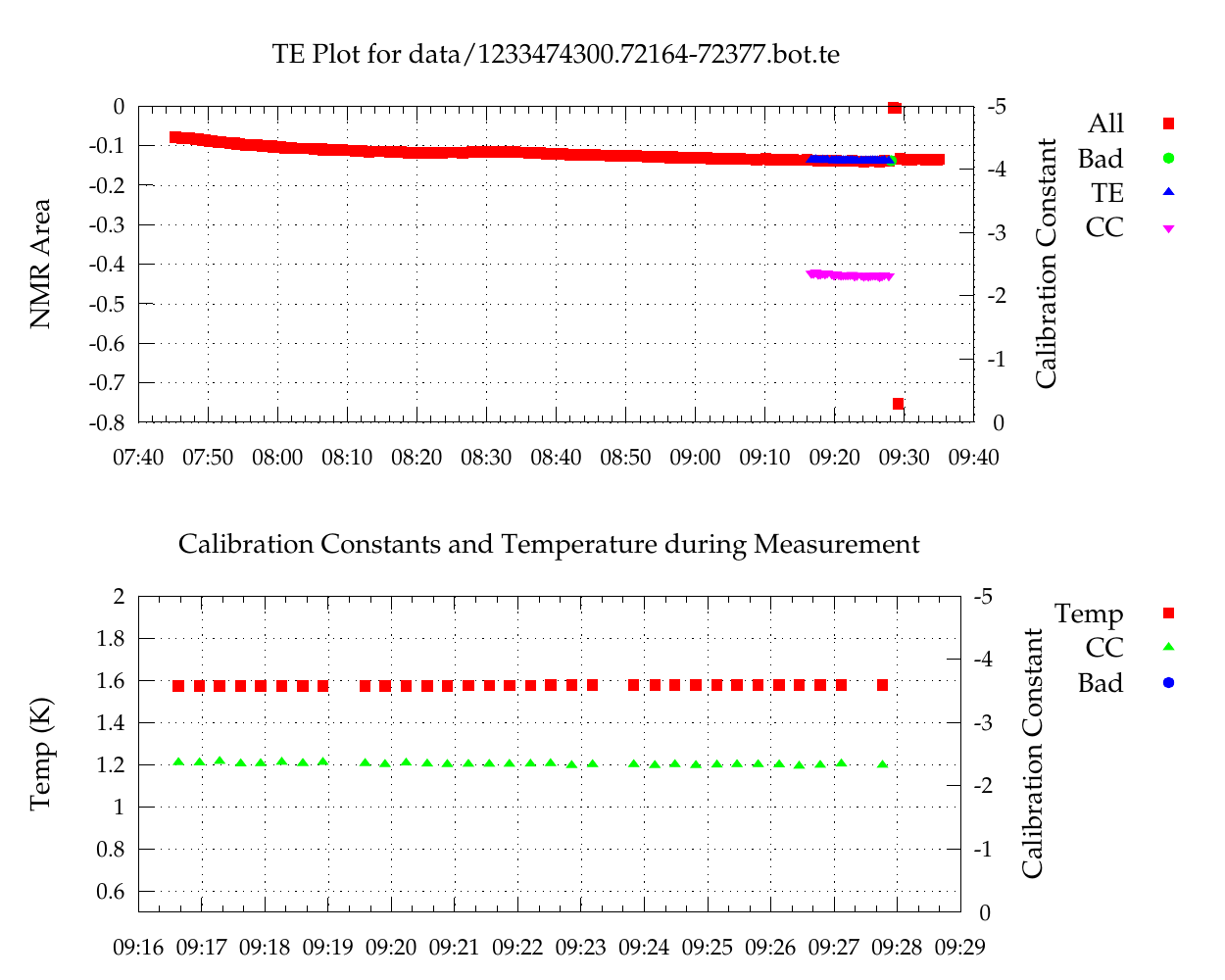}
 \includegraphics[width=2.9in]{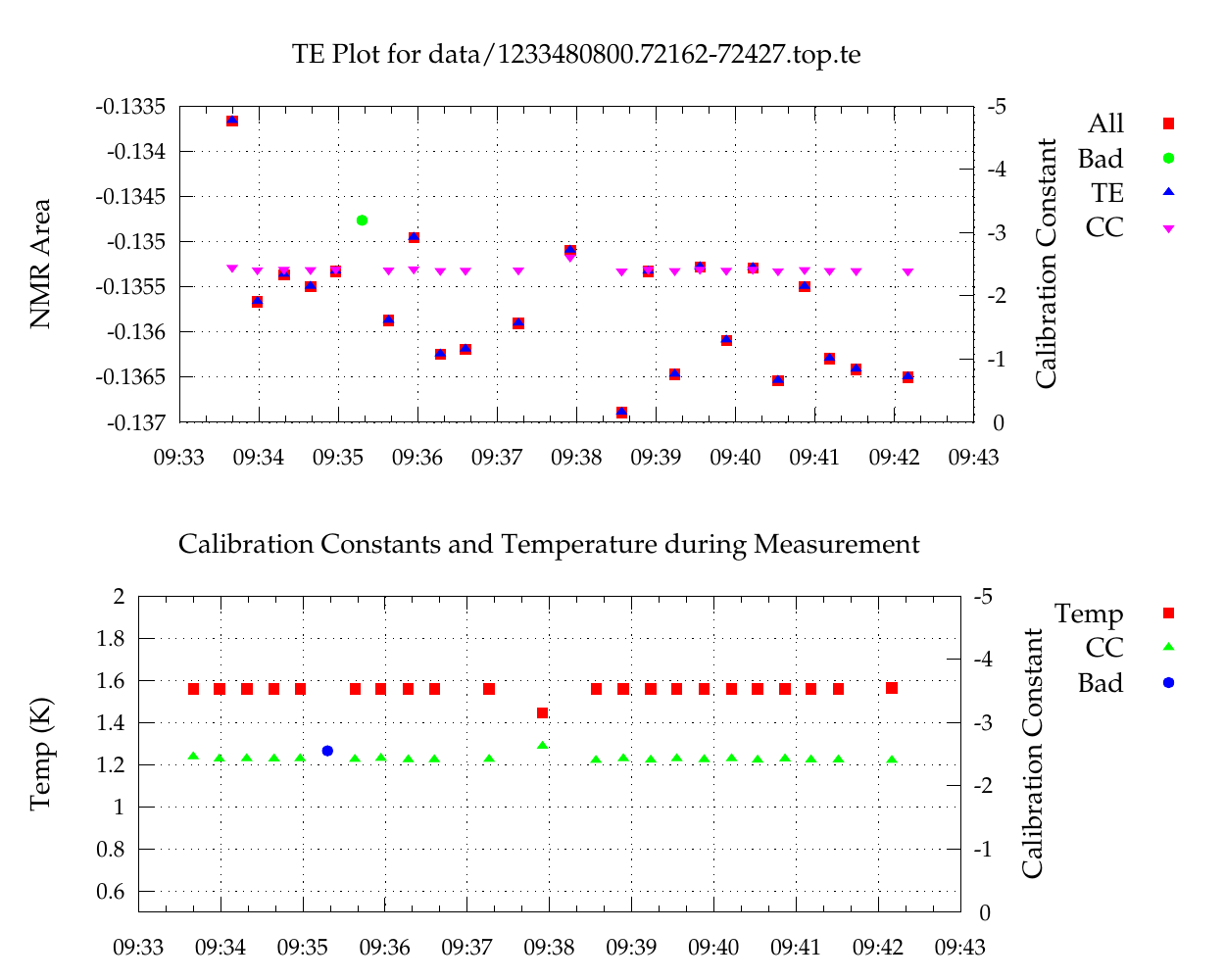}
\end{center}
\end{figure}
\begin{figure}[p]		
\begin{center}
 \includegraphics[width=2.9in]{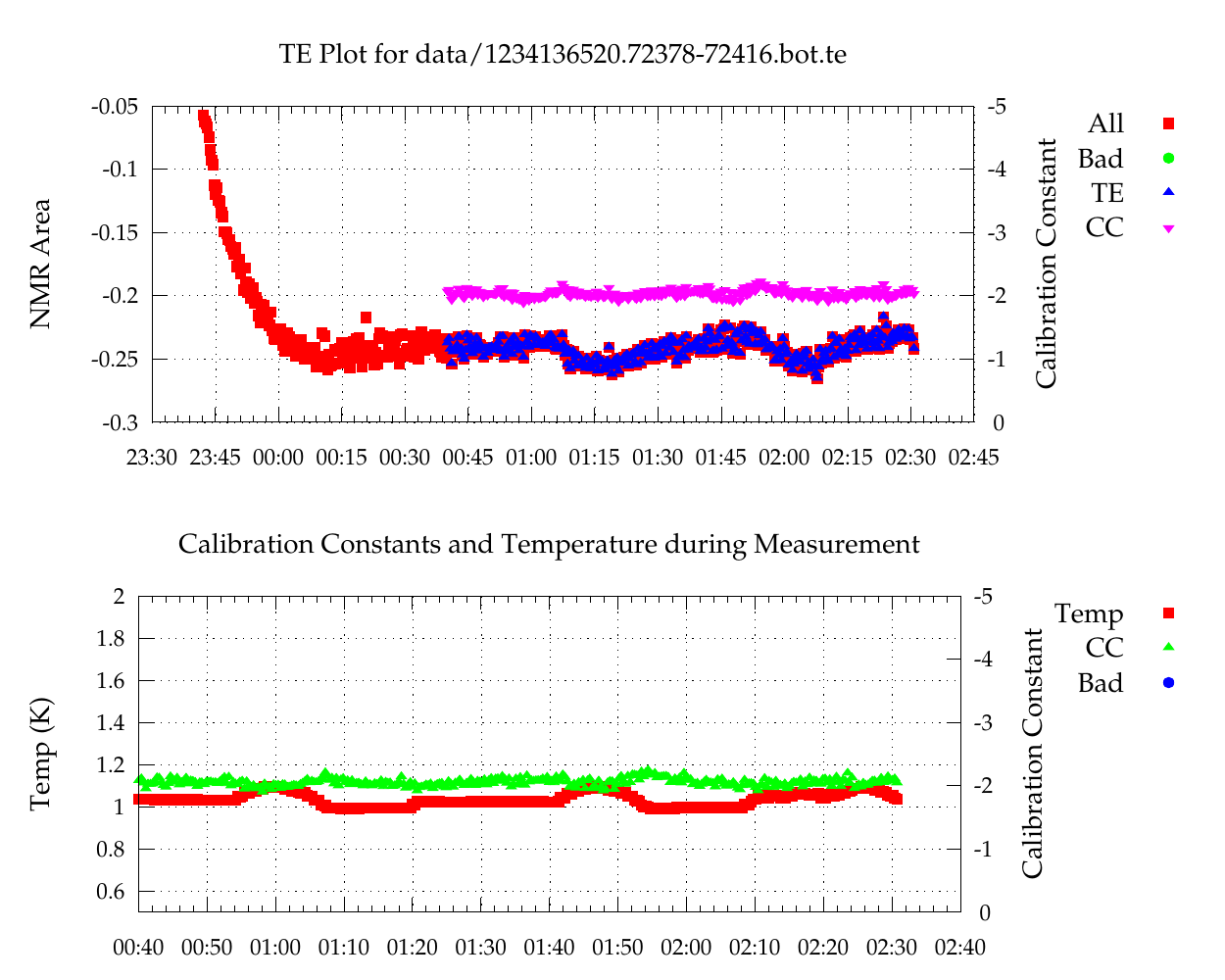}
 \includegraphics[width=2.9in]{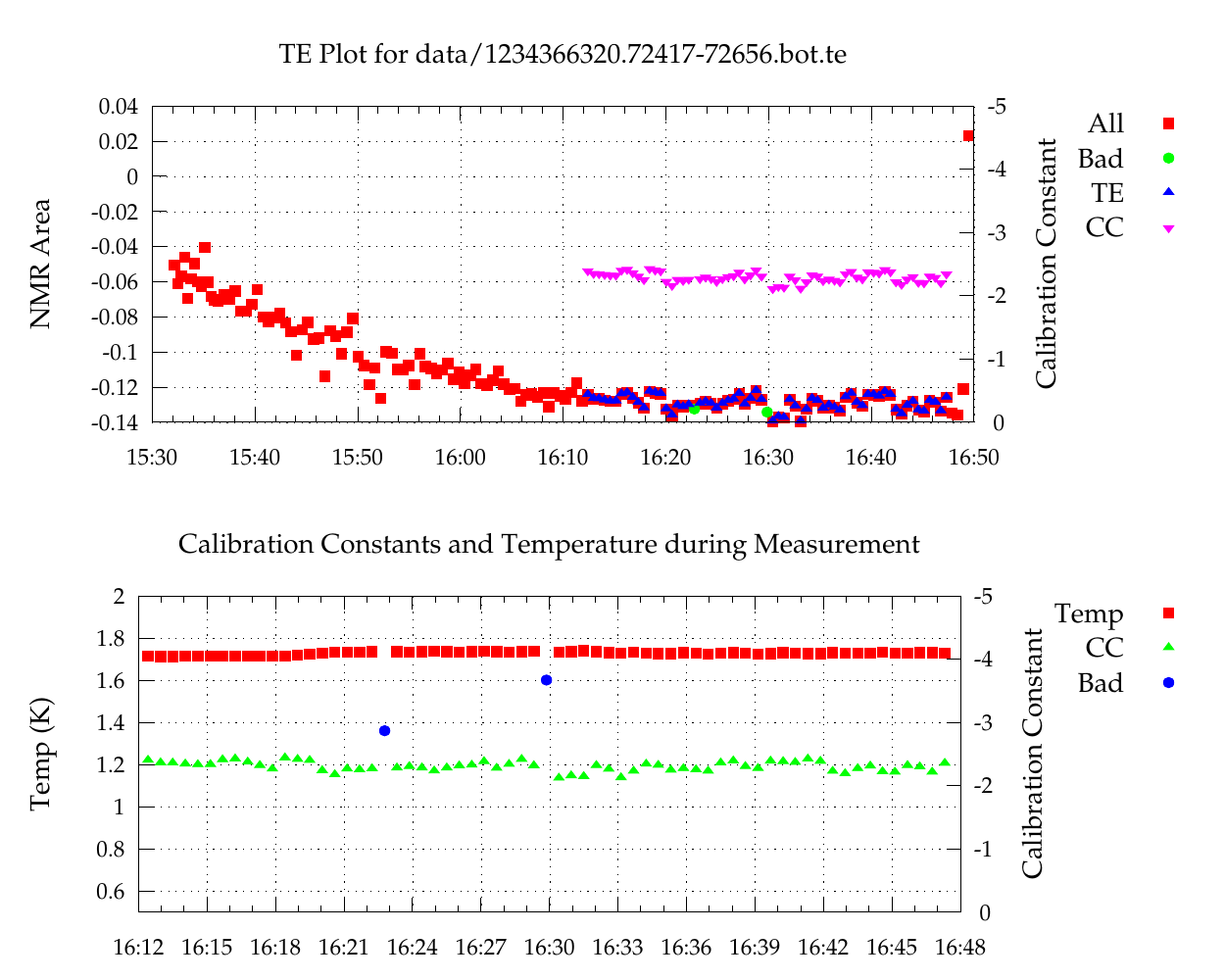}
\end{center}
\end{figure}
\begin{figure}[p]		
\begin{center}
 \includegraphics[width=2.9in]{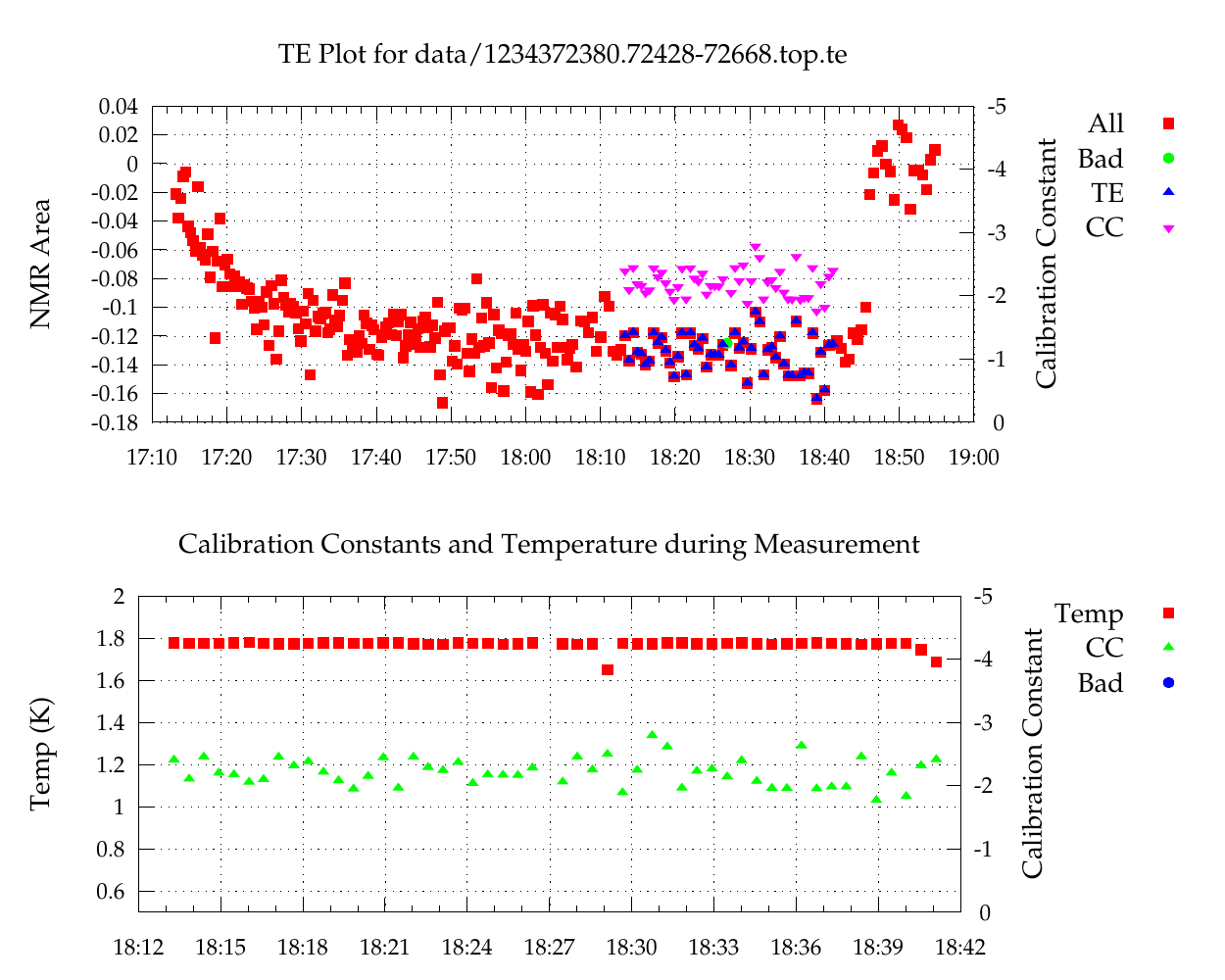}
 \includegraphics[width=2.9in]{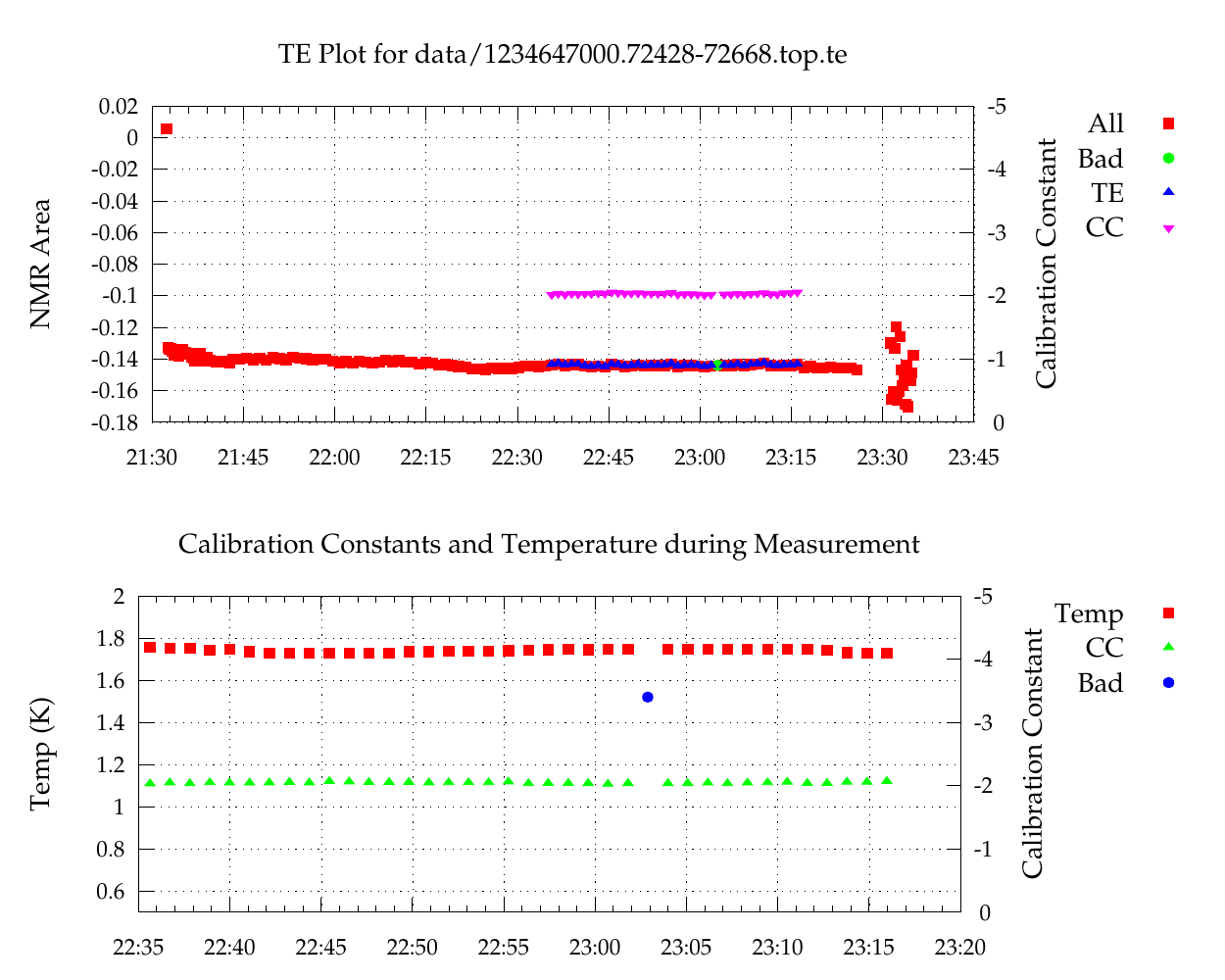}
\end{center}
\end{figure}
\clearpage 
\begin{figure}[p]		
\begin{center}
 \includegraphics[width=2.9in]{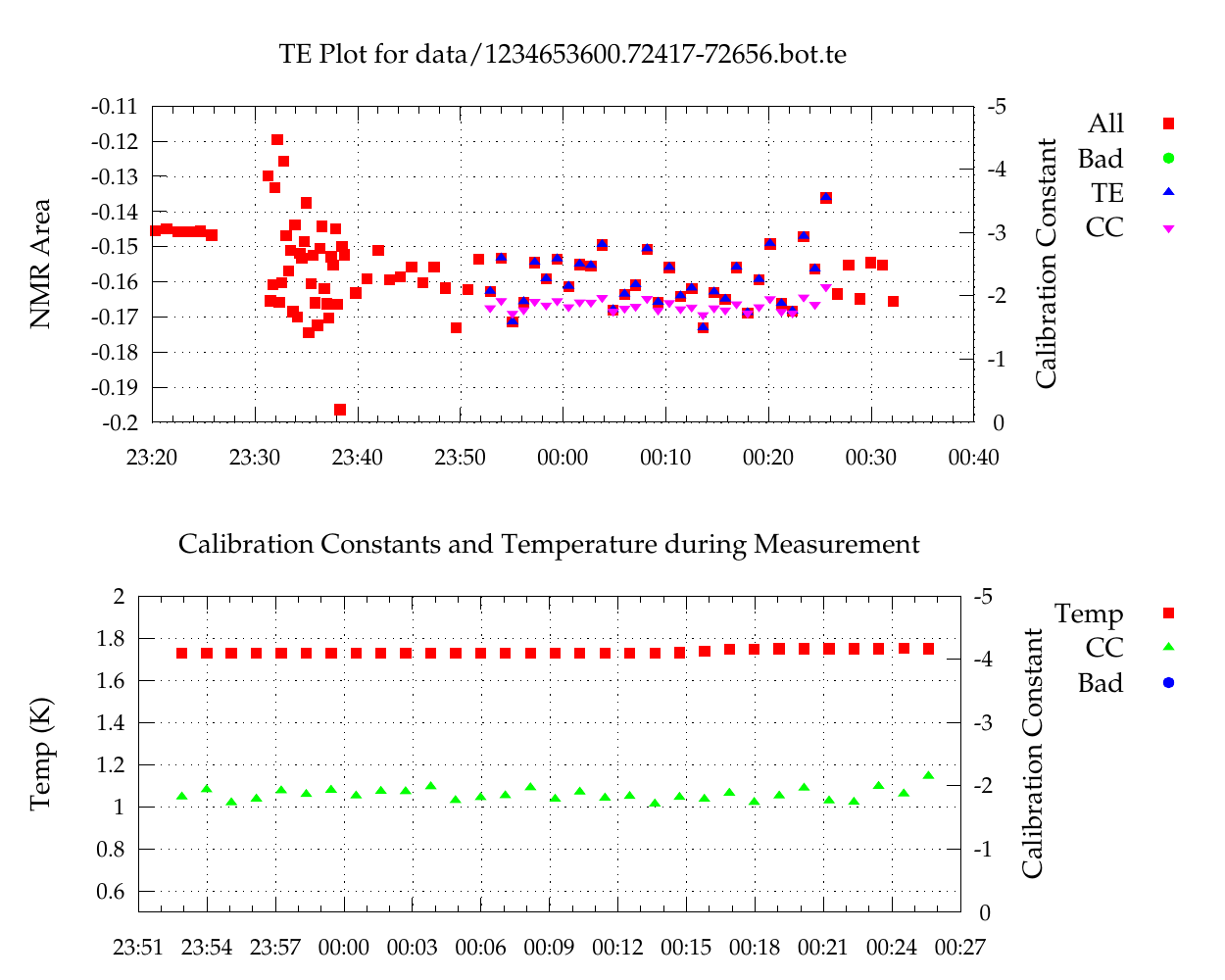}
 \includegraphics[width=2.9in]{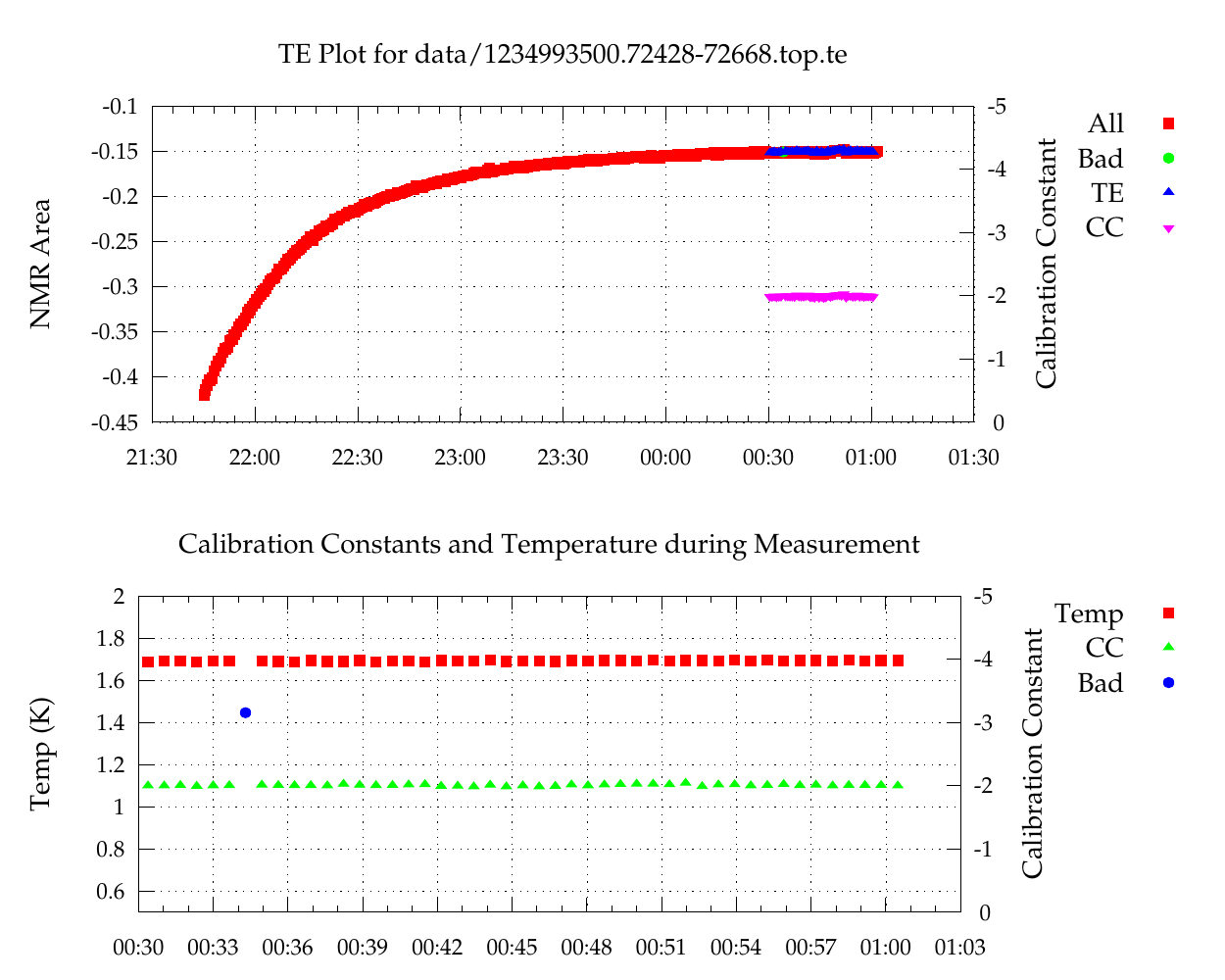}
\end{center}
\end{figure}
\begin{figure}[p]		
\begin{center}
 \includegraphics[width=2.9in]{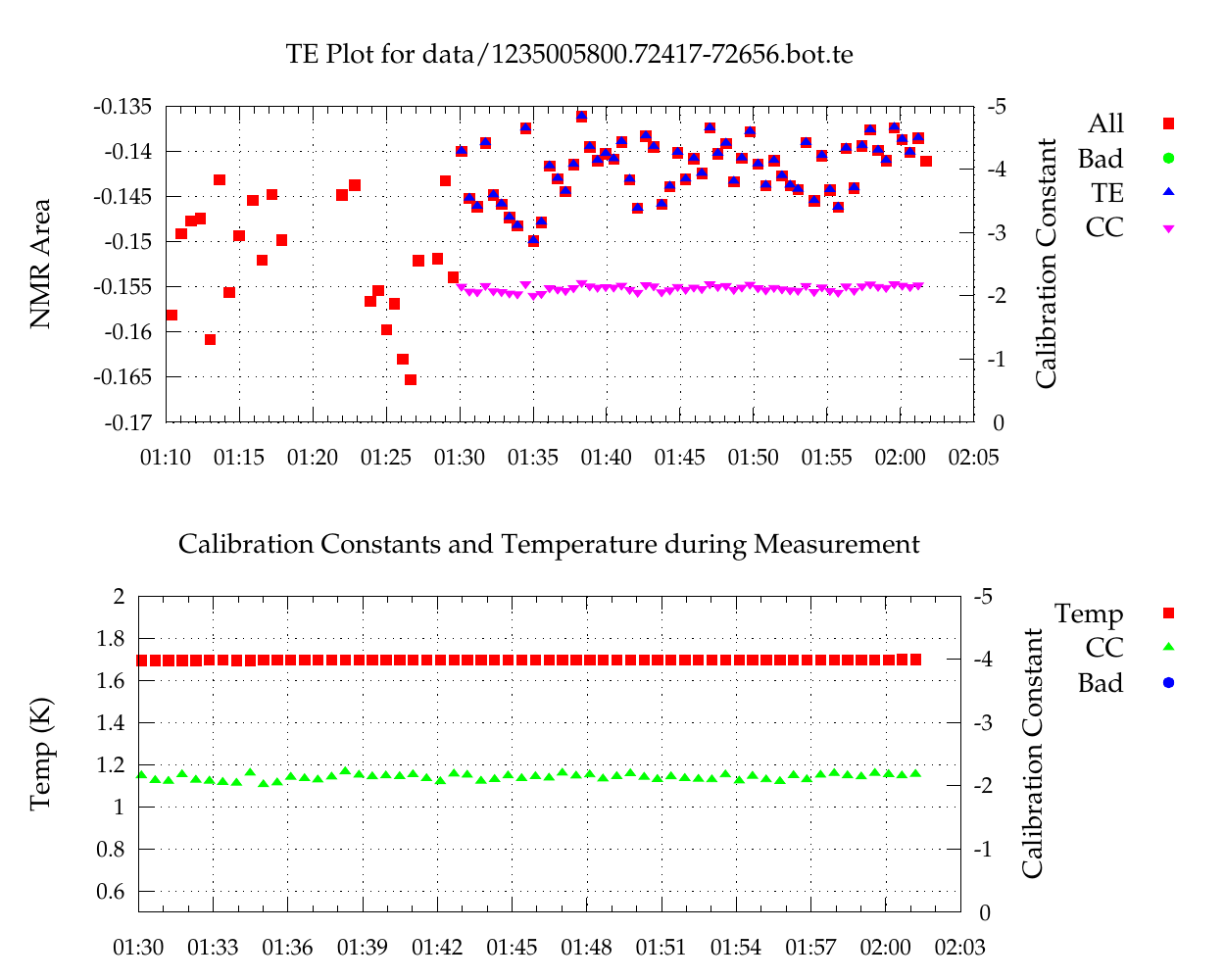}
 \includegraphics[width=2.9in]{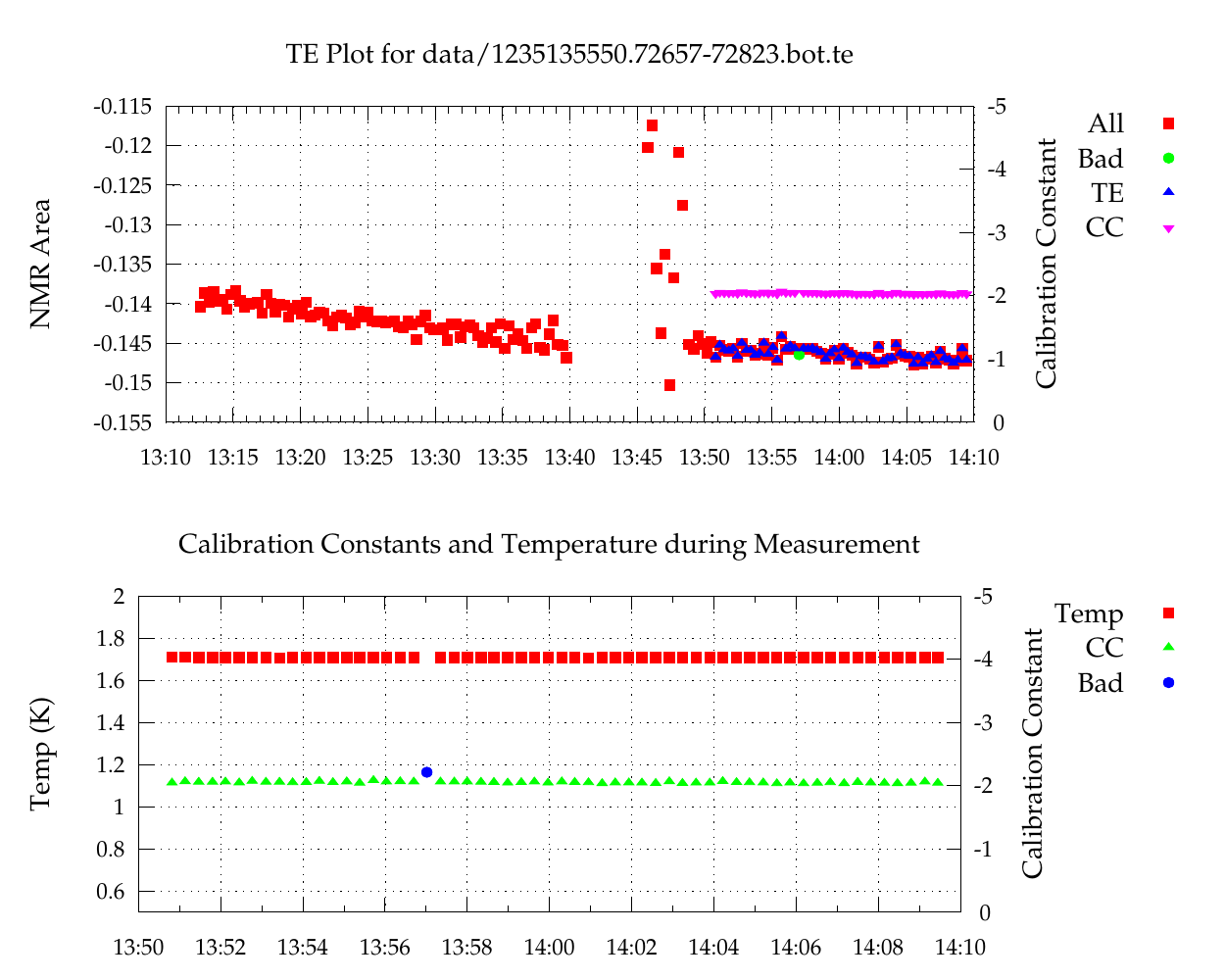}
\end{center}
\end{figure}
\begin{figure}[p]		
\begin{center}
 \includegraphics[width=2.9in]{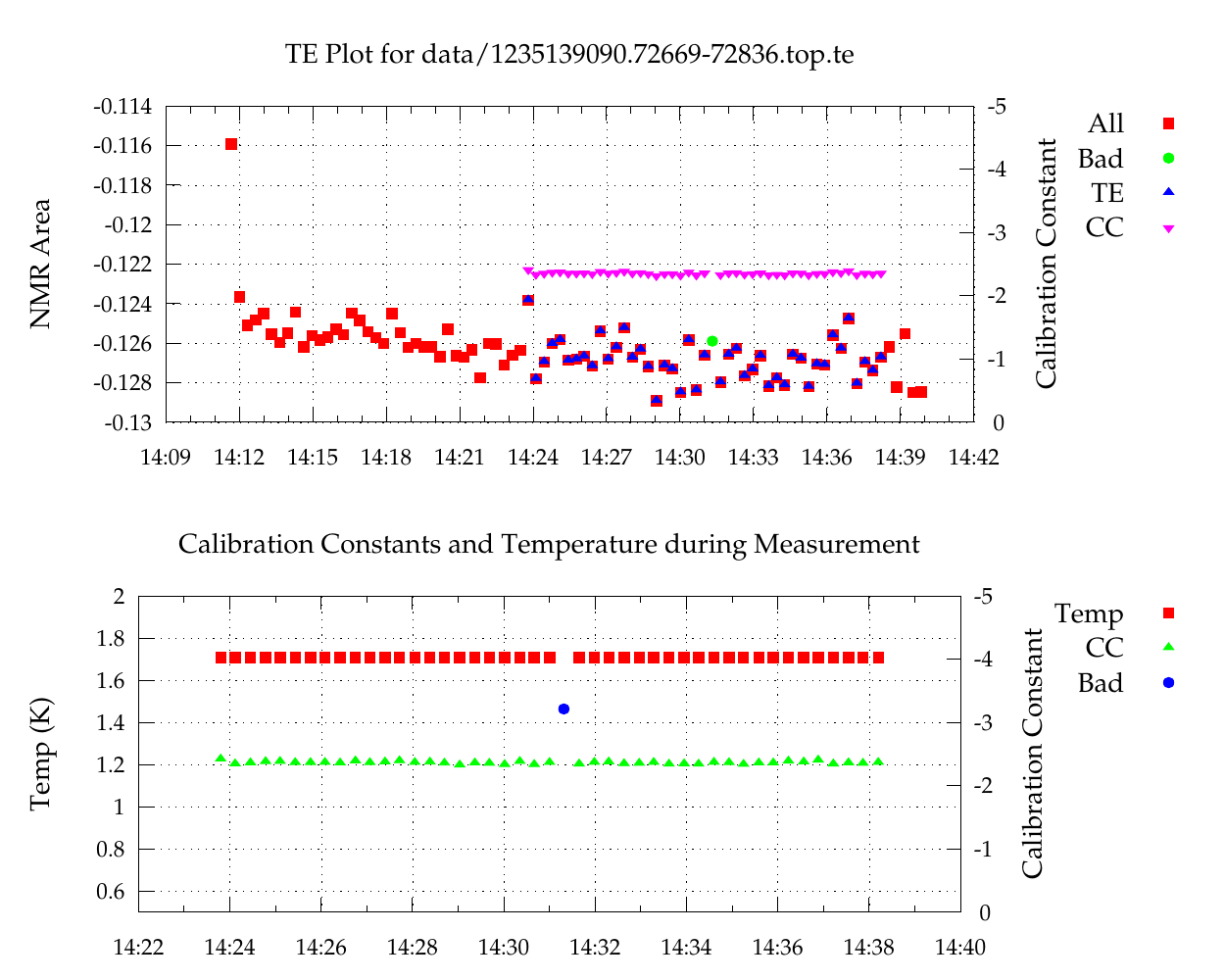}
 \includegraphics[width=2.9in]{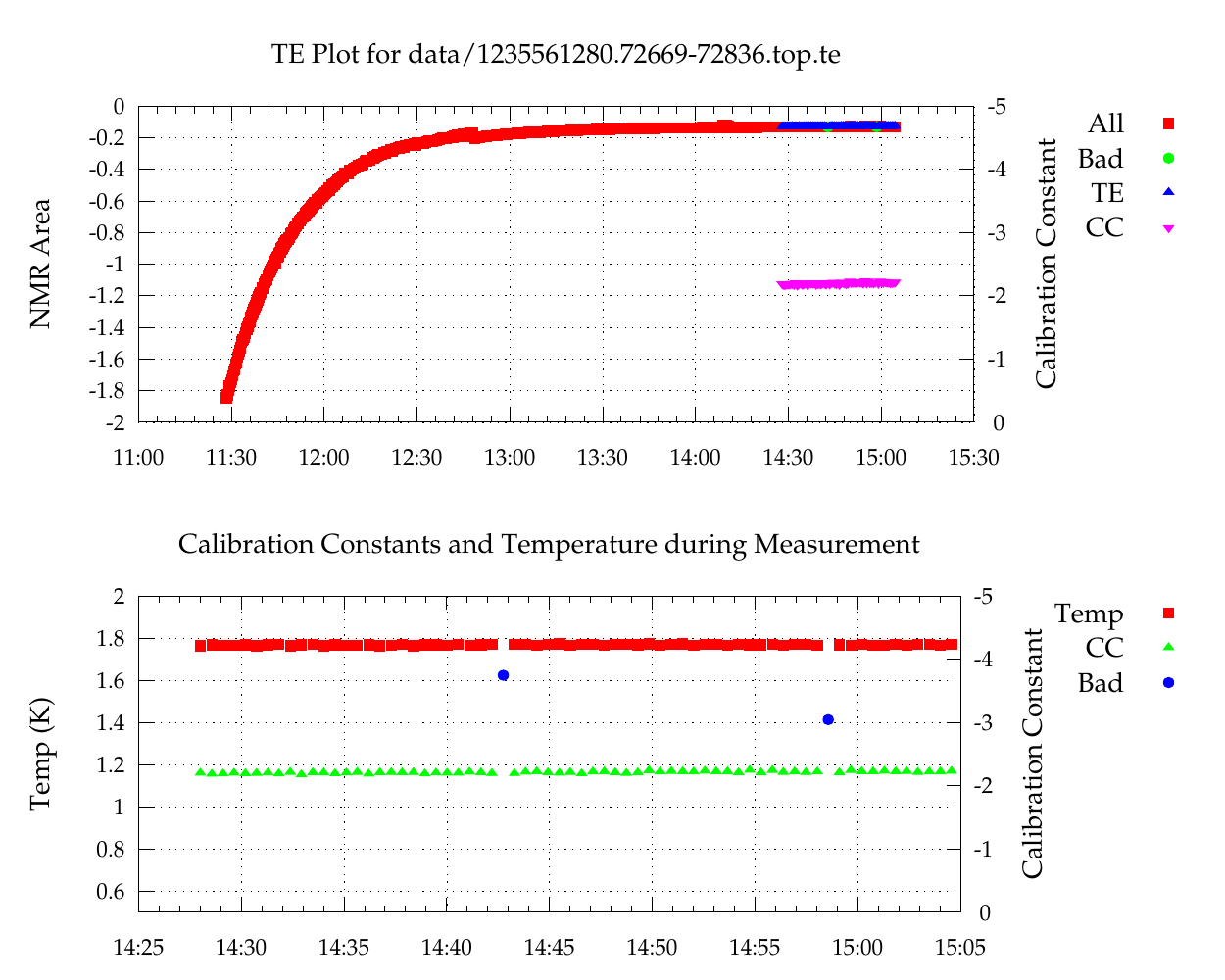}
\end{center}
\end{figure}
\clearpage 
\begin{figure}[p]		
\begin{center}
 \includegraphics[width=2.9in]{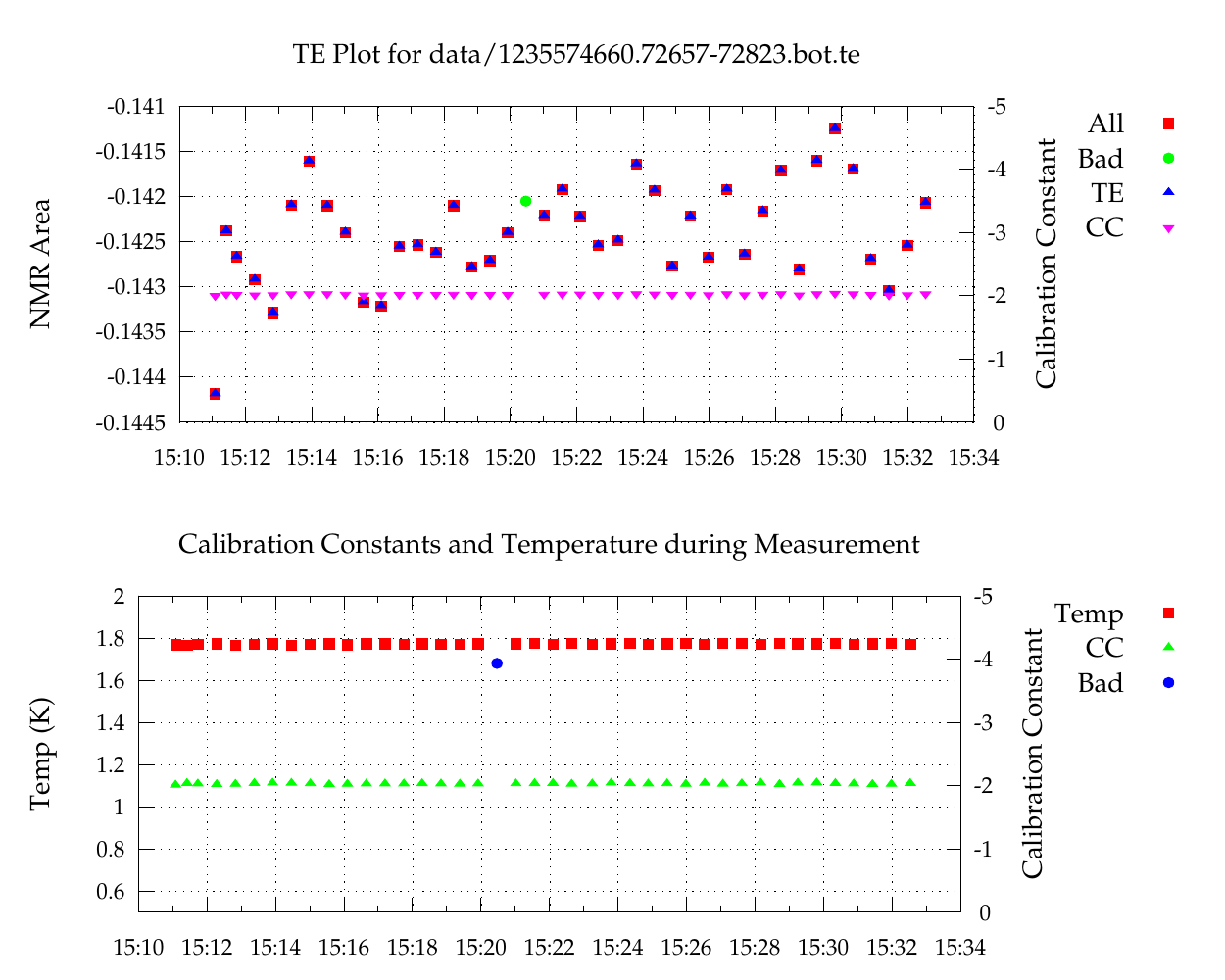}
 \includegraphics[width=2.9in]{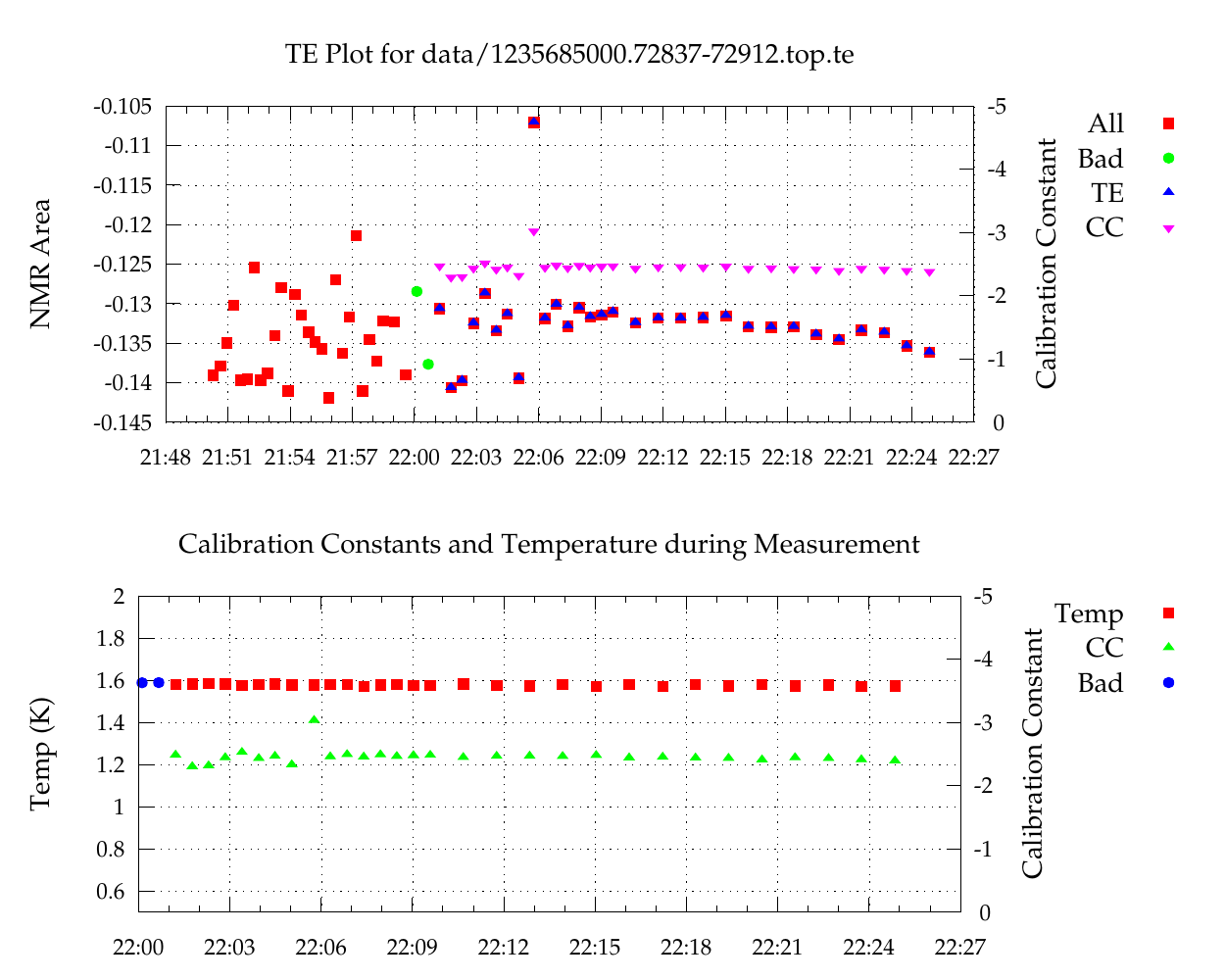}
\end{center}
\end{figure}
\begin{figure}[p]		
\begin{center}
 \includegraphics[width=2.9in]{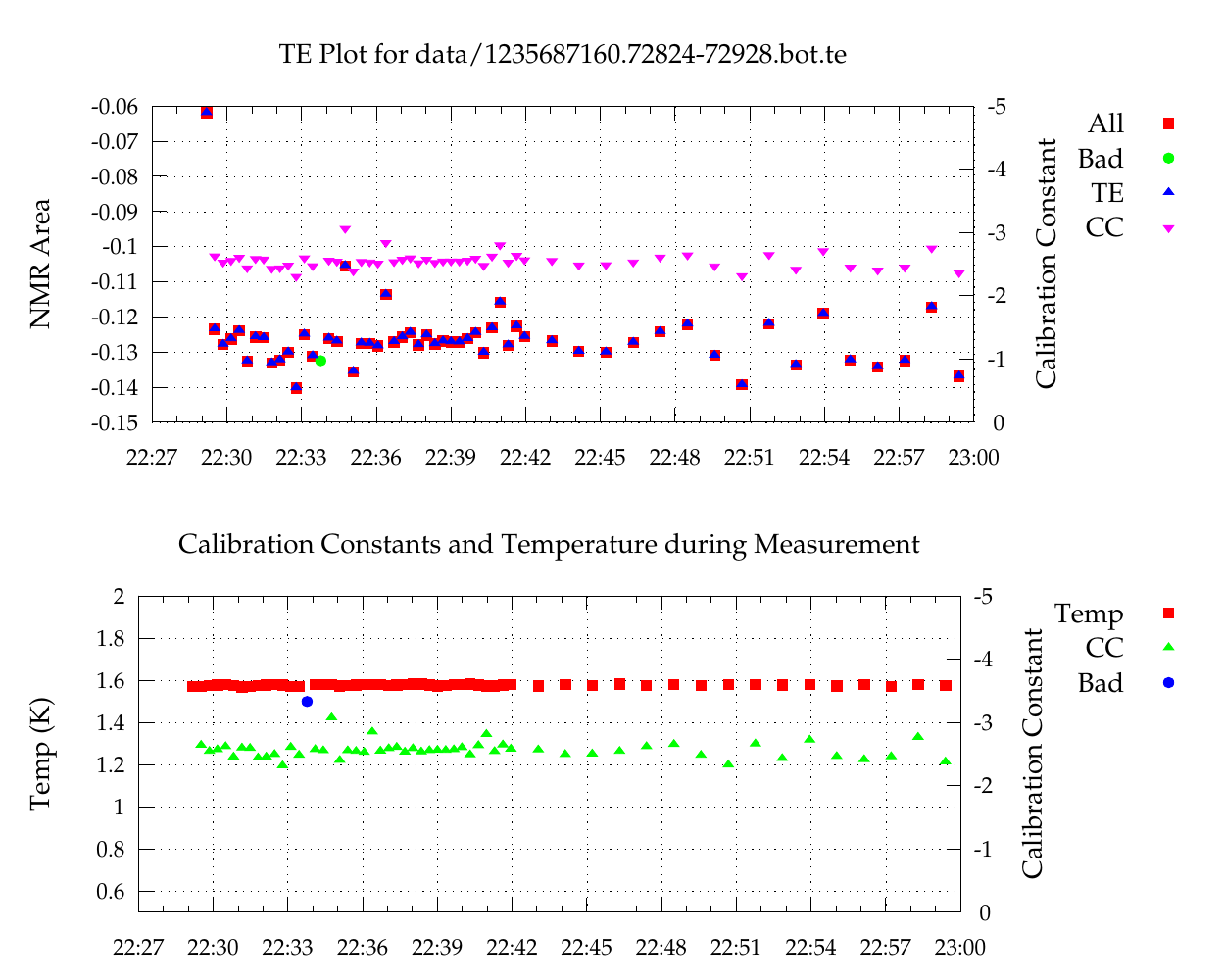}
 \includegraphics[width=2.9in]{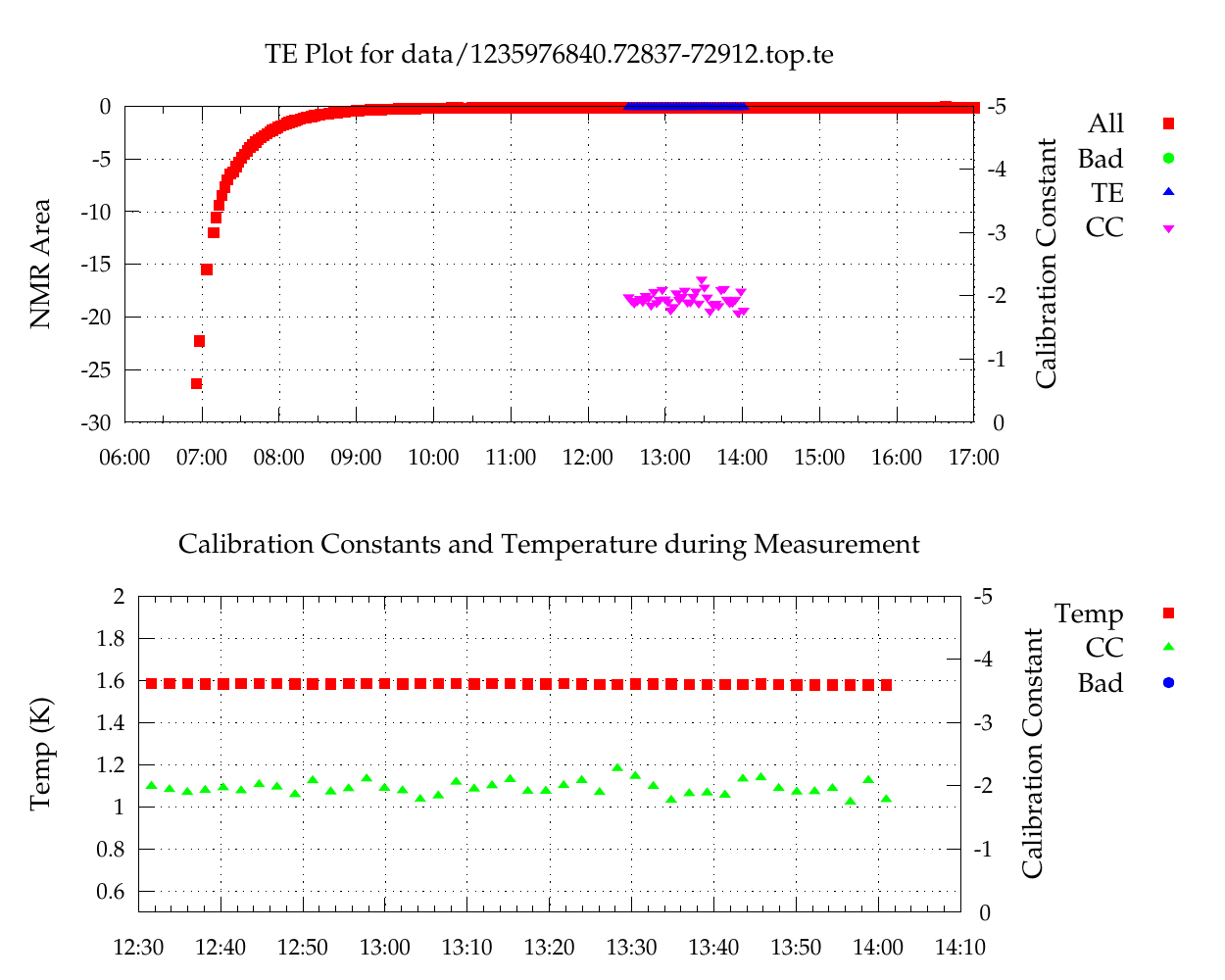}
\end{center}
\end{figure}
\begin{figure}[p]		
\begin{center}
 \includegraphics[width=2.9in]{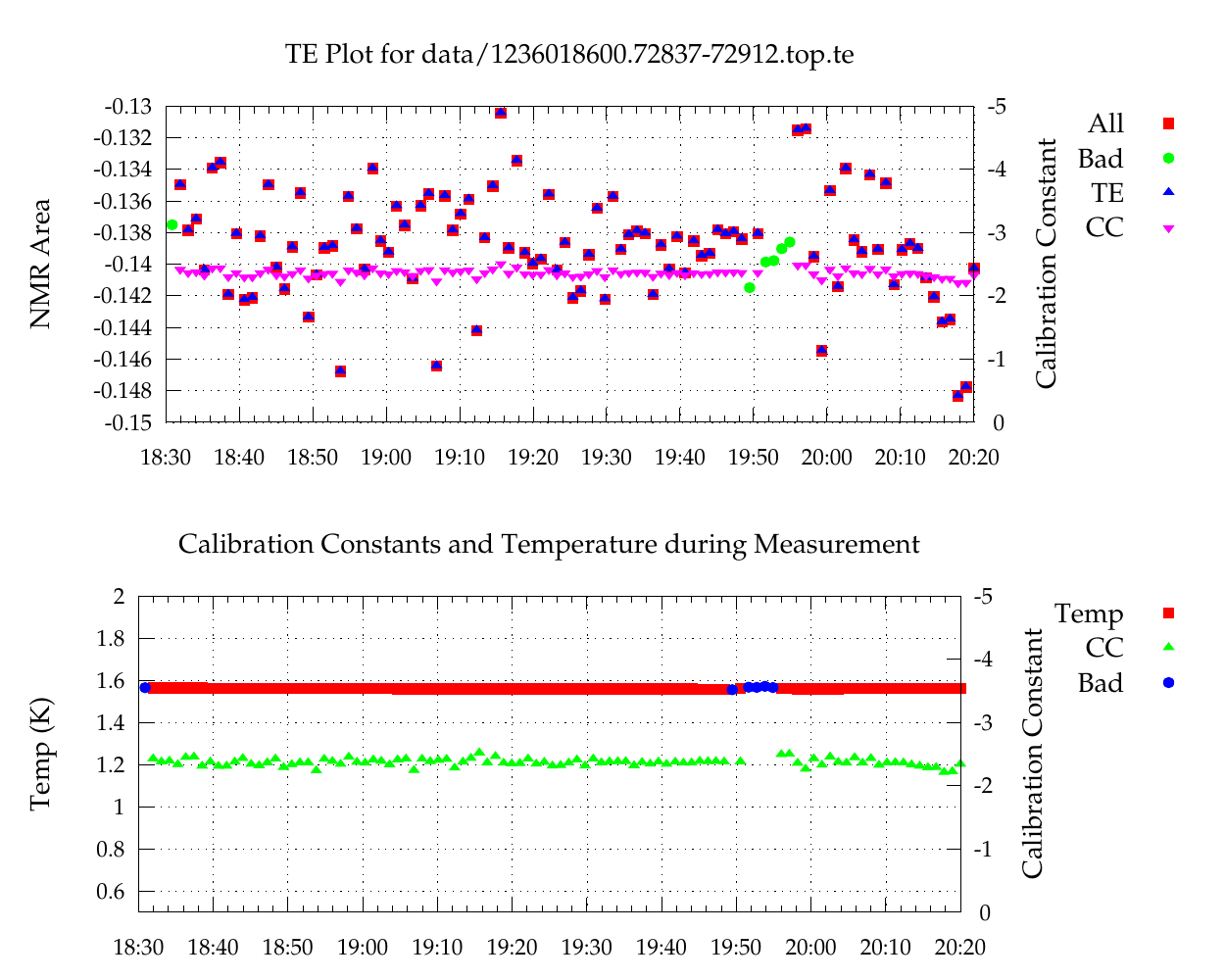}
 \includegraphics[width=2.9in]{figures/18_te.pdf}
\end{center}
\end{figure}
\clearpage 
\begin{figure}[p]		
\begin{center}
 \includegraphics[width=2.9in]{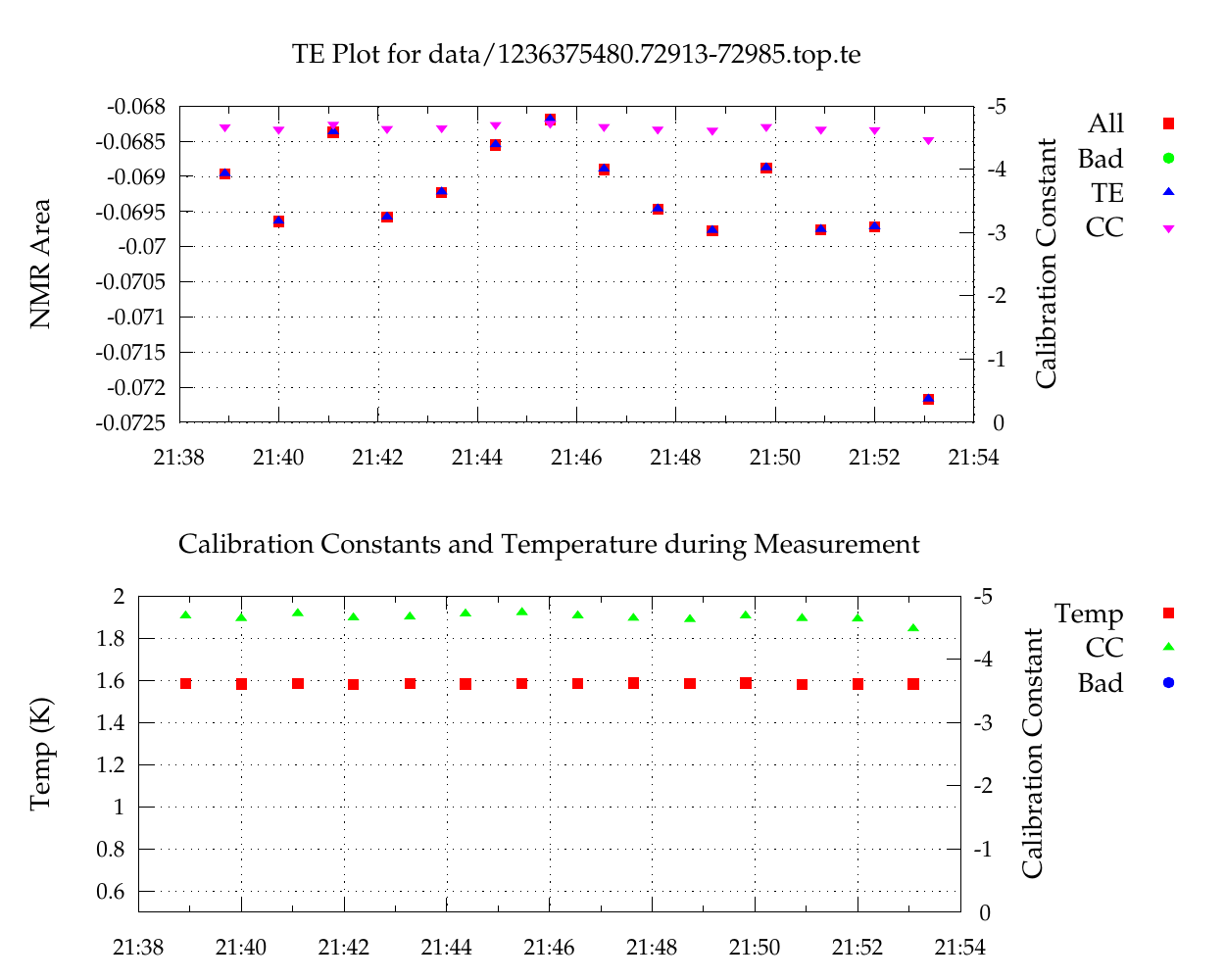}
 \includegraphics[width=2.9in]{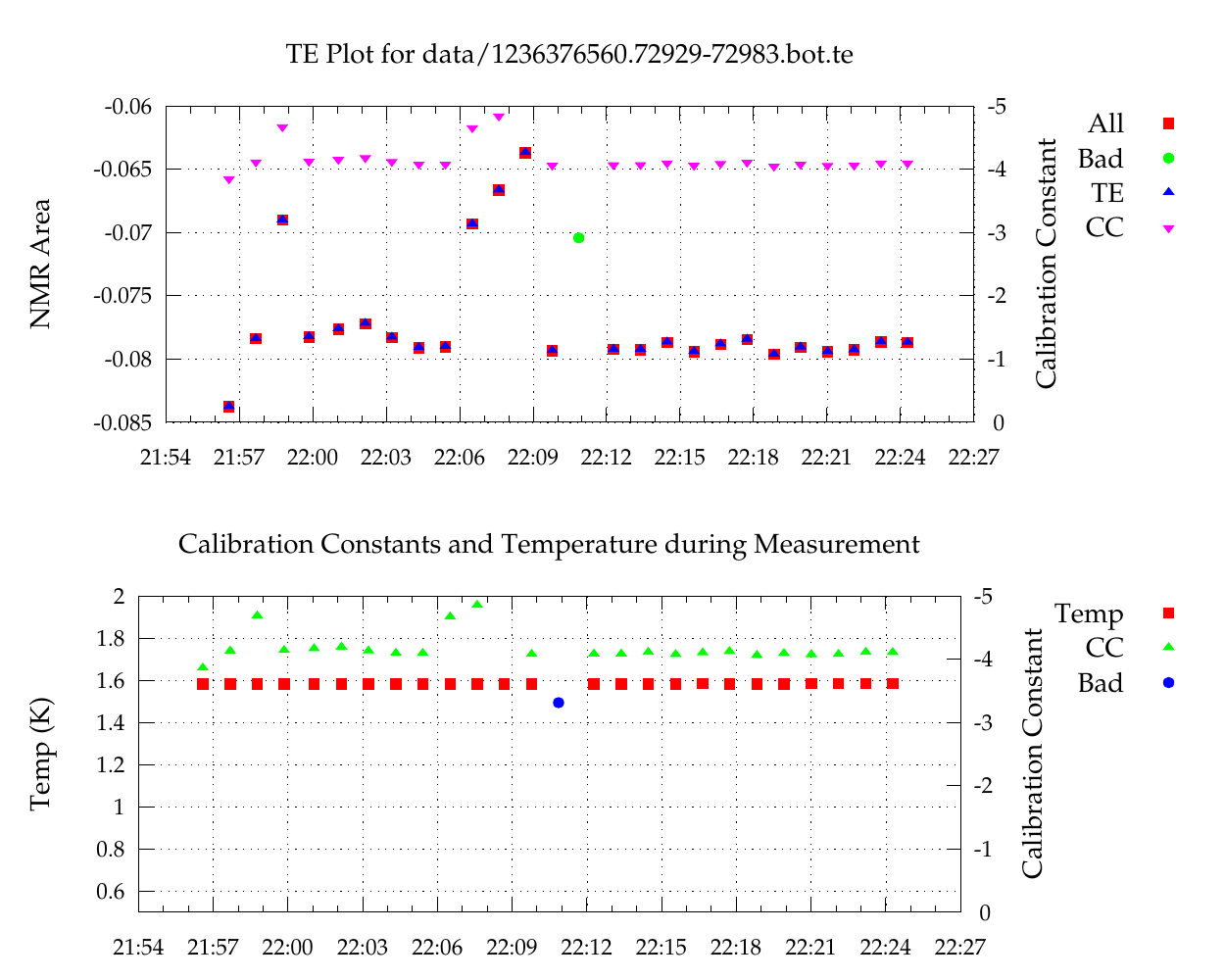}
\end{center}
\end{figure}
\begin{figure}[p]		
\begin{center}
 \includegraphics[width=2.9in]{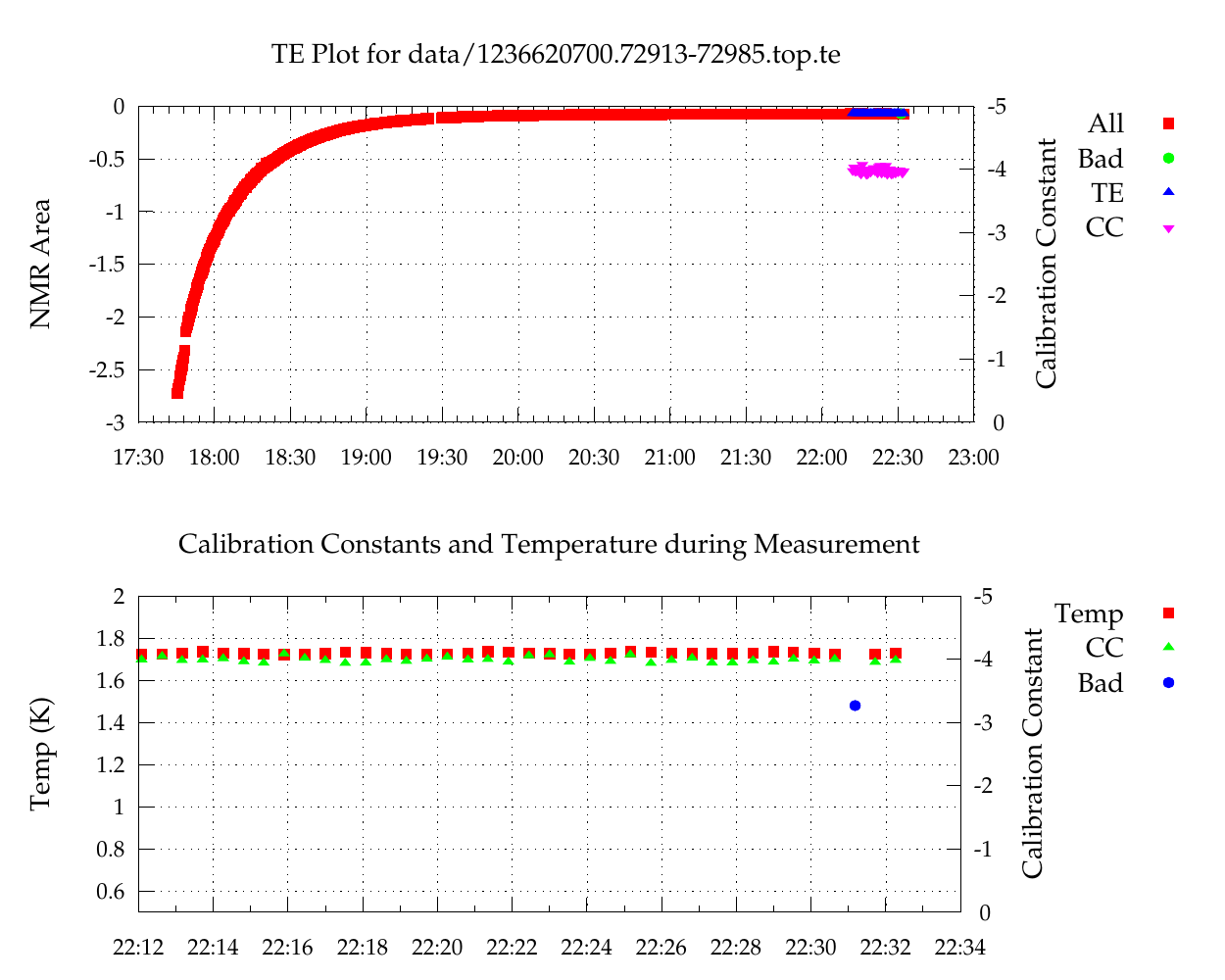}
 \includegraphics[width=2.9in]{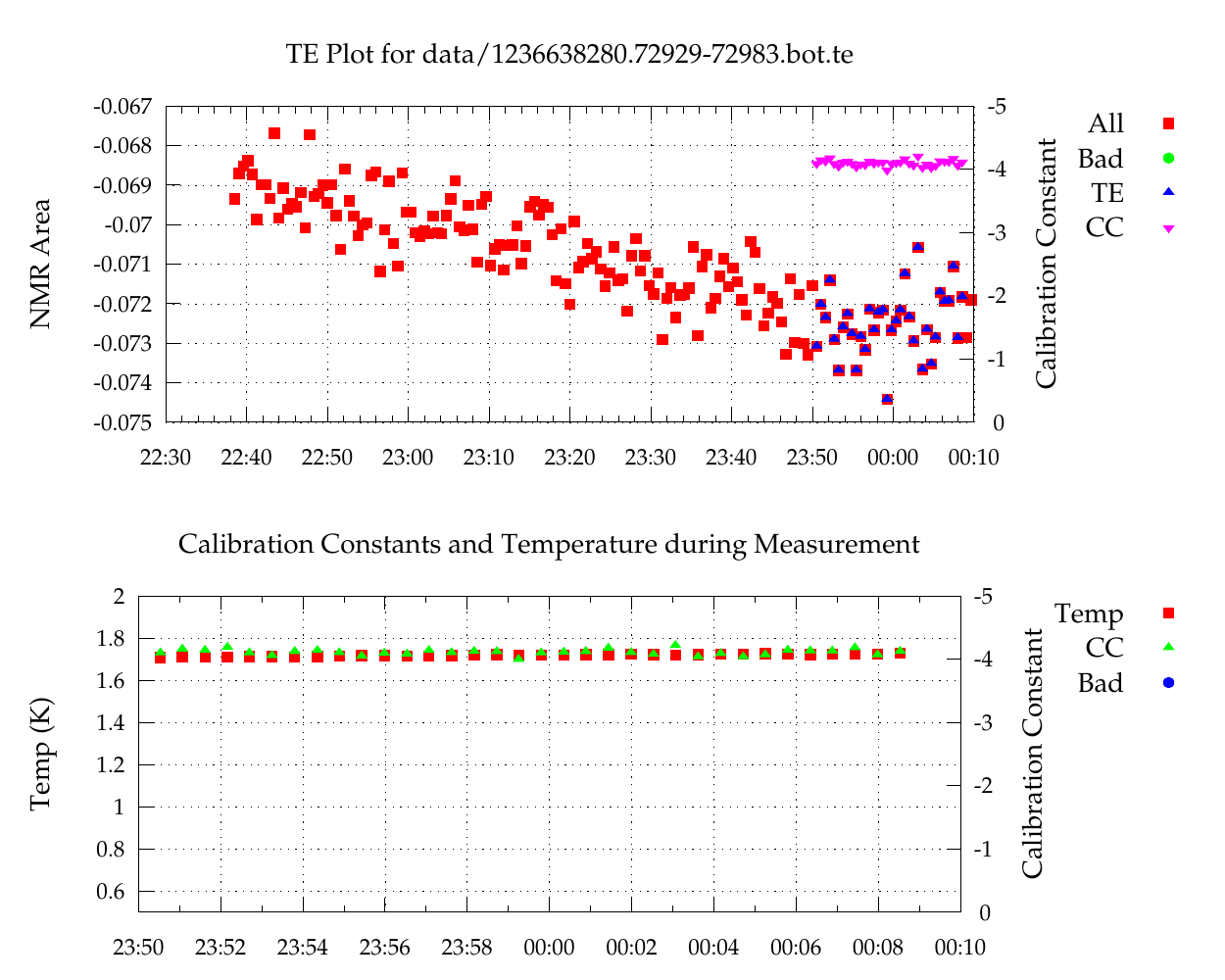}
\end{center}
\end{figure}
\begin{figure}[p]		
\begin{center}
 \includegraphics[width=2.9in]{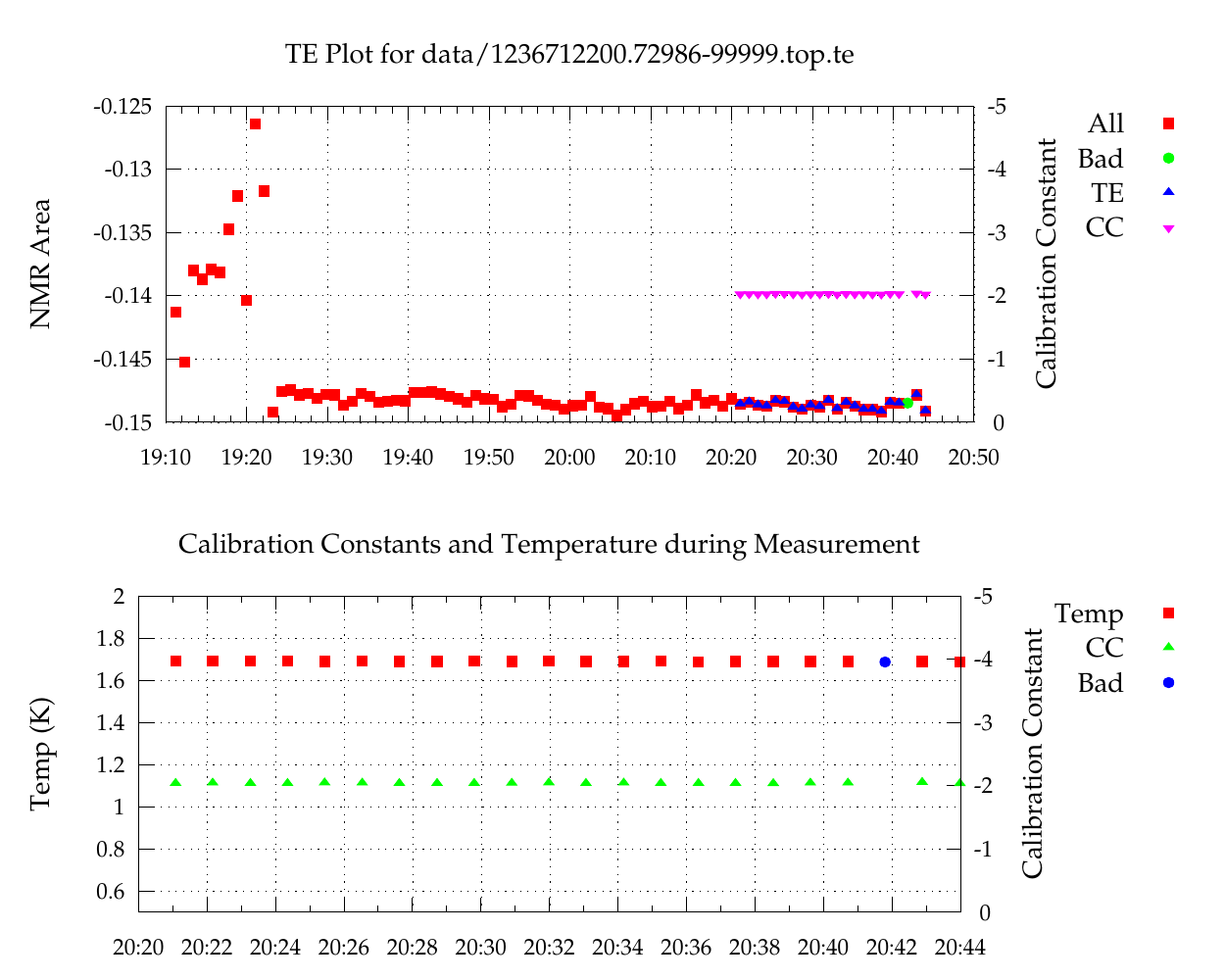}
 \includegraphics[width=2.9in]{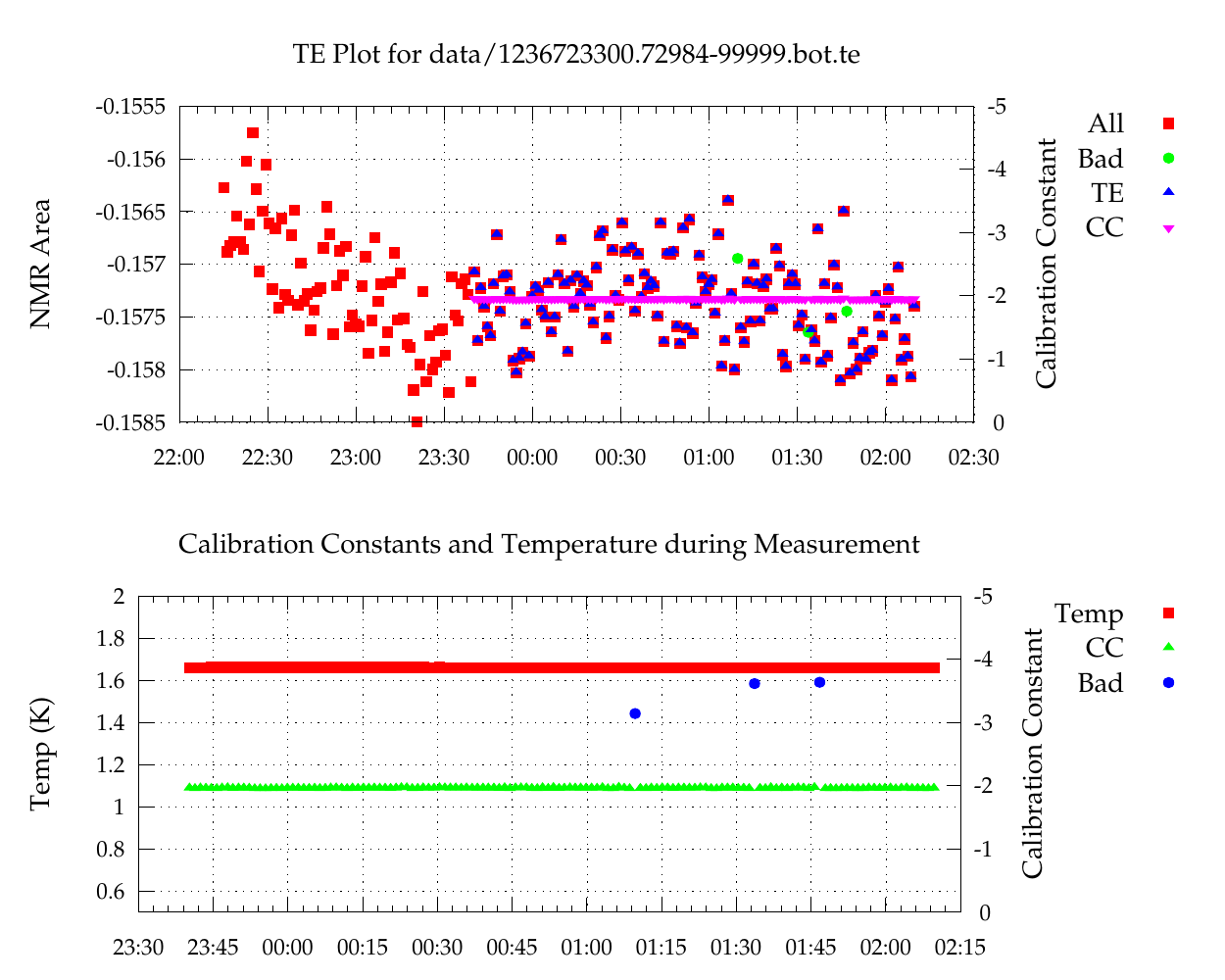}
\end{center}
\end{figure}
\clearpage

\chapter{Target Material Lifetimes}
\label{sec:applife}

As promised in section \ref{sec:life}, this appendix features plots of the polarization performance for all 13 target material samples used during SANE.  The polarizations are given over charge accumulated, in which 20$\times10^{15}$ e$^-/$cm$^2$ translates to around a week of use in the beam.  Red circles give positive polarization points, in which the polarization is aligned the target magnetic field, while blue diamonds give negative polarizations.  Vertical gold bars represent an anneal.

\begin{figure}[p]		
\begin{center}
 \includegraphics[width=5in]{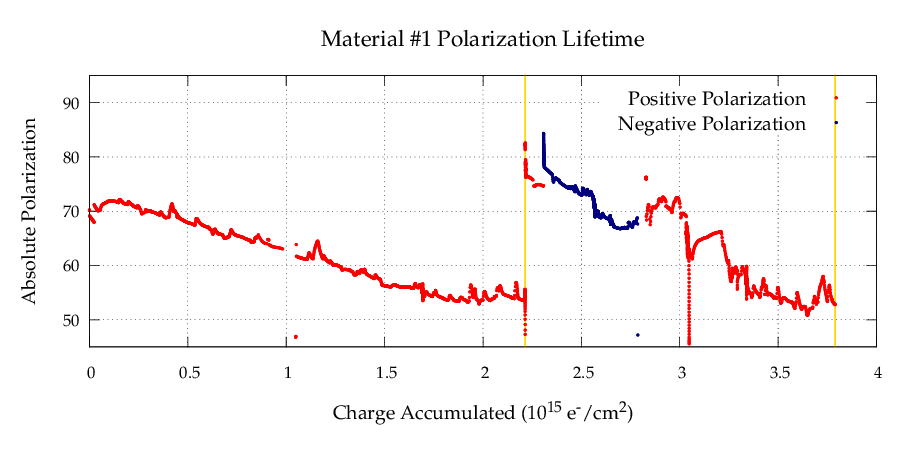}
\end{center}
\end{figure}
\begin{figure}[p]		
\begin{center}
 \includegraphics[width=5in]{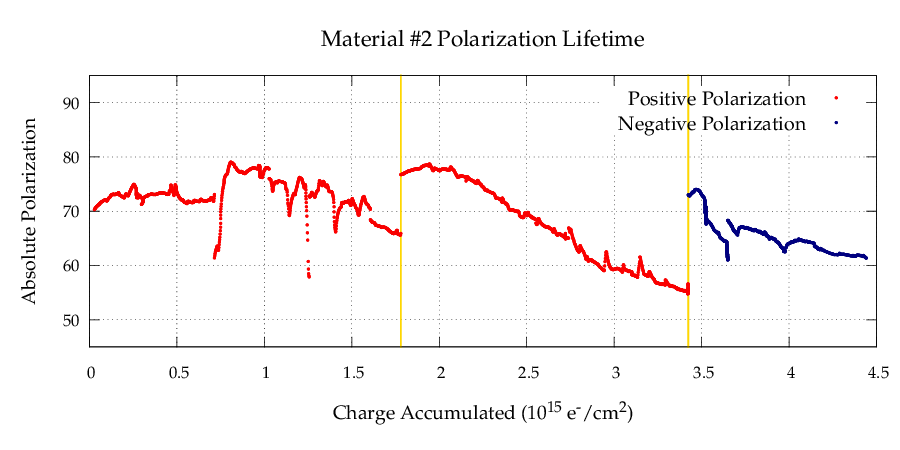}
\end{center}
\end{figure}
\begin{figure}[p]		
\begin{center}
 \includegraphics[width=5in]{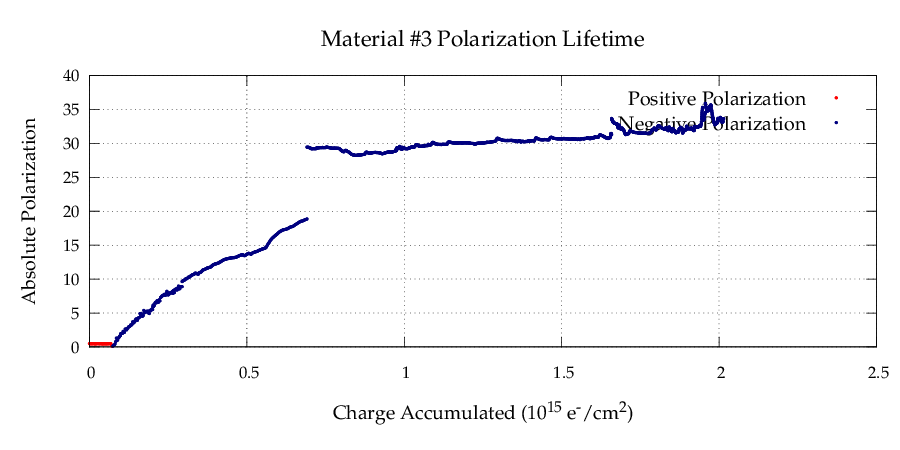}
\end{center}
\end{figure}
\clearpage 
\begin{figure}[p]		
\begin{center}
 \includegraphics[width=5in]{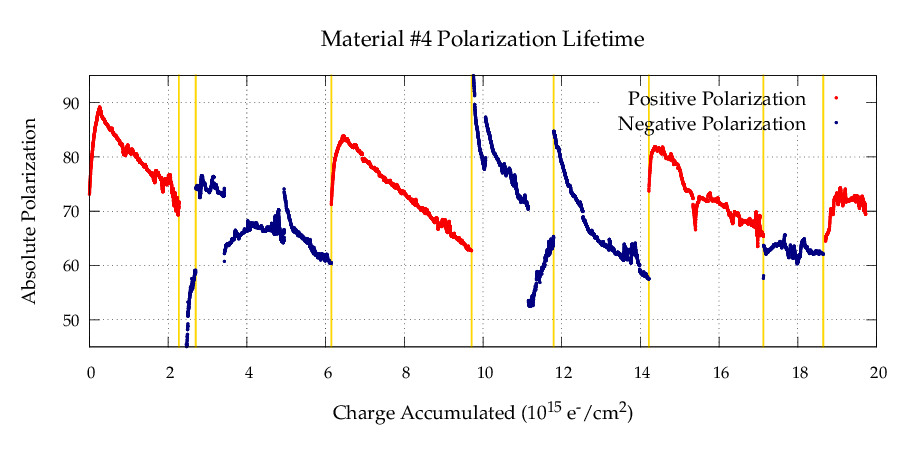}
\end{center}
\end{figure}
\begin{figure}[p]		
\begin{center}
 \includegraphics[width=5in]{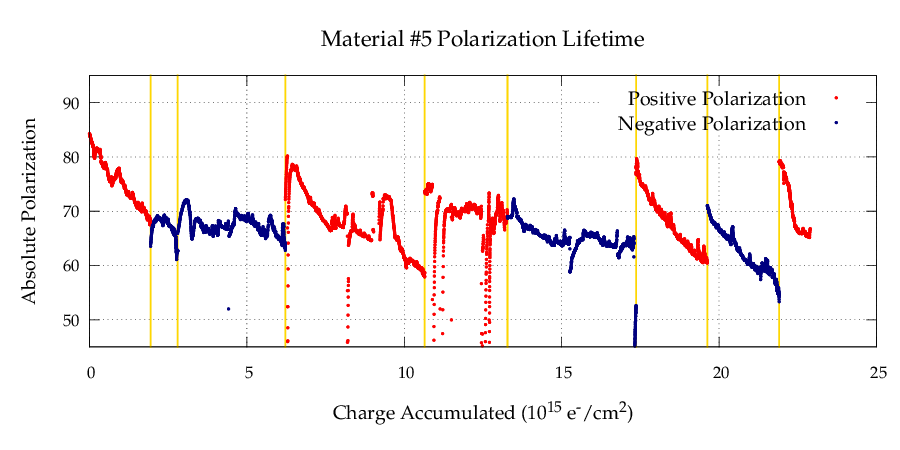}
\end{center}
\end{figure}
\begin{figure}[p]		
\begin{center}
 \includegraphics[width=5in]{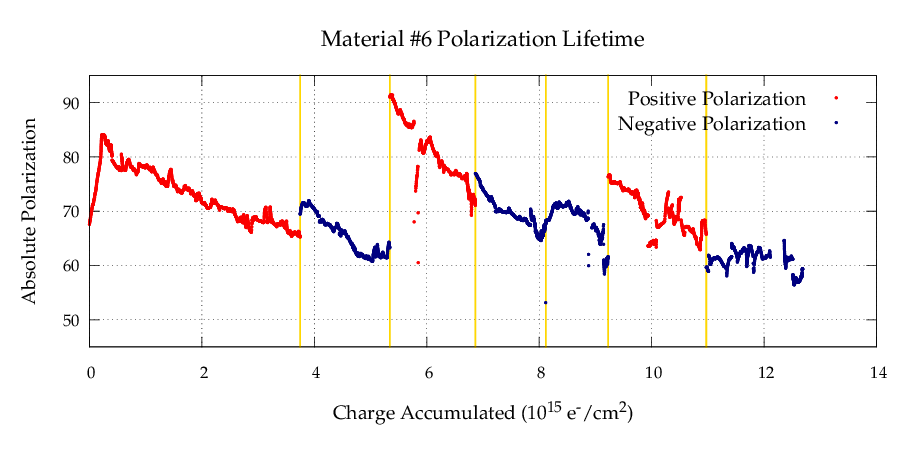}
\end{center}
\end{figure}
\clearpage 
\begin{figure}[p]		
\begin{center}
 \includegraphics[width=5in]{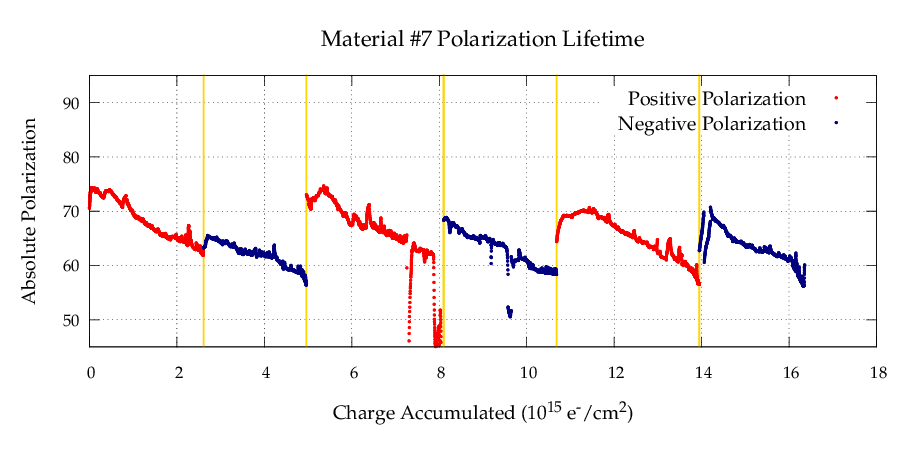}
\end{center}
\end{figure}
\begin{figure}[p]		
\begin{center}
 \includegraphics[width=5in]{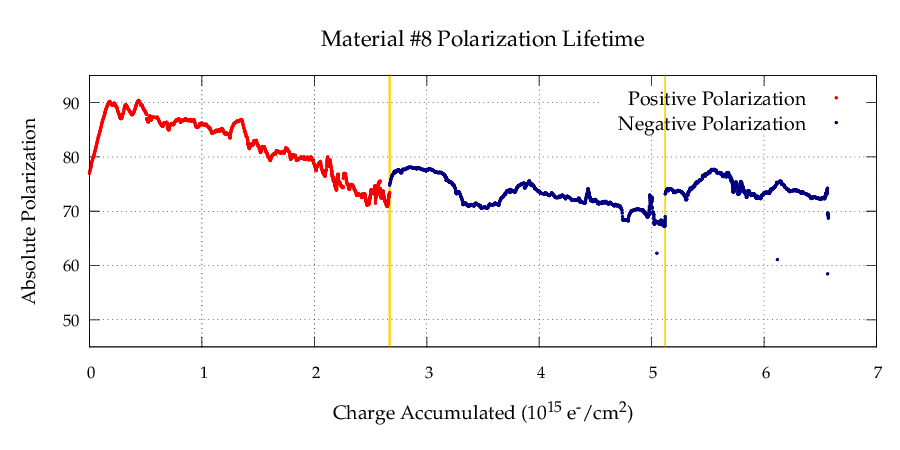}
\end{center}
\end{figure}
\begin{figure}[p]		
\begin{center}
 \includegraphics[width=5in]{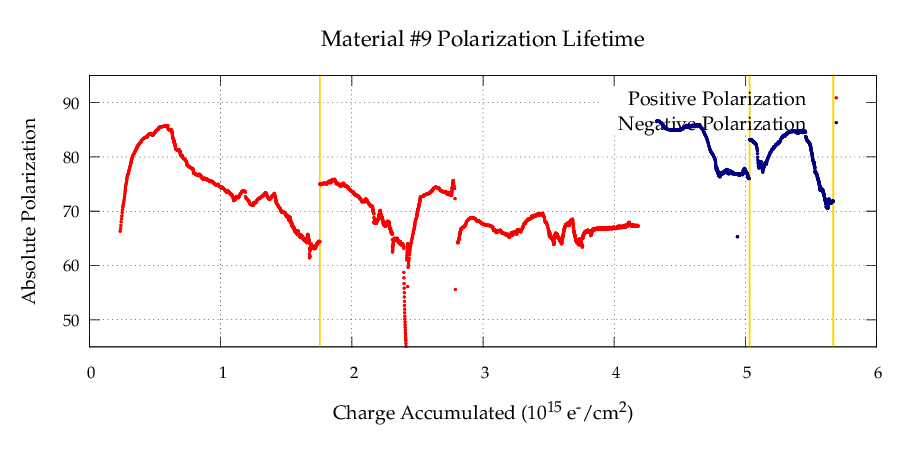}
\end{center}
\end{figure}
\clearpage 
\begin{figure}[p]		
\begin{center}
 \includegraphics[width=5in]{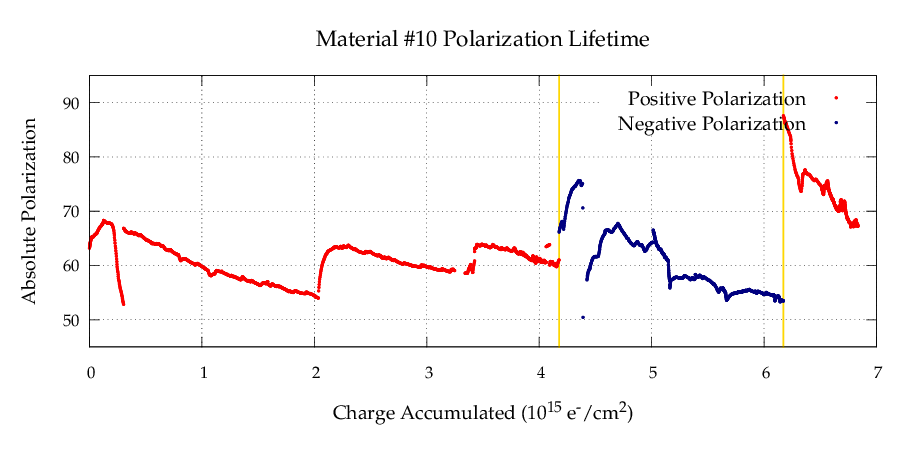}
\end{center}
\end{figure}
\begin{figure}[p]		
\begin{center}
 \includegraphics[width=5in]{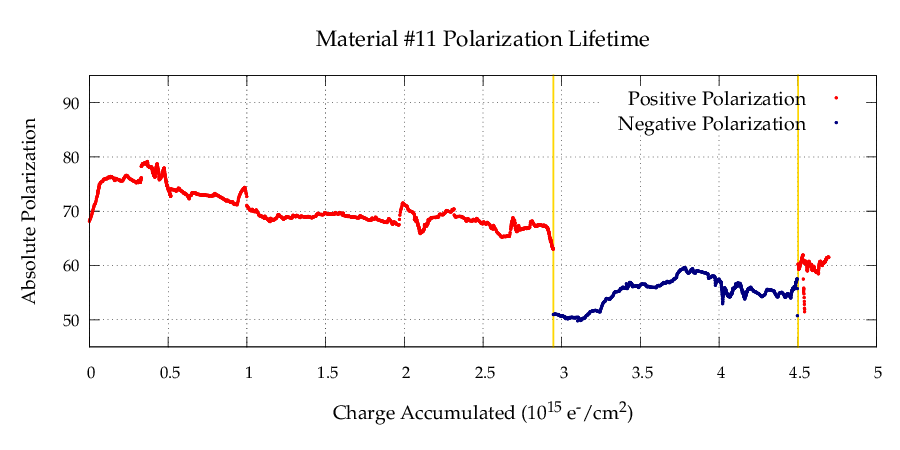}
\end{center}
\end{figure}
\begin{figure}[p]		
\begin{center}
 \includegraphics[width=5in]{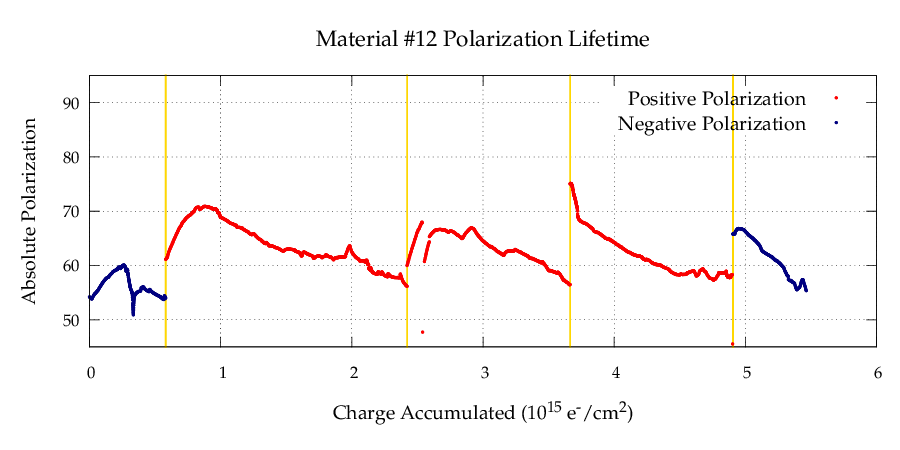}
\end{center}
\end{figure}
\clearpage 
\begin{figure}[p]		
\begin{center}
 \includegraphics[width=5in]{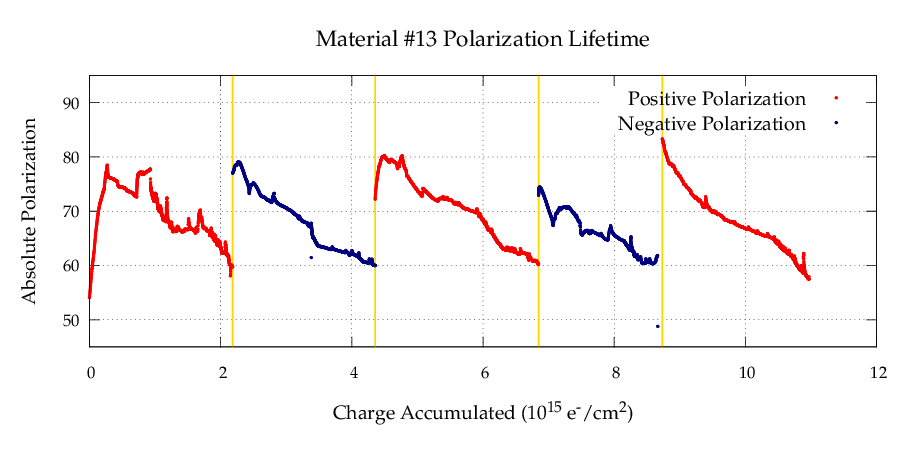}
\end{center}
\end{figure}

\chapter{Radiative Corrections}
\label{sec:radapp}

The formalism of the radiative tail correction is presented in this appendix, along with specifics of the procedure and code used to produce the corrections for SANE.  This appendix augments the discussions of section \ref{sec:radcor}.

\section{Introduction}
\begin{figure}[htb]		
\begin{center}
    \includegraphics[width=4in]{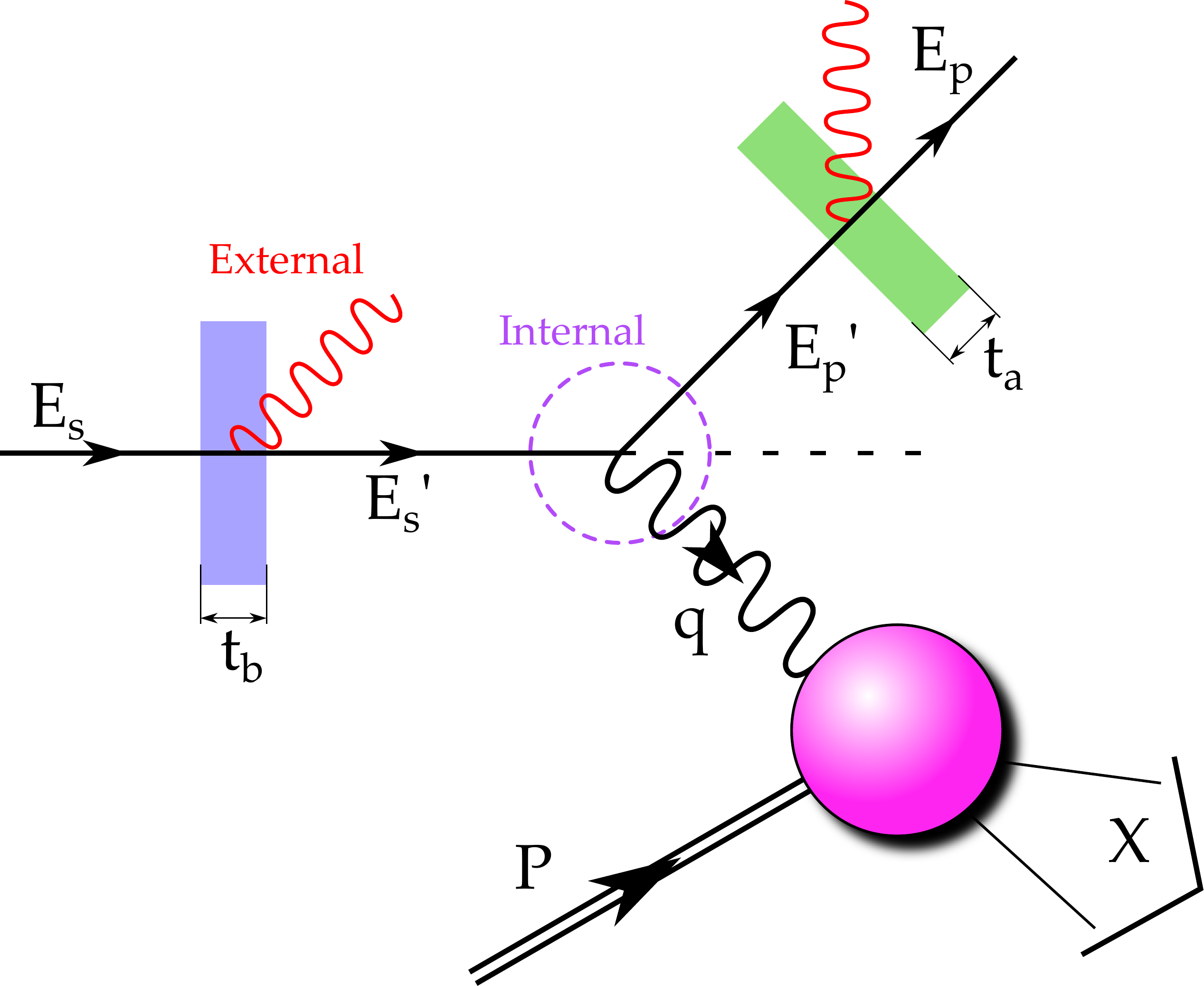}
  \end{center}
  \caption{Diagram of the mechanisms which make radiative corrections necessary.}
  \label{fig:radcorapp}
\end{figure}

SANE measured polarized asymmetries in inclusive electron scattering on a polarized target for incident energy of 5.895 and 4.725 GeV in a momentum transfer range of $2.5<Q^2<6.5$ GeV.  Observed asymmetries must be corrected for losses due to external and internal radiative processes, as seen in figure \ref{fig:radcorapp}.  External corrections are needed due to bremsstrahlung and ionization in all material transversed by the incident electron before and after the scattering process of interest occurred.  This includes aluminum beam windows, nose, helium, ammonia, etc, which can be summarized as contributing to a radiation length. Internal corrections involve vacuum polarization, vertex corrections and internal bremsstrahlung. 

The technique behind the radiative elastic tail calculation presented here will be outlined in the next section, and the section following will give a quick map of where these steps are done in the code.  The last two sections present results of this work in the form of elastic tails for RSS and SANE kinematics.  A nice overview of radiative corrections in general can be found in K. Slifer's thesis \cite{sliferthesis}.

\section{Elastic Radiative Corrections}

The formalism of radiative corrections is covered in great detail in Mo and Tsai \cite{motsai}.  The treatment by Stein \cite{stein} is a bit simpler to follow.  As the code used here mostly follows Stein, so will this description.

\subsection{External Corrections}

The first correction we will consider is the external radiative correction.  This takes into account bremsstrahlung and ionization in all material before the scattering of interest takes place; this means windows, the tail-piece, lid and even ammonia target material.  There is also a correction to apply due emission of single photons when in fact multiple soft photon radiation may be occurring.

As in figure \ref{fig:radcor}, we will follow Stein's notation where subscript $s$ denotes incident and subscript $p$ denotes outgoing quantities of the electron; $E_p$ would then be the outgoing electron energy.  Thicknesses $t_b$ and $t_a$ denote radiation lengths before and after the scattering.

From Stein's appendix A, we have the cross section contribution to the radiative tail from straggling from ionization and bremsstrahlung (equation A49):
\begin{equation}
\begin{split}
\sigma_b &= \left(\frac{d^2\sigma}{d\Omega dE_p} \right)_b \\
&= \frac{M_T+2(E_s-\omega_s)\sin^2(\theta/2)}{M_T-2E_p\sin^2(\theta/2)} \\ &\quad\times \tilde{\sigma}_{el}(E_s-\omega_s)\left[\frac{bt_b}{\omega_s}\phi(v_s)+\frac{\xi}{2\omega_s^2}\right] +\tilde{\sigma}_{el}(E_s)\left[\frac{bt_a}{\omega_p}\phi(v_p)+\frac{\xi}{2\omega_p^2}\right],
\end{split}
\end{equation}
where
\begin{equation}
\begin{split}
\omega_s &= E_s - \frac{E_p}{1-(2E_p/M_T)\sin^2(\theta/2)}\\
\omega_p &=  \frac{E_s}{1-(2E_s/M_T)\sin^2(\theta/2)} - E_p
\end{split}
\end{equation}
and 
\begin{equation}
\begin{split}
\xi &= \frac{\pi m}{2\alpha}\frac{t_b+t_a}{(Z+\eta)ln(183/z^{1/3})} , \\
v_s &= \omega_s / E_s,\\
v_p &= \omega_p/(E_p+\omega_p) , \\
\phi(v) &= 1- v + 3v^2/4,\\
\tilde{\sigma}_{el}(E) &= \tilde{F}(q^2)\sigma_{el}(E).
\end{split}
\end{equation}
Here $\sigma_{el}$ is the elastic cross section and $\tilde{F}$ is a multiplicative correction to the cross section (Stein's A44):
\begin{equation}
\begin{split}
\tilde{F}(q^2) &= (1 + 0.5772\cdot bT) + \frac{2\alpha}{\pi}\left [ \frac{-14}{9} +  \frac{13}{12}\ln \frac{Q^2}{m^2}\right]\\ &\quad + \frac{\alpha}{\pi} \left [ \frac{1}{6}\pi^2 - \Phi(\cos^2\frac{\theta}{2}) \right].
\end{split}
\end{equation}

As Karl mentions in his thesis, the first term of $\tilde{F}$ is a normalization factor from the bremsstrahlung expression, the second term is the sum of the vacuum polarization and vertex corrections.  A third term in Stein's equation A44 has been removed as it deals with the peaking approximation which we aren't using here. The last term is at most a half percent correction, and contains the Spence function
\begin{equation}
\Phi (x) = \int^x_0 \frac{-\ln \vert1-y\vert}{y}dy.
\end{equation}

We can correct the cross section for single-photon emission to account for multiple-soft-photon emission by multiplying by
\begin{equation}
F_{soft} = \left(\frac{\omega_s}{E_s}\right)^{b(t_b+t_r)} \left(\frac{\omega_p}{E_p+\omega_p}\right)^{b(t_a+t_r)},
\end{equation}
for $t_r = b^{-1}(\alpha/\pi)[\ln(=q^2/m^2)-1]$, the thickness of an ``equivalent radiator'' to account for internal effects.

\subsection{Internal Corrections}

The exact calculation of the internal correction is an integral of kinematic factors and elastic structure functions $W^{el}_1(q^2)$ and  $W^{el}_2(q^2)$, and accounts for one-photon exchange and single-photon emission.  Here we now have four vectors $s$, $p$, $t$ and $k$, for referring to the incident electron, outgoing electron, target particle and real photon emitted, respectively, as well as $u=s=T-p$ and $P_f = u-k$.  This is Mo and Tsai equation B.5 or Stein equation A24:

\begin{equation}
\begin{split}
\sigma_{exact} &= \left(\frac{d^2\sigma}{d\Omega dE_p} \right)_{exact} = \frac{\alpha^3}{2\pi}\left(\frac{E_p}{E_s} \right) \int^1_{-1} \frac{2M_T\omega d(\cos\theta_k)}{q^49u_0-\vert\vec{u}\vert\cos\theta_k)} \\
& \times \Biggl( \tilde{W}_2(q^2) \Biggl\{  \frac{-am^2}{x^3} \left[2E_s(E_p+\omega)+\frac{q^2}{2}\right] - \frac{-am^2}{y^3} \left[2E_p(E_s-\omega)+\frac{q^2}{2}\right]  \Biggr. \Biggr.\\ 
& \quad - 2+2v(x^{-1}-y^{-1}) \left[ m^2(sp-\omega^2)+sp \left(2E_sE_p - sp + \omega (E_s+E_p) \right) \right] \\
&  \quad  + x^{-1}\left[2(E_sE_p+E_s\omega+E_p^2)+\frac{q^2}{2} - sp - m^2\right]\\
& \Biggl.   \quad  + y^{-1}\left[2(E_sE_p+E_p\omega+E_s^2)+\frac{q^2}{2} - sp - m^2\right] \Biggr\}\\
&+\tilde{W}_1(q^2)\Biggl\{\left(\frac{a}{x^3}+\frac{a}{y^3}\right)m^2(2M^2+q^2)+4+4v(x_{-1}+y_{-1})sp(sp-2m^2)\Biggr.\\
&\Biggl.\Biggl.\quad +(x_{-1}+y_{-1})(2sp+2m^2-q^2)\Biggr\} \Biggr)
\end{split}
\end{equation}
which includes numerous kinematic factors, such as $a$, $b$, $v$, $x$, and $y$, which are Stein's equations A25 through A41.

\section{Code}
This section addresses where the above calculations are handled in the code.  The code is Karl Slifer's, as compiled and altered from various other sources, and his tech notes can be consulted to determine the origination of different subroutines \cite{slifer1}.  

The subroutine ``sub{\_}rtail.f'' performs the calculation, taking as arguments $E_s$, $E_p$, $\theta$, $t_b$, and $t_a$, in addition to the polarization angle and flag, and returns the external and internal cross section corrections.  The polarization flag causes a different subroutine to be used for the internal corrections.
If the polarized flag is used, the output is as $\Delta \sigma /2$ instead of an unpolarized cross section correction.

After calculating a few useful factors, ``sub{\_}rtail.f'' calls ``fbar'' to calculate $\tilde{F}$ which is kept in a common block, then calls ``externl'' to calculate the external portion of the radiative tail correction.  ``Externl'' is a subroutine in the sub{\_}rtail.f file, and it follows Stein's formulation closely, with comments denoting the equation numbers from that paper. 

A function ``sigbar'' is called to compute the elastic cross section as a function of incident energy, getting form factors  from subroutine ``fmfac.''  This elastic cross section is immediately corrected by $\tilde{F}$.   The $F_{soft}$ correction is also done in externl.  Once this is done, externl returns the cross section correction xextb and xexta, the external corrections for before and after the target.

After the external corrections are done, internal corrections are calculated, integrating with the Simpson integration routine.  The integrand is put together in function ``xsect,'' and it should be noted that subroutine ``xsectp'' needs to be called for each kinematic setting before the xsect function is used.  Xsect follows Mo and Tsai's equation B.3 and B5 to produce the internal bremsstrahlung integrand.

With the internal integration done, the internal and external corrections are passed back to the main program from sub{\_}rtail.f.  From here they are printed for each kinematic setting, separately and as a sum of both internal and external.


\section{RSS Kinematics}

\begin{figure}[p]		
\begin{center}
   \includegraphics[width=4in]{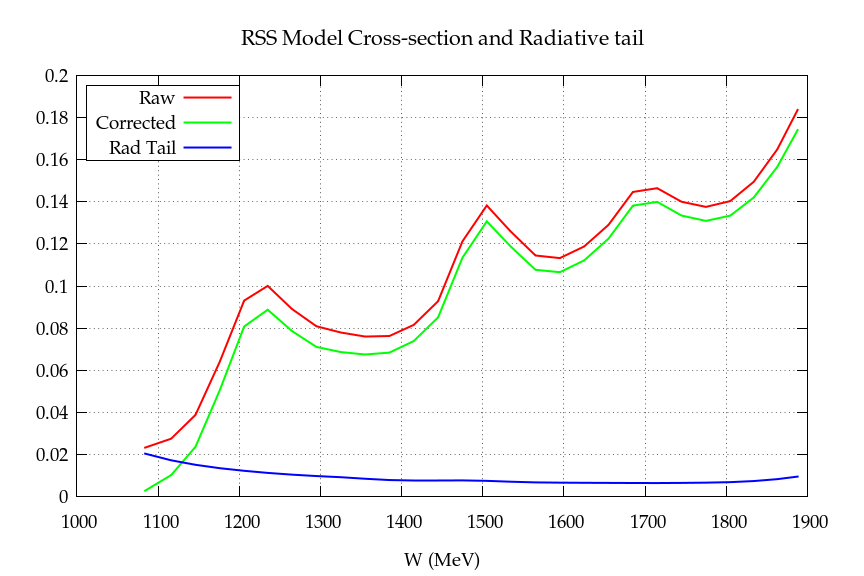}
  \end{center}
  \caption{RSS model cross sections with radiative tail.}
  \label{fig:rss}
\end{figure}

To confirm that the code is working as expected after being compiled on the 64-bit server at UVa, Twist, the radiative tail correction was reproduced for the kinematics of the RSS experiment.  Plotted in figure \ref{fig:rss} is a plot reproduced from RSS data.  The data points are based on RSS model cross sections, and give raw and radiative tail corrected data.  Subtracting the two gives the elastic radiative tail.    

\begin{figure}[p]		
\begin{center}
    \includegraphics[width=4in]{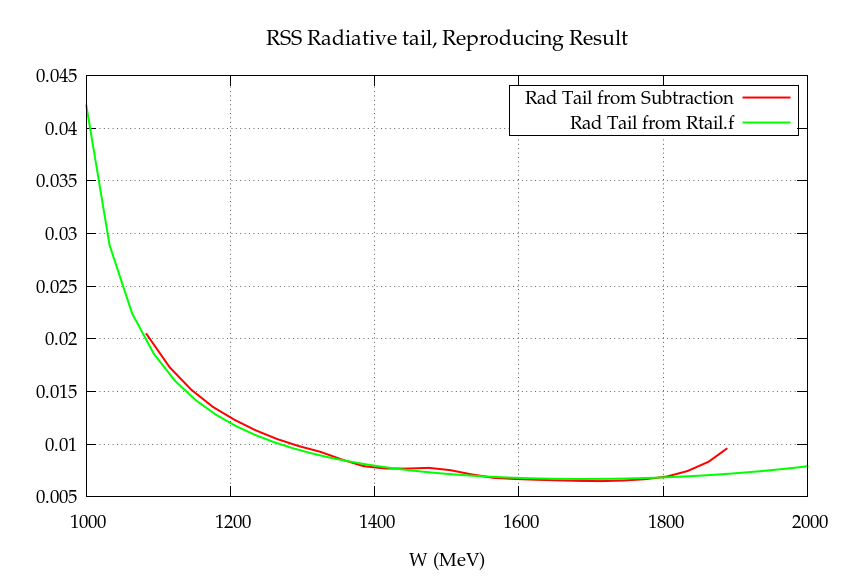}
  \end{center}
  \caption{Reproduction of RSS radiative tail, shown as produced from RSS result subtraction and from this analysis.}
  \label{fig:reprod}
\end{figure}

Shown in figure \ref{fig:reprod} is the radiative tail correction above, now shown with a radiative tail correction produced with the code running at RSS kinematics. The curves show close agreement, with the except of points at the extrema of the RSS data set with high extrapolation errors.  This at least reassures us that the code is working as it did for RSS.

\section{SANE Results}

To apply this procedure to SANE, we first calculate the thickness of the radiators before and after the target.  In table \ref{tab:thick}, the radiation thicknesses are shown for each of the components of the SANE beamline and target which are traversed by the beam.

\begin{table}[htb]
  \begin{center}
\begin{tabular}{lllc}
\toprule
Component		& Material  & Thickness (mg/cm$^2$) & $\chi_0$ (\%) \\
\midrule
Target Material & $^{14}$NH$_3$ &  1561 & 3.82\\ 
Target Cryogen & LHe &  174 & 0.18\\
Target Coil & Cu &  13 & 0.10\\
Cell Lid & Al &  10 & 0.04\\
Tail Window & Al &  27 & 0.12\\
Rad Shield& Al &  7 & 0.03\\
N Shield& Al &  10 & 0.04\\
Beam Exit& Be & 24 & 0.04\\
\multirow{2}{*}{Vacuum Chamber Windows} & Be &  94 & 0.14\\ 
 & Al & 139 & 0.58\\ 
\cmidrule{2-4}
&\multicolumn{2}{l}{Perp Total Before}  &2.98\\
&\multicolumn{2}{l}{Perp Total After}  &2.36\\
&\multicolumn{2}{l}{Para Total Before}  &2.54\\
&\multicolumn{2}{l}{Para Total After}  &2.36\\
 
 \bottomrule
\end{tabular}
\caption[Table of component thicknesses for radiative corrections]{Table of component thicknesses for radiative corrections, assuming a target material packing fraction of 0.60.  Total thicknesses before and after the center of the target are given for each configuration.}
  \label{tab:thick}
\end{center}
\end{table}

\begin{figure}[htbp]		
\begin{center}
    \includegraphics[width=2.9in]{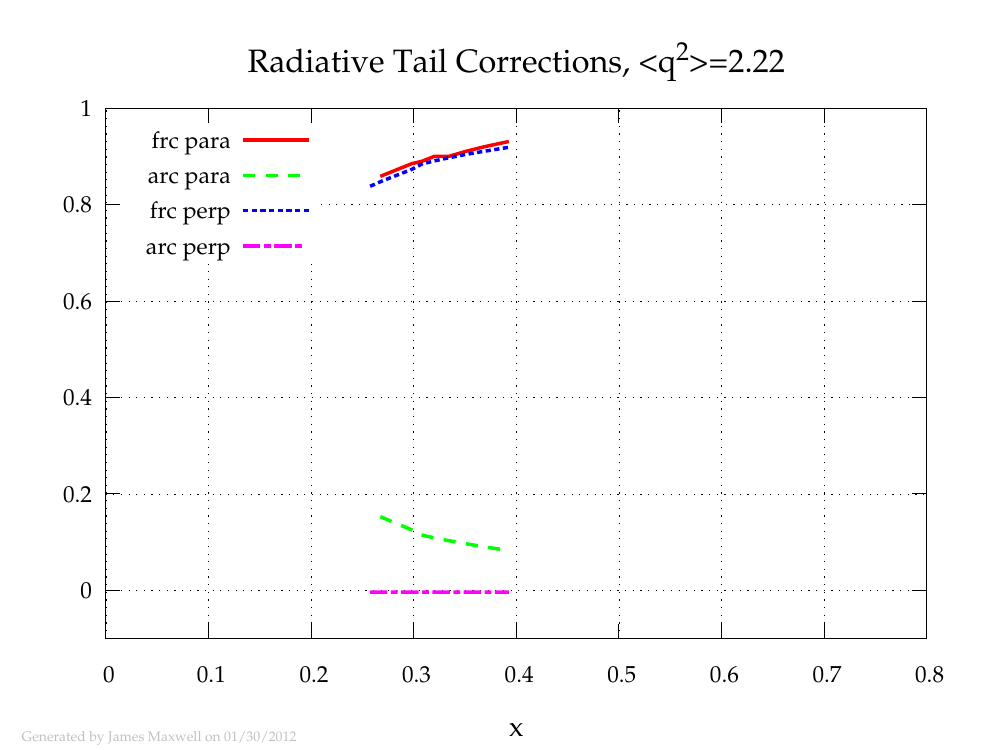}
    \includegraphics[width=2.9in]{figures/rad_tail_q2_3.pdf}
    \includegraphics[width=2.9in]{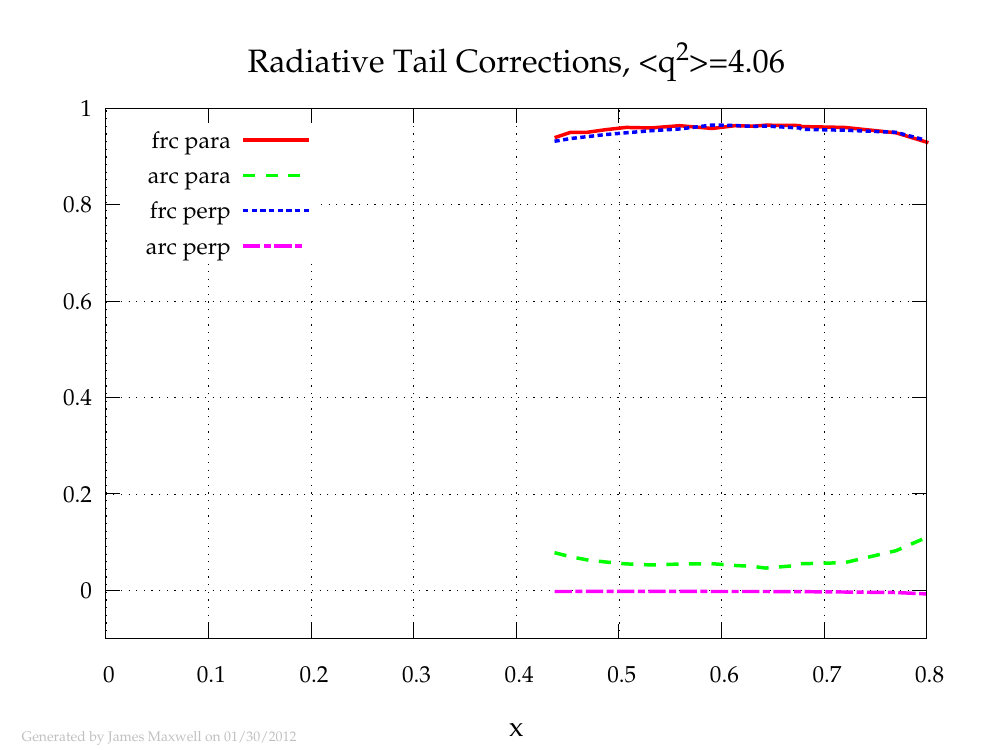}
    \includegraphics[width=2.9in]{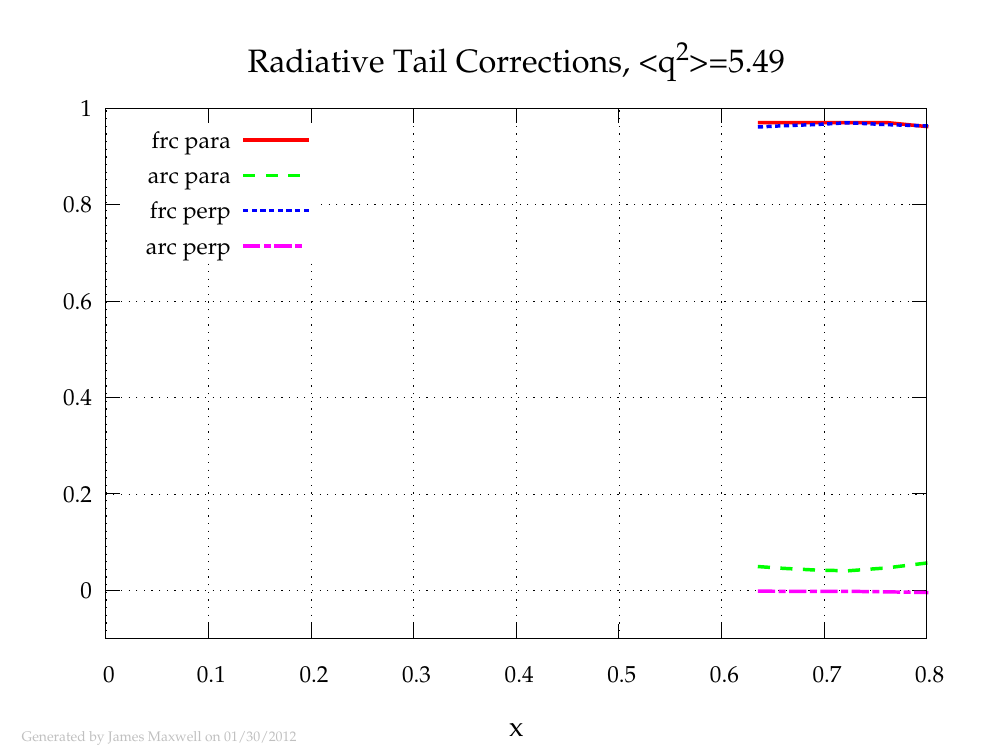}
  \end{center}
  \caption{Radiative tail corrections as applied to our results in each kinematic bin.}
  \label{fig:radres}
\end{figure}

A radiative tail correction was created for each bin per run, using the averaged $E$, $E'$ and $\theta$ of the events as input to perform the calculation.  Figure \ref{fig:radres} shows the averaged radiative tail correction factors as averaged from all SANE runs into kinematics bins.  The ``perpendicular'' results are actually calculated for 80$\degrees$.

\chapter{Target Magnet Failure}
\label{sec:magapp}

This appendix relates in great detail the events leading to and following the failure of the UVa target magnet during SANE.  The failure caused weeks of delay for repair many further hours lost to instability.   As the UVa polarized target represents a crucial slice of the spin physics program at JLab, a record of the failure and repair is worth recording here.  This appendix falls outside the purview of the spin structure measurement, and we do not intend it to be read by most. 

The UVa magnet was built by Oxford Instruments in 1991 and was used in experiments E143,
E155 and E155x at SLAC, and GEN98, GEN01 and RSS at JLab. In JLab's Hall C, in 1998, a surveying
tripod was pulled into the magnet's outer vacuum can while the magnet was energized. The implosive
loss of vacuum caused the magnet to quench, although without apparent damage to the magnet itself.
The magnet remained in storage at JLab after GEN01 until it was removed for testing before SANE.

\section{Precipitating Factors}

\subsection{Loss of Vacuum Incident}
The first difficulty occurred in the ``EEL" test lab at JLab, during a trial cool-down and magnet
energization in June of 2008, the first use of the magnet after six years of storage. When the magnet was
cold and energized, a test of the target motion control with a new design of target insert caused a tear in
the thin aluminum window separating the target cryo-refrigerator space from the outer vacuum chamber.
Helium from the cryo-fridge filled the vacuum, causing a warming and subsequent quench of the magnet,
in a similar but less violent circumstance as in 1998. Time and monetary restrictions did not allow for the
magnet to be tested again before installation in Hall C.

\subsection{Persistence Switch Quench}
The UVa polarized target was installed in Hall C, its field parallel to the beam, and cooled to
operating temperature by October 31st. That night, the magnet was energized to 77.200A without
difficulty, but a quench occurred due to a GPIB communication error with the magnet's Oxford IPS-120
power supply. The persistence switch, which allows connection of the magnet coil's leads to the power
supply, is operated by heating a length of superconducting wire until it is resistive, removing the
superconducting path to allow current through the leads (p11 manual). The Oxford power supply
contains a current source to heat this switch, but the ``on" status of this current is not an effective indicator
of the switch's status. At least 15 seconds of heating or cooling is necessary to ``flip" the switch before
ramping the current in the magnet. In this case, a buffer issue in the GPIB communications with the
power supply caused a delay long enough that the leads of the magnet were allowed to ramp down before
the switch was fully superconducting. The rapid change in current in the magnet coils due to the de-energization
of the still-connected leads caused a quench.

\subsection{Quench Ramping Down}
The following day, November 1st, the magnet was successfully energized to 77.200A and the
magnet entered persistent mode without issue. After 2 days taking beam on a CH2 target, the magnet was
to be de-energized to zero and then re-energized to be tested in the opposite current polarity. During the
de-energization process on November 3rd, the ramping was performed at the rates specified as those for a
``trained" magnet, instead of the slower rates for a newly energized system. The prescribed rates for
``training" are 1.2A/m for 0A to 60A, 0.6A/m for 60A to 72A, and 0.3A/m for 72A-78A, while the
``trained" rates are 2.0A/m for 0A to 60A, 1.5A/m for 60A to 72A, and 1.0A/m for 72A-78A. The higher
ramp rate, which was due to operator error, initiated a quench as the rate increased from 1.5A/m to 2.0A/
m at 60A.

\section{Failure and Repair}
After allowing the magnet to cool as prescribed by its manual, an attempt to re-energize the
magnet in the opposite polarity failed with a quench at -26A; the increased helium boil-off from the
magnet suggested a new resistive element in the superconducting path. Resistance measurements of the
cold magnet confirmed these fears, and repair was deemed necessary.

\subsection{Repairs}
After removing the magnet from the hall and grinding open its casing in the EEL test lab, damage
was observed upon inspection of the magnet's quench protection circuitry and coil wiring, which was set
with plasticine in a channel of the phenolic support ring of the magnet. In a bundle of ten wires which
connected coils 5a-d and 4a-b, six had fused together entirely and four had lost insulation. In addition, a
barrier diode for the protection of coil 4a-b was broken. Repairs were performed in the EEL by J.
Beaufait, a hall C technician, with the assistance of P. Brodie, a specialist from Oxford Instruments. The
offending protection diode, a MBRP30045CT, was out of production and was replaced with a
MBRP40045CT diode. The damaged wires were reconnected with 1 inch superconducting joints and 3
inch copper to copper contacts.

\subsection{Cause of Damage}
The prevailing theory concludes the protection diode's failure was the source of the thermal
damage of the superconducting wire. During the energization or de-energization of the magnet coils,
either by the power supply or more violently by a quench, an emf is induced in the coils equal to the
inductance of the coils times the rates of change of the current. The barrier diodes protect the coils of the
magnet from excessive voltage by releasing the magnet current into a parallel path with a 0.25 Ohm
resistor (manual p13). (Vc = -IdRd + Lc*dI/dt >= Vf ~0.6V at 78A, Td=78A, Rd=0.25ohms) Two diodes
are used in parallel for each coil (or pair of coils), with opposing bias, to protect from the positive voltage
induced during energization and the negative voltage induced during de-energization, while the roles of
the diodes are reversed with the current polarity.

If a protection diode does not trip at high voltage, thermal damage can occur to the coils and
support wiring, which was seen in the UVa magnet as burnt insulation and fusing of the wires the failed
diode. In this case, with the diode for the 4a-b coil being ramped ``negatively" damaged, the current
flowing through the coil protection circuits would have been diverted through the J4 joint connecting the
5a and 4b coils to the protection circuit. It was the wire carrying this current through joint J4, 4red and its
neighbors in the plasticine, that were fused.

\section{Behavior After Repair}
After the repair and new software and administrative safety measures were put in place to prevent
quenches due to human or software error, the magnet was returned to experimental Hall C. The physics
program of SANE prompted the magnet to be installed with its field 80 degrees to the incident beam. By
the 16th of December, the magnet was cooled down to 4K, and on the 18th it was energized to the
positive polarity, 77.300A, for the first time in the ``perpendicular" configuration. Due to worries about
the behavior of the superconducting switch expressed by a safety review committee, the magnet power
supply leads were left at a current of 77.300A for the remainder of the run period, although the
superconducting switch was allowed to go superconducting, thus separating the magnet current from the
power supply.

The next day, the 19th of December, while taking beam on a CH2 target, the magnet again
quenched; the causes of this quench are still unclear. The clearance for the beam to pass through the coils
of the magnet was much tighter, only +-4cm, in the perpendicular configuration, so the 2cm rastered
beam may have been creating many secondary particles by clipping some structure in the target, as could
the ``sheet of flame" produced by Bremsstrahlung radiation from the chicane beam-line magnets. The
very orientation of the magnet itself in the perpendicular configuration may also be a suspect, as the
chicane and SOS iron magnets put added stress in the magnet, perhaps enough to add instability. All of
these factors may been essential contributors to this quench.
Following the quench of the 19th, another attempt to energize the magnet was made on the 20th,
although a poor vacuum in the outer vacuum chamber of the magnet increased liquid helium consumption
to an unacceptable rate that precluded energization. The experiment was then delayed until after the
winter accelerator break of the 23rd to January 13th of the new year.

\subsection{January to March of 2009}
By the 13th, the magnet was cold and filled with liquid helium. Leaks in the target cryo-fridge
created another week-long delay. The remainder of the experiment, from the 20th of January to the 16th
of March, 2009, was plagued with over a dozen quenches. These quenches were of two categories, the
first being quenches during energization or de-energization, ``ramping quenches.” There were also a
number of quenches during a supposed ``rest" magnet state, when the magnet was in persistent mode, with
the power supply leads de-energized, and the magnet quenched spontaneously with or without beam
being delivered to the hall.

After the winter break the voltage over the magnet coils was directly measured and recorded,
when previously only the voltage in the power supply was logged, and that only once every 30 seconds.
While the coil voltage should be simply related to the power supply voltage ($V_{ps} = IR - V_c, V_c =
LdI/dt$), the added time resolution made it possible to observe new features in the voltage over time.
These measurements were used to determine the source of the ramping quenches.

\subsection{Abnormal ``Charging'' Coil Voltage}
Starting on January 18th, a strangely sporadic behavior complicated the energization of the
magnet coils up to 10A. Under normal energization, the voltage in the coils will jump quickly, as a step
function, to the expected $LdI/Dt$, usually about 4V per 1A/m, as the power supply begins energizing the
magnet. When this abnormal behavior occurred, the voltage in the coils instead would approach the
expected voltage only exponentially.
 Once the energization
was stopped at the power supply and the change in time of the current was zero, the coil voltage then
decayed to zero exponentially as well. This decay curve is significant as it shows the behavior of the
voltage to be more complicated than $LdI/dt$; when the power supply stops ramping, the current through
the leads was not changing. Unless the power supply was put into ``hold" to stop the energization as the
voltage approached the expected value of the coil voltage, the magnet would quench, or go into a ``miniquench"
state in which the voltage goes over the quench protection voltage on the power supply, but the
magnet itself remained energized at a slightly lower current than at the time of the quench.

This mysterious behavior was explained only by the ``fortuitous" malfunction of the magnet's
shim switch power supply. Two shim coils, were included in the magnet's design to allow fine tuning of
the magnetic field. During normal operation in SANE, the correct provided by the shim coils was not
used. However, if these coils are left superconducting as the main magnet coils are energized, induced
current will build up in these smaller coils and precipitate a quench. Thus a power supply remains
connected to the shim coils throughout the experiment to dissipate the induced current. Like the main
coils, the shims are connected to the power supply via a superconducting switch powered by a simple
current source set to 100mA. On January 22nd, the power supply used to open the shim switches was
found to be faulty, powered off and unable to power on. This discovery led to the suspicion of the shims
as the culprit for the ``charging" behavior.

To test the shim coils and their superconducting switches, each shim coil switch, Z1 for one half
of the magnet, Z2 for the second, was powered independently. The shim coils themselves are wired in
series. The voltage in the main magnet coils, as well as the shim coils themselves via the shim power
supply, were recorded. The main coils were energized at a rate of 0.5A/m in each case; in theory, we
expect a coil voltage of ~0.7V at this current ramp rate under normal circumstances.


With both Z1 and Z2 shim switches powered, and thus supposedly open to the power supply, a
voltage that paralleled the coil voltage over time was measured across the shims with a magnitude of
-0.033V. When the Z1 shim switch was unpowered, thus supposedly closing the Z1 shim from the power
supply, this shim voltage was unchanged. However, when the Z1 shim switch was again powered on but
the Z2 shim switch was unpowered, the voltage across the shim coils was measured as -0.002V,
indistinguishable from zero. In addition, the voltage in the main coils over time was additionally retarded
in this case. Essentially, the behavior of the shim voltage was unchanged by the powered on or off status
of the Z1 shim switch heater, whereas the power status of the Z2 shim switch affected the voltage on both
the shim and main coils. From this we conclude the Z1 shim coil was never actually connected to the
power supply as the Z1 shim switch heater did not bring the superconducting switch out of persistent
mode.

To support this hypothesis, we can see a large voltage spike, up to -4V, when the power status of
the Z2 shim switch is turned ``on", connecting the shim to the power supply, after ramping with the Z2
switch ``off." In this case, the Z2 shim coil builds induced current as the main coils are energized. When
the Z2 shim is then reconnected, this induced current is released into the power supply to be dissipated, as
seen by a voltage spike. When a similar test is attempted with the Z1 shim coil, the voltage jumps
slightly to 0.5V, as it is never actually connected to the power supply.

The ``mini-quenches" seen previously were understood in this new context to be quenches of the
shim coils, not of the main coils. As the induced current built up above the limits of the superconducting
shims, they would quench, releasing their current as heat and inducing a sudden back emf on the main
coils. This back emf created a large enough voltage in the mains coils to exceed the power supply's
quench protection voltage, as well as lower the current in the main coils as observed.

With the ``charging" ramping understood, steps were taken to prevent quenches due to this
behavior. It was hoped that by increasing the current in the Z1 shim switch heater, enough heat could be
created in the switch to bring it out of superconducting. However after increasing the current to 20\%
greater than the prescribed 100mA current, and thus increasing the power by 44\%, the switch remained
superconducting.

As was mentioned before, the ``charging" voltage behavior was sporadic. It didn't occur in several
circumstances, and only occurred below 10A when it did. During energization and if it did not quench,
the abnormal behavior would eventually reach a ``transition point" at between 5A and 8A. This transition
point involved a voltage spike in the coil voltage, after which the voltage would no longer charge upon
energization, but stepped up directly to the expected voltage . After this point in the ramp, the magnet
could continue to 77A without further complication.

Why the ``charging" behavior was not consistent, and what caused these ``transition points" are
still unknown. One possible explanation is a physical change in the shim switch above a certain current.
Perhaps the magnetic field caused a mechanical change in the switch enough to allow it to stop
superconducting.

On January 29th a large iron sheet was placed within a meter of the magnet as shielding for the
experiment's Cerenkov detector, and remained in place until the target rotation of March 5th. The
following week, a second, smaller sheet was added. After both these sheets were installed, the charging
behavior in the coil voltage did not recur. It is possible that the changed magnetic field due to such a large
amount of iron could perhaps have reproduced whatever mechanical change the shim which occurred at a
main coil current of 10A. This is unfortunately conjecture, and the cessation of the ``charging" coil voltage
behavior remains unexplained.

\subsection{Current Leak}

Throughout the 2009 run period, the magnet current dropped slowly while in persistent mode,
while before repairs, no such sag existed. Over the course of two days, a drop in the Larmor frequency of
the proton, as measured by NMR, dropped from 213.0MHz to 212.8MHz, a tenth of a percent. This loss
was not unexpected, as the superconducting joints used in the repairs were necessarily of lower quality.
Approximately every two days the magnet leads would be reconnected to lift the current to the
appropriate value.

\subsection{Quenches without Satisfactory Explanation}
The magnet quenched a dozen times between January and March in circumstances unrelated to
energization or de-energization. These quenches have not been satisfactorily explained, but common
themes offer a suggestion to the causes.

Many times quenches occurred with beam traveling through the magnet. These instances
included quenches within 5 minutes of beam being introduced, as well as quenches after several hours of
beam. All may be explained due to beam steering in the confined space of the transversely oriented
magnet, or perhaps the ``sheet of flame'' mentioned earlier. These quenches occurred on December 20th,
February 2nd, 26th, 27th, again on the 27th, and March 3rd.

Other times beam could not be blamed. Several quenches occurred when excess heat was in the
vicinity of the magnet, in the form of a target anneal on February 5th and a cryo-fridge back-fill on
February 8th. Once a quench was due to a tiny ramping to take a thermal equilibrium measurement off
field on the 17th of February, which involved shim and switch heaters. On one occasion, January 26th,
the shim heaters were on for 5 minutes and the magnet spontaneously quenched. On February 17th the
magnet spontaneously quenched 30 minutes after it had finished energizing.

The common thread is heat, although this is by no means the certain cause. It is clear from the
many quenches that the magnet was unstable after the repairs of December. Heat, beam, magnet
orientation and ferromagnetic shielding may have contributed to each quench, although the importance of
each is indefinite. It is crucial to note however, that after the target rotation of March 5th upon which the
iron plate was also removed, the magnet stayed persistent until the end of the experiment on the 16th. In
this orientation the magnet was instead energized in the negative polarization. The final de-energization
of the magnet was performed at the ``trained'' magnet ramp rates (with the exception of the highest rate,
which was 1.8A/m instead of 2.0A/m), but the magnet did not quench during the de-energization.

\addcontentsline{toc}{chapter}{Bibliography}
\bibliography{jdm_thesis}
\bibliographystyle{unsrt}
\end{document}